\pdfoutput=1

\documentclass[a4paper,11pt]{report}
\usepackage[a4paper]{geometry}
\usepackage{graphicx,rotating}
\usepackage{cite}
\usepackage{amssymb}
\usepackage{mathrsfs}
\usepackage{amsmath}
\usepackage{amsfonts}
\usepackage{multirow}
\usepackage{eurosym}
\usepackage{dcolumn}
\usepackage[pagewise]{lineno}
\usepackage{lscape}
\usepackage{url}

\makeatletter
\DeclareRobustCommand*{\bfseries}{%
  \not@math@alphabet\bfseries\mathbf
  \fontseries\bfdefault\selectfont
  \boldmath
}
\makeatother

\newcommand{\bbonu}{\ensuremath{\beta\beta0\nu}}
\newcommand{\bbtnu}{\ensuremath{\beta\beta2\nu}}
\newcommand{\qbb}{\ensuremath{Q_{\beta\beta}}}
\newcommand{\Qbb}{\ensuremath{Q_{\beta\beta}}}
\newcommand{\mbb}{\ensuremath{m_{\beta\beta}}}

\newcommand{\kgy}{\ensuremath{\rm kg \cdot y}}
\newcommand{\ckky}{\ensuremath{\rm counts/(keV \cdot kg \cdot y)}}

\newcommand{\bb}{\ensuremath{\beta\beta}}

\newcommand{\XE}{\ensuremath{{}^{136}\rm Xe}}
\newcommand{\BA}{\ensuremath{{}^{136}\rm Ba}}
\newcommand{\CS}{\ensuremath{{}^{137}\rm Cs}}
\newcommand{\GE}{\ensuremath{{}^{76}\rm Ge}}

\newcommand{\TEX}{\ensuremath{{}^{130}\rm Te}}
\newcommand{\TL}{\ensuremath{{}^{208}\rm{Tl}}}

\newcommand{\CO}{\ensuremath{{}^{60}\rm Co}}

\newcommand{\THO}{\ensuremath{{}^{232}{\rm Th}}}
\newcommand{\BI}{\ensuremath{{}^{214}}Bi}
\newcommand{\RN}{\ensuremath{{}^{222}}Rn}

\newcommand{\Gonu}{\ensuremath{G^{0\nu}(E_0, Z)}}

\newcommand{\M}{\ensuremath{M}}

\newcommand{\Xe}{\ensuremath{^{136}}Xe}

\newcommand{\Bi}{\ensuremath{^{214}}Bi}

\newcommand{\Tl}{\ensuremath{^{208}}Tl}

\newcommand{\Pb}{\ensuremath{^{208}}Pb}

\newcommand{\Po}{\ensuremath{^{214}}Po}

\long\def\symbolfootnote[#1]#2{\begingroup%
\def\thefootnote{\fnsymbol{footnote}}\footnote[#1]{#2}\endgroup}

\begin{document}

\begin{titlepage}

\begin{center}
{\large \sf CONCEPTUAL DESIGN REPORT} \\ \vspace{0.65cm}
{\Huge \bf The NEXT-100 experiment for \bbonu\ searches at LSC} \\ \vspace{0.75cm}
{(16 May 2011)} \\ \vspace{0.35cm}
\end{center}

\begin{center}
\begin{minipage}{15cm}
\begin{abstract}
We propose an EASY (\textit{Electroluminescent ApparatuS of high Yield}) and SOFT (\textit{Separated Optimized FuncTion}) time-projection chamber for the NEXT experiment, that will search for neutrinoless double beta decay (\bbonu) in $^{136}$Xe. Our experiment must be competitive with the new generation of \bbonu\ searches already in operation or in construction. This requires a detector with very good energy resolution ($\lesssim1\%$), very low background contamination ($\sim10^{-4}$ \ckky) and large target mass. In addition, it needs to be operational as soon as possible. The design described here optimizes energy resolution thanks to the use of proportional electroluminescent amplification (EL), which provides a large yield of photons as a signal; it is compact, as the Xe gas is under high pressure; and it allows the measurement of the topological signature of the event to further reduce the background contamination. The SOFT design uses different sensors for tracking and calorimetry. We propose the use of SiPMs (MPPCs) coated with a suitable wavelength shifter for the tracking, and the use of radiopure photomultipliers for the measurement of the energy and the primary scintillation needed to estimate the $t_0$.  This design provides the best possible energy resolution compared with other NEXT designs based on avalanche gain devices.

The baseline design is an Asymmetric Neutrino Gas EL apparatus (ANGEL), which was already outlined in the NEXT LOI. ANGEL is conceived to be easy to fabricate. It requires very little R\&D and most of the proposed solutions have already been tested in the NEXT-1 prototypes. Therefore, the detector can be ready by 2013.  The detector may be upgraded to a fiducial mass of 1 ton after the initial physics runs, following the successful approach of GERDA and XENON experiments. 

With our design, NEXT  will be competitive and possibly out-perform existing proposals for next-generation neutrinoless double-beta decay experiments. In this Conceptual Design Report (CDR) we discuss first the physics case, present a full design of the detector, describe the NEXT-1 EL prototypes and their initial results,  and outline a project to build  a detector with 100 kg of enriched xenon to be installed in the Canfranc Underground Laboratory in 2013.
\end{abstract}
\end{minipage}
\end{center}

\begin{center}
\pagebreak 

{\LARGE \bf The NEXT collaboration}

\vspace{0.4cm}

{\small \sc E.~G\'omez, R.M.~Guti\'errez, M.~Losada, G.~Navarro}\\
{\it Universidad Antonio Nari\~no, Bogot\'a, Colombia}

\vspace{0.3cm}

{\small \sc A.L.~Ferreira, C.A.B.~Oliveira, J.F.C.A.~Veloso}\\
{\it Universidade de Aveiro, Aveiro, Portugal}

\vspace{0.3cm}

{\small \sc D.~Chan, A.~Goldschmidt, D.~Hogan, T.~Miller, D.~Nygren, \\J.~Renner, D.~Shuman, H.~Spieler, T.~Weber}\\
{\it Lawrence Berkeley National Laboratory, Berkeley CA, USA}

\vspace{0.3cm}

{\small \sc F.I.G.~Borges, C.A.N.~Conde, T.H.V.T.~Dias, L.M.P.~Fernandes, E.D.C.~Freitas, J.A.M.~Lopes, C.M.B.~Monteiro, H.~Natal da Luz, F.P.~Santos, J.M.F.~dos Santos}\\
{\it Universidade de Coimbra, Coimbra, Portugal}

\vspace{0.3cm}

{\small \sc M.~Batall\'e, L.~Ripoll}\\
{\it Universitat de Girona, Girona, Spain}

\vspace{0.3cm}

{\small \sc P.~Evtoukhovitch, V.~Kalinnikov, A.~Moiseenko, Z.~Tsamalaidze, E.~Velicheva}\\
{\it Joint Institute for Nuclear Research (JINR), Dubna, Russia}

\vspace{0.3cm}

{\small \sc E.~Ferrer-Ribas, I.~Giomataris, F.J.~Iguaz}\\
{\it IRFU, Centre d'\'Etudes Nucl\'eaires de Saclay, Gif-sur-Yvette, France}

\vspace{0.3cm}

{\small \sc J.A.~Hernando Morata, D.~V\'azquez}\\
{\it Universidade de Santiago de Compostela, Santiago de Compostela, Spain}

\vspace{0.3cm}

{\small \sc C.~Sofka, R.~C.~Webb, J.~White}\\
{\it Texas A\&M University, College Station TX, USA}

\vspace{0.3cm}

{\small \sc J.M.~Catal\'a, R.~Esteve, V.~Herrero, \\A.~M\'endez, J.M.~Monz\'o, F.J.~Mora, J.F.~Toledo}\\
{\it I3M, Universidad Polit\'ecnica de Valencia, Valencia, Spain}

\vspace{0.3cm}

{\small \sc R.~Palma, J.L.~P\'erez-Aparicio}\\
{\it Universidad Polit\'ecnica de Valencia, Valencia, Spain}

\vspace{0.3cm}

{\small \sc V.~\'Alvarez, M.~Ball, J.~Bayarri, S.~C\'arcel, A.~Cervera, \\
J.~D\'iaz, P.~Ferrario, A.~Gil, J.J.~G\'omez-Cadenas\symbolfootnote[1]{Spokesperson. Contact email: gomez@mail.cern.ch}, K.~Gonz\'alez, \\I.~Liubarsky, D.~Lorca, J.~Mart\'in-Albo, F.~Monrabal, J.~Mu\~noz Vidal, \\M.~Nebot, J.~P\'erez, J.~Rodr\'iguez, L.~Serra, M.~Sorel, N.~Yahlali}\\
{\it Instituto de F\'isica Corpuscular (IFIC), CSIC \& Univ.\ de Valencia, Valencia, Spain}

\vspace{0.35cm}

{\small \sc J.M.~Carmona, J.~Castel, S.~Cebri\'an, T.~Dafni, H.~G\'omez,
D.C.~Herrera, I.G.~Irastorza, G.~Luz\'on, A.~Rodr\'iguez, L.~Segu\'i, A.~Tom\'as, J.A.~Villar}\\
{\it Lab.\ de F\'isica Nuclear y Astropart\'iculas, Universidad de Zaragoza, Zaragoza, Spain}

\end{center}

\end{titlepage}

\tableofcontents

\chapter{The Physics}

\section{Majorana neutrinos and double beta decay} \label{sec.Industry}
Neutrinos could be the only truly neutral particles among the elementary fermions. A truly neutral particle would be identical to its antiparticle, as proposed by Ettore Majorana more than 70 years ago \cite{Majorana:1937vz}. 
The existence of such Majorana neutrinos would have profound implications in particle physics and cosmology. For instance, they provide a natural explanation for the smallness of neutrino masses, the so-called \emph{seesaw mechanism}. Besides, Majorana neutrinos violate lepton-number 
conservation. This, together with CP-violation, is a basic ingredient to help uncover the reasons why matter dominates over antimatter in our Universe.

The only practical way to establish experimentally that neutrinos are their own antiparticles is the detection of neutrinoless double beta decay (\bbonu) \cite{Avignone:2008, Giuliani:2010}. This is a hypothetical, very rare nuclear transition that occurs if neutrinos are massive Majorana particles \cite{Schechter:1981bd,Hirsch:2006yk}. It involves the decay of a nucleus with $Z$ protons into a nucleus with $Z+2$ protons and the same mass number $A$, accompanied by the emission of two electrons: $(Z,A) \rightarrow (Z+2,A) + 2e^{-}$. The sum of the kinetic energies of the two emitted electrons is always the same, and corresponds to the mass difference between mother and daughter nuclei, \Qbb. The decay violates lepton number conservation and is therefore forbidden in the Standard Model.

The simplest underlying mechanism of \bbonu\ is the virtual exchange of light Majorana neutrinos, although, in general, any source of lepton number violation (LNV) can induce \bbonu\ and contribute to its amplitude. If we assume that the dominant LVN mechanism at low energies is the light-neutrino exchange, the half-life of \bbonu\ can be written as:
\begin{equation}
\Big(T^{0\nu}_{1/2}\Big)^{-1} = G_{0\nu} \ \Big|M_{0\nu}\Big|^{2} \ m^{2}_{\beta\beta}
\label{eq:Tonu}
\end{equation}
where \Gonu\ is an exactly-calculable phase-space factor, $|M^{0\nu}|$ is a nuclear matrix element, and \mbb\ is the effective Majorana mass of the electron neutrino:
\begin{equation}
m_{\beta\beta} = \Big| \sum_{i} U^{2}_{ei} \ m_{i} \Big|,
\end{equation}
where $m_{i}$ are the neutrino mass eigenstates and $U_{ei}$ are elements of the neutrino mixing matrix. Therefore, a measurement of the \bbonu\ decay rate would provide direct information on neutrino masses.

Neutrino oscillation measurements constrain how the effective Majorana mass changes with the absolute neutrino mass scale, defined as the smallest neutrino mass eigenstate. Currently, only upper bounds on the absolute mass scale, of about 1 eV, exist. Also, current oscillation data does not allow to differentiate between two possible mass eigenstates orderings, usually referred to as \emph{normal} and \emph{inverted} hierarchies. In the normal hierarchy ---where the gap between the two lightest eigenstates corresponds to the small mass difference, measured by solar experiments--- the effective Majorana mass can be as low as 2 meV. If the mass ordering is the inverted ---the gap between the two lightest states corresponds to the large mass difference, measured by atmospheric experiments---, \mbb\ can be as low as 15 meV. In the particular case in which the neutrino mass differences are very small compared to its absolute scale, we speak of the \emph{degenerate} spectrum. In this case, larger values for \mbb, approximately above 50 meV, can be obtained. 

Cosmological measurements provide a further constrain on the effective Majorana mass, since they measure the sum of the masses of the three neutrino flavors. The less restrictive cosmological bounds exclude values of \mbb\ larger than a few hundred meV, while the most restrictive cosmological bounds exclude values above some 100 meV \cite{GonzalezGarcia:2010un}.

\begin{figure}[tb]
\centering
\includegraphics[width=0.75\textwidth]{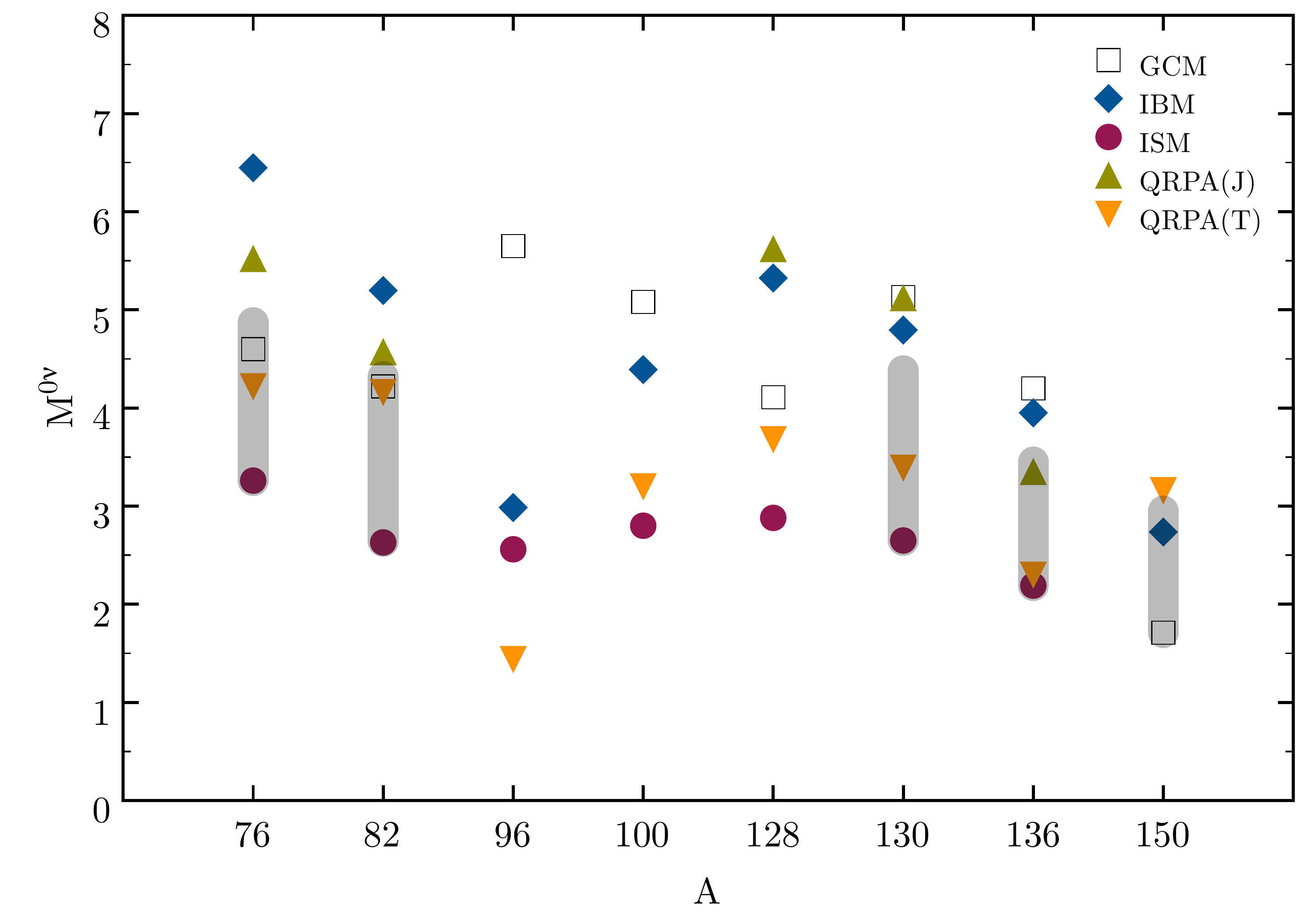}
\caption{Recent NME calculations from the different techniques (GCM \cite{Rodriguez:2010mn}, IBM \cite{Barea:2009zza}, ISM \cite{Menendez:2009di, Menendez:2009oc}, QRPA(J) \cite{Suhonen:2010ci}, QRPA(T) \cite{Simkovic:2009cu,Simkovic:2009fv,Fang:2010qh}) with UCOM short range correlations. All the calculations use $g_{A} = 1.25$; the IBM-2  results are multiplied by 1.18 to account for the difference between Jastrow and UCOM, and the RQRPA are multiplied by 1.1/1.2  so as to line them up with the others in their choice of r$_0$=1.2~fm.}
\label{fig.NME}
\end{figure}

All nuclear structure effects in \bbonu\ are included in the nuclear matrix element (NME). Its knowledge is essential in order to relate the measured half-life to the neutrino masses, and therefore to compare the sensitivity and results of different experiments, as well as to predict which are the most favorable nuclides for \bbonu\ searches. Unfortunately, NMEs cannot be separately measured, and must be evaluated theoretically. In the last few years the reliability of the calculations has greatly improved, with several techniques being used, namely: the Interacting Shell Model (ISM)
\cite{Caurier:2007wq, Menendez:2009di, Menendez:2009oc}; the Quasiparticle Random Phase Approximation (QRPA) \cite{Suhonen:2010ci, Simkovic:2008fr, Simkovic:2009cu}; the Interacting Boson Model (IBM) \cite{Barea:2009zza}; and the Generating Coordinate Method (GCM) \cite{Rodriguez:2010mn}. Figure \ref{fig.NME} shows the most recent results of the different methods.  

\section{Double beta decay experiments} \label{sec.dbd}
\begin{figure}[bthp!]
\centering
\includegraphics[width=0.5\textwidth]{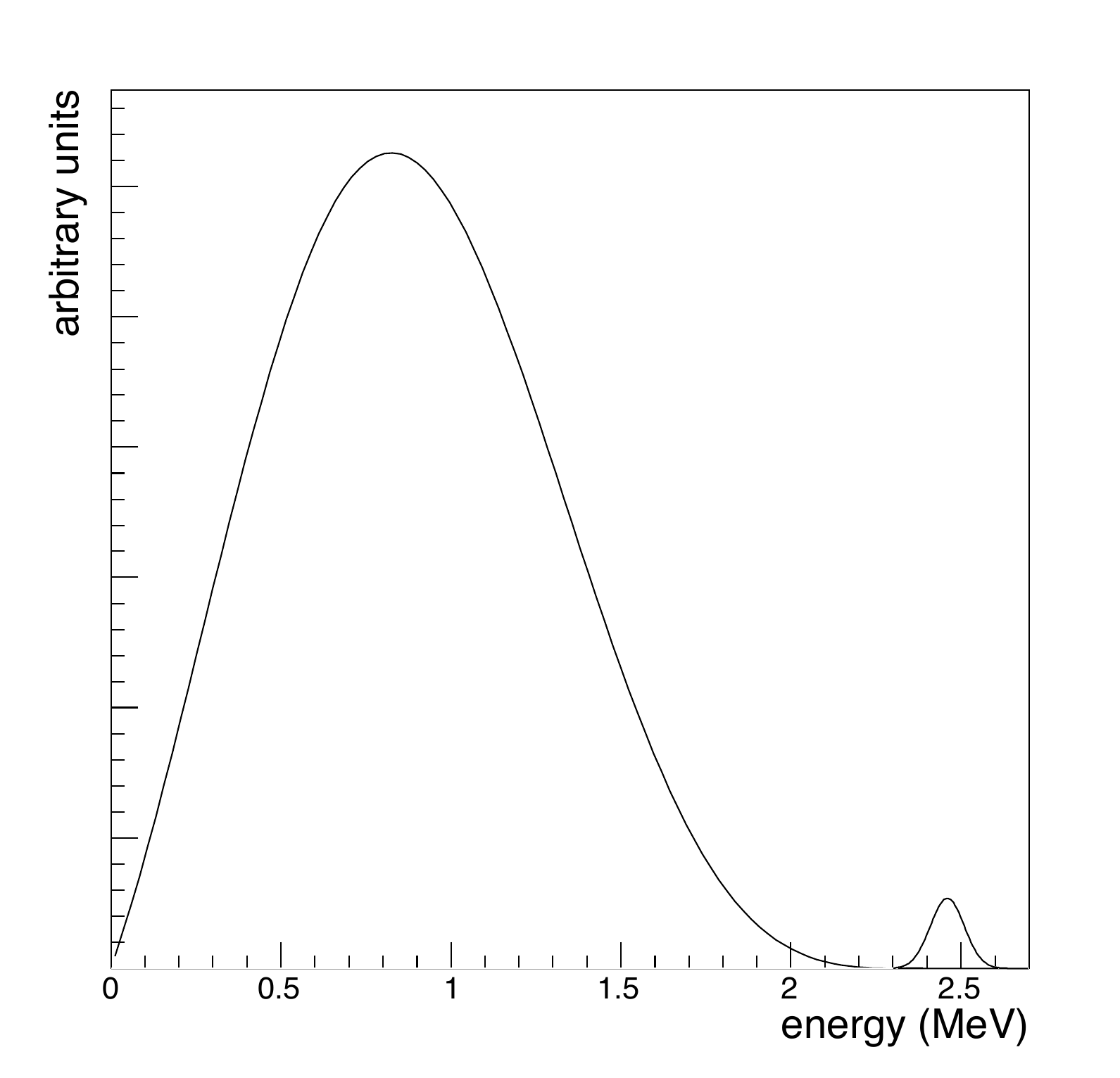}
\caption{Energy resolution is a must to exploit the golden signature in \bbonu\ searches.}
\label{fig.golden}
\end{figure}

Double beta decay experiments are designed, in general, to measure the kinetic energy of the electrons emitted in the decay. The golden signature available to a \bbonu\ experiment is the sum of such kinetic energies, which, for a perfect detector, equals \Qbb. However, due to the finite energy resolution of any detector, \bbonu\ events are reconstructed within a non-zero energy range centered around \Qbb, typically following a gaussian distribution (Figure \ref{fig.golden}). Unfortunately, any background event falling in this energy range limits dramatically the sensitivity of the experiment. Good energy resolution is therefore essential. That's why germanium-based experiments have dominated the field so far: in a \GE\ experiment, a region able to contain most of the signal ---called the \emph{region of interest} (ROI), and often taken as 1 FWHM around \Qbb--- would be only a few keV wide. 

The Heidelberg-Moscow (HM) experiment \cite{Klapdor:2000sn}, using high-purity germanium diodes enriched to 86\% in the isotope \GE, set the most sensitive limit to date: $T^{0\nu}_{1/2}(\GE) \ge 1.9 \times 10^{25}$ years (90\% CL). The experiment accumulated a total exposure of 71.7 \kgy, and achieved a background rate in the ROI of 0.18 \ckky. The energy resolution (FWHM) at \Qbb\ was $4.23 \pm 0.14$ keV. A subset of the Collaboration observed evidence for a \bbonu\ signal \cite{KlapdorKleingrothaus:2001ke}. The claim has been severely questioned \cite{Aalseth:2002dt}, but no one has been able to prove it wrong. According to it, the isotope \GE\  would experiment \bbonu\ decay with a lifetime of about $1.5\times10^{25}$ years. 

Unfortunately, energy resolution is not enough by itself: a continuous spectrum arising from natural decay chains can easily overwhelm the signal peak, given the enormously long decay times explored. Consequently, additional signatures to discriminate between signal and backgrounds are desirable. Also, the experiments require underground operation and a shielding to reduce external background due to cosmic rays and surrounding radioactivity, and the use of very radiopure materials. In addition, large detector masses, high \bb\ isotope enrichment, and high \bb\ detection efficiency are clearly desirable, given the rare nature of the process searched for. 

One way to evaluate the interest of new proposals is to compare their performance with that of the HM experiment. This can be done in terms of the three basic quantities mentioned above. The total exposure, the expected background rate in the ROI and the energy resolution.

\section{The double beta race} \label{sec.dbr}
Clearly, when considering a new \bbonu\ experiment such as NEXT, one must take into account the expected performance and schedule of the competition. However, this is not easy for several reasons: a) there are many \bbonu\ proposals, at very different stages of development; b) there are significant uncertainties in the schedules of most projects; and c) there are significant uncertainties in the expected performance of the different experiments.

On the other hand, it is still important to estimate the sensitivity to a light Majorana neutrino that can be reached by the major \bbonu\ players, in a time window that would correspond to the ``next generation'' of experiments. In
\cite{GomezCadenas:2010gs}, it was found that the four most promising 
experiments searching for \bbonu\ processes were  GERDA, CUORE, EXO and KamLAND-Zen. In this section we estimate their expected performance, extrapolated until 2020, and considering for each one two possible stages or phases as actually planned by GERDA and KamLAND-Zen. 

\begin{sidewaystable}
\begin{center}
\begin{tabular}{l c c c l c c c c}
\hline\hline
Experiment 	& Isotope 	& Resolution & Efficiency & \multicolumn{1}{c}{Phase} & Mass & Exposure & Background rate & Sensitivity \\
& & (keV) & & & (kg) & (kg$\cdot$year) & (\ckky) & (meV) \\ \hline \noalign{\smallskip}
\multirow{2}{*}{CUORE} & \multirow{2}{*}{\TEX} & \multirow{2}{*}{5} & \multirow{2}{*}{0.8} & 2015--2017 (I) & 200 & 600 & $10^{-1}$ & 140 \\
& & & & 2018--2020 (II) & 200 & 600 & $4 \times 10^{-2}$ & 85 \\ \hline \noalign{\smallskip}
\multirow{2}{*}{EXO} & \multirow{2}{*}{\XE} & \multirow{2}{*}{100} & \multirow{2}{*}{0.7} & 2012--2014 (I) & 160 & 480 & $7\times10^{-3}$ & 185 \\
& & & & (II) 2016--2020 & 160 & 800 & $5\times10^{-3}$ & 150 \\ \hline \noalign{\smallskip}
\multirow{2}{*}{GERDA} & \multirow{2}{*}{\GE} & \multirow{2}{*}{5} & \multirow{2}{*}{0.8} & 2012--2014 (I) & 18 & 54 & $10^{-2}$ & 214 \\
& & & & 2016--2020 (II) & 35 & 175 & $10^{-3}$ & 112 \\ \hline \noalign{\smallskip}
\multirow{2}{*}{KamLAND-Zen} & \multirow{2}{*}{\XE} & \multirow{2}{*}{250} & \multirow{2}{*}{0.8} & 2013--2015 (I) & 360 & 1440 & $10^{-3}$ & 97 \\
& & & & 2017--2020 (II) & 35 & 2700 & $5\times10^{-4}$ & 60 \\
\hline\hline
\end{tabular}
\end{center}
\caption{Proposals considered in the \mbb\ sensitivity comparison. For each proposal, the isotope that will be used, together with estimates for detector performance parameters --- FWHM energy resolution, detection efficiency and background rate per unit of energy, time and \bb\ isotope mass --- are given. Two possible operation phases, with estimates for the detector mass and the background rate achieved, are given for each experiment.}\label{tab:proposals} 
\end{sidewaystable}

\subsection{CUORE}
CUORE \cite{Arnaboldi:2003tu, Ardito:2005ar, Bandac:2008zz, Gervasio:2000vj, Sisti:2010zz, Arnaboldi:2008ds} is an array of TeO$_2$ bolometers. Because $^{130}$Te has a large natural abundance ($\sim$34\%), the need for enrichment is less important than in other isotopes.  CUORE can collect a large mass of isotope ($\sim 200$~kg for a total detector mass of 740 kg). The advantages and disadvantages of the technique are similar to those of \GE\ experiments with about the same  energy resolution and efficiency for the signal. However, the ratio surface to active volume is better in the \GE\ experiments.

CUORICINO measured an irreducible background of $0.153 \pm 0.006$ \ckky\ (in kilograms of detector mass). The major sources of contamination are Compton events from 2615 keV peak of \TL\ (from \THO\ cryostat contamination) and degraded alphas on copper and crystal surfaces. The more recent Three Tower Test (TTT) measures
$0.122\pm 0.001$ \ckky (detector mass), or about $4 \times 10^{-1}\ \ckky$~in isotope mass. CUORE expects to be able to reduce the background to $10^{-2}$\ckky~(per kg of detector), or about $4 \times 10^{-2}\ \ckky$ (per kg of isotope). 

CUORE is scheduled to start data taking in 2014. We assume a commissioning run of one year (2014--2015),  a physics run of 2 years (2015, 2016) at the background level of  $10^{-1}\ \ckky$~per kg of isotope. For stage II, after one year off for upgrade, we assume a
run of three years (2017-2020) with a background $4 \times 10^{-2}\ \ckky$~per kg of isotope.

\subsection{GERDA}
GERDA \cite{Abt:2004yk, Heider:2007cj, Barabanov:2009zz, Zuzel:2010gz} will search for \bbonu\ in \GE\ using arrays of high-purity germanium detectors. This is a well-established technique that offers outstanding energy resolution (better than 0.2\% FWHM at the $Q$-value) and high efficiency ($\sim$ 0.80). Its main drawback is the scalability to large masses. 
 
The first phase of GERDA will run with the same detectors used by HM and IGEX, for a total of 18 kg of isotope. Currently, the measured background rate is at the level of $5 \times 10^{-2}\ \ckky$, while the goal of the collaboration is to reduce it to $10^{-2}\ \ckky$. The second phase will add 40 kg of thick-window p-type BEGe detectors (34 kg of isotope). Such detectors have enhanced background discrimination, due to pulse shape analysis. They are also optimized to suppress surface backgrounds. The experiment expects to reach  $10^{-3}\ \ckky$.

GERDA is already commissioning the detector with depleted Ge crystals, and will probably start a physics run in 2011. We assume that their background target will be reached in 2012 and a full run of 3 years (2012-2014) for phase I, followed by one year offline for upgrade to phase II and commissioning and 5 years run (2016-2020) at their target background rate of $10^{-3}\ \ckky$.

\subsection{EXO}
The Enriched Xenon Observatory (EXO) \cite{Danilov:2000pp} will search for \bbonu\ decay in $^{136}$Xe using a liquid-xenon TPC (LXe) with 200 kg total \XE\ mass (enriched at 80\% in \XE). The use of liquefied xenon results in a relatively modest energy resolution ($\sim$ 4\% FWHM resolution at \Qbb\ \cite{Conti:2003av}). The strong point of a LXe TPC is self-self-shielding and good position resolution. It is then possible to select a fiducial volume capable to reject superficial backgrounds and multi-hit events, in such a way that only energetic gammas from \TL\ and \BI\ constitute a significant source of background\footnote{The ultimate goal of the EXO Collaboration, whose benefits are not considered here, is to develop the so-called \emph{barium tagging}. This technique would allow the detection of the ion product of the $^{136}$Xe decay, and thus eliminate all backgrounds but the intrinsic \bbtnu}. 

We have performed simulations of the response of a LXe TPC to the \BI\ and \TL\ backgrounds. Our results,
in spite of the fact that we consider a rather idealized detector are more pessimistic that those quote by the EXO collaboration. While they find a background rate of the of  $10^{-3}\ \ckky$~we obtain $7 \times \sim10^{-3}$. We believe that this discrepancy can be explained, at least partially, by taking into account that the recently measured value of \XE\ end-point \cite{Redshaw:2007un} is considerably closer to the \BI\ peak that believed a few years ago.  
EXO-200 is commissioning its detector in 2011. We consider one year to bring the detector to the desired level of background (2012) and two phases, as detailed in table \ref{tab:proposals}. In the second phase we assume that the mass is increased to 400 kg and the background reduced to  $5 \times \sim10^{-3}$, at the cost of efficiency (40\% rather than 60\%).

\subsection{KamLAND-Zen}
The KamLAND-Zen (Zen for Zero Neutrino double beta decay) \cite{Terashima:2008zz, Koga:2010ic} experiment plans to dissolve 400 kg of xenon enriched at 90\% in \XE\ in the liquid scintillator of KamLAND. Xenon is relatively easy to dissolve (with a mass fraction of more than 3\% being possible) and also easy to extract. The major modification to the existing KamLAND experiment is the construction of an inner, very radiopure  and very transparent (to the liquid scintillator emission wavelength, 350--450 nm) balloon to hold the dissolved xenon. The balloon, 1.7 meters in radius, would be shielded from external backgrounds by a large, very pure liquid scintillator volume. While the energy resolution at \Qbb\ is poor, of the order of 10\%, the detection efficiency is much better ($80\%$) due to its double envelope. 
The estimation of the background rate is affected, among other things by the radiopurity of the balloon and the liquid scintillator, as well as the lifetime of the \bbtnu\ process. 
A reasonable estimation for phase I of the experiment (400 kg) is a background rate of $10^{-3}\ \ckky$. The detector expects to start operation in 2012. We assume one year commissioning (2012--2013), and a physics run of 3 years for phase I. For phase II  the collaboration plans to purchase 1 ton of xenon and optimization of shielding could
lead to an improvement in the background rate to a level of about $5 \times 10^{-4}\ \ckky$.

\section{The NEXT challenge} \label{sec.nextc}
NEXT will search for neutrinoless double beta decay in \XE\ using a high-pressure gaseous xenon time-projection chamber. As it will be discussed with great detail later, the experiment aims to take advantage of both good energy resolution and the presence of a \bbonu\ topological signature for further background suppression. As a result, the background rate is expected to be one of the lowest of the new generation of \bbonu\ experiments.

Nevertheless, one should not forget that CUORE, EXO and GERDA are commissioning already their detectors, and KamLAND-Zen may start commissioning in 2012. NEXT must, therefore, be in business as soon as possible. As it will be shown in this report, the detector can be built by 2013 and commissioned in 2014. This would allow a 5-years run until 2020. This schedule is aggressive but feasible, since it is based in a technology amply demonstrated by the NEXT-1 prototypes, and will benefit of the experience in radiopurity available in the Collaboration and elsewhere.

The construction of NEXT-10 ---an intermediate, radiopure ``demonstrator'' of about 10 kg---, as initially foreseen in our LOI \cite{Granena:2009it}, would introduce, we believe, an unacceptable delay. NEXT-10 could only be built
after the final technological solutions are specified. Building and commissioning the detector would take one year, and another year would be needed to analyze the data. Irrespectively of the additional costs, the construction of a fully-fledged NEXT-10 detector would push the start date of NEXT to circa 2017, leaving only a 3 years run before 2020.

We examine now the various parameters that define the detector:
\begin{enumerate}
\item{\bf Mass:}  The LSC has already procured 100 kg of xenon, enriched at 90\% in \XE. We argue that, if the initial NEXT run (2015 and 2016) is successful, the mass could be increased to 1 ton. This would require a new vessel and more instrumentation, but would benefit from many of the infrastructures available for NEXT100, including the gas system and the shielding. 
\item{\bf Resolution:} The intrinsic energy resolution (FWHM at \Qbb) that can be achieved by an EL xenon detector operating at high pressure is about 0.3\% \cite{Bolotnikov:97,Nygren:2009zz}. Bolozdynia \cite{Bolozdynya:96} and White \cite{White:2009} have demonstrated resolutions of around 0.5\% in large systems equipped with 19 and 7 PMTs respectively. The resolution of the PMTs initially considered for the NEXT detector has also been measured \cite{Fernandes:2010gg} to be about 0.4\%. The intrinsic resolution of micro-bulk micromegas have also been measured in the NEXT collaboration to be about 3\% at 10 bar \cite{Balan:2010kx}. It follows that an EL TPC is the best option in terms of resolution. Our initial results from the NEXT1-LBNL prototype show a resolution of 0.8--0.9 \% at \Qbb.
\item{\bf Background rate:} The background rate depends on three factors. Good shielding from external backgrounds (this can be achieved by standard shielding techniques, such as installing the chamber inside a water tank); low activity of the construction materials (which in turn requires low activity sensors and a radiopure metal to build the vessel and field cage rings, as well as isolating any electronics near the fiducial volume and carefully measuring the radioactive budget of any component in the chamber), and detector performance (which depends on the energy resolution, as well as the topological rejection factor).
\end{enumerate}


To motivate our choices, we start by defining a reference scenario (R), with a total mass of 90 kg \XE, a fiducial  efficiency of 20\% (that takes into account that some \XE\ is wasted  since our design considers only a single gas volume) and a 6-years run (2015 to 2020). We assume an energy resolution of 0.8\%, as measured by Bolozdynya \cite{Bolozdynya:96}, and consistent with our initial results described later in this document. Our detailed simulations of the detector, also described later yield a background rate of $2 \times 10^{-4}$ \ckky.

We also define a ``slow scenario'' (S), which differs from the previous one in a shorter run (3 years) and an ``aggressive
 scenario'' (A) which assumes a 2 years run with 90 kg of xenon and the ANGEL baseline and 3 years run with 1 ton  and an improved detector with a somewhat lower background rate ($10^{-4} \ckky$).
 
 
\begin{table}[t!b!]
\begin{center}
\begin{tabular}{c D{.}{.}{3.0}}
\hline \hline
Proposal  & \multicolumn{1}{c}{Sensitivity} \\
& \multicolumn{1}{c}{(meV)} \\
\hline
A & 38 \\
R  & 89 \\
S & 115 \\
\hline \hline
\end{tabular}
\end{center}
\caption{Sensitivity of NEXT under different scenarios (see text for further details).}
\label{tab.SNEXT}
\end{table}%

Table \ref{tab.SNEXT} summarizes the sensitivity of NEXT under the different scenarios discussed above. We argue that the aggressive scenario is entirely possible and would result in a sensitivity capable to outperform even that of KamLAND-Zen. NEXT is still very competitive in our reference scenario, at the level of GERDA and CUORE. The slow scenarios is, obviously, less interesting.    

The lessons are clear: given the current situation, the challenge that the NEXT collaboration must face is to build a state of the art detector, with the best possible resolution, and  as fast as possible. There is no room in the current race for a prototype of 10 kg. Rather, we should keep in mind that the 100 kg detector can and should be upgraded to 1 ton in a not-too-distant future. In spite of the fierce competition, NEXT has the chance to be among the  leaders of the next generation of \bbonu\ experiment.

%
%
\chapter{EASY \& SOFT}

\section{Energy resolution in xenon detectors} \label{sec.nextc}
\begin{figure}[tbh]
\centering
\includegraphics[width=0.65\textwidth]{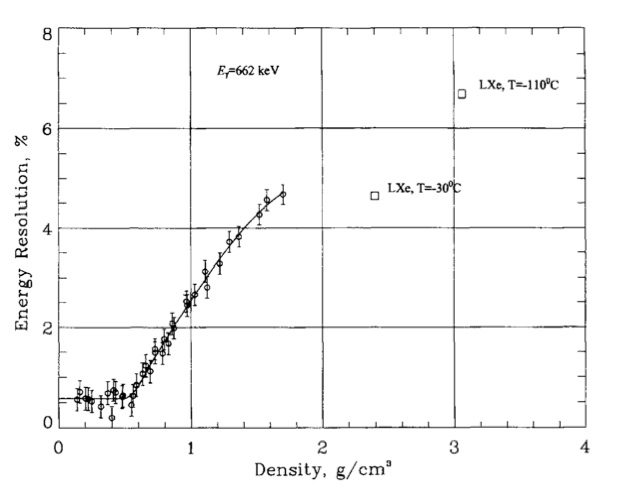} 
\caption{The energy resolution (FWHM) is shown for $^{137}$Cs 662 keV gamma rays, as a function of xenon density, for the ionization signal only. Reproduced from \cite{Bolotnikov:97}.} \label{fig:bolotnikov} 
\end{figure}

Excellent energy resolution is a crucial ingredient for a  \bbonu\ experiment. Indeed, physics allows such resolution to be attained in  a high-pressure gaseous xenon chamber (HPGXe) (instead, those very same intrinsic physics processes appear to limit the performance in a liquid xenon chamber, LXe). This is clearly seen in Figure \ref{fig:bolotnikov}, reproduced from Bolotnikov and Ramsey (1997) \cite{Bolotnikov:97}. The resolutions displayed were extracted from the photo-conversion peak of the 662 keV gamma ray from the $^{137}$Cs isotope. Only the ionization signal was detected. A striking feature in Figure~\ref{fig:bolotnikov} is the apparent transition at density 
$\rho_t \sim 0.55$ g/cm$^3$. Below this density, the energy resolution is approximately constant:
\begin{equation}
\delta E/E = 6 \times 10^{-3} {\rm ~FWHM}.
\label{eq:intrinsic}
\end{equation}
For densities greater than $\rho_t$, energy resolution deteriorates rapidly, approaching a plateau at LXe density.

The most plausible explanation underlying this strange behavior is the appearance, as density increases, of two-phase xenon (see \cite{Nygren:2007zz} and references therein). In contrast, given the xenon critical density, the intrinsic resolution in the gas phase is very good up to pressures in the vicinity of 50 bar, at room temperature, although practical and technical issues dictate operation at smaller pressures, in the range of 10 to 20 bar.  

Extrapolating the observed resolution in Figure \ref{fig:bolotnikov} as $\sqrt{E}$ to the \XE\ Q-value (2458 keV), a naive energy resolution is predicted:
\begin{equation}
\delta E/E = 3 \times 10^{-3} {\rm ~FWHM}.
\label{eq:res}
\end{equation}

Based on ionization signals only, the above energy resolution reflects an order of magnitude improvement relative to LXe. For densities less than $\rho_t$, the measured energy resolution in Figure~\ref{fig:bolotnikov} matches the prediction based on Fano's theory \cite{FANO}. The Fano factor $F$ reflects a constraint, for a fixed energy deposited, on the fluctuations of energy partition between excitation and the ionization yield $N_I$. For electrons depositing a fixed energy $E$, the rms fluctuations $\sigma_I$~ in the total number of free electrons $N_I$ can be expressed as:
\begin{equation}
\sigma_I = \sqrt{F\ N_I}.
\label{eq:fano}
\end{equation}
For pure gaseous xenon (GXe), various measurements \cite{Nygren:2007zz} show that:
\begin{equation}
	F_{GXe} = 0.15 \pm 0.02
\label{eq:gas-fano}
\end{equation}
In LXe, however, the anomalously large fluctuations in the partitioning of energy to
ionization produce an anomalous Fano factor:
\begin{equation}
	F_{LXe} \sim 20,
	\label{eq:liquid-fano}
\end{equation}
larger than the one corresponding to GXe by about two orders of magnitude.

A second advantage of gas relative to liquid is the ability to exploit the topological signal of a \bbonu\ event, that is the tracks left in the gas by the two electrons produced in the \bbonu\ decay. At 10 bar the track length of the electrons is of the order of 30 cm and can be easily imaged in a TPC. Such a topological signature is not available in LXe detectors, due to the high density of the liquid phase. This two advantages (good energy resolution and topological signature) are the key reason why NEXT can be competitive with a LXe experiment such as EXO and a Xenon-liquid-scintillator experiment such as KamLAND-ZEN. 
 
 Next we examine the various signals available in an HPGXe and how they can be exploited to search for \bbonu\ events.
 
\subsection{Scintillation}

Two processes are produced in xenon, as a response to the passage of charged particles. The first one is {\em ionization} of the gas (call $W_I$~to the average energy spent in the creation of one electron-ion pair);  the second is {\em emission of scintillation light} (call $W_s$~to the average energy spent in the creation of one primary scintillation photon). The detection of primary scintillation allows the measurement of the start-of-the-event $t_0$, needed to place
an event properly in 3-D space in a TPC. 

A recent measurement, within the context of the NEXT R\&D \cite{Fernandes:2010gg} yields:
\begin{equation}
	W_s = 76 \pm 6 ~\mathrm{eV}
\label{eq:Ws}
\end{equation}
Since the end-point of the $\XE \rightarrow \BA$~transition is $\Qbb=2457.83$~keV, we obtain that this translates in
\begin{equation}
	N_s = 32342 \pm 2551 ~{\rm photons.}
\label{eq:Ns}
\end{equation}

These photons are emitted isotropically and need to be readout with photosensors capable to count single photoelectrons, that is, photomultipliers. Furthermore, it implies the use of pure xenon, since primary scintillation signals are quenched by common molecular additives such as nitrogen, hydrogen or methane.

\subsection{Ionization}
Ionization can be used to measure both the energy of the (\bbonu) event and to track the two signature electrons. In order to do so, electrons must be first drifted towards the anode. This, in turn, requires a suitable electric field. 

The longitudinal diffusion of electrons drifting towards the anode has a minimum in xenon given by:
\begin{equation}
	E/p = 0.03 {\rm V/cm Torr}
\label{eq:long-diff}
\end{equation}
or 375 V/cm at 15 bar, with a drift velocity of about $1~\mathrm{mm}/\mu\mathrm{s}$.

At E/p above that given by \ref{eq:long-diff}, the drift velocity changes slowly. A low drift velocity is not necessarily a disadvantage in a low-rate experiment such as NEXT, assuming that attachment (whose effect increases with drift time) is kept under control. Indeed, too a high electric field can result in unwanted systematic effect \cite{Tavora:2004}. Finding the optimum electric field, that maximizes the resolution and minimizes systematic effects is one of the goals of the NEXT-1 program. However, we expect it not to be very different from the value defined by Eq.~\eqref{eq:long-diff}. Indeed, the initial results from our prototypes show that 
resolution of the photoelectric Cs-137 peak stays constant for values of the drift field between 400 V/cm and about 2 kV/cm. 

 The diffusion depends on both electric field and electron temperature. For a drift of 1 m and an electric field of 375 V/cm (at 15 bar) the transverse diffusion is of the order of 1 cm. This value is large, but seems acceptable. At 15 bar the track of the two electrons produced in a \bbonu\ is of the order of 20 cm, and thus a pitch of 1 cm allows to track the event comfortably. On the other hand, the large value of the diffusion implies that there is no obvious advantage on a too fine grain pitch, since the intrinsic resolution is dictated by the physics. 
 
  
For pure gaseous xenon, various measurements \cite{Nygren:2007zz} show that:
\begin{equation}
	W_I  = 24.8 ~\mathrm{eV}\,.
\label{eq:WI}
\end{equation}
This results in a number of primary electrons at \Qbb\ of:
\begin{equation}
	N_I = 2457.8/24.8 = 99112
\label{eq:NI}
\end{equation}
or, roughly, $10^5$~primary electrons for a \XE\ \bbonu\ event. 

\subsection{Intrinsic energy resolution}

For electrons depositing a fixed energy $E$,
the (rms) fluctuations $\sigma_l$ in the total number of free electrons $N_I$
can be expressed as:
\begin{equation}
\sigma_l = (F N_I)^{1/2} = (0.15 \times 10^5)^{1/2}= 122 {\rm~rms~electrons}
\label{eq:rms-fano}
\end{equation}

The intrinsic energy resolution (FHWM) can be obtained as:
 \begin{equation}
\delta E/E = 2.35\ \sigma_l /N_I = 2.35\times 122/10^{5} \sim 3 \times 10^{-3} \ {\rm FWHM}
\label{eq:res-int}
\end{equation}
which corresponds to the value found in \cite{Bolotnikov:97}.

Of course, there are many factors that can spoil this very good intrinsic resolution. Among these, we mention the following:
\begin{enumerate}
\item Losses of drifting electrons due to electronegative impurities, volume recombination, grid transparency, etc., represented by a factor $L = 1 - \epsilon$, where $\epsilon$~is the overall electron collection efficiency. 
\item Gain processes such as avalanche multiplication, which multiply the signal by $m$~and introduce fluctuations in the detected signal, represented by a variance $G$.
\item Electronic noise, in electrons RMS at signal processing input, represented by $n$.
\end{enumerate}

In addition there are other important sinks of resolution, such as fluctuations associated to Bremsstrahlung losses, channel equalization, non-linearities, etc. However, an analysis of the previous list is sufficient to understand the main issues to be addressed to approach the intrinsic resolution. Assuming that all the above-mentioned sources are gaussian and uncorrelated, we can combine them in quadrature: 
\begin{equation}
\sigma_n^2 = (F + G + L)N_I + \frac{n^2}{m} .
\label{eq:total-sigma}
\end{equation}
where $\sigma_n$~is the total number of electrons (rms), due to fluctuations in all sources. Then:

\begin{eqnarray}
\delta E/E &= &\frac{2.35 \ \sigma_n}{N_I\epsilon} \\
&=& 2.35 \ \left[\frac{F + G + L + \frac{n^2}{mN_I}}{N_I\epsilon^2}\right]^{1/2},
\label{eq:deltae}
\end{eqnarray}

The challenge in NEXT is to minimize the factor $L$~ (this can be achieved with a very clean gas that minimizes attachment) and the gain fluctuation factor $G$.

The gain factor turns is considerably larger than $F$~in gas proportional counters involving avalanche multiplication. On the other hand, $G$~can be made at least as small as $F$~using electroluminescence.

\begin{figure}[tb!]
\centering
\includegraphics[width=0.85\textwidth]{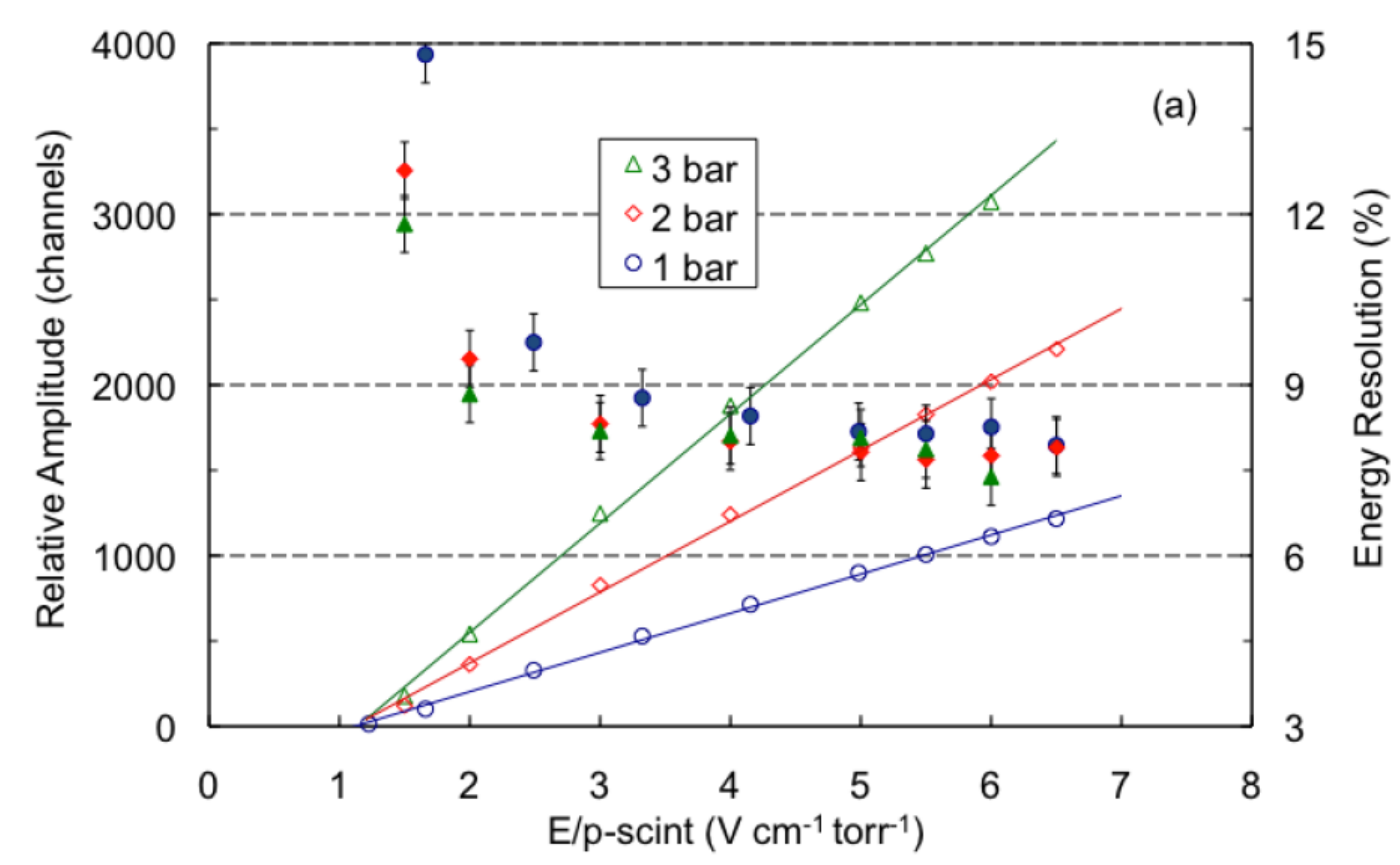} 
\caption{\small Pulse amplitude (open symbols) and energy resolution (full symbols) for 5.9 keV X-rays absorbed in a NEXT-0 prototype as a function of: E/p-scint, the reduced electric field in the scintillation region. Notice that the resolution for E/p above 3 is about the same for all pressures, near 8\%. This extrapolates to better than 0.5\% at \Qbb.}
\label{fig.nr}			
\end{figure}

A recent measurement done in the context of the NEXT R\&D (\cite{Fernandes:2010gg}), shows, indeed, that the use of EL allows to achieve resolutions close to intrinsic. Figure \ref{fig.nr} shows pulse amplitude (open symbols) and energy resolution (full symbols) for 5.9 keV X-rays absorbed in a NEXT-0 prototype as a function of the reduced electric field in the scintillation region. Notice that the resolution for E/p above 3 is about the same for all pressures, near 8\%. This extrapolates to about 0.4\% at \Qbb.

\begin{figure}[tb!]
\centering
\includegraphics[width=0.85\textwidth]{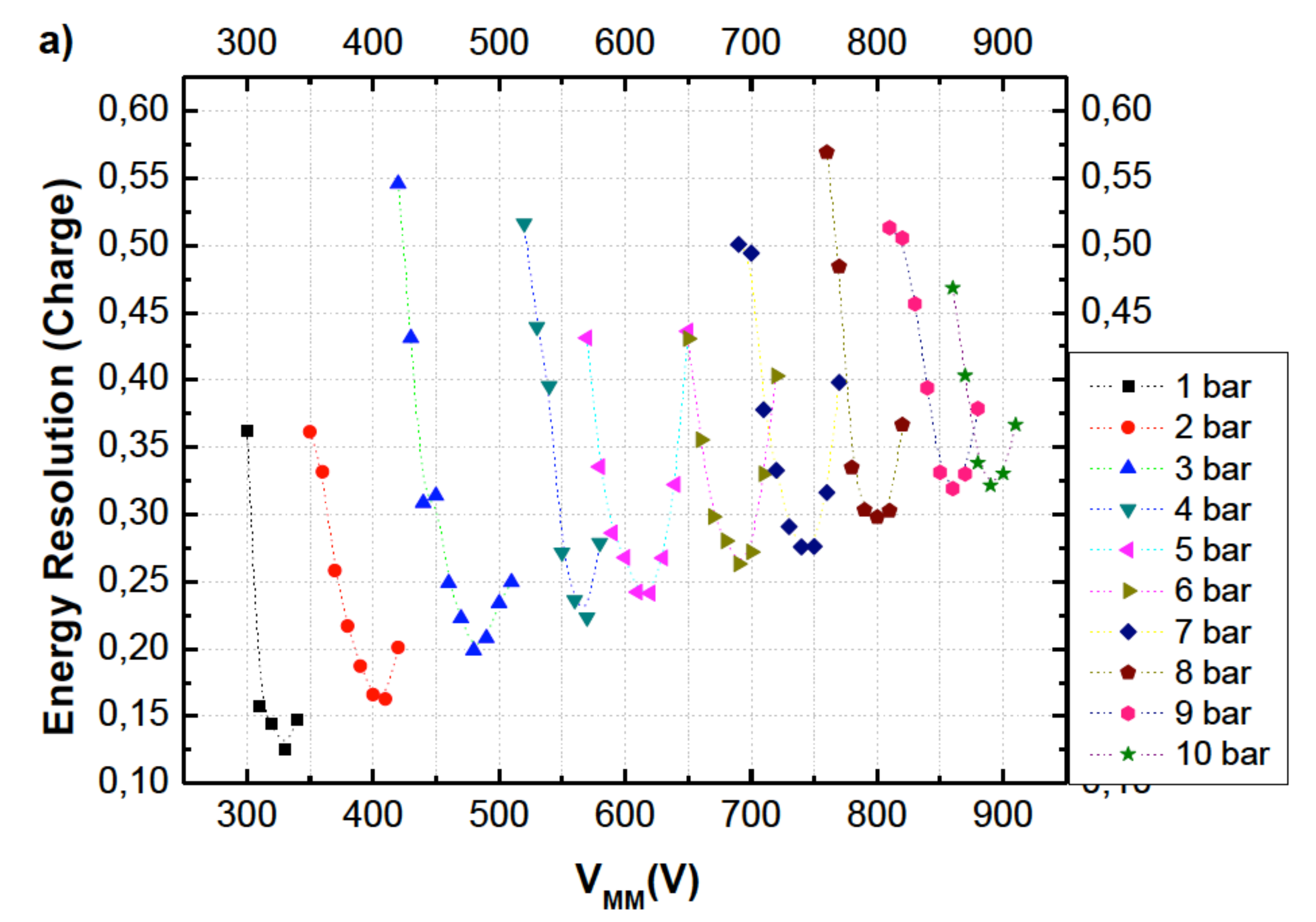} 
\caption{\small Resolution of a micro-bulk Micromegas (the last generation of micromegas devices) as a function
of the pressure for 22.1 keV photons. The resolution varies between 12\% at 1 bar ($\sim$ ) about 1\% at \Qbb, to
 32\% at 10 bar, 3\% at \Qbb. The results were measured as a part of the NEXT collaboration R\&D
 \cite{Balan:2010kx}. }
\label{fig.mm}			
\end{figure}

In the next section we will describe the physics of electroluminescence in more detail. Notice, on the other hand, that
intrinsic resolution is not the only merit of using linear amplification rather than avalanche gain. As we have seen,
the use of conventional molecular additives (quenchers) seems prohibitive in a HPGXe detector. But stable operation of avalanche-based devices requires normally the use of such quenchers to stabilize the gas and avoids sparks.

However, some of the most robust gain devices, such as the Micromegas  can perform well even in pure xenon and at high pressures. Nonetheless, their resolution appear to degrade with increased pressure, as demonstrated in  \cite{Balan:2010kx}. Figure \ref{fig.mm} shows that the resolution attainable at \Qbb\ by the last-generation micro-bulk micromegas at 10 bar would be 3\%, to be compared with that of 0.4\% found in (\cite{Fernandes:2010gg}, Figure \ref{fig.nr}), using electroluminescence. Both measurements were carried out with very small setups, in close-to-ideal conditions, and therefore can be taken as reflecting the intrinsic performance of the devices under study. 

Thus, the choice of an avalanche-gain device such as micromegas does not appear to be optimal in terms of energy resolution. 
On the other hand, the micro-bulk micromegas has been measured to be very radiopure and it could, conceivably, improve the topological signature in NEXT if a mixture capable to reduce the diffusion could be found. In this case one could take advantage of a reduced pitch, and the use of such devices as a trackers could be interesting. Thus, within the context of our R\&D for a detector upgrade, we will keep exploring the possibility of using micromegas as tracking devices compatible with an EL TPC.

\section{Electroluminescence} \label{sec.el}
\subsection{The Gas Proportional Scintillation Chamber}

\begin{figure}[tbhp!]
\begin{center}
\includegraphics[width=0.8\textwidth]{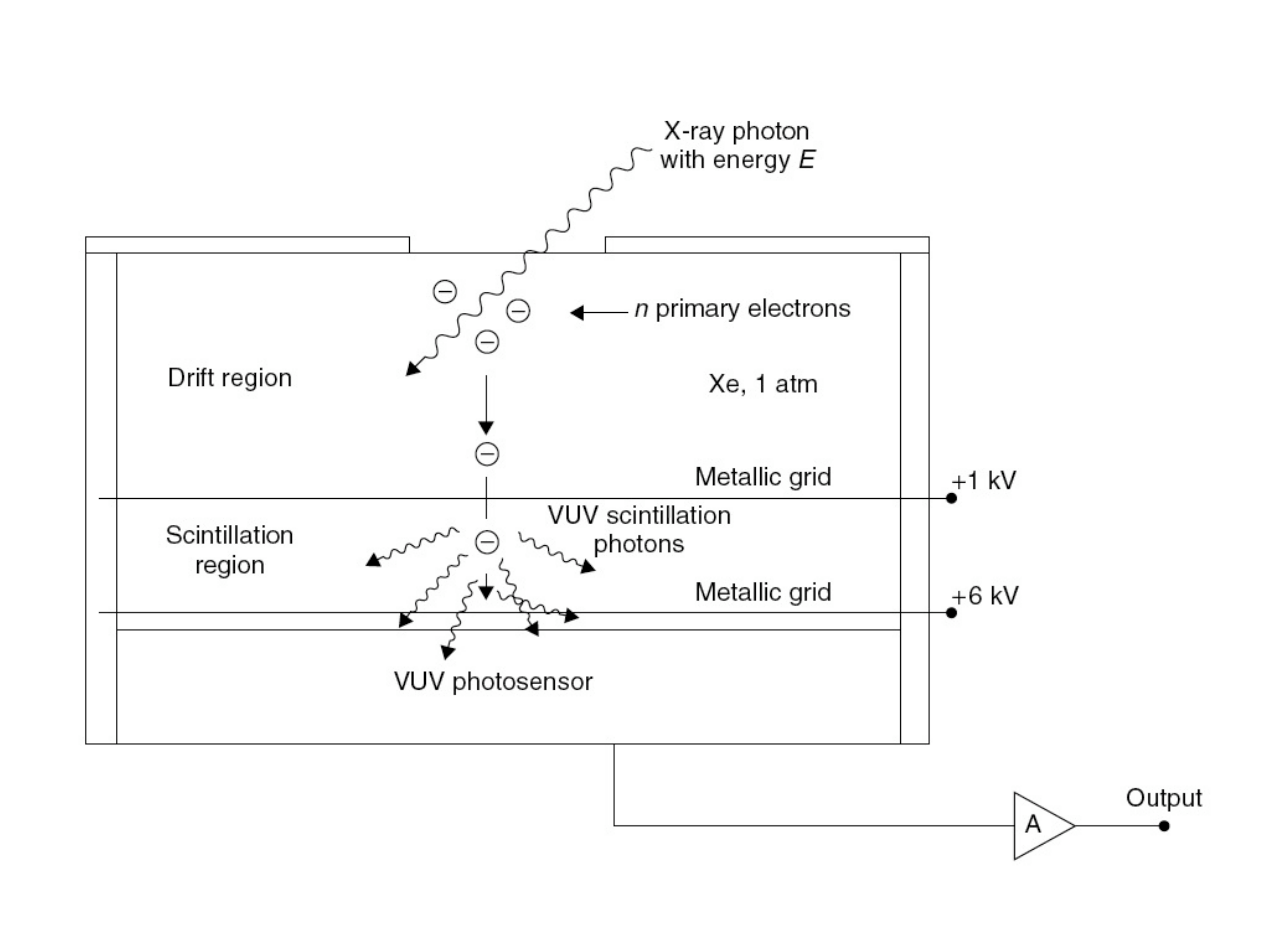}
\includegraphics[width=0.8\textwidth]{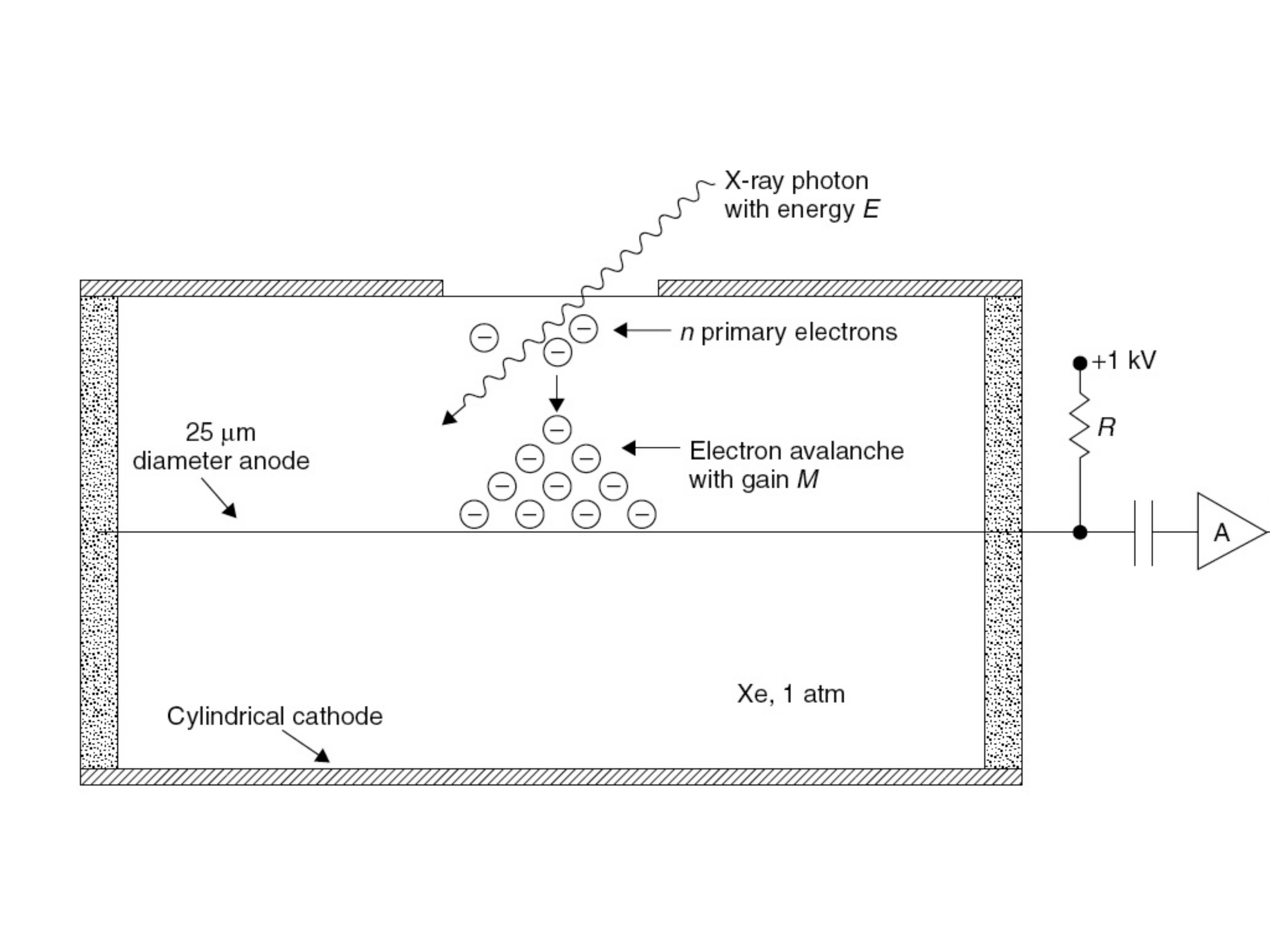} 
\end{center}
\caption{\small Top: principle of a Gas Proportional Scintillation Counter. Bottom: principle of a Gas Proportional Counter with avalanche gain (from \cite{Conde:04}).}
\label{fig:GPSC} 
\end{figure}
 
Figure \ref{fig:GPSC} (top), illustrates the principle of a Gas Proportional Scintillation Chamber (GPSC) \cite{Conde:67,Charpak:75}. An x-ray enters through the chamber window and is absorbed in a region of weak electric field
($>0.8~{\rm kV ~cm^{-1}~ bar^{-1}}$) known as the drift region. The ionization electrons drift under such field to a region of moderately high electric field (around $3-4 ~{\rm kV ~cm^{-1}~ bar^{-1}}$ range), the so-called scintillation
or EL region. In the scintillation region, each electron is accelerated so that it excites, but
does not ionize, the gas atoms/molecules.
The excited atoms decay, emitting UV light (the so-called
secondary scintillation), which is detected by
a photosensor, usually a photomultiplier tube. The intensity of the secondary scintillation light is
two or three orders of magnitude stronger than
that of the primary scintillation. However, since
the secondary scintillation is produced while the
electrons drift, its latency is much longer than that for the primary scintillation, and its rise time is much slower
(a few $\mu$s compared to a few ns). For properly chosen electric field strengths and EL region spatial widths, the number
$n_{ph}$ of secondary scintillation photons produced by
a single primary electron is nearly constant and can
reach values as large as a few thousand photons per electron.

The average total number, $N_t$, of secondary
scintillation photons produced by an X-ray photon
is then $N_t = n_{ph}\cdot N_I$, (recall that $N_I$~is the number of primary 
ionization electrons) so the photosensor signal
amplitude is nearly proportional to E, hence the
name of gas proportional scintillation counter
(GPSC) for this device.

What made the devices extraordinarily attractive was their improved energy resolution compared with conventional Proportional Chambers (PC) --- Figure \ref{fig:GPSC} (bottom). In a PC
the primary electrons are made to drift
towards a strong electric field region, usually
in the vicinity of a small diameter (typically
25~$\mu$m) anode wire. In this region,
electrons engage in ionizing collisions that lead to
an avalanche with an average multiplication gain
M of the order of 10$^3$ to 10$^4$. If M is not too
large, space charge effects can be neglected, and
the average number of electrons at the end of the
avalanche, $N_a = M\cdot N_I$, is also proportional to
the energy E of the absorbed X-ray photon (hence
the name proportional (ionization) counter given
to this device).
 However, for PC detectors, there are
fluctuations not only in $N_I$ but also in M; for GPSCs, since
the gain is achieved through a scintillation process
with almost no fluctuations, only fluctuations in
$N_I$ and in the photosensor need to be considered. Thus a better energy
resolution was achieved in the latter case; typical values for 5.9 keV X-rays were 8\%
for GPSC and 14\% for PC. 
 
 The Scintillation Drift Chamber (SDC) was invented in 1975~\cite{Charpak:75}.  An SDC
is a TPC with EL readout instead of charge gain by electron avalanche
multiplication in gas. A large SDC with 19 PMTs ~\cite{Bolozdynya:96} demonstrated excellent energy resolution at high pressure (9 bar), and for high energy X-rays. However, for mainstream particle physics, EL has had application primarily in only one technique: two--phase LXe detectors aimed at direct detection of WIMPs \cite{Aprile-book}. In that
very successful application, the enabling asset of EL is not excellent energy
resolution (limited, as we have seen, by the anomalous Fano factor in liquid), but the capability to detect single electrons.

\subsection{Xenon atomic energy structure}
In the EL region the drift electrons are accelerated and collide with the gas atoms.
If the electric field is not too large the electrons collide elastically with the atoms and can also excite but not ionize them. In the case that an excitation collision happens, the atom can stay in one of several excited levels. Since, at high pressures (above few hundreds of torr \cite{Tanaka2}), the time intervals between collisions of an excited atom with other atoms of the gas are much smaller than the atomic radiative lifetimes \cite{OliveirauE} the main channel of de-population of these excited atoms is through the formation of excimers--electronically excited molecular states. Excimers, $R_2^{**}$, are formed through three-body collisions between one excited atom, $R^*$, and two atoms in the ground state, $R$:
\begin{equation}
R^*+2R \rightarrow R_2^{**} + R
\label{eq:excimerformation}
\end{equation}

The excimers responsible for the VUV electroluminescence, $0_u^+\left(^1\Sigma_u^+\right)$, $\left(\nu=0\right)^1\Sigma_u^+$ and $\left(\nu=0\right)^3\Sigma_u^+$, are formed from $1s_4$ and $1s_5$ atomic levels ($J=3/2$) \cite{Mulliken} or by  radiative transitions from higher excited molecular levels which are formed from atoms with an energy higher than $1s_4$ and $1s_5$ \cite{Koehler}. The excimers represented with ``$\left(\nu=0\right)$'', $R_2^*$, are vibrationally relaxed through two-body collisions between the unrelaxed excimers, e.g. $0_u^+\left(^1\Sigma_u^+\right)$, $R_2^{**}$, and ground atoms, $R$:
\begin{equation}
R_2^{**}+R \rightarrow R_2^{*} + R
\label{eq:vibdecay}
\end{equation}

Vibrational unrelaxed excimers can decay to the repulsive ground state, $^1\Sigma_g^+$, emitting a VUV photon:
\begin{equation}
R_2^{**} \rightarrow  2R + h\nu
\label{eq:excimerdecay1}
\end{equation}
as well as vibrational relaxed excimers:
 \begin{equation}
R_2^{*} \rightarrow  2R + h\nu
\label{eq:excimerdecay2}
\end{equation}

In the case that the radiative decay is from $R_2^{**}$ the energy of the VUV photon is slightly higher than if the decay is from $R_2^{*}$. In this way it is, at low gas pressures, observed a continuum emission spectrum with two peaks, usually called ``first continuum'' - at higher frequencies - and ``second continuum'' - at lower frequencies. The reason why, at high pressures, one usually only observe the ``second continuum'' is because process (\ref{eq:vibdecay}) is preferable to (\ref{eq:excimerdecay1}) due to the increasing in the number of atom collisions.

\subsection{Simulation of EL in NEXT}
NEXT software includes a platform to simulate EL \cite{OliveirauE},  based in Garfield \cite{garfield} and Magboltz \cite{magboltz1,magboltz2}. We have used it to study the expected EL yield and energy resolution. 

Primary drift electrons were allowed to drift a distance of $d=5$ mm under the influence of an uniform electric field created by two infinite parallel planes (except in the edges this arrangement simulates well the large EL grids in the NEXT detector). It was considered that the gas is at a pressure $p=10$ bar and at a temperature of $293\textrm{K}$. A set of $N_e=10.000$ primary electrons was used for each value of the potential applied between the parallel planes, $V$. 

\subsection{Electroluminescence yield}

\begin{figure}[tbhp!]
\centering
\includegraphics[width=1.0\textwidth]{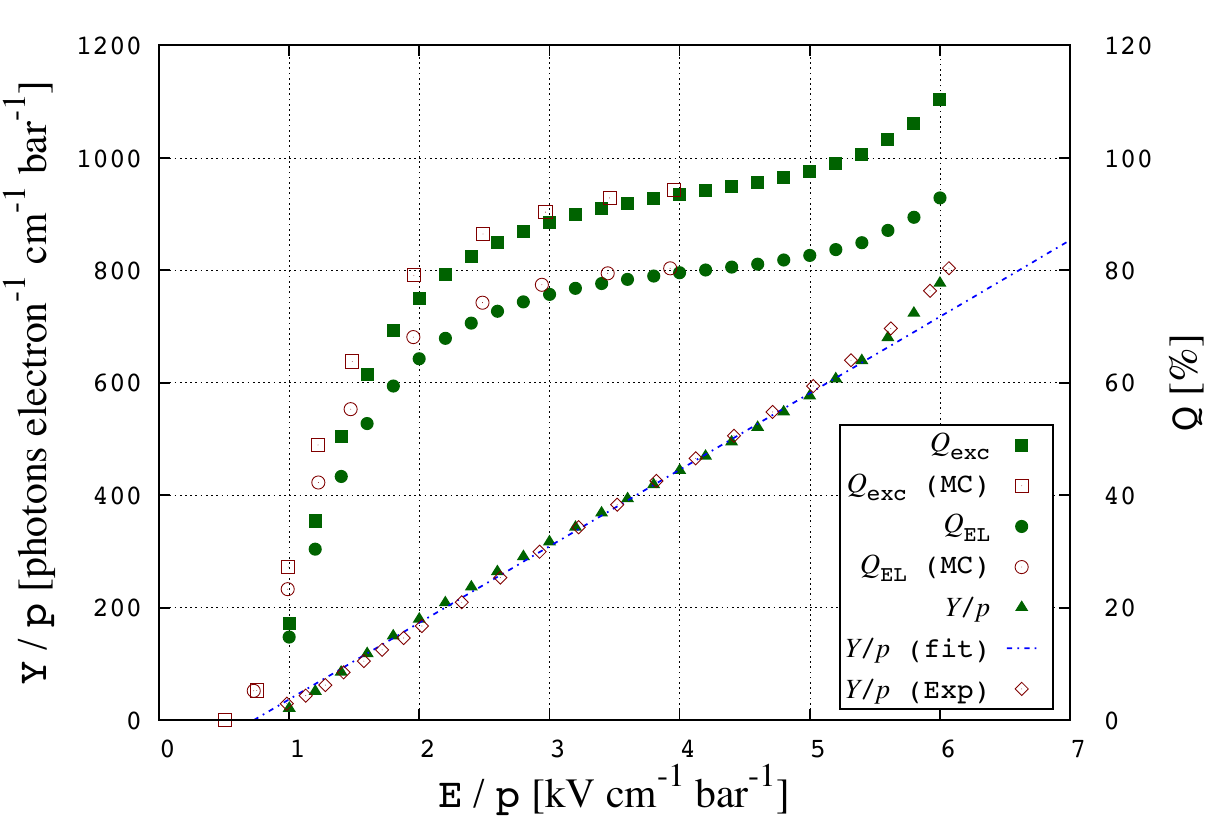}
\caption{Reduced electroluminescence yield, $\left(\frac{Y}{p}\right)$, as a function of the reduced electric field (pressure units), $\left(\frac{E}{p}\right)$. Excitation efficiency, $Q_{{\rm exc}}$, and electroluminescence efficiency, $Q_{{\rm EL}}$, as a function of the reduced electric field. are also shown. Full symbols are results of this
work. Former Monte Carlo results of $Q_{{\rm exc}}$ and $Q_{{\rm EL}}$ \cite{3d} as well as experimental measurements of
the reduced EL yield \cite{Monteiro:2007} are included (open symbols) for comparison.}
\label{fig:yop_q}
\end{figure}

\begin{figure}[tbhp!]
\centering
\includegraphics[width=1.0\textwidth]{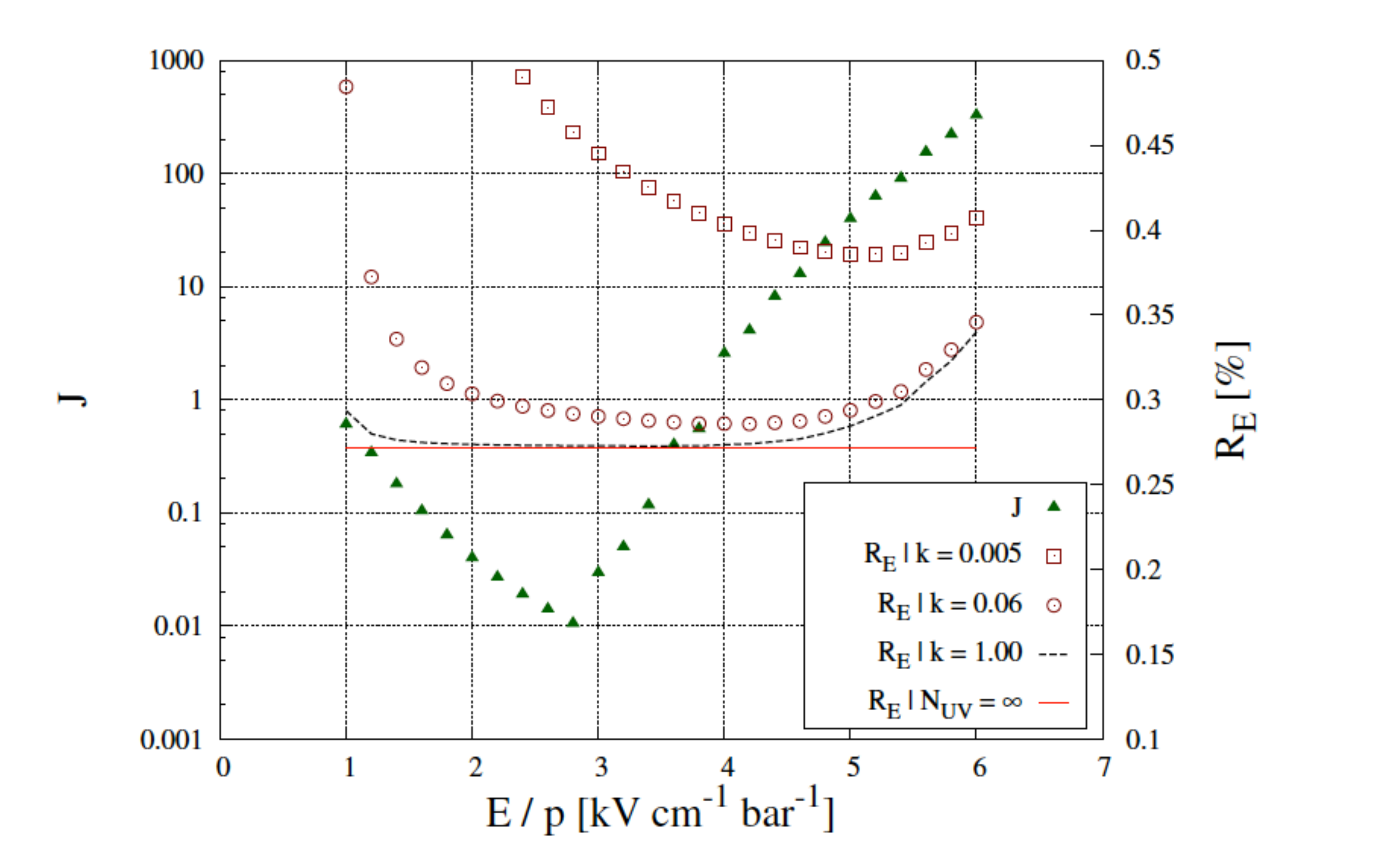}
\caption{Relative variance in the number of emitted EL photons as a function of the reduced electric field. Energy resolution, $R_{E}$, as a function of the reduced electric field for three different scenarios: a) an ideal detector that detects all EL photons; b) an detector with 50\% effective PMT coverage and an effective detection efficiency per PMT of 10\%  ($k=0.5 \times 0.1 = 0.05$); and c) a detector with 5\% PMT coverage and 10\% detection efficiency ($k=0.005$).  The value of $R_E$ if only the fluctuations in the production of primary charge contributed is also shown (red line). Notice that the resolution is always better than 0.5\% even in case c).}
\label{fig:j}
\end{figure}

In Figure \ref{fig:yop_q} it is shown the reduced electroluminescence yield, $\left(\frac{Y}{p}\right)$, as a function of the reduced electric field, $\left(\frac{E}{p}\right)$. The reduced electroluminescence yield is defined as being the number of photons emitted per primary electron and per unit of drift length divided by the number density of the gas, $N$. The behavior of $\left(\frac{Y}{p}\right)$ with $\left(\frac{E}{p}\right)$ is approximately linear even when the actual ionization threshold is achieved at $\left(\frac{E}{p}\right)\sim3~\textrm{kV~cm}^{-1}\textrm{bar}^{-1}$. In Figure \ref{fig:j} it can be easily seen that this threshold is achieved since, for higher values of the electric field, the fluctuations in the secondary charge production, which are bigger than in the electroluminescence, start to dominate. The EL yield keeps its linear behavior while the probability of ionization is low.

Performing a linear fit to the obtained points we obtain the dependence:
\begin{equation}
\left(\frac{Y}{p}\right)=\left(130\pm1\right)\left(\frac{E}{p}\right)-\left(80\pm3\right) \left[\textrm{photons electron}^{-1}\textrm{ cm}^{-1}\textrm{ bar}^{-1}\right]
\label{eq:yopfit}
\end{equation}

Consider, for example, the number of photons produced in a TPC operating at 15 bar pressure, with a EL region
of 5 mm and $E/p$~set to 3.5. One then produces 2800 EL photons per primary electron. Since there are about $10^5$ primary electrons in a \bbonu\ event, we end up with $\sim 3 \times 10^8$~EL photons.

%

Consider now the variance $G$~of the gain:

\begin{equation}
G = 1/Y+ (1+ \sigma_{pd}^2)/n_{pe}
\label{eq:G}
\end{equation}

The contributions to the gain resolution $G$ must include fluctuations in:
\begin{enumerate}
\item the EL gain Y;
\item $n_{pe}$, the number of photo-electrons per incident electron;
\item the gain process in the photo-detector per single photo-electron, whose fluctuation we express by $\sigma_{pd}$.
\end{enumerate}

The first term in (\ref{eq:G}) is much smaller than the second, since Y is large, while the limited photon detection efficiency results in a smaller number for $n_{pe}$. Assuming $\sigma_{pd}^2= 0.5$ (most PMTs will do better than that) and setting G = 0.15 (so that it contributes no more than the Fano factor) one obtains:
\begin{equation}
n_{pe}^{EL} \ge 10
\label{eq:npe}
\end{equation}
Thus, in order to optimize the resolution is necessary a device capable to detect at least 10 photoelectrons per primary electron. We will revisit this condition when discussing the NEXT design.

Using our detailed simulation we can estimate the energy resolution attainable by NEXT for the \bbonu\ events.  Figure \ref{fig:j} shows energy resolution $R_{E}$ curves as a function of the reduced electric field for three different scenarios: a) an ideal detector that detects all EL photons; b) an detector with 50\% effective PMT coverage and an effective detection efficiency per PMT of 10\%  ($k=0.5 \times 0.1 = 0.05$); and c) a detector with 5\% PMT coverage and 10\% detection efficiency ($k=0.005$). $J$~is the parameter that describes the
fluctuations relative to the electroluminescence production, defined as the relative variance in the number of emitted VUV photons per primary electron, $N_{EL}$:

\begin{equation}
\displaystyle J=\frac{\sigma^2_{N_{EL}}}{\bar{N}_{EL}}
\label{eq:j}
\end{equation}
%

To summarize, we have a detailed understanding of the EL process, which is fully simulated within the NEXT software framework. Our predictions for both the EL yield and the EL resolution are consistent with data available in the literature and with our own measurements within the NEXT R\&D and confirm the possibility to reach a resolution at \Qbb\ which can be as low as 0.4\%, and in any case, at the level of our target resolution of 1\%. 

\section{The SOFT concept} \label{sec.soft}

A HPGXe TPC design for \bbonu\ searches must capture true events with high efficiency while rejecting backgrounds to the greatest extent possible. From this, three main challenges emerge:
\begin{enumerate}
\item Determination of the total energy of each candidate event with near-intrinsic resolution. Our target goal is to measure the energy with a resolution better than 1\% FWHM at \qbb. This energy resolution goal can be met using secondary scintillation or electroluminescence (EL).
\item Determination of the complete topology of each event in 3-D, based on energy-sensitive tracking of the \bb\ decay electrons and identification of ``satellite'' deposits of energy. The 3-D localization requires efficient detection of the primary scintillation light to accurately define the start-of-the-event time, $t_0$.
\item Fabrication materials of sufficient radio-purity such that the background rejection capabilities of the HPGXe TPC provide the desired sensitivity. 
\end{enumerate}

\begin{figure}[tbhp!]
\begin{center}
\includegraphics[width=1.0\textwidth]{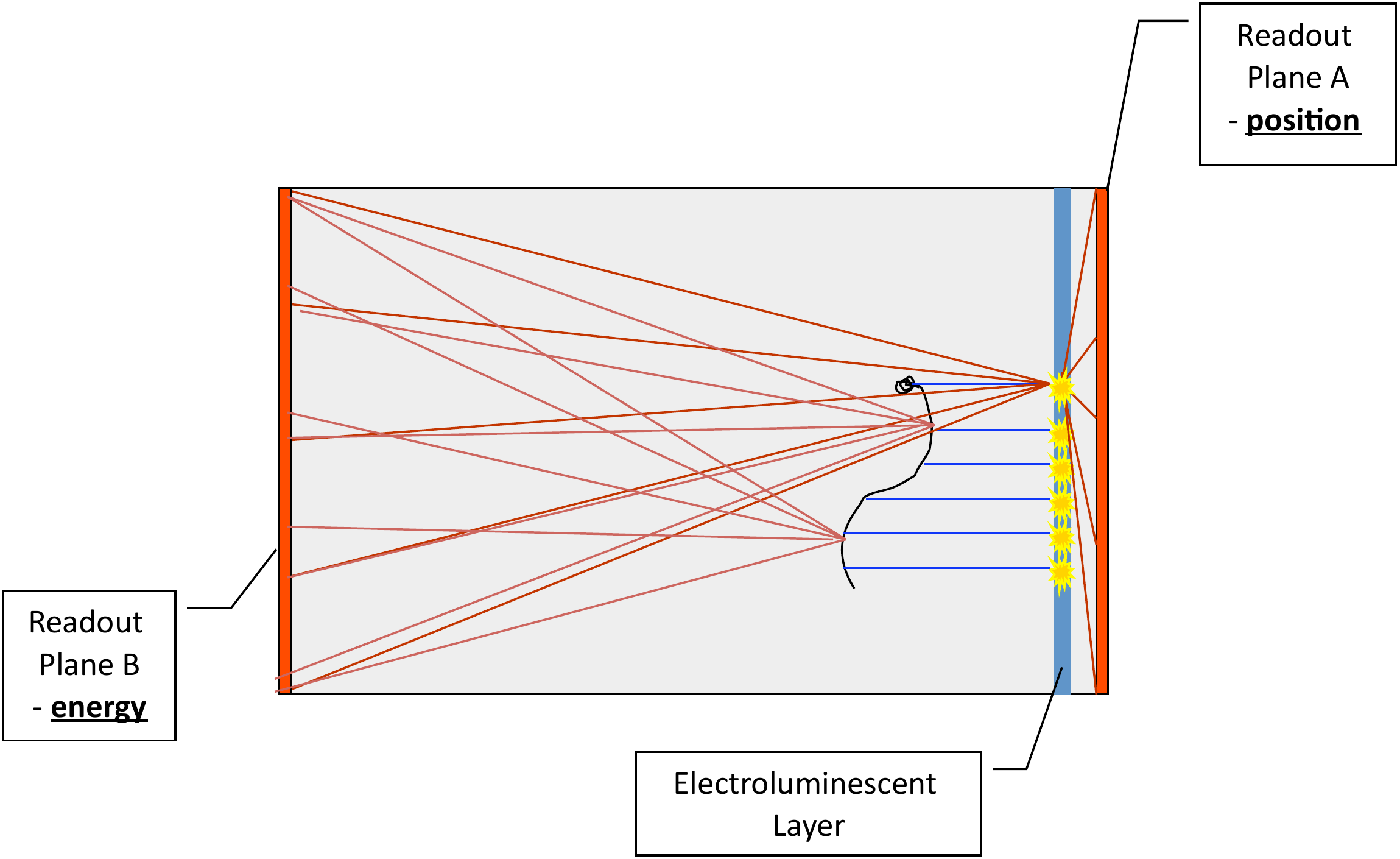} 
\end{center}
\vspace{-0.5cm}
\caption{\small The SOFT concept. EL light generated at the anode is recorded in the photosensor plane right behind it and used for tracking. It is also recorded in the photosensor plane behind the transparent cathode and used for a precise energy measurement.}
\label{fig:stpc-soft} 
\end{figure}

The concept of a Separated Function TPC was proposed by Nygren in \cite{Nygren:2007zz} and extended by Nygren and G\'omez-Cadenas in the NEXT LOI document
\cite{Granena:2009it}, proposing that tracking and energy measurements were carried out by different sensors. 
Figure \ref{fig:stpc-soft} illustrates an asymmetric TPC with Separated Optimized Functions. An event, shown as a wiggly track, generates primary scintillation recorded at both planes (this is called the $S_1$~signal, following the
slang used by the experiments searching for direct detection of Dark Matter). EL light generated at the anode ($S_2$) is recorded in the photosensor plane right behind it and used for tracking. It is also recorded in the photosensor plane behind the transparent cathode and used for a precise energy measurement.

To understand the advantages of the SOFT approach consider first the case in which both the cathode and the anode are instrumented with PMTs. Recording S1 at any point in the TPC requires PMTs optimizes for detecting single photoelectrons (e.g, high gain). Instead, recording the light pattern that defines a track with the anode PMTs requires, in general, lower gain, to avoid saturation (since a lot of light is produced very near the PMTs). Furthermore, the energy of the event can be measure with relatively sparse coverage, while a detailed tracking may require a denser coverage. If the TPC is designed as non SOFT (e.g, a symmetric TPC with the same instrumentation in anode and cathode) it is difficult to reconcile this conflicting requirements. Instead the SOFT paradigm would prescribe to instrument the anode with a dense array of small, low-gain PMTs and the cathode with a sparse array of large, high-gain PMTs. It follows that the separation of functions allows optimizing the TPC performance. The next step may be to decide that the anode PMTs can be replaced by other sensors. In our design SiPMs are chosen on the grounds of low cost, low expected radioactivity and good sensor response.

\begin{figure}[tbhp!]
\begin{center}
\includegraphics[width=1.0\textwidth]{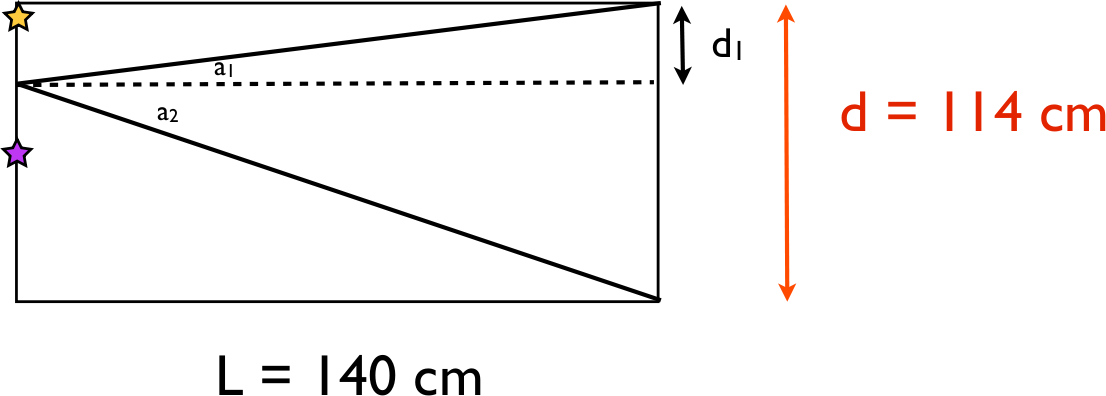} 
\end{center}
\vspace{-0.5cm}
\caption{\small The soft dependence of solid angle in the soft concept.}
\label{fig:solid-soft} 
\end{figure}

An important question is the radial dependence of the light collected in the PMT plane. The notion is illustrated in
Figure \ref{fig:solid-soft}. A photon impinging in the center of the chamber spans a solid angle:
\[
a_1 = a_2 = \arctan(\frac{d/2}{L})\,.
\]
As the track moves along the radius the fraction of the solid angle described by $a_1$~and $a_2$~changes according to:
\begin{eqnarray*}
a_1 &=& \arctan{\frac{d_1}{L}}\\
a_2 &=& \arctan{\frac{d-d_1}{L}} \\
\end{eqnarray*}

Taking $L=140$~cm, $R=114$~cm, one obtains $a_1 = a_2 = 0.387$, $a_1+a_2=0.774$ in the center of the
chamber. Taking a point near the border, $d_1 \sim 5$~cm, $a_1 = 0.036, a_2 = 0.66$, $a_1+a_2=0.757$. 
The solid angle span from the center is 3.7\% of $4\pi$~to be compared with the solid angle span from the corner,
3.0\% of $4\pi$. The ratio between both points is 81\%. 

\begin{figure}[tbhp!]
\begin{center}
\includegraphics[width=0.8\textwidth]{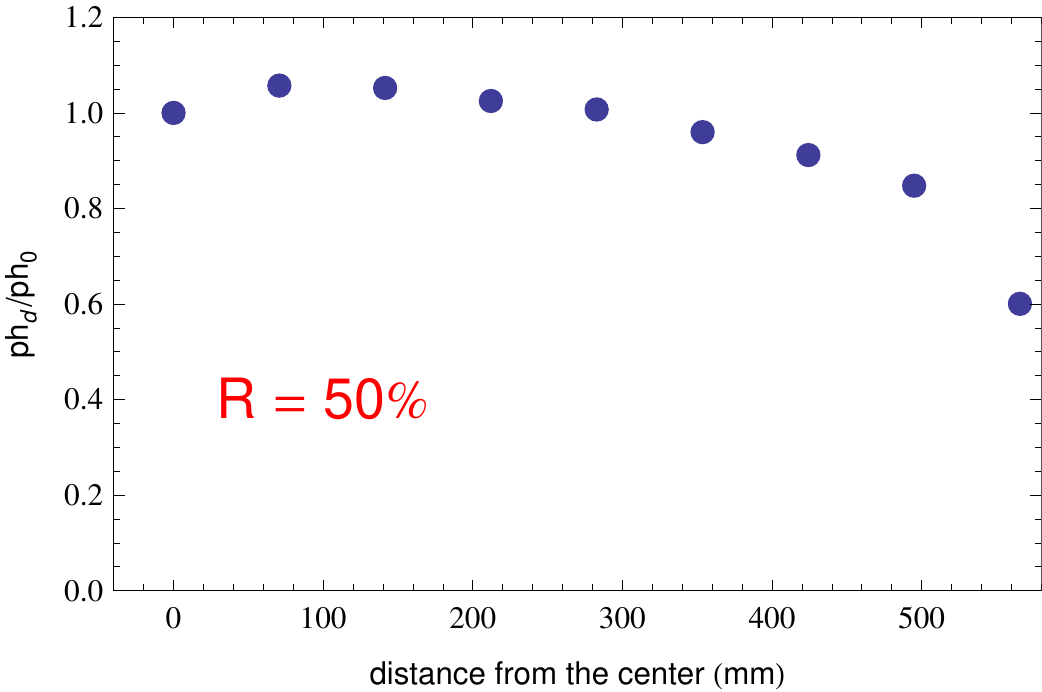} 
\includegraphics[width=0.8\textwidth]{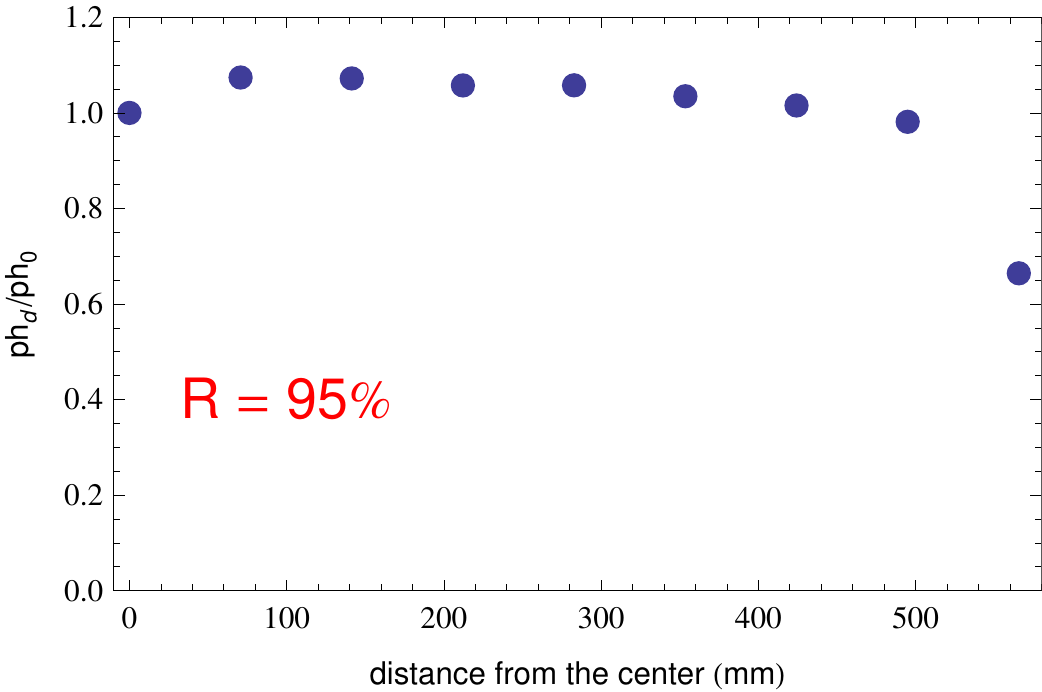} 
\end{center}
\vspace{-0.5cm}
\caption{\small Radial dependence in ANGEL. The points show the fraction of light collected in the cathode from light generated in the anode at a distance $d$ from the  the axis of the chamber (${\rm ph}_d$), relative to the light  collected from light generated on the axis (${\rm ph}_{0}$). Results are shown for a light tube of 50\% reflectivity and a light tube of 95\% reflectivity.}
\label{fig:light-curve} 
\end{figure}

Figure \ref{fig:light-curve} shows the light curves for tubes of reflectivity 50\% and 95\%. As it can be seen, the radial dependence is rather mild even for the poor reflectivity corresponding to the case of uncoated PTFE, and almost flat (except in the edge of the detector) for the target reflectivity of the ANGEL design (see chapter 3). The radial dependence can be fully corrected by mapping the ratio between light generated in the EL grids and light detected in the PMT plane. In addition to light maps generated by Monte Carlo simulation, low-energy photons of energies $\sim$100 keV can be used to obtain the correction from the data themselves. A demonstration of this technique will be exposed when we discuss the NEXT-1 prototypes. 

%
%
\chapter{The ANGEL design} \label{sec.Angel}
Our baseline design for the NEXT-100 detector is the \emph{Asymmetric Neutrino Gas EL Apparatus} (ANGEL), an asymmetric high-pressure gas xenon TPC with separated-optimized functions (SOFT). The energy function is provided by PMTs
located behind a transparent cathode and sealed inside pressure-resistant individual housings. The tracking function is provided by SiPMs (also known as MPPCs) located behind transparent EL grids and coated with TPB. The fiducial mass is chosen to be 100 kg, and the operative pressure 15 bar. 

ANGEL has been chosen as our baseline following the principle of designing the simplest TPC that can result in a competitive physics program in the shortest possible time. As a consequence, we have opted for solutions that require little or no additional R\&D on top of the  program already under way with the NEXT-1 prototypes. 

Since the number of parameters that define the detector is large, it is useful to describe the design examining the detector from different angles. This are: the choices relatives to the target; those related with the
 pressure vessel; those related with the field cage, high-voltage and EL grids; the energy plane; and the tracking plane.  

%

\section{Source mass} 
The source mass of the NEXT experiment is gas xenon enriched in the \XE\ isotope. In the ANGEL design the xenon is also used as electric insulator. As such, an annulus of 4 cm radius around the field cage is needed to insulate the high voltage. This choice has the drawback that a significant amount of valuable enriched xenon is wasted. It could be avoided by enclosing the fiducial volume inside a xenon vessel, and using an additional, cheaper gas (such as nitrogen) as electric buffer. However, the use of such a vessel implies a considerable complication from the engineering point of view. The first problem is the choice of the construction material for the vessel, which must be non-conducting and radiopure. Kevlar or acrylic glass are among the possibilities. The latter is more radiopure, but xenon diffusion into the buffer gas may be an issue. The enriched xenon and the buffer gas must be kept at the same pressure (in fact the buffer gas must track the pressure of the xenon, to avoid a transient that could break the xenon vessel). Two gas systems are necessary. Safety systems must be duplicated. If the optical sensors (PMTs and SiPMs) are to be placed outside the xenon vessel one needs transparent interfaces (acrylic coated with TPB could be used here). 

While all the above is feasible, it requires considerable R\&D and increases both the cost of the detector, the time needed to build it, and the uncertainties on rapid commissioning. Consequently we have opted for the simplest system for the baseline, assuming a fiducial mass of xenon of nearly 100 kg, for a total mass of almost 120 kg (see Table \ref{tab.AngelTarget}). The nominal operative pressure is 15 bar and all systems will be designed accordingly. Nevertheless, if we need to operate in the initial run with a total of 100 kg (about 70 kg in the fiducial region), the operating pressure will be reduced accordingly to about 12.5 bar. 

The LSC has procured already a 100 kg of xenon enriched at $\sim90\%$ in the isotope \XE\ from russian suppliers, taking advantage of the coordination provided by JINR (a group of this laboratory is part of NEXT). The purchase conditions are extremely favorable at present, and could be kept if a further order is placed in a short period. The NEXT collaboration will actively seek for additional funds to purchase as much enriched xenon as possible in as short term. Ultimately a large mass (of the order of 1 ton) is necessary to compete with the other xenon-based experiments, EXO and KamLAND-Zen.

\begin{table}[tb!]
\begin{center}
\begin{tabular}{lD{.}{.}{-1}}
\hline \hline
Fiducial Radius (cm) &  53.0 \\
Fiducial Length (cm) &  130.0 \\
Total Chamber Radius (cm) & 57.0 \\
Total Chamber Length (cm) & 135.0 \\
Fiducial Volume (m$^{3}$) & 1.15 \\
Fiducial Source Mass (kg) & 99.14 \\
Total Source Mass (kg) & 119.0 \\
\hline\hline
\end{tabular}
\end{center}
\caption{Source mass in the ANGEL design.} \label{tab.AngelTarget}
\end{table}%

\section{The energy plane}

In ANGEL the energy measurement will be provided by the detection of EL light via PMTs, which will also record the scintillation light needed for $t_0$. Those PMTs will be located behind a transparent cathode.

\begin{figure}[p]
\begin{center}
\includegraphics[width=0.75\textwidth]{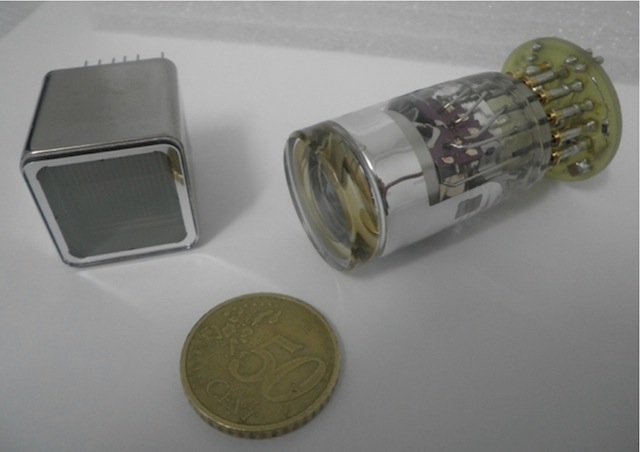}		
\end{center}
\caption{On the left, the Hamamatsu R8520-406 PMT. This is a radiopure PMT, sensitive to VUV, that can take up to 5 bar pressure. On the right, the PMT used in the NEXT-1 prototypes, model R7378A. This phototube is also sensitive to VUV and can resist pressure up to 20 bar, but is not radiopure.}
\label{fig:PMT}
\end{figure}

\begin{figure}[p]
\centering
\includegraphics[width=0.85\textwidth]{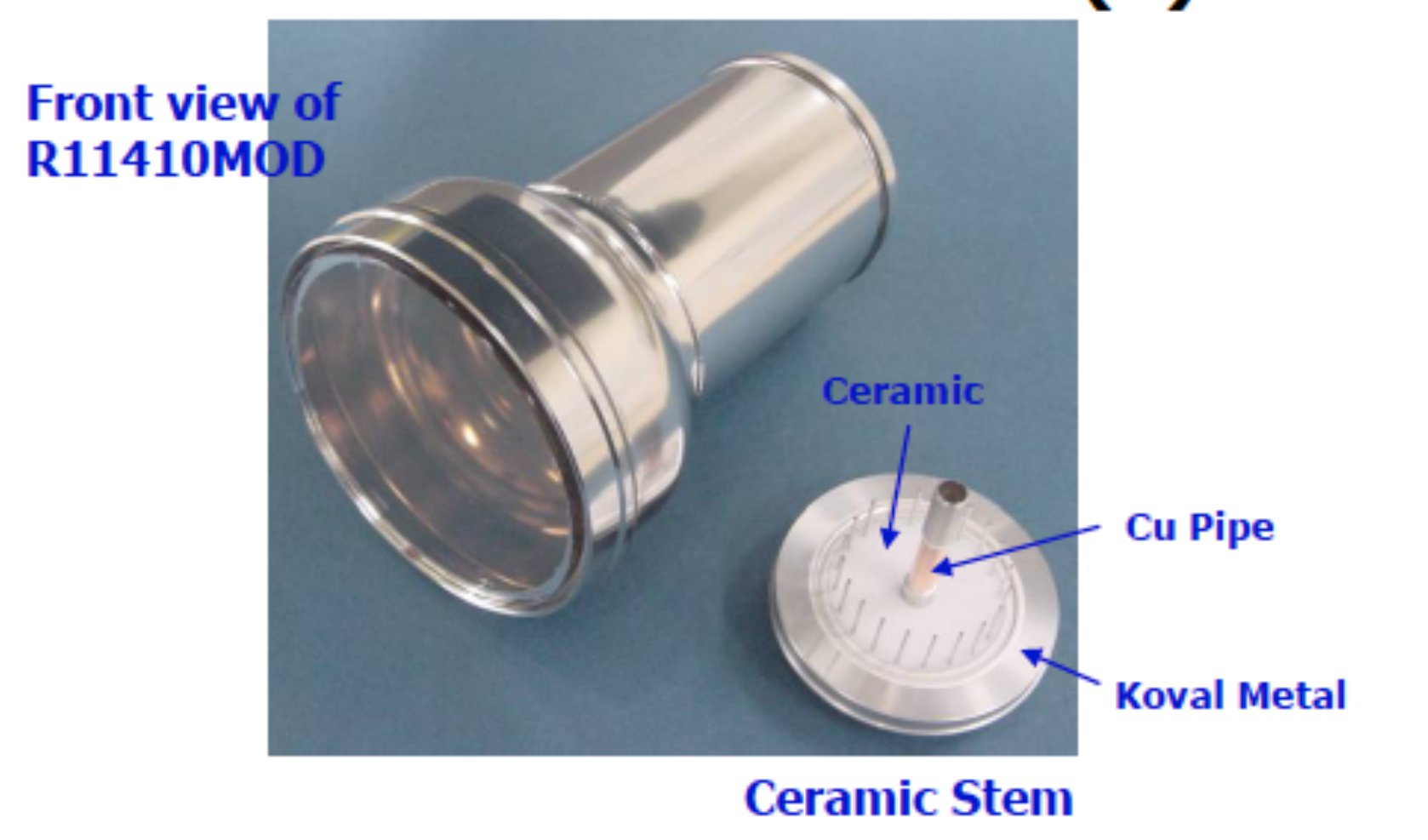}		
\caption{The Hamamatsu R11410MOD phototube. This is a large PMT, 3'' in diameter, with an average radioactivity of 3 mBq for the U chain and 2 mBq for the Th chain.}
\label{fig.R11}
\end{figure}

\begin{figure}[p]
\centering
\includegraphics[width=0.95\textwidth]{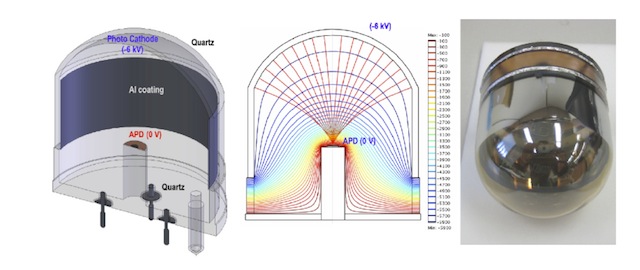}		
\caption{Left and center: principle of operation of the QUPID. Right: a picture of the actual device. Each QUPID has in
average a radioactivity of 0.5 mBq for the U chain and 0.5 mBq for the Th chain.}
\label{fig.QUPID}
\end{figure}

The PMTs used for the NEXT-1 prototypes are Hamamatsu R7378A. This small sensors (1 inch diameter) are sensitive to the VUV emitted by xenon and can resist pressure up to 20 bar. Unfortunately, they are quite radioactive, about 50 mBq per unit of the uranium and thorium chains. Hamamatsu has another small PMT (1'') which is both radiopure (0.5 mBq per unit) and sensitive to xenon VUV light, the R8520-406. This PMT, square in shape, can take up to 5 bar. Both PMT models can be seen in Figure \ref{fig:PMT}. The bigger (3-inch diameter) R1141MOD from Hamamatsu (Figure \ref{fig.R11}) has levels of \BI\ and \TL\ activity per unit of area smaller than that of the R8520. It has been developed for use in cryogenic noble liquid detectors for dark matter searches and cannot resist high pressure. A more ambitious possibility ---suitable perhaps for an upgrade of NEXT--- would be the QUPID (Figure \ref{fig.QUPID}), which features high QE (33\%) and even smaller background, 0.5 mBq for the U chain and 0.5 mBq for the Th chain.

\subsection{PMTs and pressure}
The R8520 cannot be used at our baseline pressure of 15 bar, and is probably risky to use it at its nominal pressure of 5 bar, due to the possibility of failure after several cycles of vacuum-pressure (we have observed this phenomenon in the R7378A, in spite of the fact that the PMT has not operated at a pressure higher than 10 
bar). 

\begin{figure}[p]
\begin{center}
\includegraphics[width=0.6\textwidth]{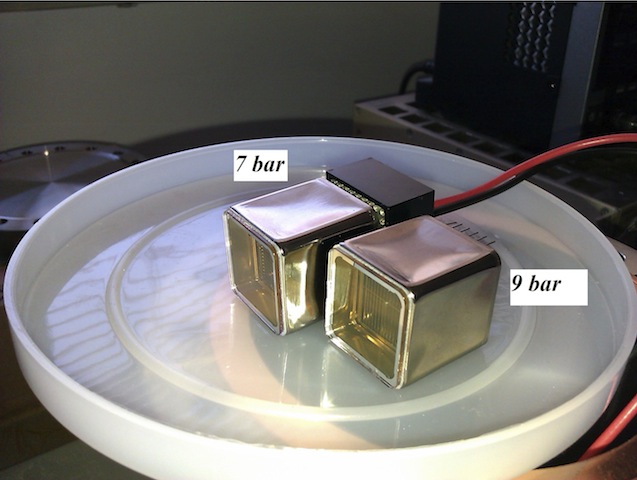}
\end{center}
\caption{PMTs after the pressure resistant tests. The R8520 starts to shrink at 7 bar. The deformation becomes large at 9 bar. After few hours at this pressure the PMTs stop working.} \label{fig.PMTunder}
\end{figure}

The collaboration has studied a number of PMTs reinforced by Hamamatsu. These samples were placed under high pressure in argon, and single photoelectron data was taken. The tests show that the metal body shrinks for pressures above 6 bar. At 7 bar the effect becomes visually apparent, and at 9 bar the shrinking is such that the side surface moves more than 1mm inside the PMT volume (see Figure \ref{fig.PMTunder}). After being few hours at 9 bar, the PMT stops functioning. It appears as if the PMT sealing is still good, but the shrinking causes a short circuit inside the device. Further tests will be conducted during the next few months. However it appears unlikely that the R8520 can operate reliably at high pressure during long periods.

\begin{figure}[p]
\centering
\includegraphics[width=0.65\textwidth]{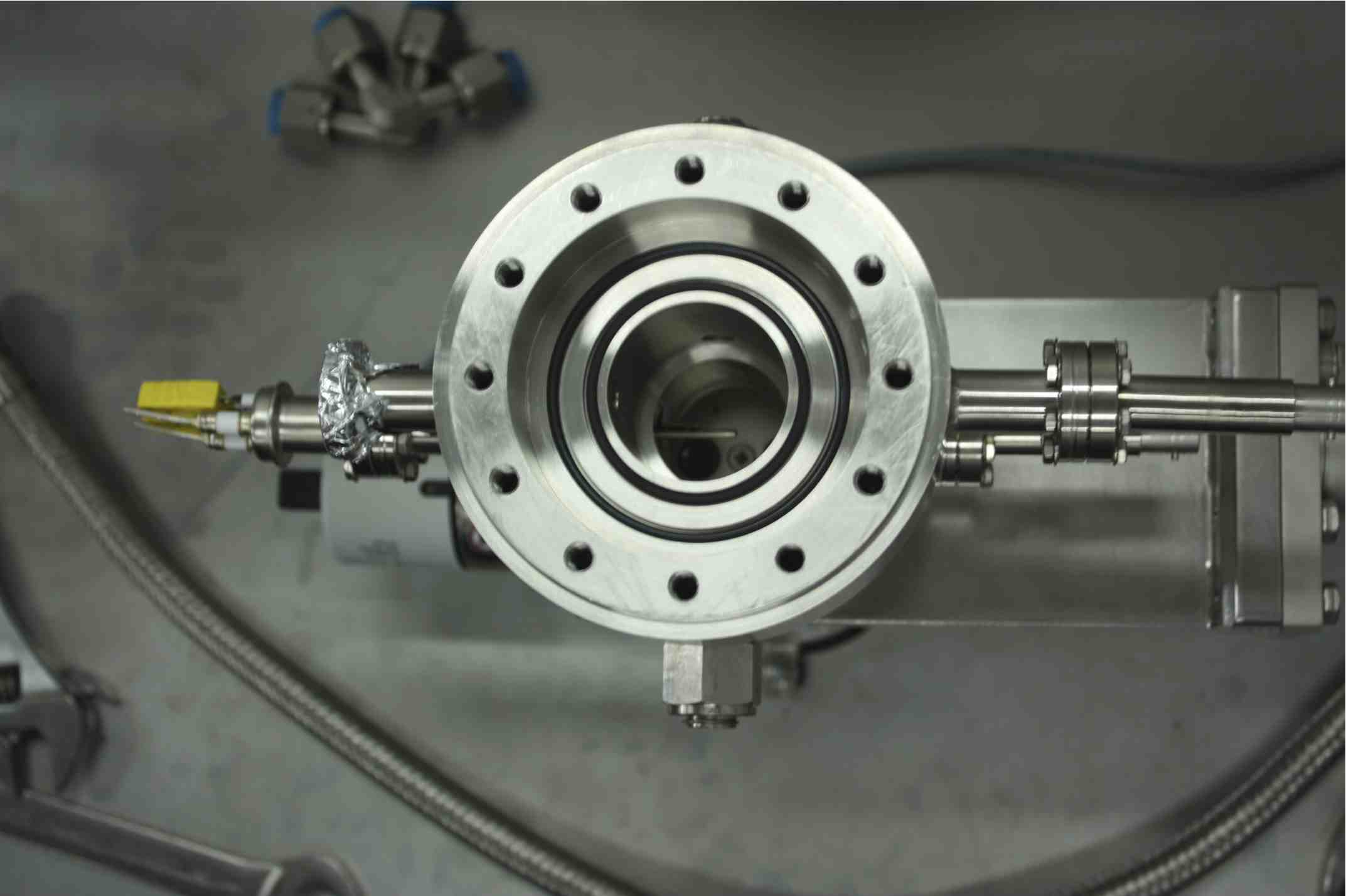}		
\caption{The NEXT-0 detector at IFIC has tested the concept of sealing quartz windows to separate the pressure atmosphere from the PMT.} \label{fig.NEXT0}
\end{figure}

If the PMTs cannot be directly operated under pressure, the obvious solution is to leave the devices outside of the pressurized volume, viewing the chamber through a transparent, pressure-resistant window. This concept was tested in the IFIC's NEXT-0 detector (Figure \ref{fig.NEXT0}), where a quartz (fused silica) window, 15 mm thick, that can resist pressure up to 15 bar, seals the fiducial volume. Clear EL signals were observed in a R8520 PMT optically coupled to the window.

Sapphire is a better material than quartz to build large windows (it has a 10 times higher tensile strength). While UV-grade sapphire is available, it is also possible to coat the windows with TPB, that will shift the VUV light to blue.

With the PMTs  protected in their houses it is possible to use either the R8520 or the R11410MOD. In the first case, one would
house 4 small PMTs in each individual housing. At present we have not yet made a final decision about which type of PMT to use although it appears likely that the first run uses the R8520, which appears to be more readily available than the R11410MOD.

\begin{figure}[p]
\centering
\includegraphics[width=0.8\textwidth]{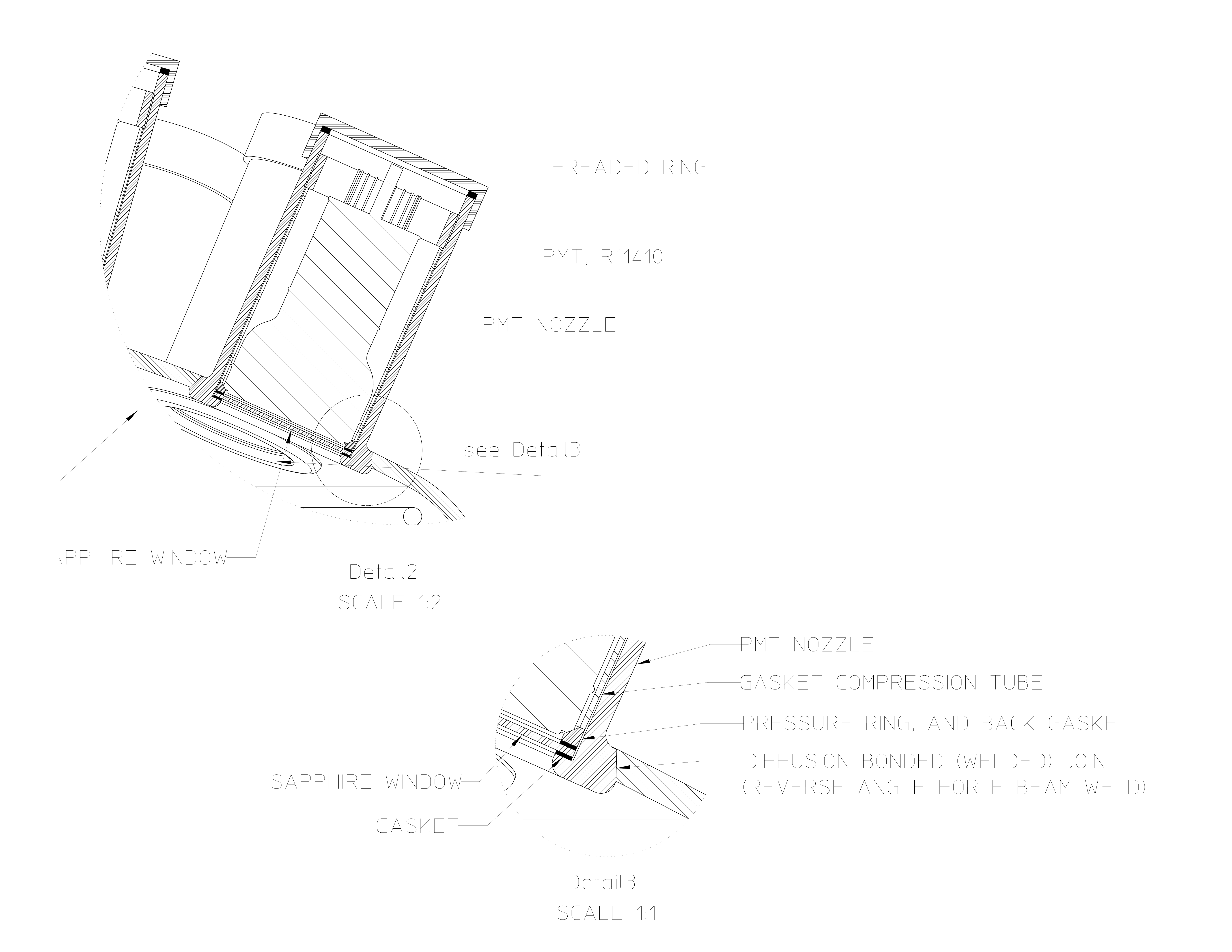}		
\caption{A detail of the housing protecting the PMTs in ANGEL.}
\label{fig.housings}
\end{figure}

\begin{figure}[t!b!]
\centering
\includegraphics[width=0.8\textwidth]{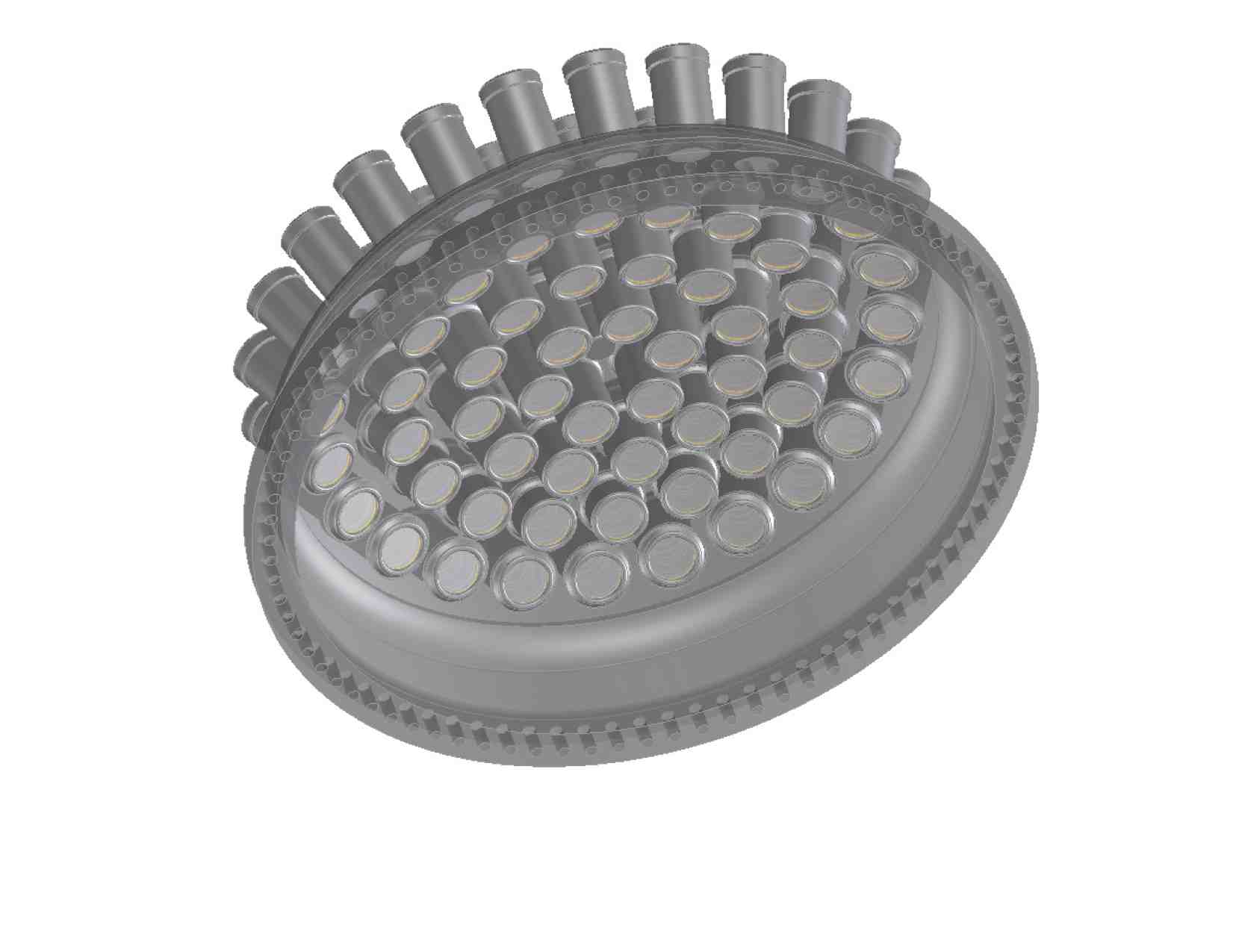}		
\caption{The ANGEL torispheric head, showing housing for  60 PMTs (30\% coverage).}
\label{fig.housings2}
\end{figure}


\subsection{How many PMTs?}

To answer this question let's first consider the detection of scintillation light. The number of photons that arrive to the PMT housing windows depend of the properties of the reflector as well as the transparency of the EL grids. Our simulation shows that a light tube of 50\% reflectivity (which could be made of uncoated PTFE) transfer 3\% of the photons produced in the EL grids to the cathode. A light tube of 90\% reflectivity (made of PTFE coated with TPB) will transfer 9\% of the photons.

Assume now that an event is produced near the EL grids (the worst scenario for the detection of primary scintillation light with the
cathode PMTs). Recall from Eq.~(\ref{eq:Ns}) that 13158 scintillation photons are produced per MeV.  Then the number of photoelectrons (pes) detected by the PMTs at the cathode is:
\[
(13158\ \mathrm{photons/MeV}) \times C \times T_R \times T_W \times QE
\]
where $C$~is the cathode coverage, $T_R$~is the reflector transfer function, $T_W$~is the housing window transfer function and $QE$~is the PMT quantum efficiency. Monte Carlo simulation yields $T_W=0.75$~for a sapphire  window coated with TPB. Then, setting $C=0.15, R=0.03$~(R=50\%) and $QE=0.3$~we obtain:
\[
(13158\ \mathrm{photons/MeV}) \times 0.15 \times 0.03 \times 0.75 \times 0.3 \sim 13~(pes/MeV)
\]
While for a light tube of 90\% reflectivity we have a factor 3 more, $\sim 40$~pes/MeV. Thus a reflector of $R=0.9$~allows to detect S1 in the full chamber range up to energies of about 100 keV. 
 This is important, not only to study the lower part of the
\bbtnu\ spectrum, but also to trigger in low energy gammas sources for detector calibration.

Consider now EL light. According to Eq.~\ref{eq:npe}, we need at least 10 pes per primary electron to optimize resolution. ANGEL optical gain is near $3 \times 10^3$. The number of pes per electron for 15\% coverage and 50\% reflector is:
\[
3 \times 10^3\ \mathrm{(photons/electron)}\ 0.15\times 0.03 \times 0.75 \times 0.3 \sim 3\ \mathrm{(pes/electron)}
\]
Here one can immediately see that, in order to fully exploit the potential of the PMTs we need a extra factor of 3 that can come either from a 90\% reflector or doubling the coverage to some 30\%.

\section{The tracking plane}

In ANGEL the tracking function is provided by a plane of photo-sensors operating as a light-pixels and located behind the transparent EL grids. In addition to position information tracking pixels need to provide a rough measurement of the energy (since we are interested in measuring the energy per unit length of the track), but they can have much less resolution than PMTs. In exchange, they must be smaller, since physics dictates a pitch of around 1 cm (while the PMTs have a diameter of 7.5 cm). They also must have much less radioactivity and cost per unit, since they are needed in large numbers (5,000 to 10,000 depending on the pitch).

\subsection{MPPCs}

\begin{figure}[tbhp!]
\begin{center}
\begin{tabular}{c}
\includegraphics[width=0.4\textwidth]{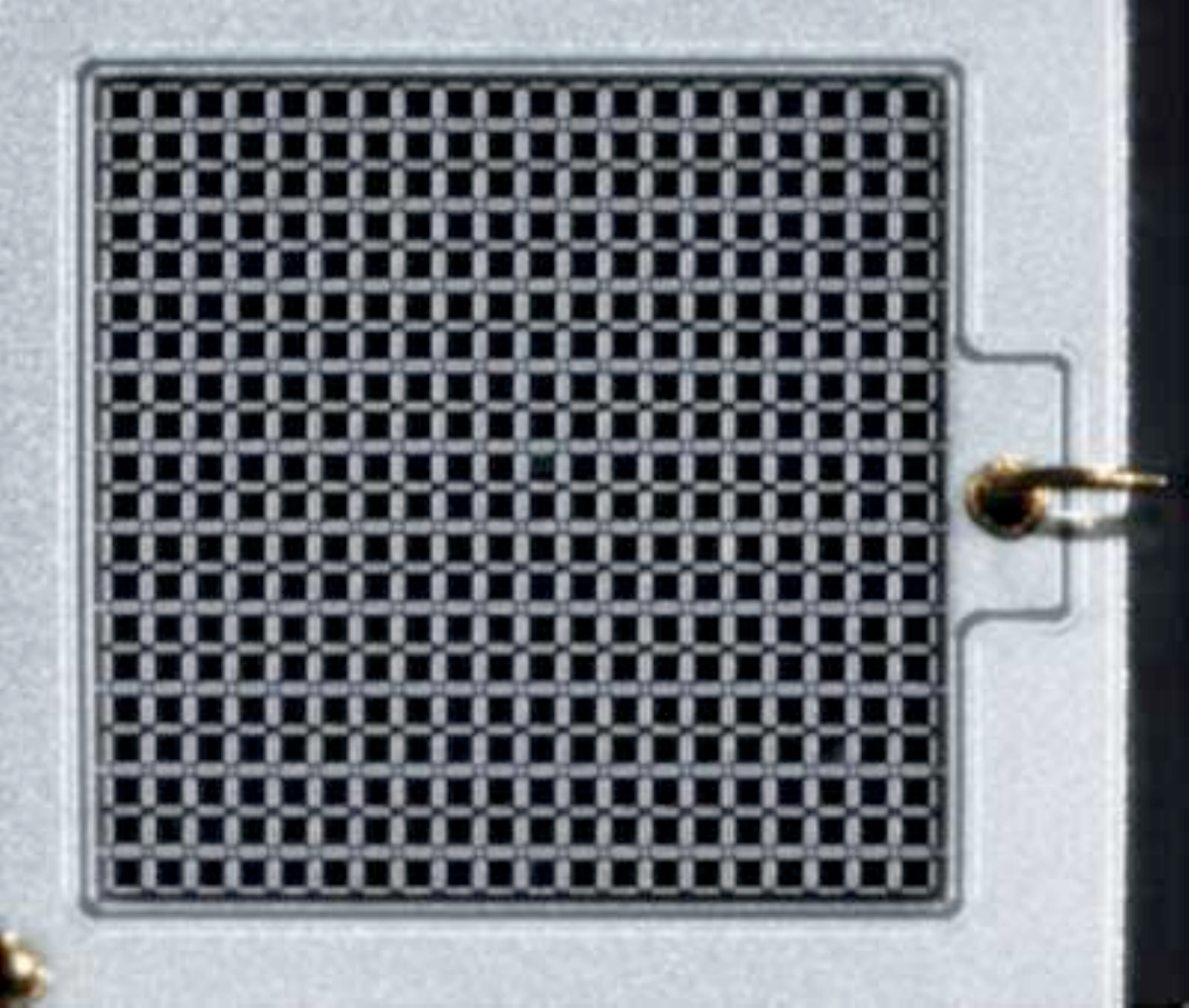} 
\includegraphics[width=0.4\textwidth]{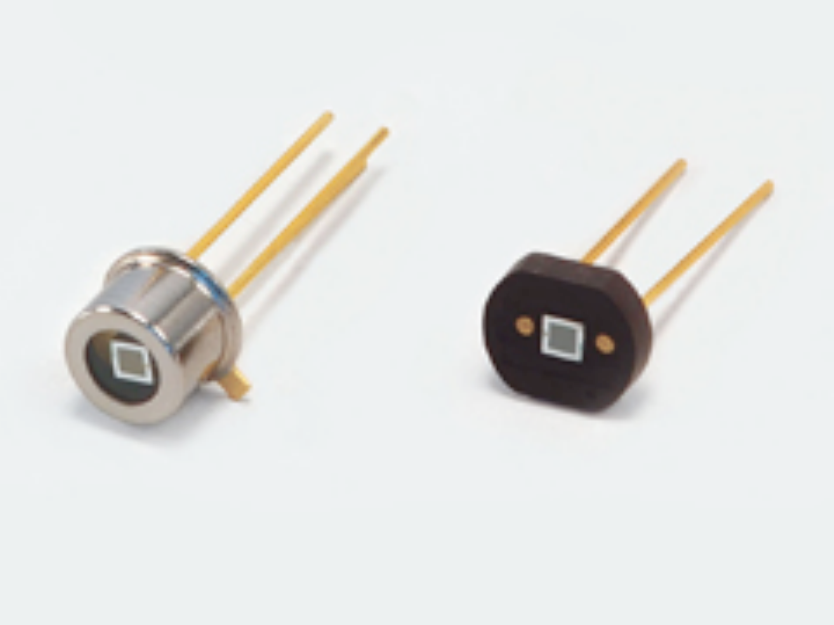}
\end{tabular}
\end{center}
\vspace{-0.75cm}
\caption{\small Left: detail of a multi-pixel photon counter (MPPC), showing the multiple APD pixels composing the MPPC. Right: two MPPC detectors (1 mm$^2$) from Hamamatsu Photonics.}
\label{fig:mpcc} 
\end{figure}

\begin{figure}[h]
\centering
\includegraphics[width=0.95\textwidth]{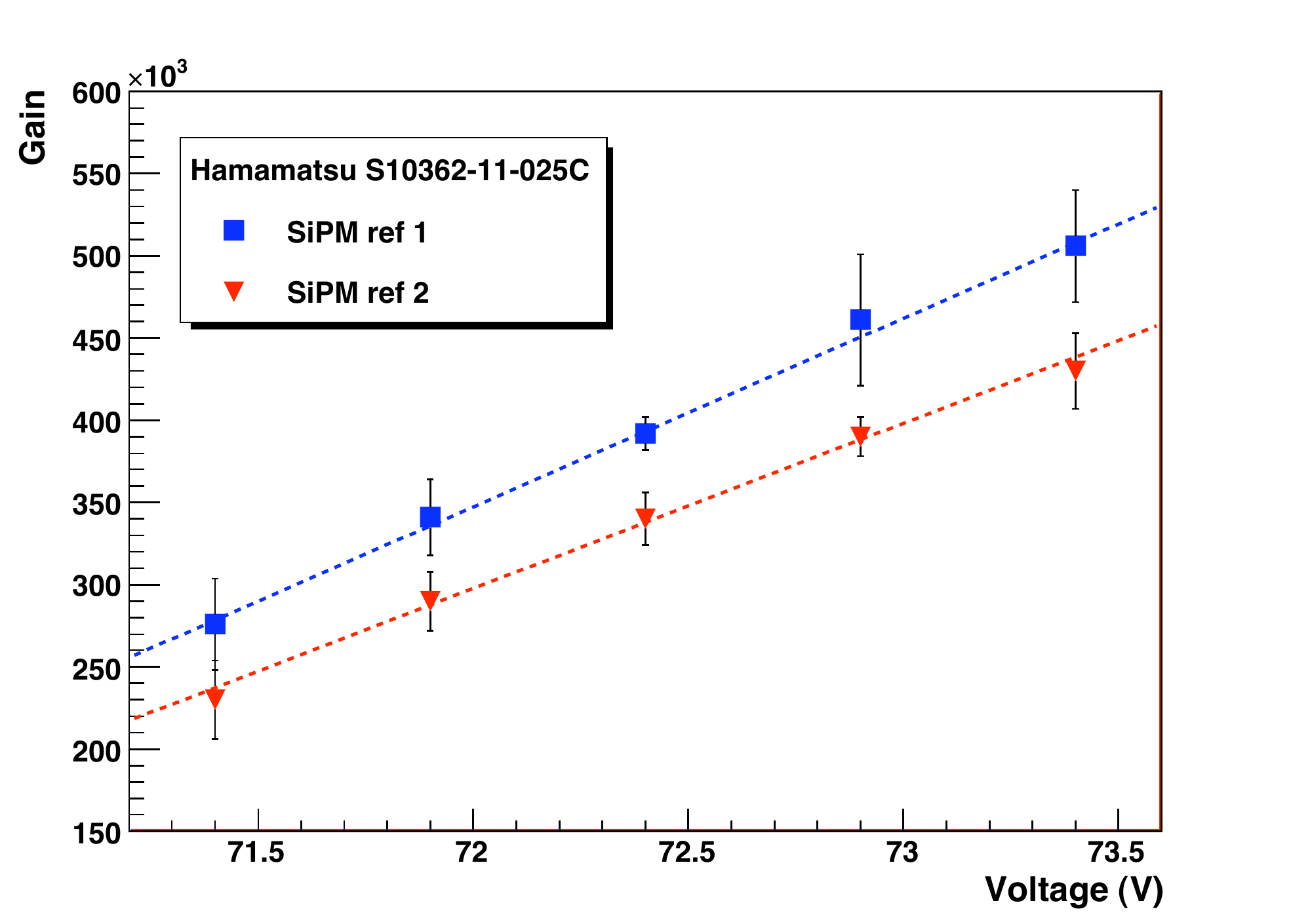}		
\caption{Gain of two SiPMs Hamamatsu S10362-11-025C measured as a function of the operating voltage.}
\label{Fig: Gain_vs_HV}
\end{figure}

%
The best option for light pixels at present appears to be the so-called Silicon Photomultipliers (SiPMs), also called
Multi-Pixel Photon Counter (MPPCs) by Hamamatsu (Figure \ref{fig:mpcc}). 
MPCCs are cheap when delivered in large quantities (about 10 euro per unit), have large gain (close to $10^6$), Figure \ref{Fig: Gain_vs_HV}) and  very low levels of radioactivity (a MPPC is a 1 mm$^2$ piece of silicon). 

\subsection{Coating MPPCs with TPB}

\begin{figure}[tbhp!]
\centering
\includegraphics[width=0.85\textwidth]{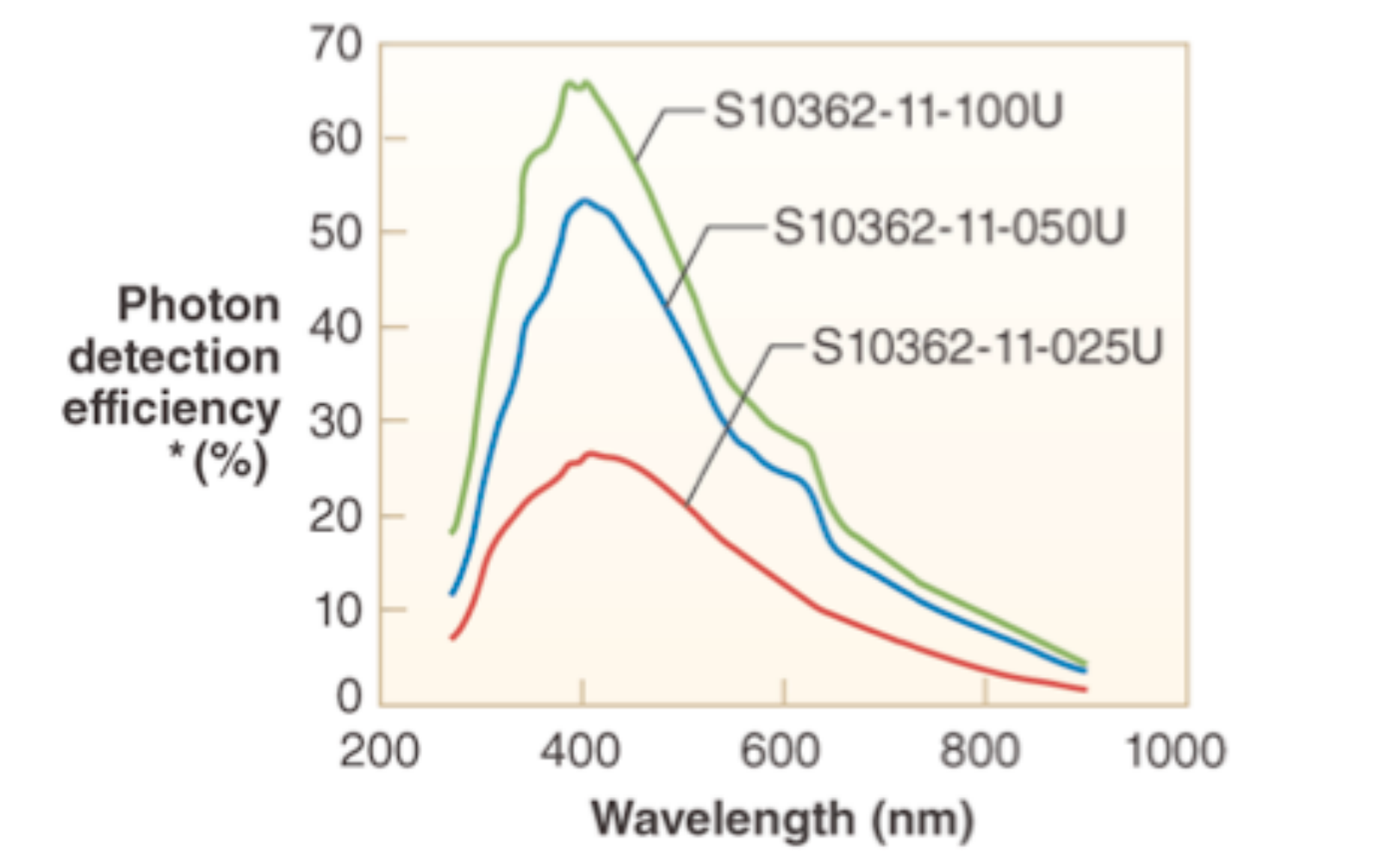}		
\caption{The Photon Detection Efficiency as a function of the wavelength of the incident light for the two SiPM models considered in NEXT.}
\label{Fig:PDE}
\end{figure}

\begin{figure}[tbhp!]
\centering
\includegraphics[width= 6.5cm]{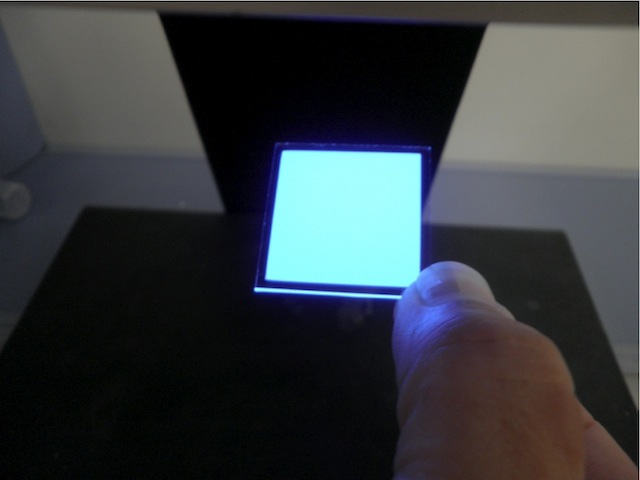}	
\includegraphics[width= 6.cm]{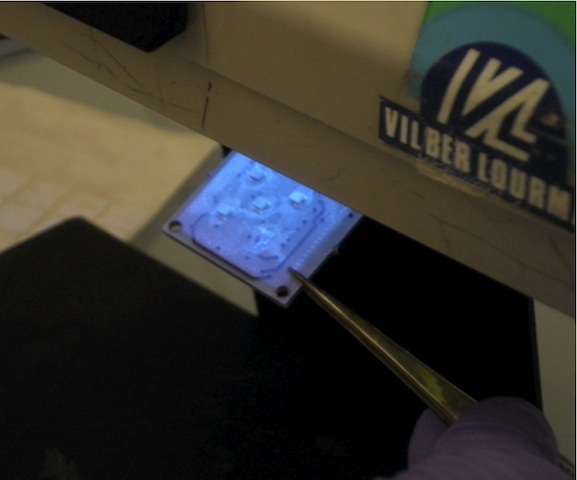}		
\caption{Illumination with  UV light  of a glass-slice (left) and a 5-SiPM board (right) both coated with TPB.}
\label{Fig:UV_illumination}
\end{figure}

\begin{figure}[h]
\centering
\includegraphics[width=0.95\textwidth]{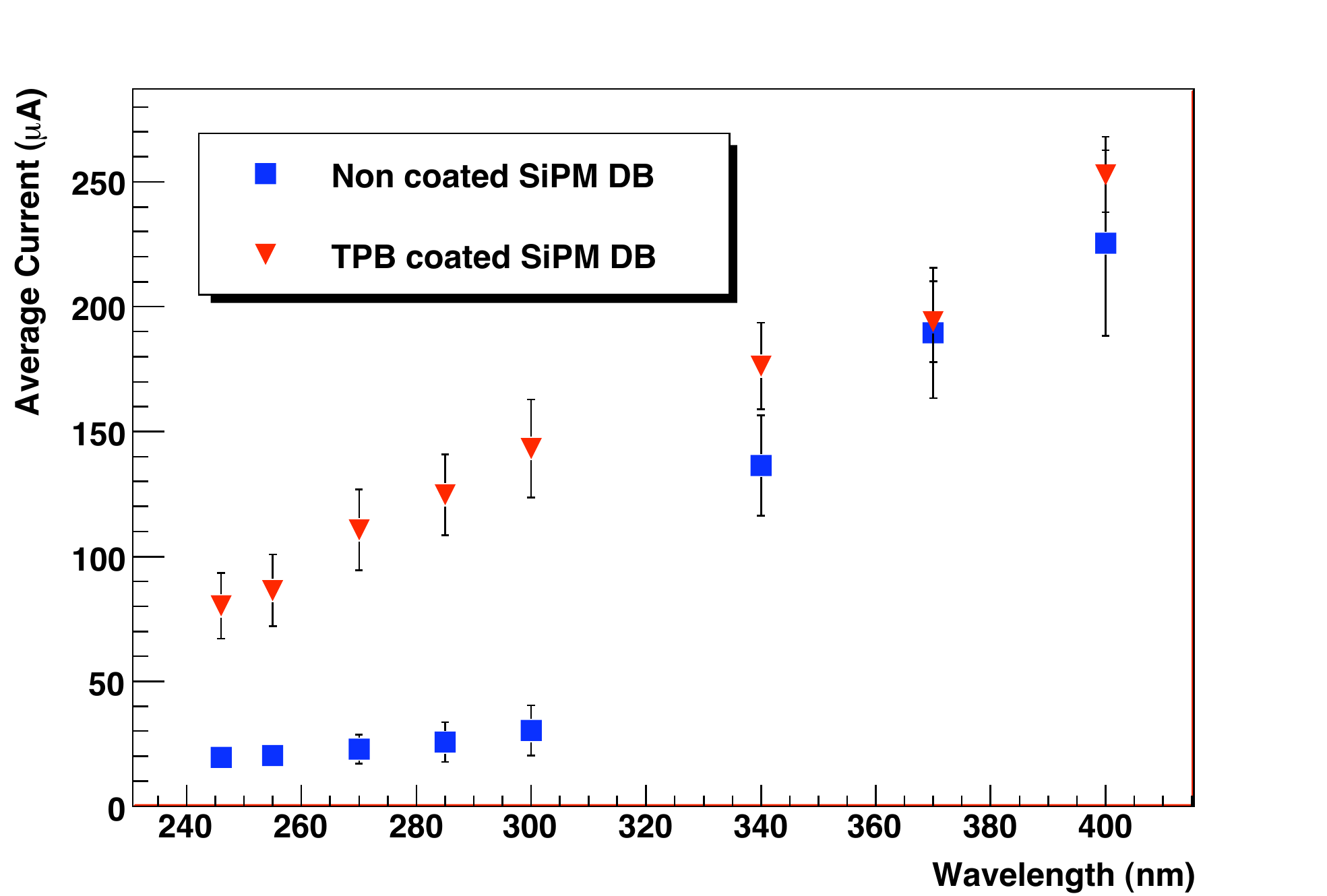}			
\caption{Average current in the TPB coated SiPMs DB compared to the average current of a non coated SIPM DB.  Both SiPM DB are illuminated by the same Xenon Lamp coupled to a Monochromator to select the input wavelength.}
\label{Fig:Xenon_lamptest}
\end{figure}

The main drawback of SiPMs is that they are not sensitive to VUV light (their particle detection efficiency, PDE, peaks near the blue, as shown in Figure \ref{Fig:PDE}) and therefore it is necessary to coat them with a WLS, such as terphenyl-butadiene (TPB) to shift the light to blue, where SiPMs are most sensitive. The procedures is illustrated in Figure \ref{Fig:UV_illumination}, which shows how a TPB-coated glass (and a TPB coated SiPM board) glows with blue light (emitted by the TPB) when illuminated with VUV light. Figure \ref{Fig:Xenon_lamptest} shows the response
of TPB-coated SIPMs compared with that on non-coated devices. Notice that the response of the uncoated SiPMs at low wavelengths is about a factor 10 lower than that in the blue region. With coating, however, the response increases by a factor 3, making light detection with these detectors feasible. 

%

\subsection{Implementation of the tracking plane}

%
%

\begin{figure}[htbp!]
\centering
\includegraphics[width= 6cm]{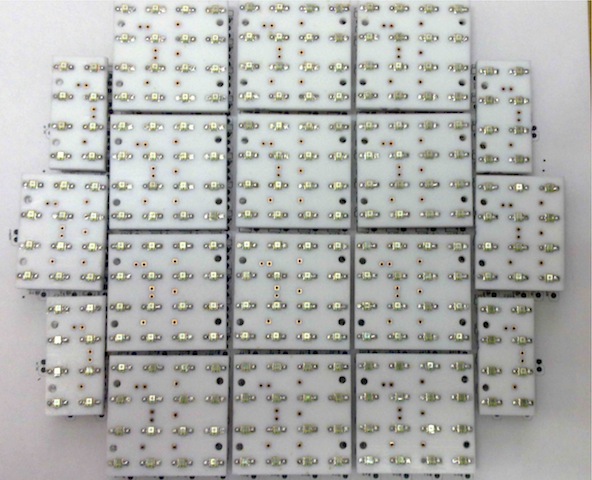}	
\includegraphics[width= 6cm]{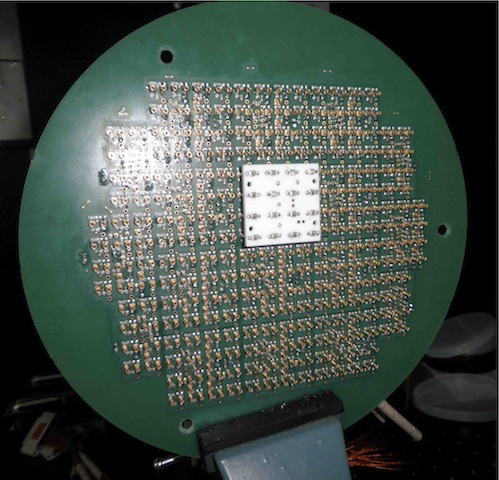}	
\caption{Left:  SiPM Daughter-Boards containing each $4\times4$ SiPMs mounted on a Cuflon support for the central ones and $2\times4$ or $3\times4$ SiPMs for the ones on the external edges of the plane. Right: Mother-Board with the front-end electronics on which center one of the Daughter-Boards is connected}
\label{Fig:Tracking_plane}
\end{figure}

Figure \ref{Fig:Tracking_plane} shows the implementation of the tracking plane
in the NEXT-1-IFIC prototype. 
The SiPMs are mounted in daughter boards (DB), made of cuflon, (PTFE fixed to a copper back plane). Most DB mounts 16 SiPMs. The DB are connected to a Mother Board (MB) that distributes signals and power to the SiPMs.
Each DB is coated with TPB in a facility available at ICMOL, an institute near IFIC before installation. A total of 248 SiPM are mounted in the MB, located 2 mm behind the anode grid.

The ANGEL design will be a larger version of the NEXT-1 tracking plane. We plan to build DB holding 64 SiPMs each in an array of 8$\times$8. About 160 such DB (assuming a pitch of 1 cm and 10,000 SiPMs) will need to be mounted and coated with TPB. Each DB will have its own biasing and the signals from the SiPMs will be amplified and serialized inside the chamber, before being dispatched via optical link (see chapter 7).


\subsubsection{Tracking with MPPCs}

SiPMs generate nearly uniform single photoelectron (spe) pulses. However, the uniformity of the SiPMs noise pulses and the large intensities of tracking signals allows to set a digital threshold (at about 5 p.e)  high enough to eliminate almost all noise, without degrading the spatial resolution.  

Each primary electron entering the meshes produces EL light for a time interval given by the gap size divided by the drift velocity.  For E/p $\sim$ 3.5 kV/cm bar, this time interval is about 3 $\mu$s. Typically, 600--1200 electrons contribute to the track imaging at any moment. Tracks less parallel to the TPC axis contribute the higher number of electrons within the EL gap. With so many primary electrons per mm, the statistical contribution to spatial resolution is $\sim$1 mm rms, even for the maximum possible diffusion within the chamber.  

With an EL gain of around 3000 and a track population of $\sim$1000 electrons within the EL meshes, the total EL luminosity is in the range of  $3 \times 10^6$ photons per $\mu$s (counting only the forward-going photons). A detection element of 1 mm$^2$ at a distance of 5  mm from the luminous region will subtend a solid angle fraction of $\sim$0.003.  Hence, about 3,000 photons per $\mu$s will impinge on such a 1 mm$^2$ detection area. From the measured response of the coated SiPMs (Figure \ref{Fig:Xenon_lamptest}) we take a transmittance of 30\% at 170 nm relative to the peak transmittance in the blue region. The PDE in the blue is about 50\%. Therefore one expects to record about 500 p.e.\ per $\mu$s, plenty for good tracking.  

\begin{figure}[tbhp!]
\begin{center}
\includegraphics[width=0.8\textwidth]{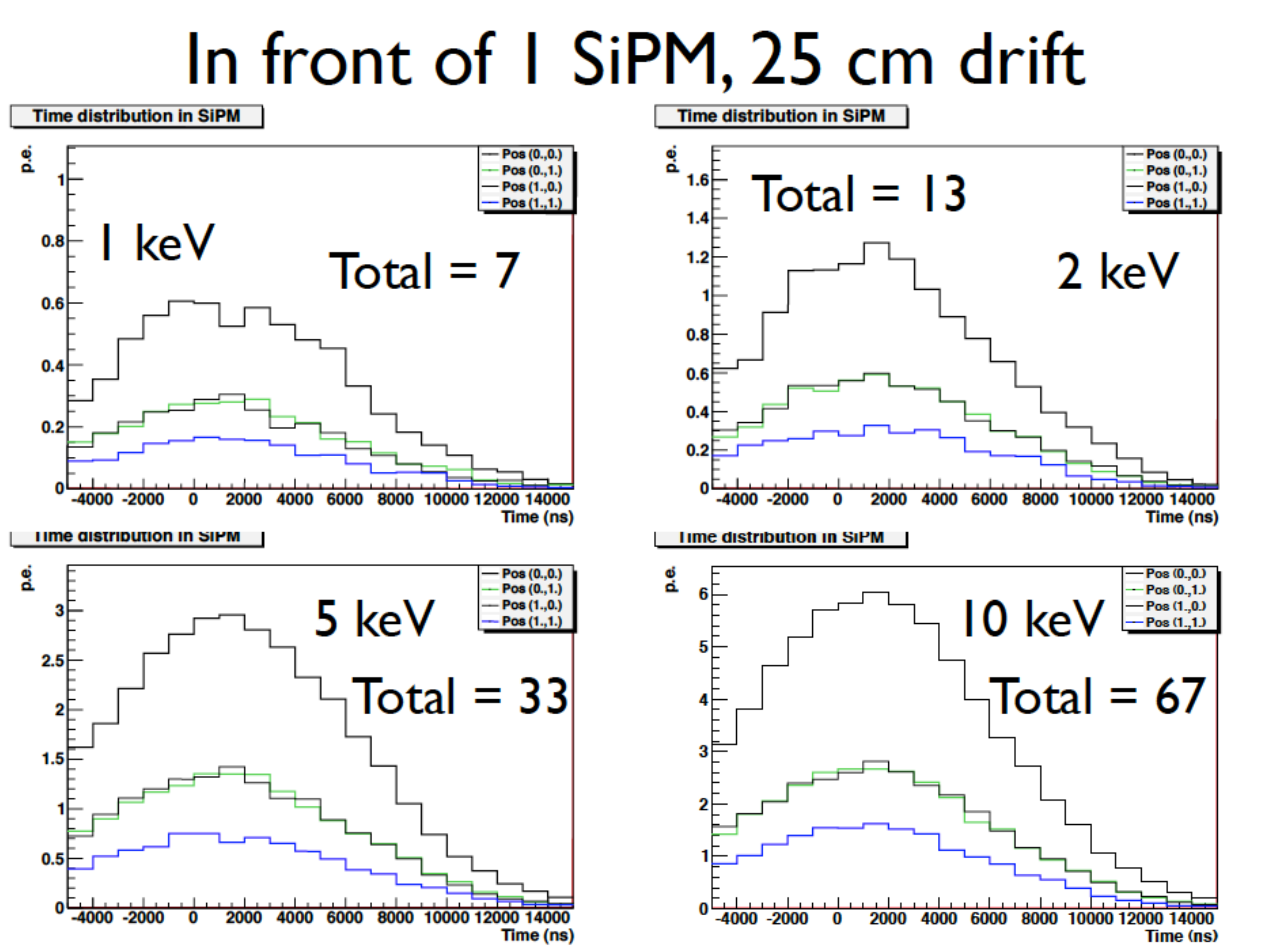} 
\end{center}
\vspace{-0.5cm}
\caption{\small Light received in an array of 4 SiPMs (labeled (0,0), (0,1), (1,0) and (1,1)) when gammas of
various energies produce light in an EL region of 5 mm at E/P of 4. In each case the signal is shown for the four
SiPMs of the array as a function of time. A suitable cut to get rid of the dark current in a SiPM is 3--4 p.e., thus, 
with a count of 7 p.e, for the 1 keV case, the SiPMs are sensitive to small energy deposits.  }
\label{fig:eltrk22} 
\end{figure}

In a more quantitative way, Figure \ref{fig:eltrk22} shows the Monte Carlo simulation of the light received in an array of 4 SiPMs (labeled (0,0), (0,1), (1,0) and (1,1)) when gammas of various energies produce light in an EL region of 5 mm at $E/p$ of 4. In each case the signal is shown for the four SiPMs of the array as a function of time. A suitable cut to get rid of the dark current in a SiPM is 3--4 p.e., thus,  with a count of 7 p.e, for the 1 keV case, the SiPMs are sensitive to small energy deposits.

\begin{figure}[tbhp!]
\begin{center}
\includegraphics[width=0.8\textwidth]{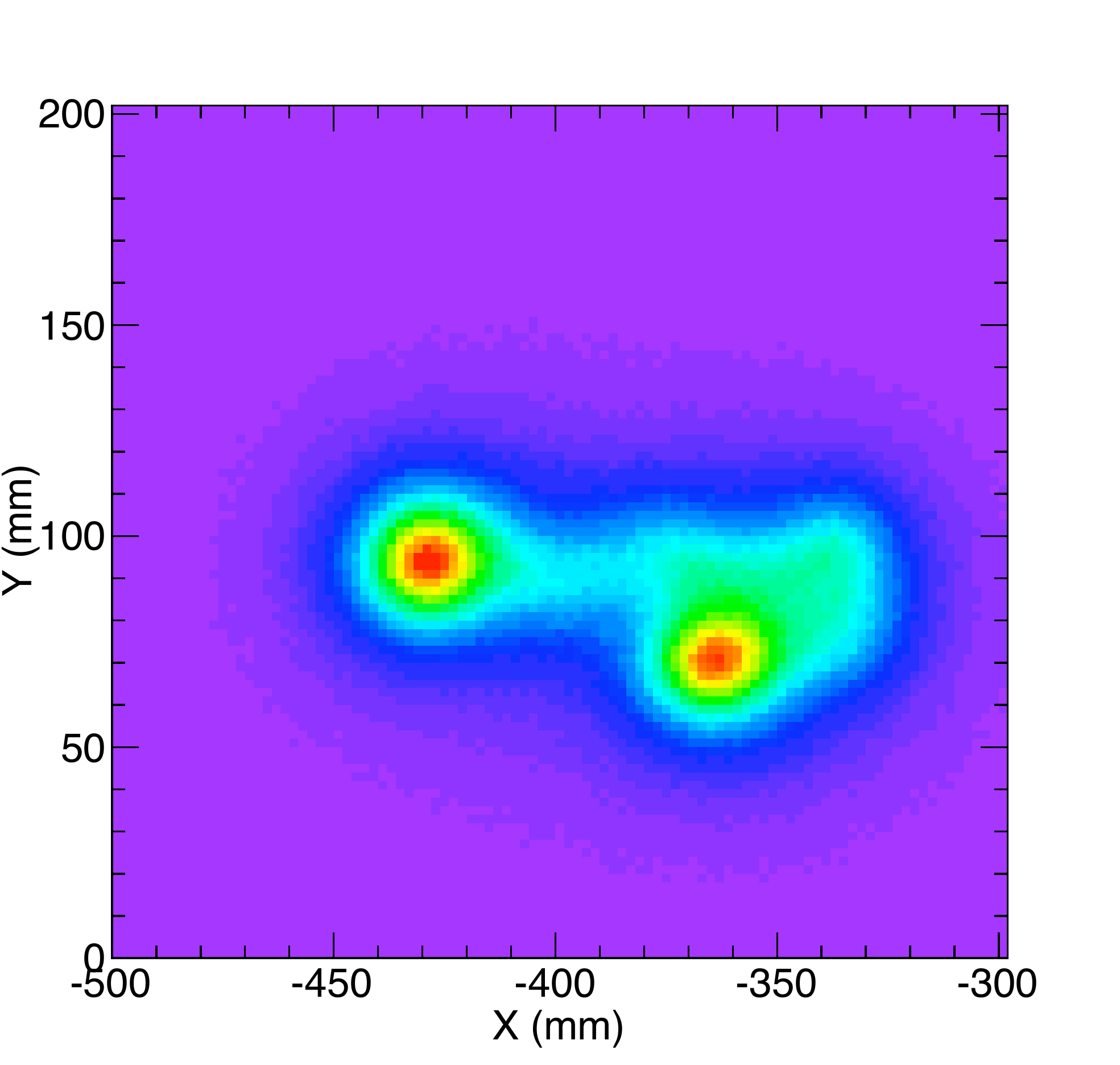} 
\end{center}
\vspace{-0.5cm}
\caption{\small Monte Carlo simulation of the image of \bbonu\ event in a plane of SiPMs.}
\label{fig:eltrk2} 
\end{figure}

Figure \ref{fig:eltrk2} shows a Monte Carlo simulation of an event tracked by SiPMs. The light background collected by cells outside the track is at most 10$^{-4}$ of the total track, and 10$^{-5}$--10$^{-6}$ for most cells.

\section{Pressure Vessel}
The construction material chosen for the pressure vessel is pure titanium, ASTM grade 2 or grade 3 (depending on the final results of the material screening now being carried out at the LSC). The choice of titanium over pure copper (ASTM oxygen-free copper C11000), material traditionally used in low background experiments, is motivated by engineering reasons. Although finite element simulations indicate that it is possible to build a vessel made of ultra-radiopure copper, the material is not contemplated  in the ASME code and standards, which will be followed through in the construction of the vessel. The use of ASME, one of the most widely accepted standards, in particular for the construction of pressurized vessels is a must, in particular to guarantee successful performance in a risk analysis, compulsory for the underground operation of the experiment.  
%

\begin{table}[t!b!]
\begin{center}
\begin{tabular}{clrr}
\hline \hline
& & \multicolumn{1}{r}{Titanium} & \multicolumn{1}{r}{Copper} \\ \hline \noalign{\smallskip}
\multirow{3}{*}{Thickness (mm)}& Cylindrical Shell & 	5  & 	30  \\
& Torispherical head & 12 &	50 \\
& Flange & 90 & 140 \\ \hline
\multirow{4}{*}{Mass (kg)} & Cylindrical Shell &	 227 & 947 \\
& Torispherical head	 & 58 & 	490 \\
& Flanges ($4\times$) & 401 & 1241 \\
& \emph{Total} & 1932 &	7412 \\
\hline
\multicolumn{2}{l}{Activity (counts/year)} & $2.4 \times 10^6$ & $2.4 \times 10^6$ \\
\hline \hline
\end{tabular}
\end{center}
\caption{Pressure vessel in the ANGEL design: titanium versus copper. Calculations based on the ASME Code, Section VIII, Division 2.}\label{tab.TiCu}
\end{table}%

As an example, table \ref{tab.AngelPV} collects the dimensions and weight of a pressure vessel made of grade 2 titanium and of OHFC copper, a material accepted by the ASME norm. Notice that the level of activity that one obtains for both materials is about the same (the specific activity of titanium grade 2 is taken from recent measurements at LBNL to be 200 $\mu$Bq/kg, to be compared with 50 $\mu$/kg for the OHFC copper). On the other hand, the titanium vessel material thickness, in particular for the flanges and the total weight are much more comfortable.

\begin{table}[tb]
\begin{center}
\begin{tabular}{clr}
\hline \hline
\multirow{5}{*}{Dimensions (mm)} 	& Cylindrical can, inner radius & 570 \\
									& Cylindrical can, length & 1350 \\
									& Cylindrical can, thickness & 5 \\
									& Torispherical head, thickness & 12 \\
									& Flange, thickness & 90 \\ \hline
\multirow{4}{*}{Mass (kg)} 	& Cylindrical can & 227 \\
							& Torispherical head & 58 \\
							& Flanges (4$\times$) & 401 \\
							& \emph{Total} & 1932 \\ \hline
\multicolumn{2}{l}{Activity in fiducial (counts/year)} & 2.4$\times10^{6}$\\
\hline \hline
\end{tabular}
\end{center}
\caption{Dimensions and weight of the titanium (grade 2 or 3) vessel in the ANGEL design. Calculations based on the ASME Code, Section VIII, Division 2.}\label{tab.AngelPV}
\end{table}%



\section{Field cage, high voltage and electroluminescence grids}

\begin{table}[t!b!]
\begin{center}
\begin{tabular}{lc}
\hline\hline
Parameter & Value \\
\hline
        E/P  &   3.50 $kV \cdot cm^{-1}\cdot bar^{-1}$ \\
        Drift voltage &    0.50 kV/cm \\
        Pressure &  15.00 bar \\
        EL grid gap &    0.50 cm \\
        Drift length &  130 cm \\
        Grid voltage &   26.25 kV \\
        Cathode voltage &   91.25 kV \\
        Optical Gain &  2800 photons/e \\
\hline\hline
\end{tabular}
\end{center}
\caption{Angel EL}\label{tab.AngelEL}
\end{table}%

Table \ref{tab.AngelEL} describes the main parameters of the EL detector. We have chosen a drift voltage of 0.5 kV/cm, near the minimum of diffusion, and $E/p$ of 3.5 which gives large optical gain and an EL gap of 5 mm. This 
results in a grid voltage of 26.25 kV and a cathode voltage of 91.25 kV.

\begin{figure}[tbhp!]
\centering
\includegraphics[width=0.99\textwidth]{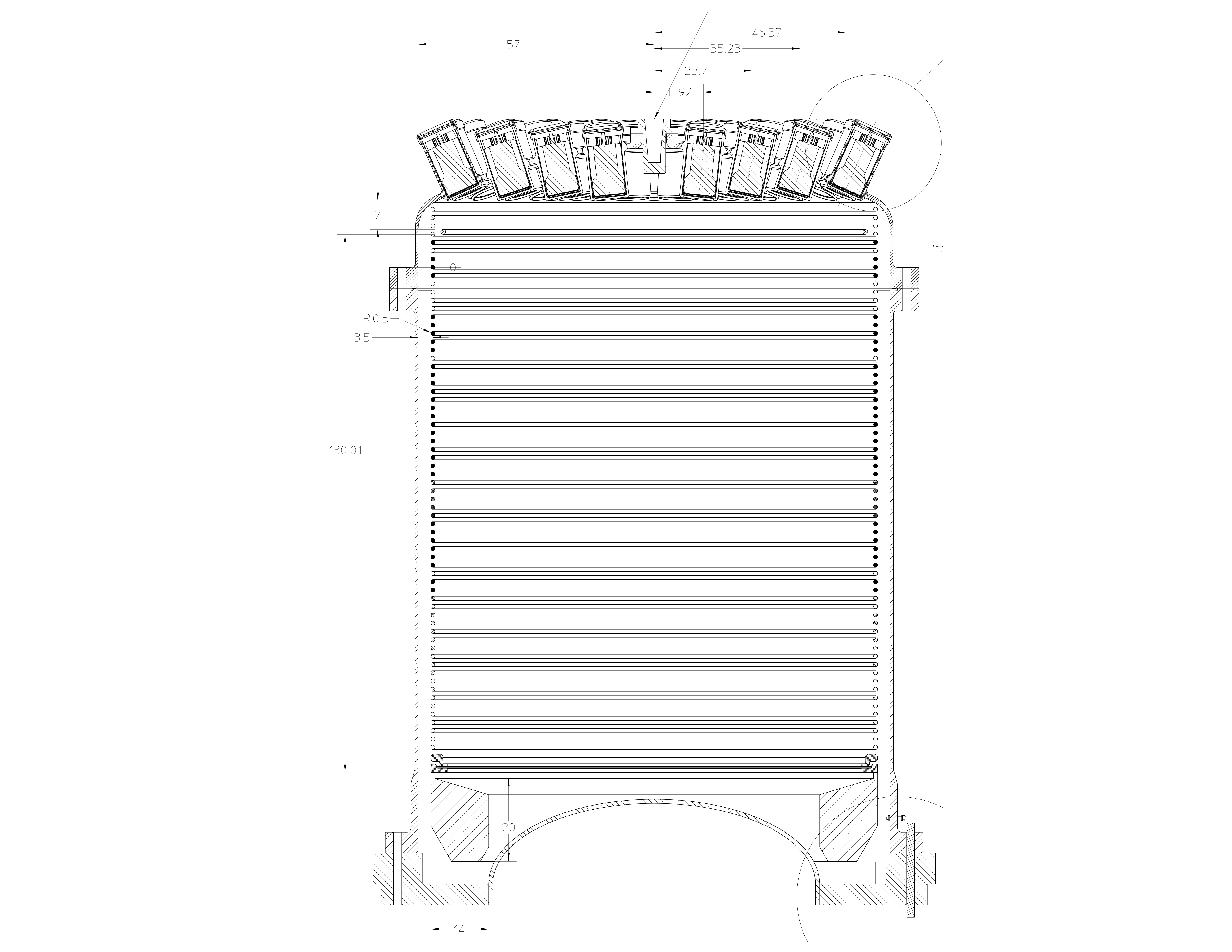}
\caption{A section of the ANGEL detector, showing the field cage.}\label{fig.FC}
\end{figure} 

\begin{figure}[ptbh!]
\centering
\includegraphics[width=0.95\textwidth]{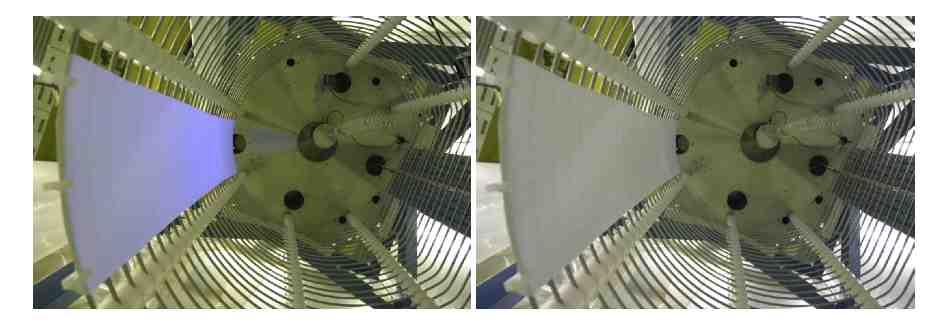} 
\caption{The ArDM field cage with a sheet of 3M+TTX coated with TPB. The left picture shows the blue emission when the sheet is illuminated with a UV lamp. The emission disappears (right) when the UV lamp is turned off. From \cite{Boccone:2009kk}.} 
\label{fig.ttx}
\end{figure}

A section of the detector, showing the field cage (FC) can be seen in Figure \ref{fig.FC}. The FC will be made
of copper rings connected by low background resistors. The light tube will consist of thin sheets of Tetratex$^{\rm TM}$ (TTX), fixed over a 3M$^{\rm TM}$ substrate, following the approach of the ArDM experiment (see for example \cite{Boccone:2009kk}) that will also operate at the LSC. We plan on a joint development in this and other aspects (e.g, low background PMTs) between both collaborations. 

TPB can efficiently absorb the VUV radiation emitted by xenon and re-emit with an spectrum that peaks
in the blue (Figure \ref{fig.ttx}). TPB can be easily deposited in the 3M+TTX sheets by vacuum evaporation. The
ArDM collaboration has measured (\cite{Boccone:2009kk} ) a reflectance coefficient at 430 nm close to 97\% for a wide range of coating thicknesses. In addition the light yields were measured at different time intervals, showing no evidence of aging in the time interval of 3 months.

\begin{figure}[t!b!]
\begin{center}
\includegraphics[width=0.85\textwidth]{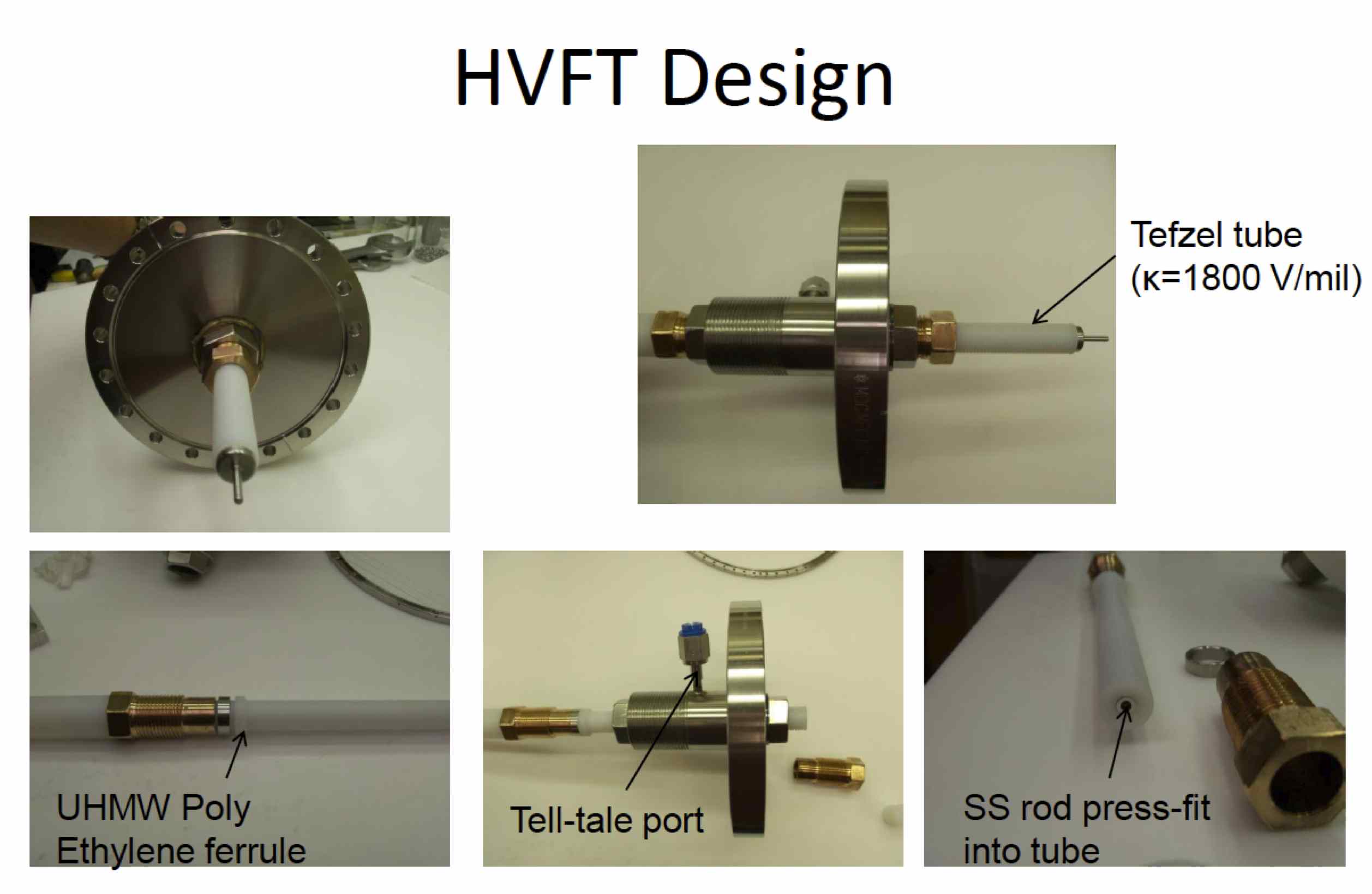} 
\end{center}
\caption{High-Voltage feedthroughs designed and built by Texas A\&M for NEXT-1-IFIC.} \label{fig.hvft}
\end{figure}

High-voltage feedthroughs (HVFT) will be constructed using a compression seal approach as illustrated in Figure \ref{fig.hvft}. A metal rod is pressed into a plastic tube (Tefzel or FEP, which have high dielectric strength) which is then clamped using plastic ferrules from both the pressure side and air side. A sniffer port is placed between the seals to assure that xenon is not leaking. This approach has been used in NEXT-1 where a cathode voltage of 40 kV has been achieved. A small prototype of the NEXT-1 feed-through was tested to 100 kV in vacuum and 70 kV in nitrogen at 3 bar. It has been demonstrated to be leak-tight at 10 bar xenon and 10$^{-7}$ mbar vacuum. The design will be scaled up for NEXT-100 as needed.

\begin{figure}[t!b!]
\begin{center}
\includegraphics[width=0.75\textwidth]{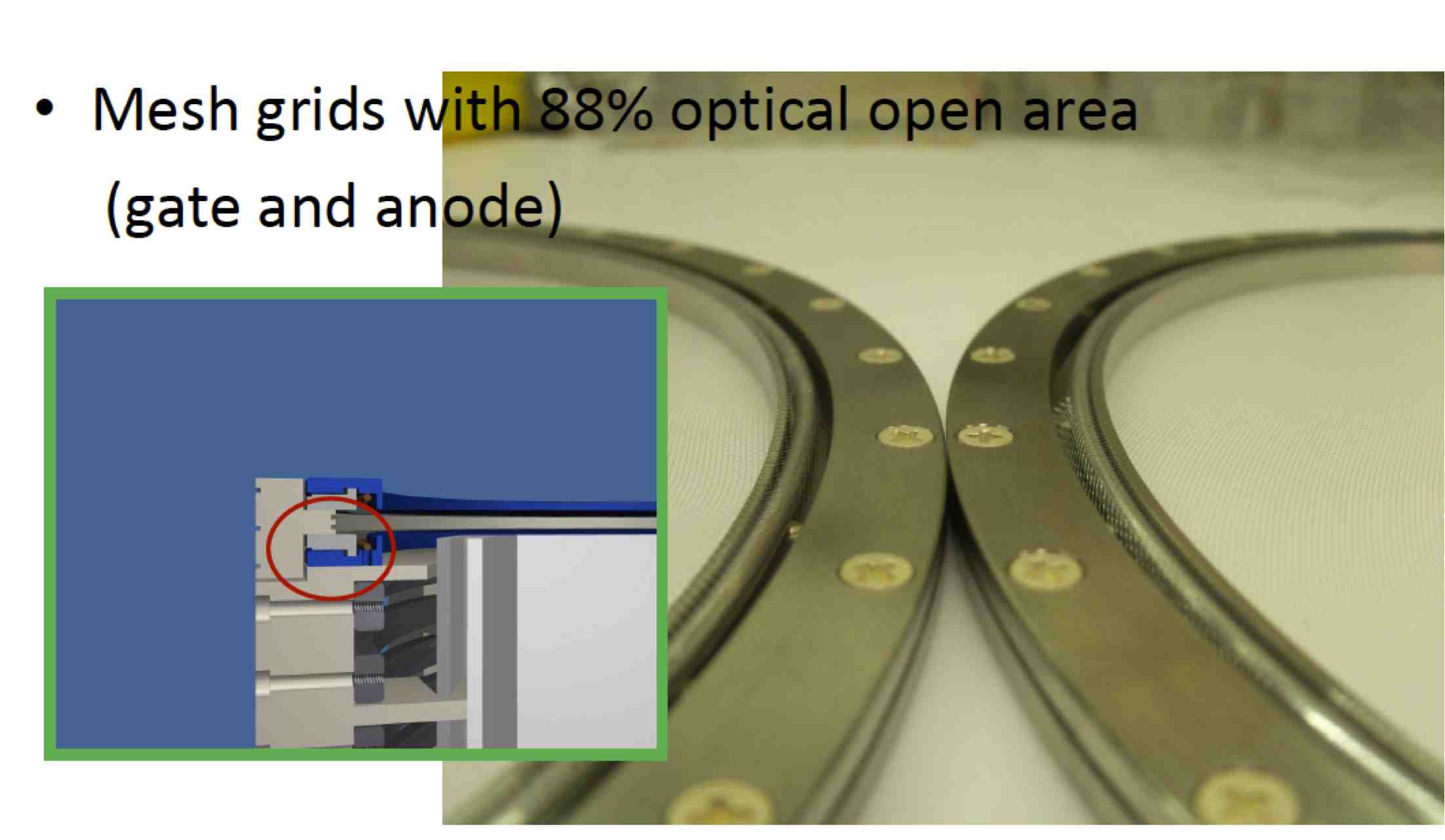} 
\end{center}
\caption{NEXT-1-IFIC EL meshes designed and built by Texas A\&M.}
\label{fig.grids}			
\end{figure}

Figure \ref{fig.grids} shows the EL grids built for the NEXT-1-IFIC prototype. 
The  grids were constructed using stainless steel mesh with pitch 0.5 mm and wire diameter 30 microns, which results in an open area of 88\%. The grids are formed by clamping in a ring with a tongue and a groove to hold the mesh and using a tensioning ring that is torqued with set-screws to achieve the optimum tension. There is considerable experience with this approach since the grids are similar to ones built for ZEPLIN II, LUX, and a number of other test chambers. The cathode and PMT shield will be wire grids with pitches of 0.5 to 1 cm to maximize open area. Again, the design will be based on previously constructed grids that are well understood. One important issue is that for the large diameter required in ANGEL, preliminary estimates show that the EL grids will bow as much as 1 mm given the modulus of elasticity of the mesh and required voltage. If this is verified and if the bowing introduces a systematic effect in the energy resolution that cannot be corrected (Monte Carlo studies are under way to assess this problem) several alternatives are possible. One possibility is to pre-stress the ground grid following the measured curvature of the HV grid. Another possibility is a design where the anode is composed of acrylic that is metalized with a transparent coating such as Indium-Tin-Oxide (ITO) and coated with TBP. In this case, the acrylic can be bowed to match the deflection of the gate grid to maintain a uniform gap. If this plate is coated on both sides, the anode side can be at positive HV and the back side grounded so that the SiPMs can still be close to the anode. 

%
\chapter{The NEXT-1 prototypes} \label{sec.next1}
The main aim of the NEXT1 prototypes built at LBNL and IFIC is demonstrating NEXT energy resolution at reasonably high energies (e.g,  using positron annihilation 511 keV gamma rays, \CS\ 660 keV gamma rays and
\CO\ 1.1 MeV gamma rays). Some of the challenges in designing an electroluminescence TPC with near optimal energy resolution are:
\begin{enumerate}
\item The total signal is distributed over many PMTs. Therefore accurate relative calibrations are needed for integration.
\item Signal duration and amplitude depends on event shape.
\begin{enumerate}
\item Low drift velocity $\sim$1 mm/$\mu$s.
\item Diffusion spreads out track signal: 0.3 mm/$\sqrt{{\rm cm}}$ longitudinal, 0.8 mm/$\sqrt{{\rm cm}}$ transverse.
\item Signals spread over time: minimum for tracks parallel to luminescence plane, maximum for tracks normal to luminescence plane (2.5 MeV electron track length $\sim$16 cm).
\end{enumerate}
\item Large electroluminescent gain and light collection/photo-efficiency is required to reduce the statistical error on photo-electron signals.
\item Light collection efficiency depends somewhat on track radial position.
\item Electroluminescent gain must be stable and uniform over the EL surface.
\item Contaminants that quench signals must be controlled (gas purification).
\item Xenon light emission is at 173 nm, in the VUV region; this requires quartz-window (VUV grade) photodetectors, or coated windows.
\item Energy resolution is measured with high-energy gammas:
\begin{enumerate}
\item Compton events are predominant, leading to distributed energy.
\item Photoelectric events from gamma absorption are frequently ($\sim$85\%) accompanied by fluorescent x-rays in the 30 keV range.
\end{enumerate}
\end{enumerate}

%

Furthermore, the NEXT1 prototypes will study with detail the functioning of a SOFT TPC, with a PMT energy plane (both NEXT1-LBNL and NEXT1-IFIC) and a SiPM plane (NEXT1-IFIC). Last but not least,
the detectors are intended as an R\&D program to test the solutions proposed by the ANGEL design. 

\section{The energy function in the NEXT-1 prototypes}

\begin{figure}[tb!]
\begin{center}
\includegraphics[width=0.8\textwidth]{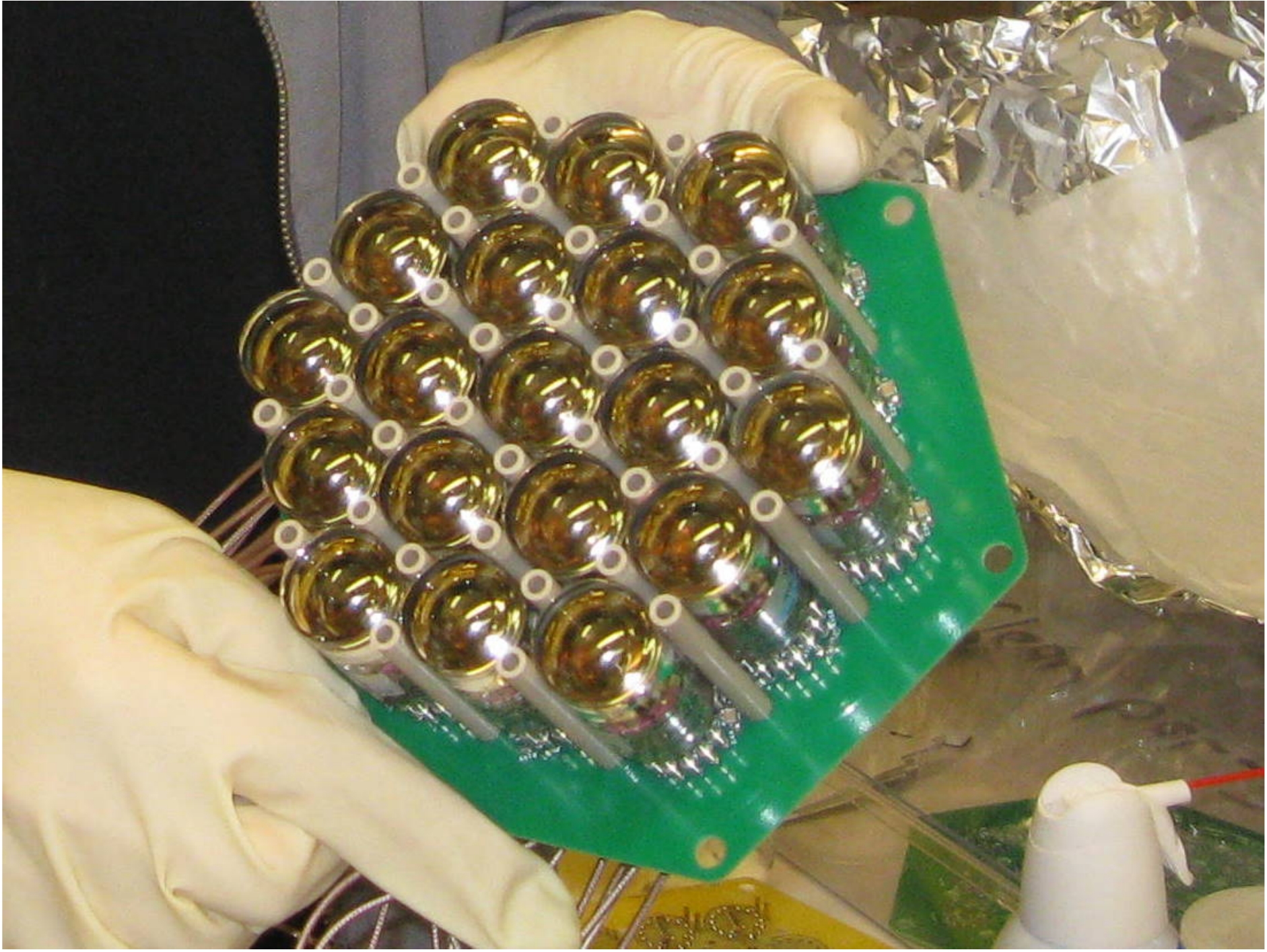} 
\end{center}
\vspace{-0.5cm}
\caption{\small The R7378A  array. This small, rugged, 1'' PMTs, are tested up to 19 bars and capable of detecting 173 nm xenon light.}
\label{fig.lbl.PMTArray} 
\end{figure}

\begin{figure}[ptbh!]
\centering
\includegraphics[width=0.85\textwidth]{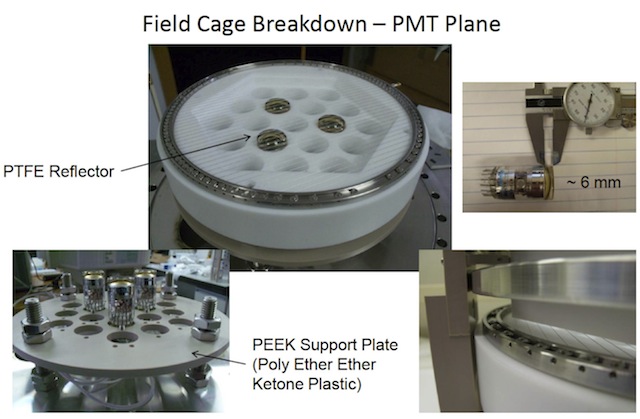} 
\caption{\small The NEXT-1 PMT plane. The lower support of the energy plane PMTs is made of peek and the upper one of PTFE (which has more reflectivity to the light). The energy plane is located behind a transparent cathode.}
\label{fig.pp}			
\end{figure}

In both the LBNL and the IFIC prototypes the energy function is provided by a plane of 19 pressure resistant R7378A PMTs, capable to detect directly the 173 nm VUV light emitted by xenon. The PMTs also measure the event start time ($t_0$).  
Figure \ref{fig.lbl.PMTArray} shows the energy plane fully assembled for the NEXT1-LBNL prototype, and 
Figure \ref{fig.pp} shows details of the NEXT1-IFIC plane. 

\section{FE readout electronics for the PMTs }

\begin{figure}[ptbh!]
\centering
\includegraphics[width=0.85\textwidth]{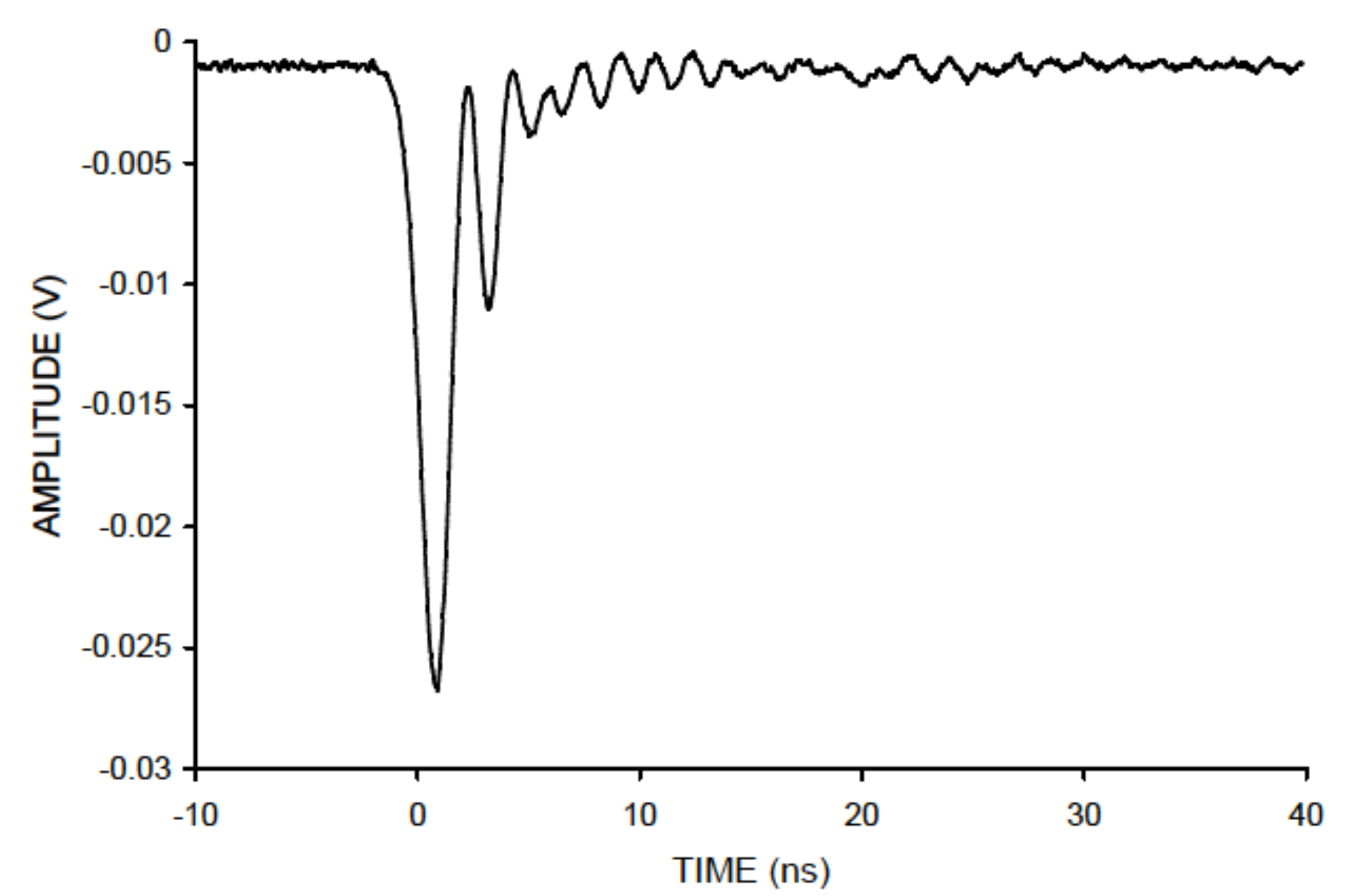} 
\caption{\small Typical Pulse from the NEXT1 prototypes PMT Plane (average of 200 pulses).}
\label{fig.ring}			
\end{figure}

The FE electronics for the PMTs is almost identical for both systems. The first step is to shape and filter the fast signals produced by the PMTs (less than 5 ns wide) to match the digitizer and eliminate the high frequency noise
which produces an unwanted ringing. The ringing, schematically shown in Figure \ref{fig.ring}, is shown 
superimposed to the fast PMT pulses. 

\begin{figure}[ptbh!]
\centering
\includegraphics[width=0.85\textwidth]{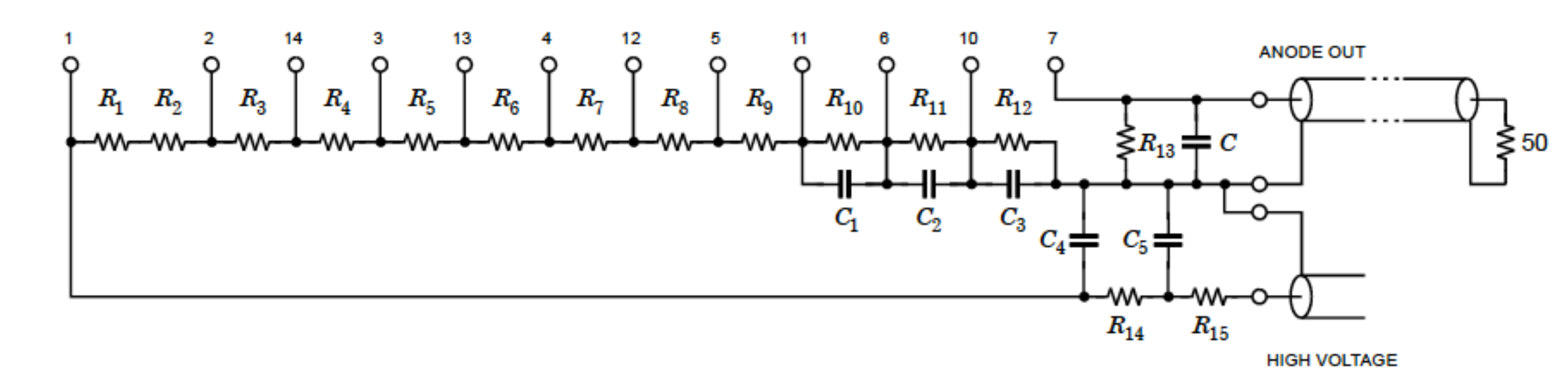} 
\caption{\small Charge integration can be carried out by adding a capacitor to the PMT base.}
\label{fig.chargei}			
\end{figure}

The key concept to eliminate ringing is to realize that the phenomenon does not change the total charge, which should be measured rather than a fast voltage pulse. An integrator can be implemented by simply adding a capacitor and a resistor (50 $\Omega$~in our case) to the PMT base (Figure \ref{fig.chargei}). The pulse rise time is
formed by the width of the PMT current pulse.
The decay time is simply:

\begin{equation}
\tau = R_L C = 50 C > \frac{1}{f_{sample}}
\end{equation}

\begin{figure}[ptbh!]
\centering
\includegraphics[width=0.85\textwidth]{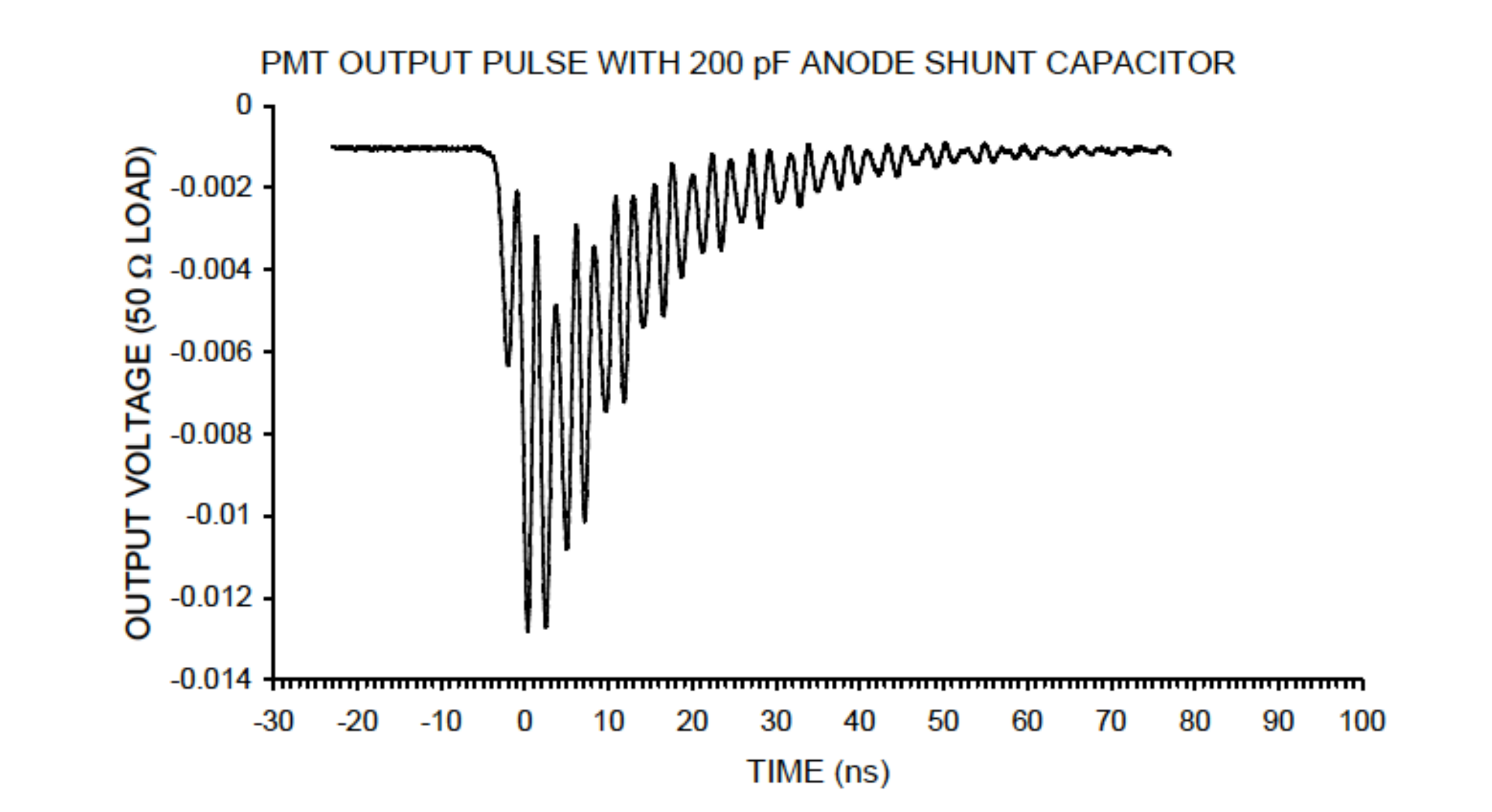} 
\caption{\small PMT output pulse with 200 pF anode shunt capacitor.}
\label{fig.shunt}			
\end{figure}

The effect of the charge integration capacitor shunting the anode can be seen in Figure \ref{fig.shunt}. It lengthens the pulse and reduces the primary signal peak voltage to:

\begin{equation}
V = \frac{Q}{C}
\end{equation}

However the amplitude of the ringing component is not reduced since it does not affect the signal current!

\begin{figure}[ptbh!]
\centering
\includegraphics[width=0.85\textwidth]{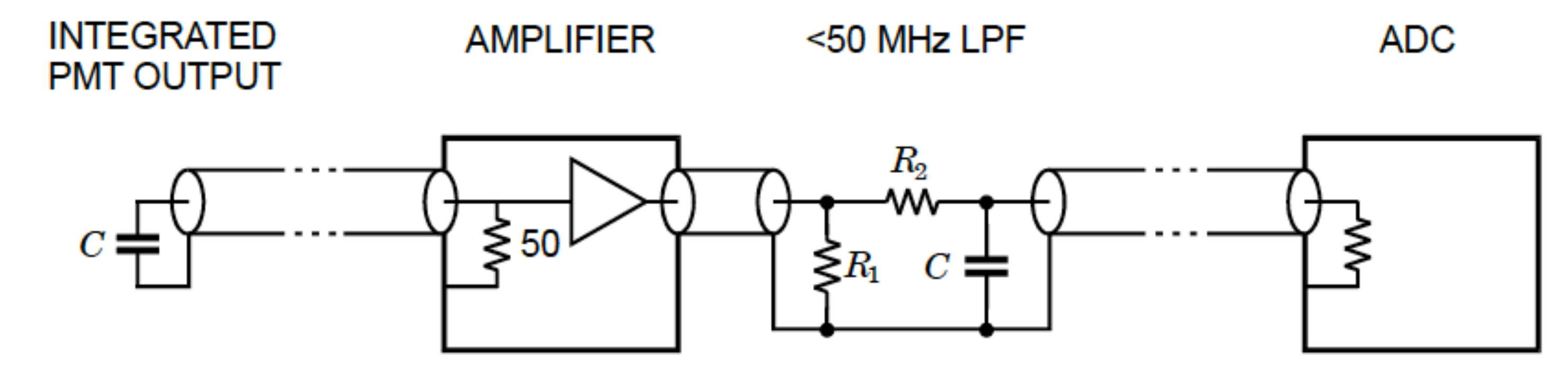} 
\caption{\small The PMT output is fed to an a chain amplifier + low pass filter.}
\label{fig.filter}			
\end{figure}

Instead of feeding the ADC directly, the integrated signal is fed to an amplifier with much
lower noise (Figure \ref{fig.filter}). Currently we are using Lecroy 660 A modules rented from the CERN pool. The signal-to-noise ratio of the single photoelectron PMT pulse sets the capacitance C and 
a low-pass filter is inserted at the output of the amplifier.
This limits the signal bandwidth to $<0.5 \times ADC$ sampling rate and reduces the total noise, since the amplifier has a greater bandwidth.

\begin{figure}[ptbh!]
\centering
\includegraphics[width=0.85\textwidth]{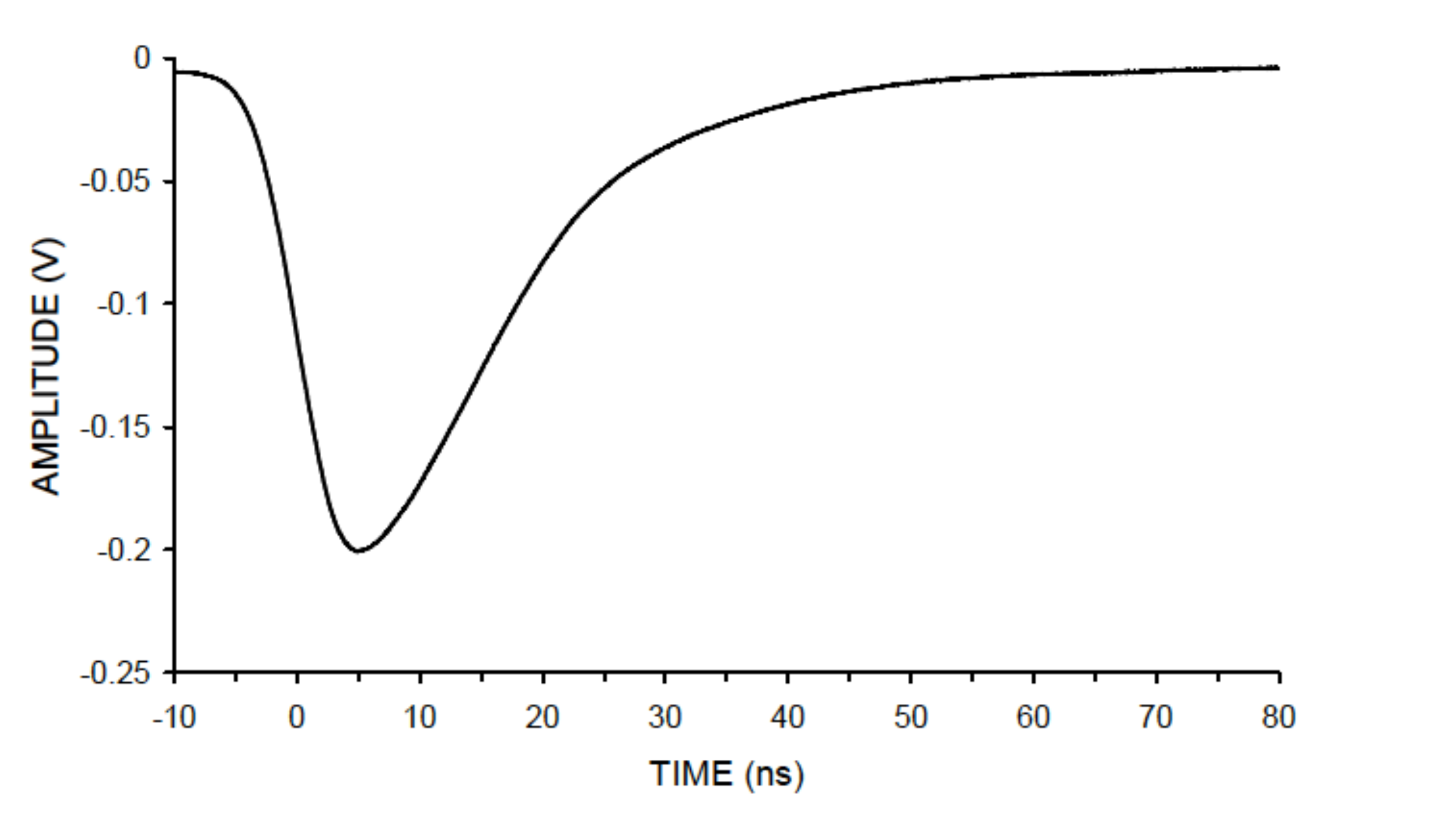} 
\caption{\small The bandwidth limit of the amplifier together with a 50 MHz low-pass filter at its
output attenuates the PMT ringing components to form a clean pulse fed to the ADC, which
has a 10 ns sampling time.}
\label{fig.filter2}			
\end{figure}

The low-pass filter is configured to provide the desired load resistance to the amplifier and
attenuation to match the ADCs maximum input voltage.
Furthermore, the amplifier provides additional attenuation at high frequencies. The result can be
seen in Figure \ref{fig.filter2}.

\section{The NEXT1 prototype at LBNL}

\begin{figure}[tbhp!]
\begin{center}
\includegraphics[width=0.9\textwidth]{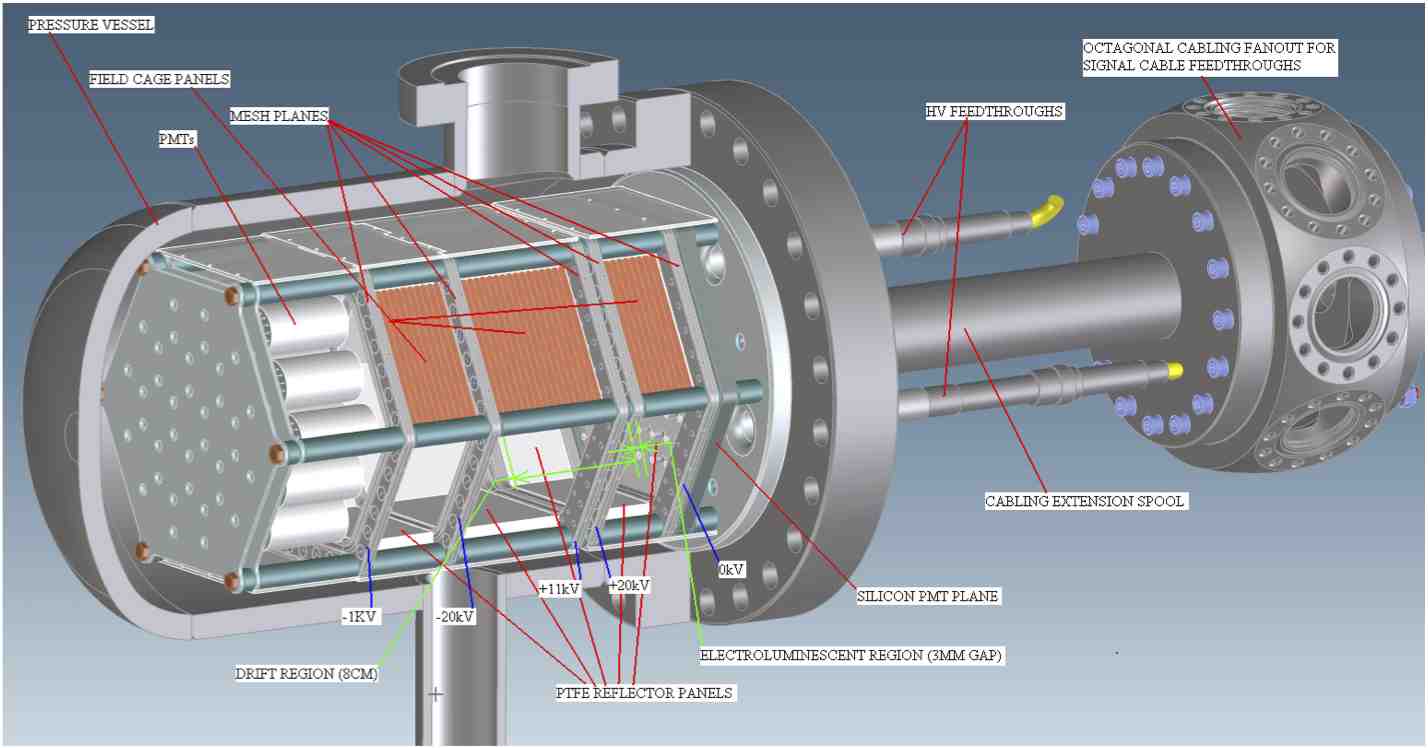} 
\end{center}
\vspace{-0.5cm}
\caption{\small Design of NEXT-1-LBNL chamber.}
\label{fig.LBNL.design} 
\end{figure}

The LBNL chamber (Figure \ref{fig.LBNL.design}) has been optimized to operate at the largest pressures considered for NEXT (20 bar). Since our aim was to provide a quick benchmark for physics, the chamber was designed to be as simple and compact as possible. Thus, only an energy plane was constructed, using pressure-resistant PMTs but no tracking plane was built.  The system is operational at LBNL since December 2010. 

\subsection{Gas system design and construction}

The gas system, capable of handling up to 300 psig operating pressure, was designed with safety as a primary criterion. Engineering controls in the form of relief valves ensure that no overpressure hazard can occur in case of an operator error.  Administrative controls, in the form of step-by-step detailed procedures, prevent the accidental loss/venting of the valuable xenon gas.   

\begin{figure}[tbhp!]
\begin{center}
\includegraphics[width=0.8\textwidth]{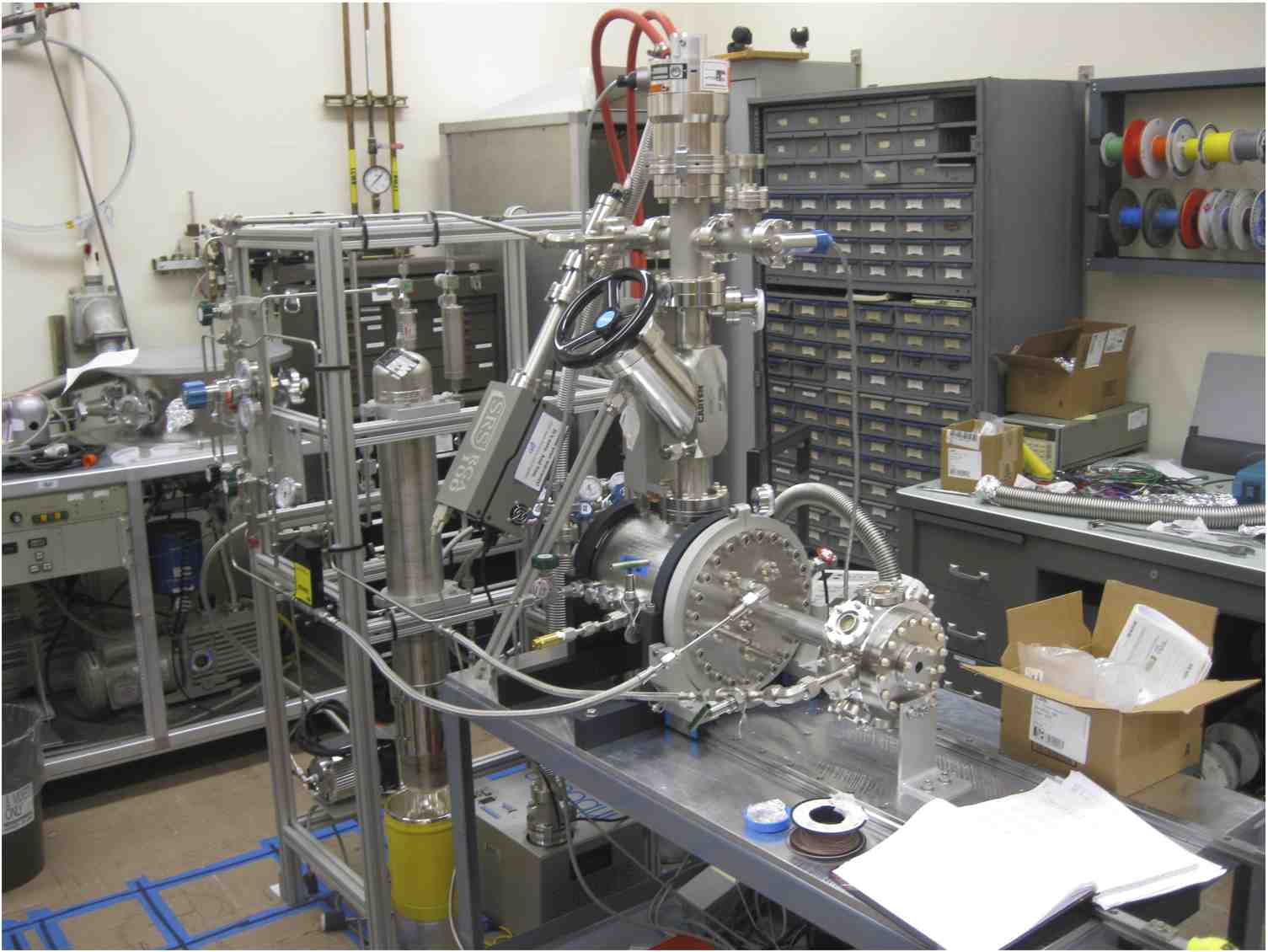} 
\end{center}
\vspace{-0.5cm}
\caption{\small The full system, currently fully operational and certified for operation to 17 bar.}
\label{fig.lbl.gasSystem} 
\end{figure}

Figure \ref{fig.lbl.gasSystem} shows the full TPC. Most of the gas system is mounted on a separate structure adjacent to the TPC table. Above the TPC a large valve (rated for vacuum and high pressure) separates the TPC from the vacuum systems.
The gas system is fully operational and is routinely operated near its maximum rated pressure of 17 bar.

\subsection{Electroluminescent TPC}

\begin{figure}[tbhp!]
\begin{center}
\includegraphics[width=0.6\textwidth]{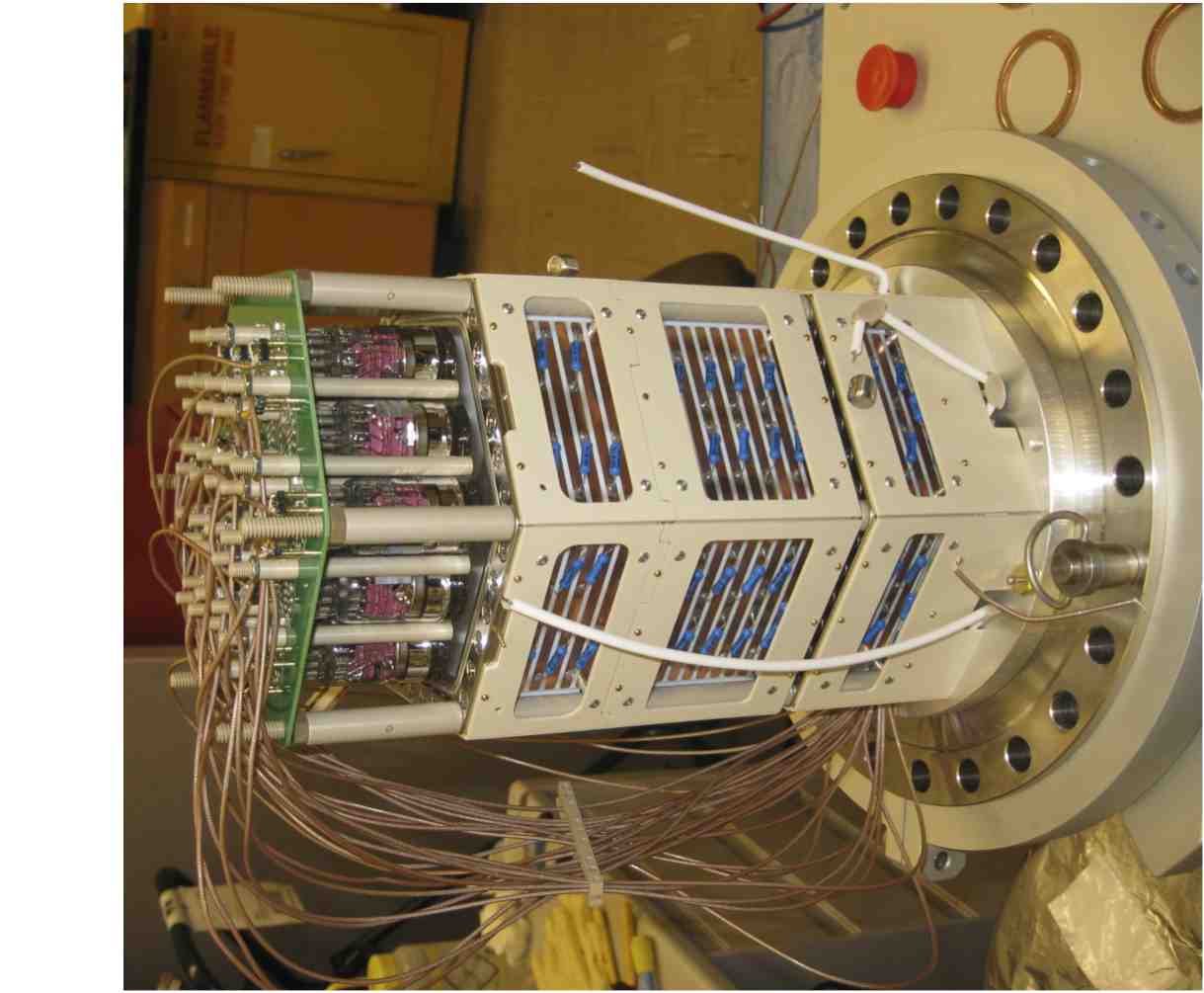} 
\end{center}
\vspace{-0.5cm}
\caption{\small Assembly process of the EL TPC. The side panels of the TPC are PTFE reflectors in the inside and field cage on the outside (copper stripes). The entire PMT and TPC assembly is held by the main vessel flange for ease of installation.}
\label{fig.lbl.FieldCage} 
\end{figure}

Five conductive meshes establish the electric potentials in the TPC. The walls of the TPC are made of PTFE, bare, on the internal side to reflect UV photons and with conductive stripes (at intermediate potentials) on the outside to establish uniform electric field lines along the TPC main axis (Figure \ref{fig.lbl.FieldCage}).
 
All HV and signal cables come out through the main large flange for ease of assembly.  Large HV for the drift and EL region come out through dedicated SHV feedthroughs directly on the main flange.  PMT HV and signal cables come out through the central tube and through multipin feedthroughs on the sides of an 8-port vessel.  The PMT array connects to a base-plane, with receptacles for the PMT pins and with electronic components for the PMTs’ voltage dividers. Capacitors on the last few PMT dynodes provide voltage stability for the long multi-photoelectron pulses expected from event size passing into a large-gain EL region. Differential signal sensing, with isolated grounds, has been implemented to avoid ground-loop noise.
%

Previous measurements by Bolotnikov et al. (using a small ionization chamber) showed a large dependence of the energy resolution in high pressure xenon on the drift 
electric field that is not fully understood, although a case can be made that the effect is an artifact of that setup. In the NEXT1-LBNL TPC the drift region is 8 cm 
long and the maximum HV on the dedicated feedthroughs is both $\pm$20 kV. It then allows us to measure the energy resolution for drift fields from 0 to 4 kV/cm 
(thus spanning the previously measured range) while maintaining a constant EL gain. 

\subsection{DAQ design, implementation and PMT measurements}

The DAQ design for the 19 PMT array begins with low-noise amplifiers, close to the detector, as described above. These are connected to Struck digitizers by low-pass circuits that shape the fast PMT pulses (less than 5 ns wide) to match the digitizer sampling rate of 100 MHz. The Struck digitizer modules have genuine 13-14 bit resolution, and are read out through a VME controller card with USB output directly into a PC running Linux. 

Currently we are using standard NIM modules for the amplification stage but plan to transition to a custom made solution based on commercially available preamp integrated circuits.  The Struck 8-channel digitizer modules are controlled by provided firmware to provide trigger generation and determine the number of samples to capture per trigger, the amount of pre-trigger, and other functions. 

In normal operation the trigger is formed by the logical-or of all the digitizer channels (one OR signal per 8-channel group). For each channel an amplitude or charge threshold can be used. A threshold at a few photo-electron level will be efficient for the EL signal and should have a very low noise-trigger rate. In calibration mode we use tagged Na-22  (two back to back 511 keV gamma rays.  One of the gammas is tagged by a dedicated NaI scintillator/PMT detector) with a low threshold on the TPC signals; the other gamma is highly likely to have entered the active volume of the detector.

\section{The NEXT1 prototype at IFIC}

The NEXT1 prototype at IFIC 
is the second prototype of the NEXT-1 EL series. It is intended to fully test and demonstrate the baseline
ANGEL design. Accordingly, NEXT1 is an asymmetric (non radiopure) SOFT TPC, with a SiPM tracking plane and a PMT energy plane. It extends both in size and in instrumentation the design of NEXT1-LBNL.

The detector has been built and commissioned, and data analysis has started. NEXT1 is conceived as an evolving prototype/demonstrator, that will allow us to fully test most of the solutions proposed for ANGEL. Specifically, we foresee the following stages (henceforth referred to as `Runs").
\begin{itemize}
\item {\bf Run I (March--September 2011):} Commissioning and initial data taking with two PMT planes, one behind the cathode, the other behind the EL grids. Each PMT plane has 19 pressure resistant R7378A. The energy function is carried out by
the cathode plane. The anode plane provides an additional measurement of the energy, important to study in
detail the SOFT concept and coarse tracking. 
\item {\bf Run II (October--December 2011):} The SiPM plane has already been built and is being tested in an independent setup with several light sources, including a xenon lamp. In this way, the tracking plane will be fully debugged before being installed in the chamber in early July. Run II will fully demonstrate tracking with SiPMs as well as study the stability of TPB coating in the sensors. In Run II we will also take data with the final FE electronics for PMTs and
the final DAQ that will be used by NEXT-100. 
\item {\bf Run III (January-June 2012):} The energy plane will be replaced, changing the R7378A inside the gas by a set of PMTs outside the chamber. The PMTs will be optically coupled to sapphire windows and will allow us to fully test the ANGEL design. In addition, the light tube will be changed from uncoated PTFE to TTX+3M foils, following the ANGEL design. Run III will show energy resolution in a system with housed PMTs will continue studying tracking with SiPMs and will study the properties and stability of a high reflectivity light tube.
\item {\bf Run IV (June 2012-December 2012):} For Run IV the SiPM electronics will be located inside the chamber.  
\end{itemize}

In our LOI we considered the possibility of a radiopure detector (NEXT-10) with a mass of about 10 kg as an intermediate detector between NEXT-1 and NEXT-100. However, such a detector appears today prohibitive in terms of costs,
man power and, very importantly, time. Instead, NEXT-1 can be evolved as described above to provide not only a full bench test of the ANGEL design, but also the needed operational experience. Along this line some of the Runs of NEXT-1 (probably Run IV but perhaps also Run III) will, very likely, be taken at the LSC. In this way, we will gain experience and prepare the underground operation of NEXT-100.  

\subsection{Design and construction }


\begin{figure}[bhtp!]
\centering
\includegraphics[width=0.6\textwidth]{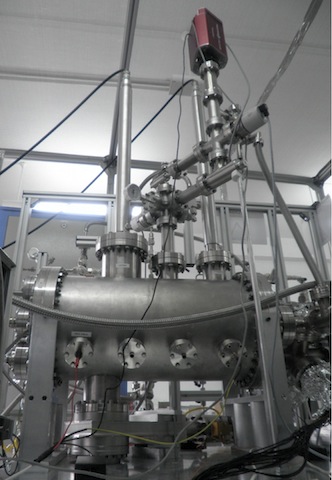}
\caption{View of the NEXT1 pressure vessel. Also visible, the gas system, mass spectrometer
and the long tubes to pass the high voltage. }   
\label{fig:can}
\end{figure}

NEXT1-IFIC is presently the largest operational HPGXe TPC in the World. It can operate up to 15 bars, although the optimal operating pressure is 10 bar. The TPC fits inside a steel chamber 600 mm long and 300 mm inner diameter (Figure \ref{fig:can}), fabricated by TRINOS Vacuum Systems, a company located in Valencia.  The fiducial  volume consists of a hexagonal cross-section, defined
by PTFE reflector panels, 160 mm across the diagonal and 300 mm drift region. The electroluminescence region is made
of two parallel grids separated by 5 mm. The maximum designed drift field is 1 kV/cm and the maximum electroluminescence field is 40 kV/cm. The energy plane (and the tracking plane during the commissioning phase of the detector) use Hamamatsu R7378A
photomultiplier tubes (capable of resisting up to 20 bar pressure). 

\subsection{The gas system}

%

The materials used for the vessel and the readout place outgas electro-negative impurities, which degrade the performance of the detector, into the Xe gas. 
The role of the gas system is to remove these. This is achieved by continuously re-circulating the Xe gas through a SAES Getters (MC500). All the gas piping,
save for the inlet gas hoses and Getter fittings, are $1/2$ inch diameter with VCR fittings. The re-circulation
loop is powered by a KNF diaphragm pump with a nominal flow of 100 standard liters per minute. At 10 bar this translates
to an approximate flow of 10 liters per minute. 

Considerable experience has been gained after the successful commissioning of three different large gas systems at LBNL, Zaragoza and IFIC. As a consequence, the CDR includes a full technical design of the gas system for NEXT-100 (See chapter on NEXT-100). 


\subsection{Field cage and light tube}

\begin{figure}[ptbh!]
\begin{center}
\includegraphics[width=0.85\textwidth]{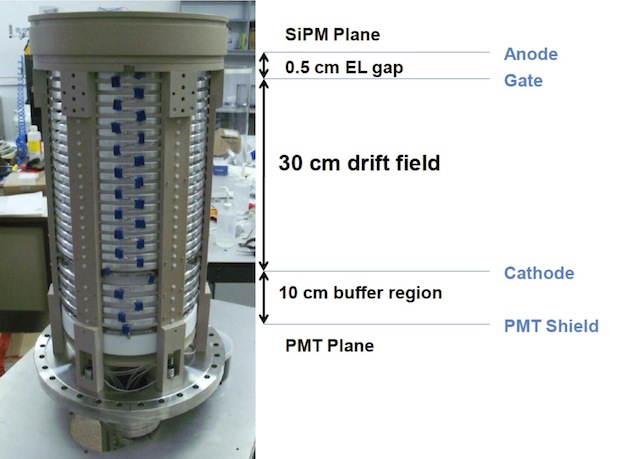}
\end{center}
\caption{The NEXT-1 field cage.}
\label{fig.fc}			
\end{figure}

\begin{figure}[ptbh!]
\begin{center}
\includegraphics[width=0.65\textwidth]{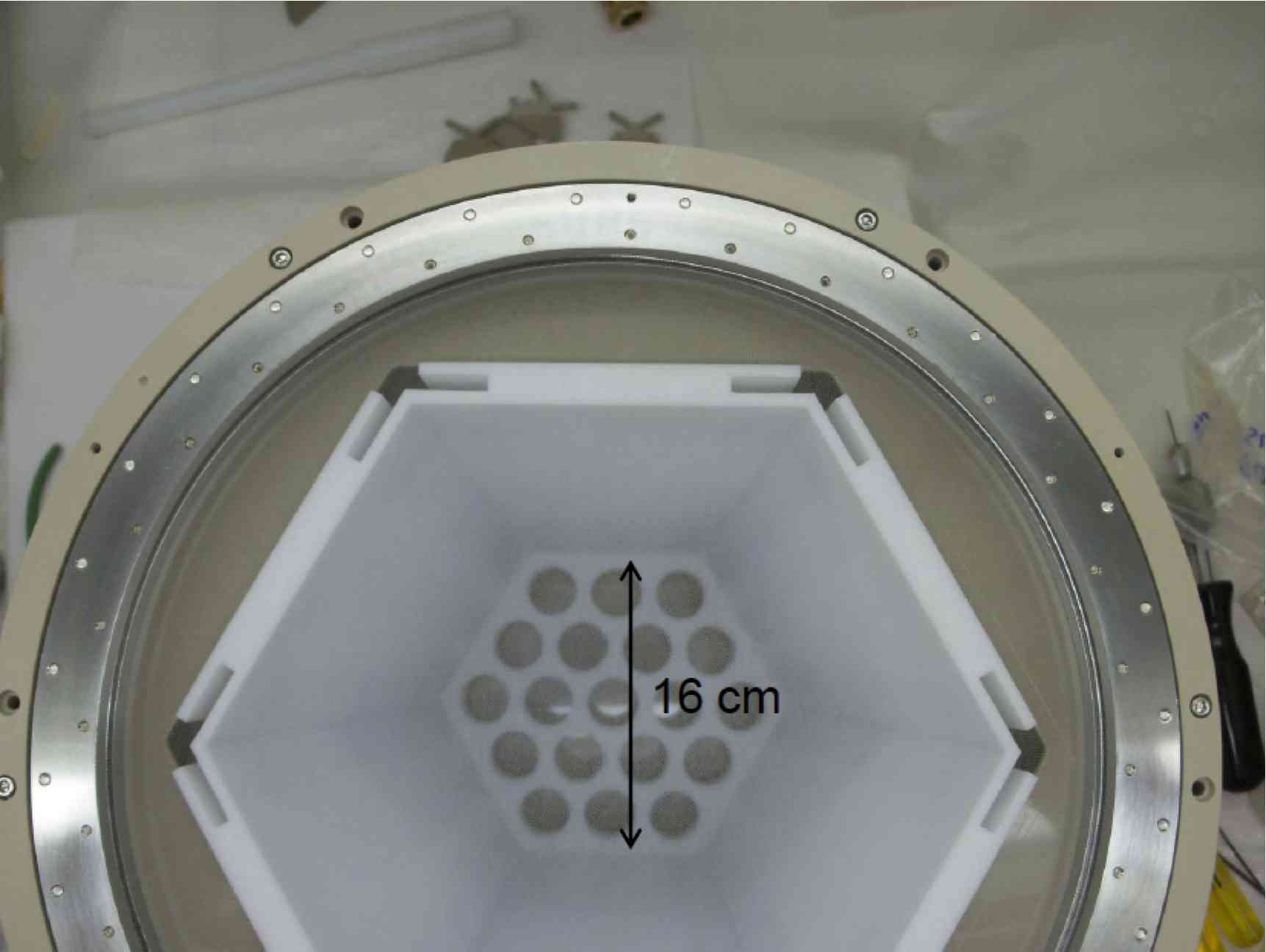} 
\end{center}
\caption{A view from the top of the NEXT1 field cage, showing the light tube, made of very reflective Teflon slabs. The energy plane honeycomb can clearly been seen through the transparent grids.} \label{fig.lt}
\end{figure}

The field cage (Figure \ref{fig.fc}) has a skeleton of peek, a very rigid, clean, non-degassing and non radioactive material. The field shaping rings of the field cage are made by cutting and machining aluminum
pipe. Figure \ref{fig.lt} shows the teflon light tube, made of PTFE.

\subsection{High voltage and feedthroughs}


The High Voltage is supplied to the Cathode and the Gate, in the electroluminescence region, through custom made high-voltage feed throughs (HVFT), described in the previous chapter. These have been tested to high vacuum and 100 kV without leaking or
sparking. The HVFT designed and built by Texas A\&M (as the field cage), are shown in
Figure \ref{fig.hvft}. 

The side of the chamber contains 8 CF40 size nipples. One set is located in the horizontal plane while the other
$135^\circ$. these contain radioactive source ports used for calibration of the TPC. The ports are made by welding a 0.5 mm blank
at the end of a 12 mm liquid feedthrough. The radioactive source is then located on the outside of the detector.

%
%
%
%
%

\subsection{The tracking plane}

\begin{figure}[bthp!]
\centering
\includegraphics[width=0.6\textwidth]{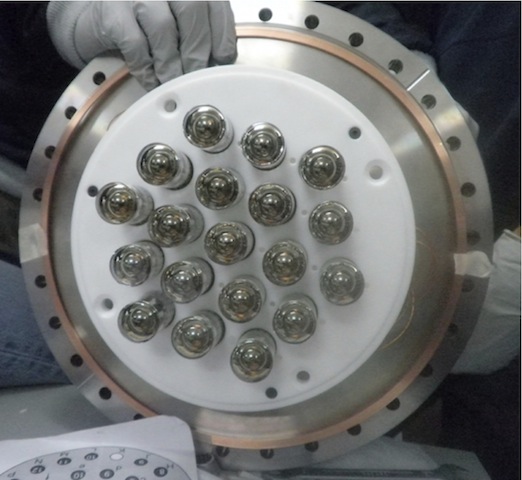}		
\caption{PMT Tracking plane installed at the TPC Anode in the commissioning of NEXT1 operation.}
\label{Fig:PMTs_trackingplane}
\end{figure}

The tracking plane
is responsible for the tracking and provides topological information on the events.
It is located closely behind the EL mesh-grids and provides a 2-D pixelization.
Figure \ref{Fig:PMTs_trackingplane} shows the PMT plane that gives the tracking function for Run I. 
\begin{figure}[htbp!]
\begin{center}
\includegraphics[width=0.85\textwidth]{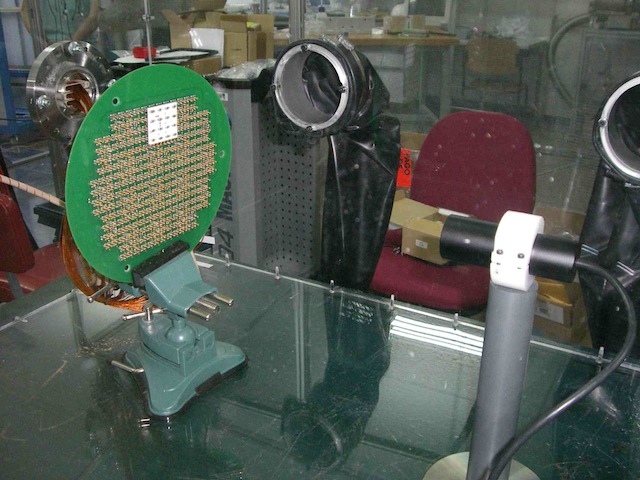}
\end{center}
\caption{The SiPM plane inside a test box, made of acrylic.}
\label{Fig:box}
\end{figure}
\begin{figure}[bthp!]
\begin{center}
\includegraphics[width=0.75\textwidth]{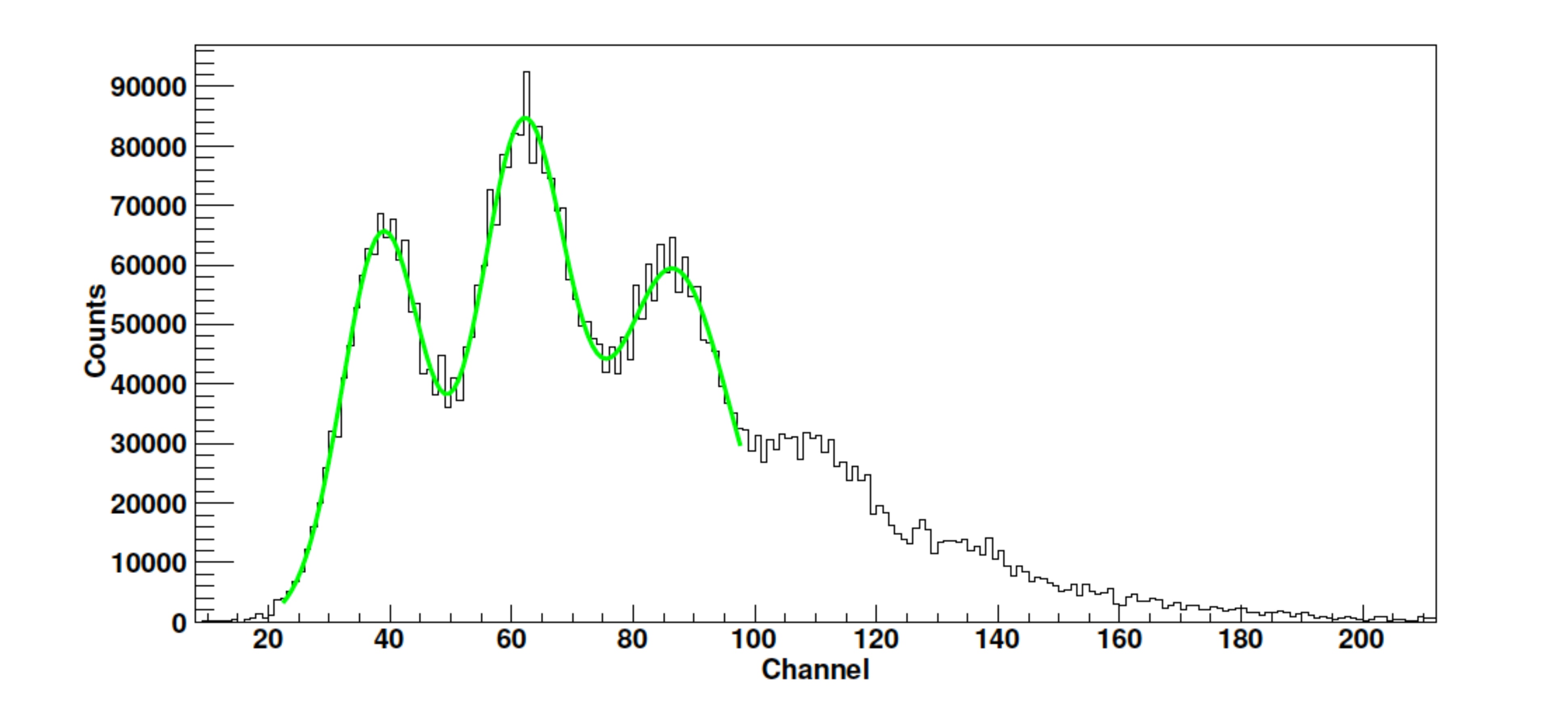}		
\end{center}
\caption{Photoelectron spectrum of the SiPMs connected to the DB in the SiPM plane.}
\label{Fig:phe}
\end{figure}
Figure \ref{Fig:box} shows the SiPM plane inside a test box, made of acrylic. The box is filled with nitrogen, that does not absorb the 172 nm produced by the VUV xenon lamp. The SiPM will be fully tested before installation in the chamber.  

\begin{figure}[htbp!]
\centering
\includegraphics[width=0.85\textwidth]{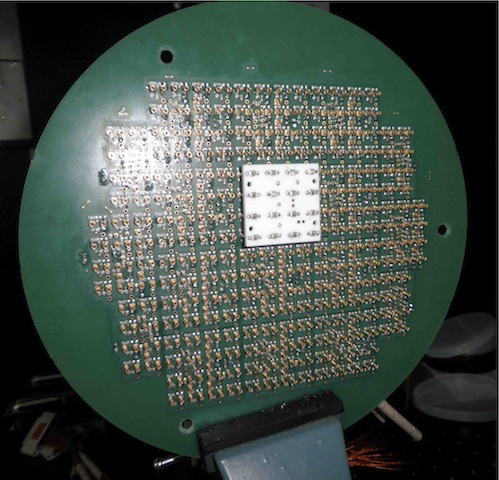}		
\caption{Mother Board with one SiPM Daughter board connected. }
\label{Fig:motherboard}
\end{figure}

Preliminary results are already available. Figure \ref{Fig:phe} shows the few-photoelectrons spectrum of one SiPM connected to the DB shown in Figure \ref{Fig:motherboard}. The peaks follow a gaussian distribution. One
can see clearly $N_\gamma=0,1,2...$. Notice that calibration of SiPMs is straight forward. The difference
between the centroid of two subsequent peaks is exactly the charge generated by one photon (e.g, one pe). In other words, the difference between peak centroids is a measurement of the gain.

\begin{figure}[htbp!]
\centering
\includegraphics[width=0.85\textwidth]{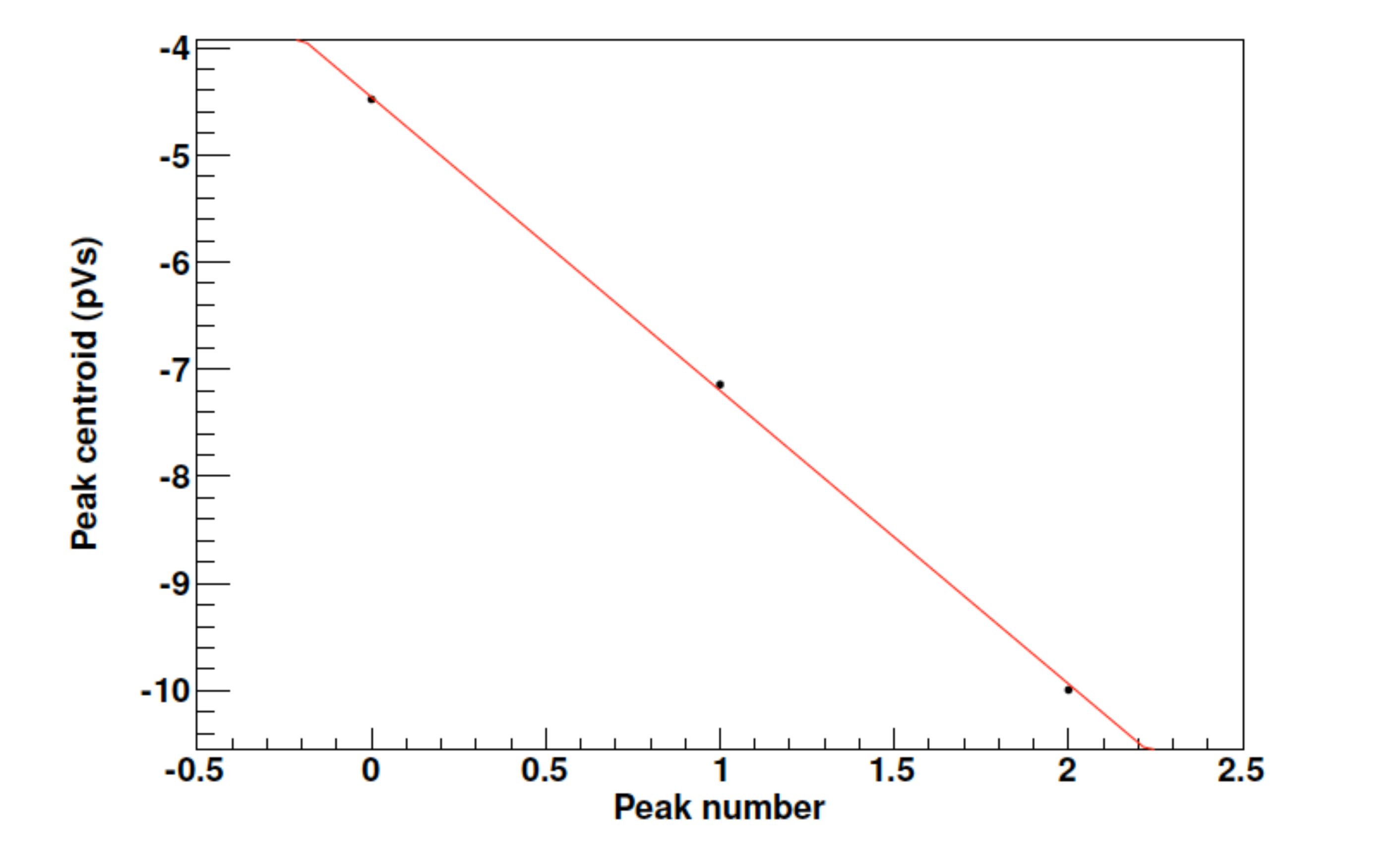}		
\caption{Charge generated by the simultaneous detection of 1,2... photons in SiPMs connected to the
DB in the SiPM plane.}
\label{Fig:gain}
\end{figure}

Figure \ref{Fig:gain} shows a linear fit to the centroid of the peak position versus the peak number. Writing:
\[
Q = G N_\gamma + Q_0
\]
Where Q is the charge corresponding to $N_\gamma=0,1,2...$, $Q_0$~ is the charge corresponding to 
$N_\gamma=0$ and $G$~is the gain, equal to the slope of the straight line fit shown in Figure~\ref{Fig:gain}:
\[
G=(3.417 \pm 0.002) \times 10^5
\]

\subsection{FE readout electronics for SiPMs}

\begin{figure}[phbt!]
\begin{center}
\includegraphics[width=0.75\textwidth]{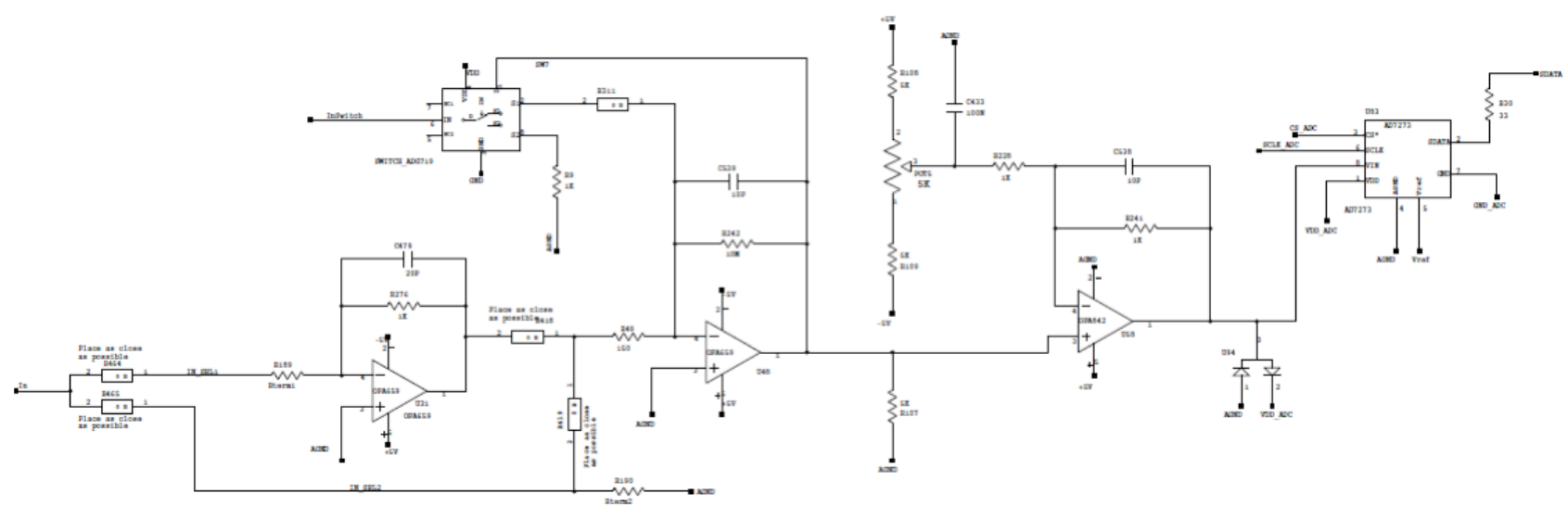}
\end{center} 
\vspace{-0.5cm}
\caption{\small Analog path of an ADC circuit for the SiPM front-end. From left to right, I/V converter, gated integrator, voltage amplifier and single-channel ADC. }
\label{fig:chip} 
\end{figure}

The readout electronics of the tracking plane poses a greater challenge than that of the PMTs (identical to the PMT electronics described for the LBNL prototype) due to the much higher number of channels. Figure  \ref{fig:chip} shows the analog path of an ADC circuit for the SiPM front end. It includes I/V converter, gated integrator, voltage amplifier and single-channel ADC. In NEXT1, this FE electronics is implemented as a part of a special card {\em outside} the chamber, and the signals are passed through flat cables and multi-pin feedthroughs. 

However, ten thousand signal wires across the TPC vessel, as needed for NEXT-100 appear as rather challenging. This leads us to the final improvement to the baseline design, which is to  install the front-end electronics, digitizers and some sort of data multiplexers inside the vessel in order to reduce the number of feedthroughs.

A possible solution consists in using 64-ch ASICs that include the analog chain (amplifier + integrator in 
Figure \ref{fig:chip}), the digitizers and the data multiplexer. A commercial serializer/deserializer chip interfaces the ASIC to an optical transceiver, using a single optical fiber for a full-duplex data connection to the DAQ stage (FECs equipped with specific add-in cards). With this multiplexing factor, 10k channels require just 157 optical links across the vessel. Alternatively to the use of an ASIC it is possible to develop small scale discreet electronics that will also fit
in the chamber (recall that the ANGEL design includes a thick ultra-pure copper ring to shield the electronics from the main gas volume).

\subsection{Commissioning}
%

The NEXT1 prototype has been designed and built in less than one year. The detector has been operating since January 2011. During the commissioning phase we have introduced a number of improvements to the original chamber, including new grids that withstand higher voltages. Drift voltages close to 50 kV and anode voltages close to 25 kV have been achieved. The detector has been operating routinely for many hours in a very stable way. A significant incident, however, has been the rupture, during the month of April, of the recirculation pump diaphragm, for causes yet unknown. The pump is being repaired and the system continues running with a spare, smaller pump. Additional safety systems will be build in the next few months to close automatically the system in the event of a similar failure, avoiding the loosing of valuable xenon. In addition, a system of slow control is being developed. This incident brings home the need of a triple diaphragm pump (as well as additional emergency devices) for the NEXT-100 gas system (see chapter 7).

The electronic system of the noise was originally quite high due to electronic pickup. Addition of RC filters, as described above has almost eliminated the problem. There remains a small pickup introduced in the power line of the PMTs that will be cured in the next few week.

\chapter{Initial results from the NEXT-1 prototypes} \label{sec.next1r}
\section{First results from the LBNL prototype}
\noindent 
The LBNL NEXT prototype went into operation in December 2010. Since then more than 2 Terabytes of data have
been taken at various pressures (10, 11 and 15 bar), drift electric fields (0.5 kV/cm through 2 kV/cm), E/P electroluminescent gains 
(1.0 though 2.6 kV/(cm bar)), trigger configurations, glass flow rates and with different calibration sources: $^{22}$Na, $^{137}$Cs, $^{241}$Am, $^{60}Co$ and cosmic ray muons.

\subsection{Setup and trigger}
\begin{figure}[p]
  \centering
  \includegraphics[scale=0.5]{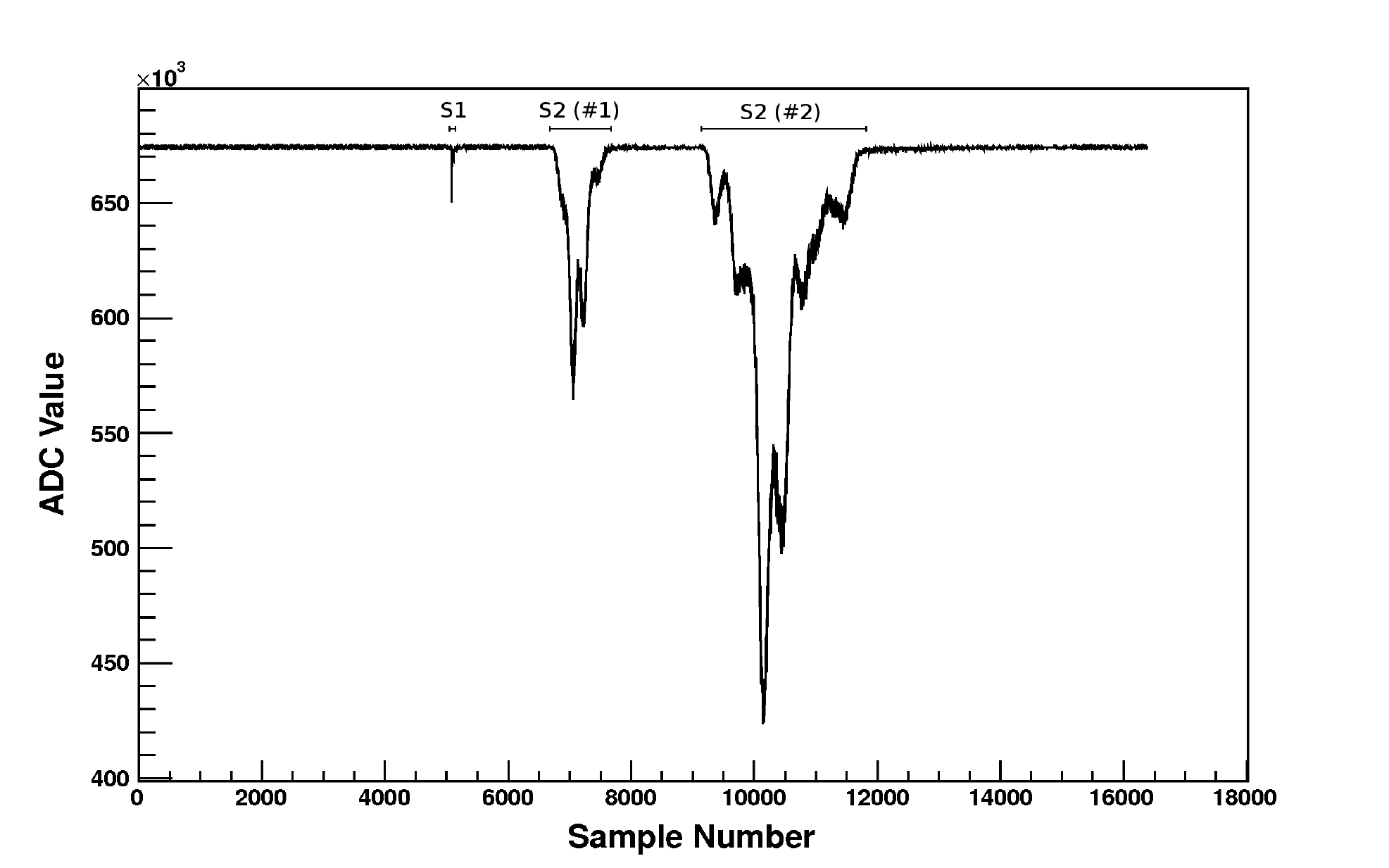}
  \caption{\label{run254_evt_5_0427_waveform_annotated} \textbf{Pulse finding and integration algorithm:} 
Shown is a typical 19-PMT summed waveform a $^{60}$Co source (1.1 and 1.3 MeV gammas). The S1 pulse is required to be a small width pulse. S2 pulses are separately identified and integrated if there is enough baseline level waveform samples between them. }
\end{figure}

\begin{figure}[p]
  \centering
  \includegraphics[scale=0.5]{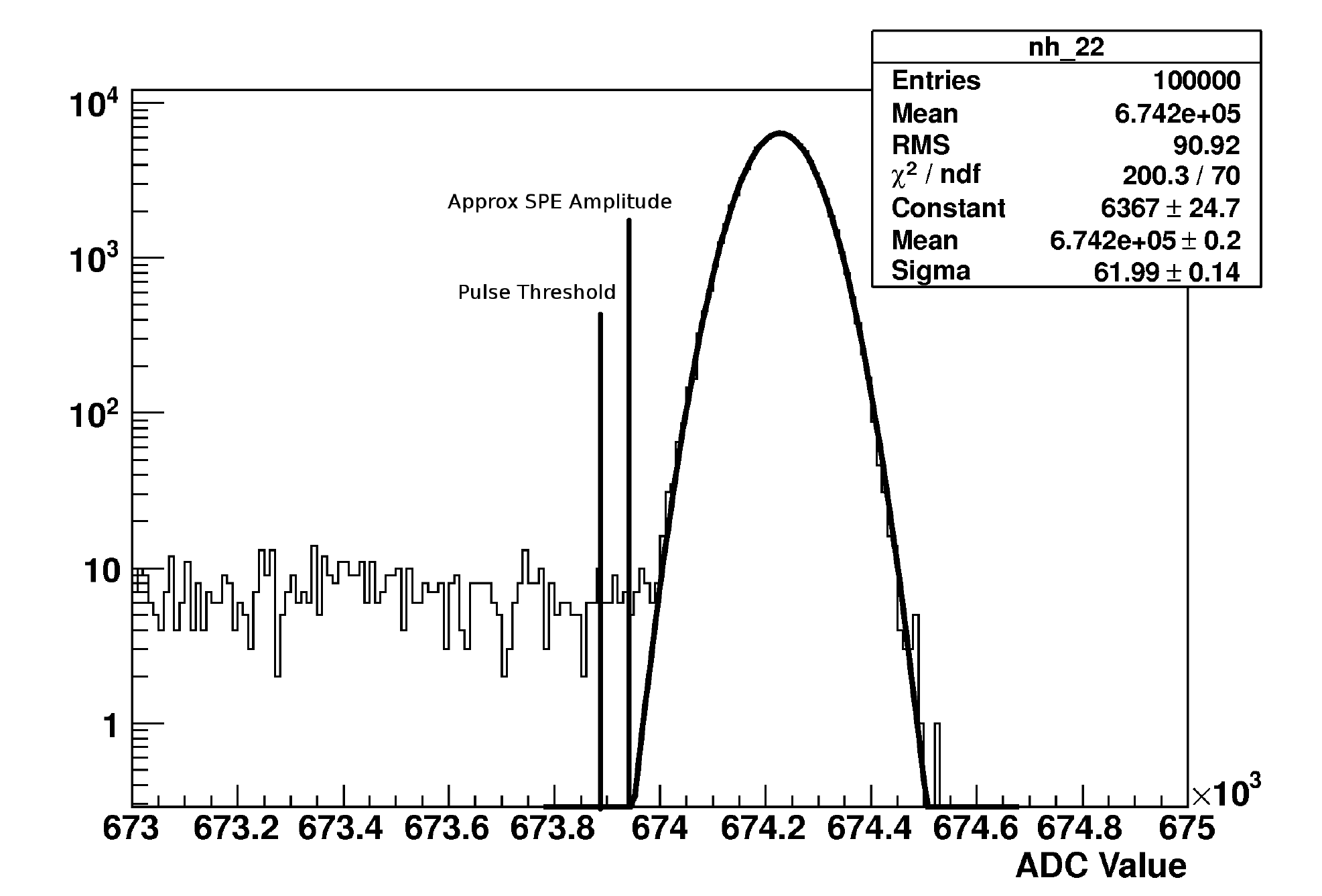}
  \caption{\label{nh22_run234_inc_range_annotated} \textbf{Noise and analysis threshold:} Histogram of pretrigger samples
of the 19-PMT summed waveforms with a fit to a gaussian of the central noise band. Below the main noise peak are samples from genuine PMT pulses. Shown also are the 5$\sigma$ analysis threshold to find S1 and S2 pulses as well as the typical amplitude of a single photoelectron. Thus the effective offline threshold is highly efficient. Of note also is the lack of outliers in the positive side indicating the absence of extraneous non-gaussian noise sources. }
\end{figure}

In the current implementation the TPC has 19 PMTs on the energy measurement side and no dedicated tracking sensor.
The 19 PMTs are 13 cm away for the 3mm EL gap region and the active drift volume is 8 cm long in the drift direction and
has a 14 cm transversal span. Typical high voltage differences across the EL gap are 8 to 15 kV depending on the pressure and 
typical voltage differences across the drift region are 4 to 16 kV. The PMTs are typically operated at a gain of 1.4x10$^6$.   
 
The trigger configuration has evolved over the 4 months since first operation. The most successful configurations involve a fast 3-fold coincidence with a few photoelectron threshold from groups of PMTs which initiates a 5-100 $\mu s$ gate during which a single channel S2 energy threshold is required to fire. The S2 energy threshold is built from a 20$\mu s$ peaking charge amplifier.
 
Signals from the 19 PMTs are then digitized with the 16-bit Struck digitizers for 160 $\mu s$ leaving typically 60 $\mu s$ of pretrigger and 100 $\mu s$ of signal 
region including the S1 and S2 signals. Figure \ref{run254_evt_5_0427_waveform_annotated} shows a typical waveform resulting from the sum of the 19 PMTs signals. 
Valid events are selected to have a narrow S1 signal followed by one or more S2 wide signals.  As seen in Figure \ref{nh22_run234_inc_range_annotated} the summed 
waveforms have small noise permitting single photoelectron efficient 5$\sigma$ analysis thresholds.

\subsection{The Cs-137 analysis}
In order to demonstrate the capabilities and performance of the LBNL NEXT prototype we will describe the analysis steps and data features for a single run taken with a 
1 mCi $^{137}$Cs 662 keV gamma ray source  highly collimated and on the TPC axis entering the pressure vessel through a 2 mm thick stainless steel window on a 
reentrance port. This run taken at 10 bar pressure had 
-10.6 kV applied to the mesh that defines the start of the drift region, +2.7 kV to the first EL mesh and +10.6kV to the second EL mesh. 

\begin{figure}[tb]
  \centering
  \includegraphics[scale=0.5]{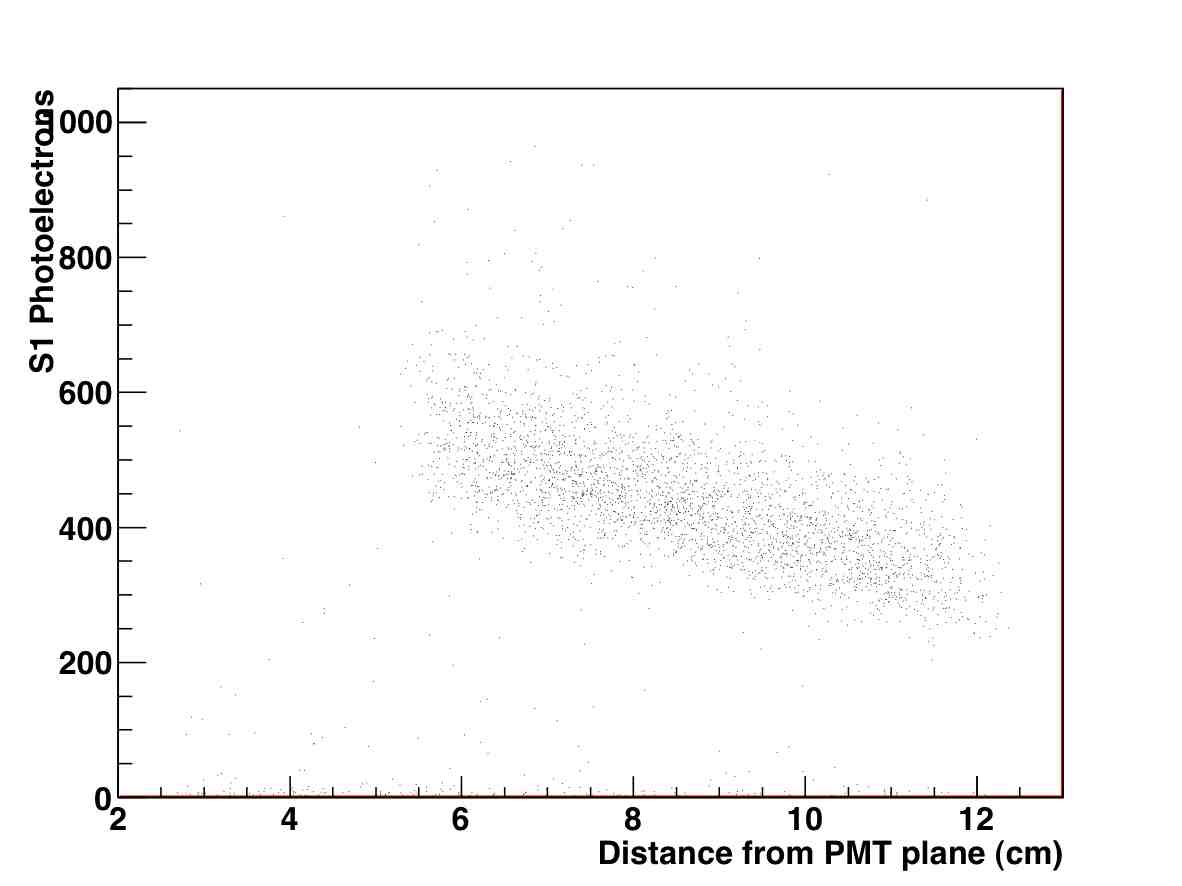}
  \caption{\label{fig_cs137_run234_s1_vs_distance} \textbf{S1 signal vs position:} 
The approximate number of primary scintillation S1 photoelectrons measured for 662 keV depositions 
is shown as a function of the event distance from the PMT plane (computed from the drift time). 
The large dependence on position is due to the large solid angle variation with position. Still, the S1 
signal is strong in all cases (hundreds of photoelectrons) leading to a very efficient and low noise 
$t_0$ measurement. A good S1 measurement may facilitate an event-by-event recombination estimation.}
\end{figure}

The analysis starts with the identification of S1 signal, whose primary function is to determine the $t_0$ for the TPC drift. In addition, the S1 signals are useful 
for identifying pileup events and could shed light on the amount of recombination. As seen in figure \ref{fig_cs137_run234_s1_vs_distance} hundreds of photoelectrons 
are measured in the S1 of full energy 662 keV events thus reliable S1 signals can be obtained for energy depositions as low as 10 keV. The large position dependence 
of the S1 signal is mainly a solid angle effect at play due to the limited reflectivity of the PTFE walls. 

\begin{figure}[tb]
  \centering
  \includegraphics[scale=0.5]{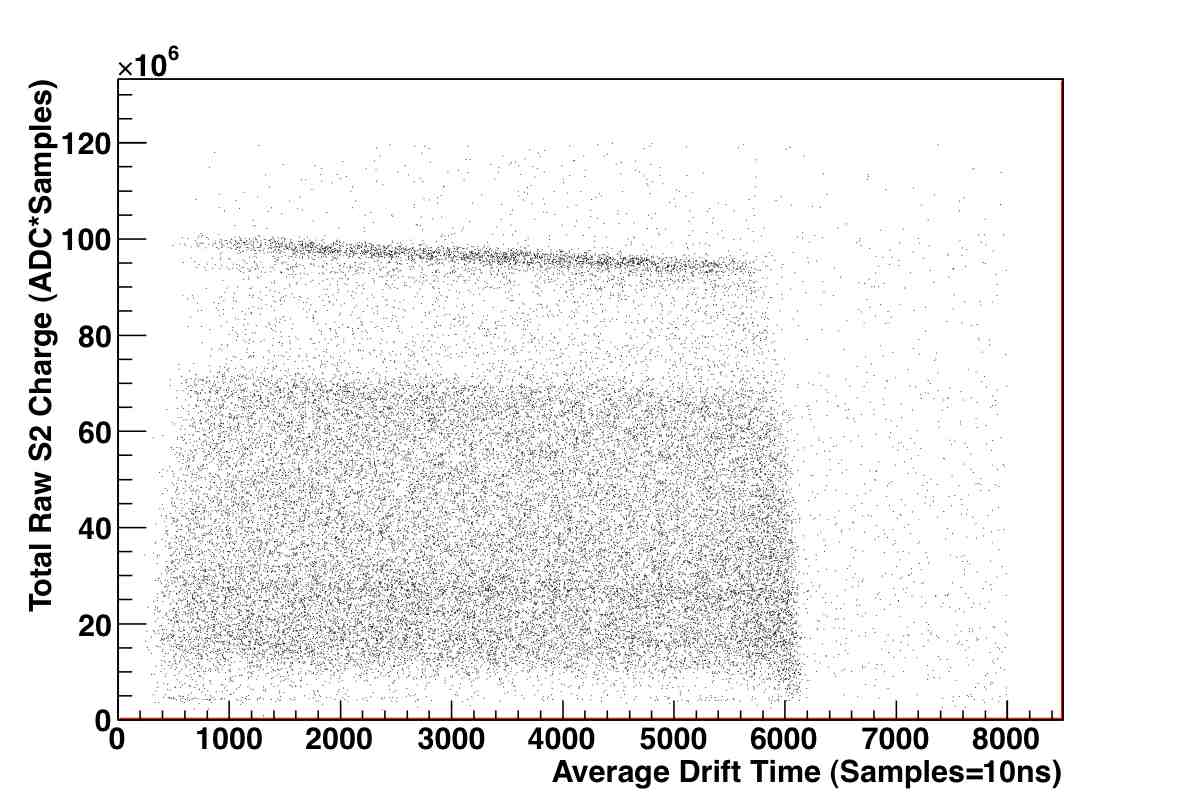}
  \caption{\label{fig_cs137_run234_s2_vs_drift} \textbf{Raw S2 vs Drift Time:} Total charge from the 
electroluminescent (S2) pulse versus the drift time with a 1 mCi $^{137}$Cs 662 keV gamma ray source 
highly collimated and on the TPC axis. The band slopes are the result of attachment impurities in the 
xenon. 
These data taken at 10 bar pressure show an electron lifetime of 900 microseconds. This, according
 to a Magboltz simulation, corresponds to O$_2$ impurities at the 0.05 part per million in the drift region.}
\end{figure}

Figure \ref{fig_cs137_run234_s2_vs_drift} shows the integrated S2 charge versus the drift time for valid events in the run with an S1 and one or more S2 pulses. A highest narrow energy band corresponds to full energy 662 keV depositions. The drift time span of the events corresponds to the maximum drift time, given by the maximum 8 cm drift length and the drift velocity. The slope in the energy bands is due to electron attachment by O$_2$ molecules in the drift region. 

\begin{figure}[!tbp]
  \centering
  \includegraphics[scale=0.5]{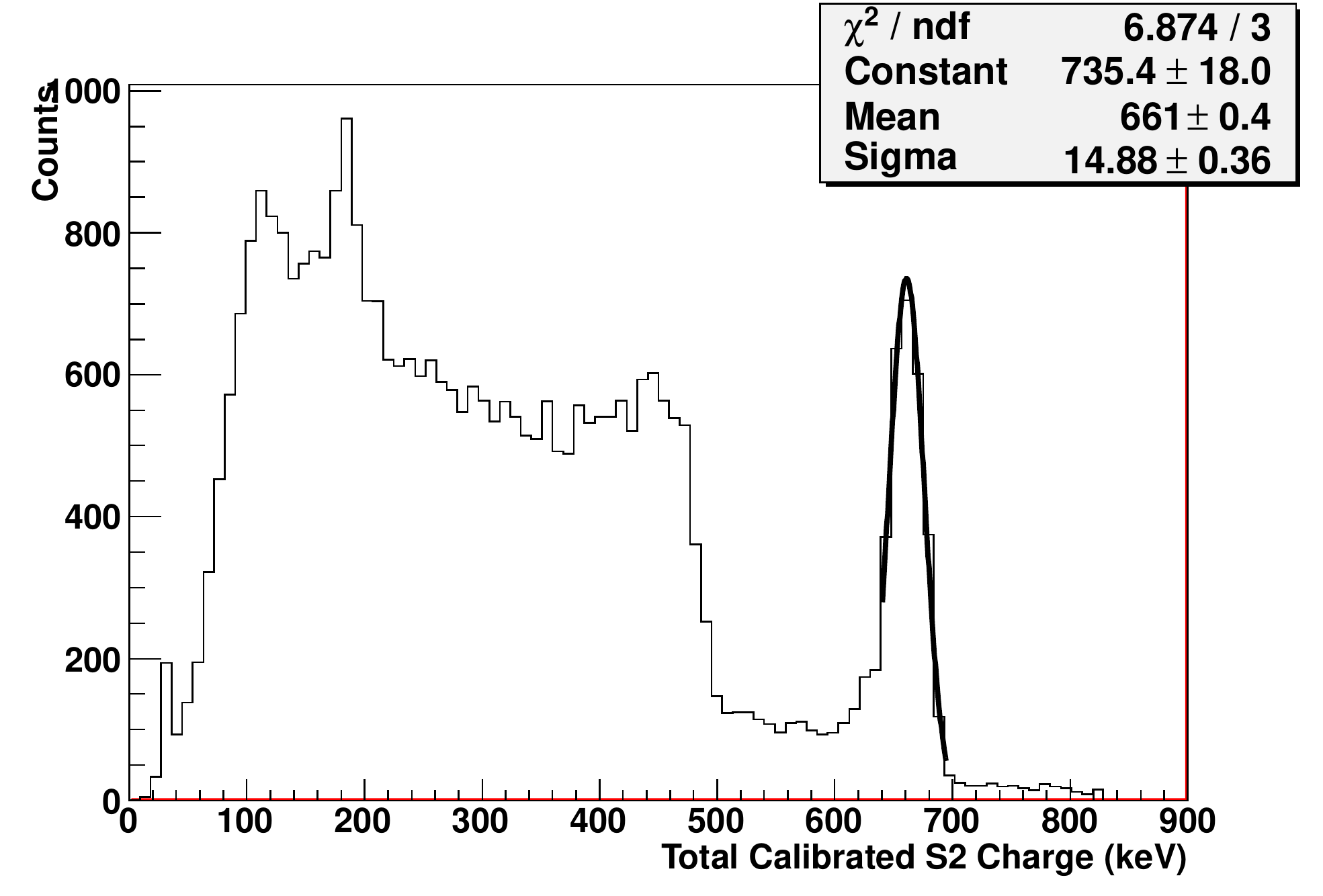}
  \caption{\label{fig_cs137_run234_s2_spectrum} \textbf{Calibrated S2 charge spectrum:} Shown is the S2 
spectrum calibrated using the full energy peak of $^{137}$Cs. In this run at 10 bar and $E/P$ of 2.56 
$kV/(bar cm)$ in the EL region about 300,000 photoelectrons are measured for events in the full energy 
peak. The Compton edge from single scatters with the loss of the scattered gamma, expected at 477 keV, 
is clearly visible. The peak from backscattered gamma rays, expected at 184 keV, is also visible above 
the normal Compton spectrum. 
The rising edge near 100 keV is due to the hardware trigger threshold. The peak near 
30keV is due to xenon x-rays (29-36 keV) following photoelctric absorption of gamma rays.
 The flat region between 500 and 600 keV is due to multiple Compton scattering events with escape of 
scattered gamma/s. The uncorrected energy resolution is 5.3\% FWHM for 662 keV.}
\end{figure}

The charge spectrum for this run shown in figure \ref{fig_cs137_run234_s2_spectrum}
shows a clear full energy peak where the distribution is calibrated, a Compton edge at the expected ~480 keV 
and a Compton backscatter peak at the expected ~180 keV. Even in the presence of the hardware trigger threshold in the 
~100 keV region, a narrow 30 keV peak can be discerned caused by events in which only the xenon x-rays (29-36 keV) following photoelectric interactions entered the 
active region of the TPC.

\begin{figure}[!tbp]
  \centering
  \includegraphics[scale=0.5]{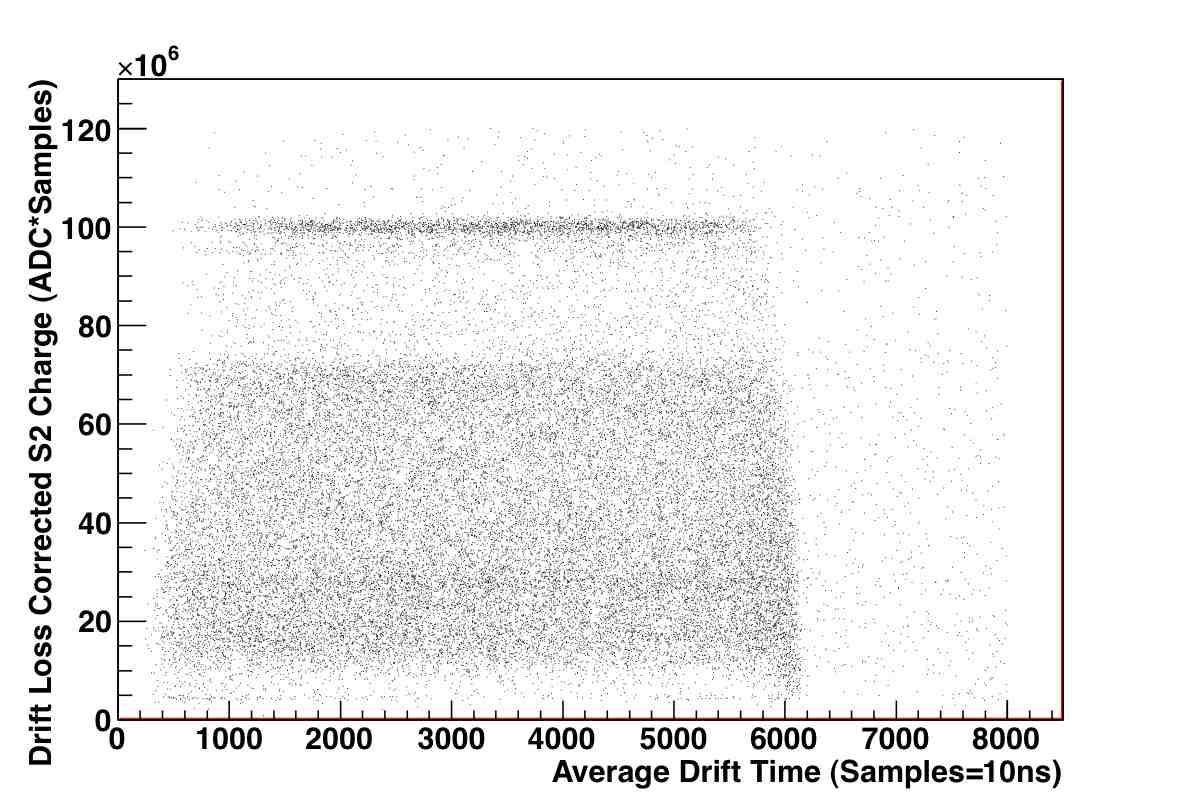}
  \caption{\label{fig_cs137_run234_s2driftcor_vs_drift} \textbf{Attachment corrected S2 versus Drift Time:} 
The drift determined from the difference in time between the primary scintillation (S1) pulse 
and the EL (S2) pulse was used to correct the S2 charge for the attachment losses resulting in straight 
horizontal bands for the energy peak/features. The drift velocity of 1.29 $mm/\mu s$ is obtained from the 
measured maximum drift time, in this case 62 $\mu s$, and the drift region length of 8 $cm$.  }
\end{figure}

\begin{figure}[!ptb]
  \centering
  \includegraphics[scale=0.5]{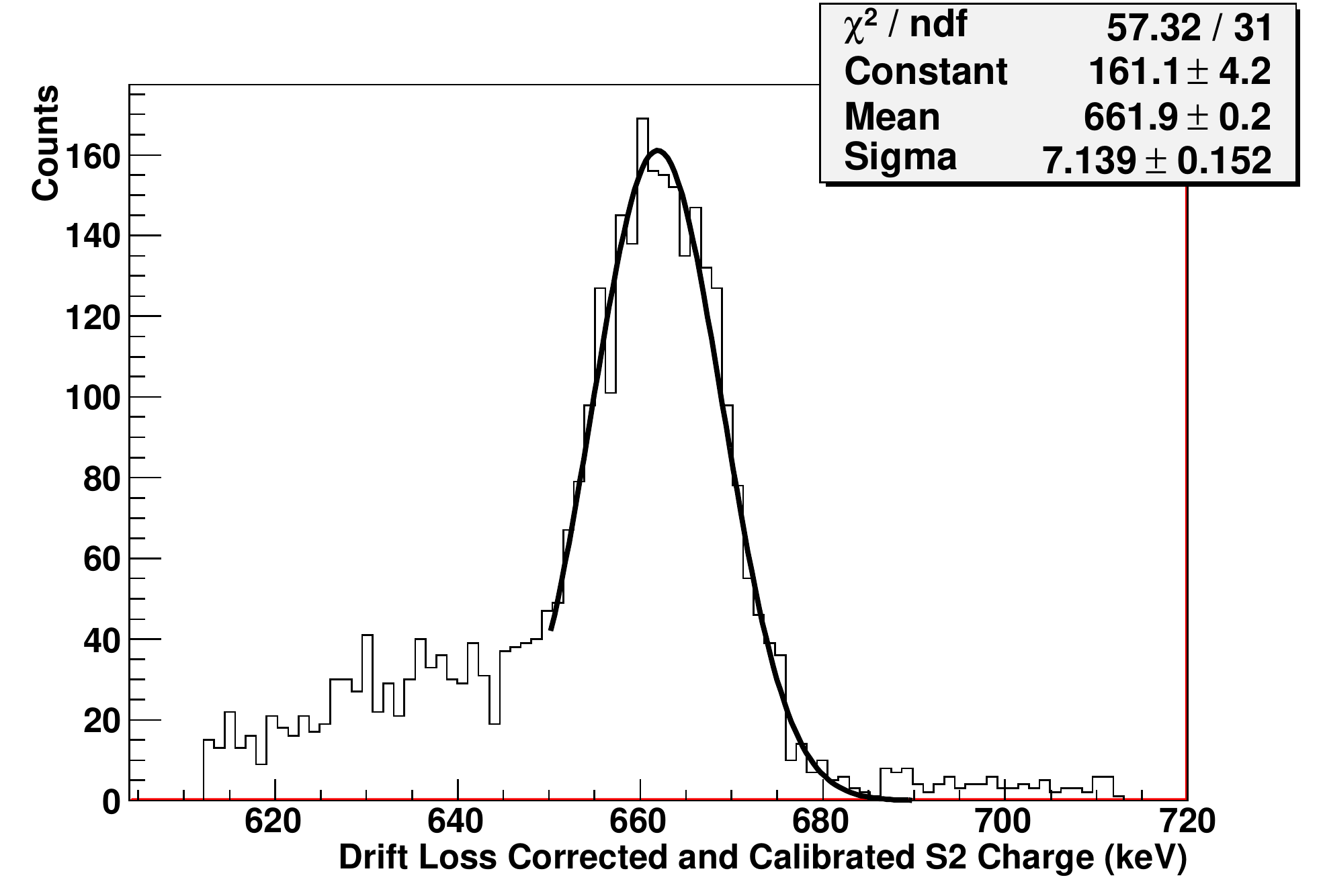}
  \caption{\label{fig_cs137_run234_s2driftcor_peak} \textbf{Attachment corrected Energy Resolution:} 
The attachment corrected energy resolution is 2.5\% FWHM for 662 keV.}
\end{figure}

The electron attachment energy correction for the exponential charge loss versus drift time is applied to the data a posteriori through the use of charge moments constructed 
from the taylor expansion of an $e^x$ weighting function. Figure \ref{fig_cs137_run234_s2driftcor_vs_drift}
shows the attachment corrected data showing horizontal energy bands. Figure \ref{fig_cs137_run234_s2driftcor_peak} 
shows a fit to the attachment corrected spectrum with a 2.5\% FWHM for the 662 keV line. 
About 300,000 photoelectrons are measured for full energy peaks. This corresponds to about 10 photoelectrons (PE) measured per ionization electron. At this EL gain 
and photoelectron yield the PE statistics have a similar contribution to the resolution function as the Fano factor, about 0.6\%.

\begin{figure}[!ptb]
  \centering
  \includegraphics[scale=0.5]{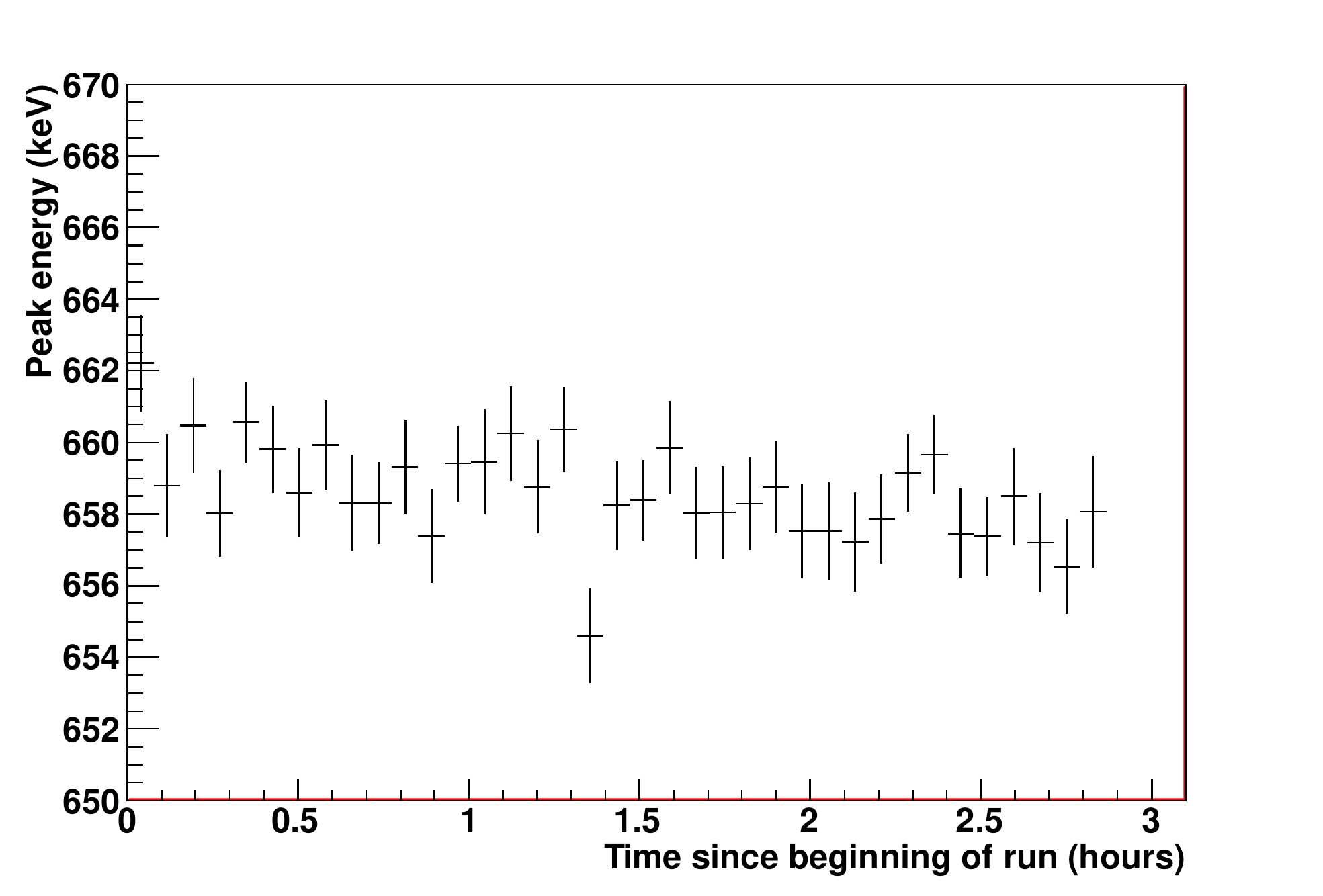}
  \caption{\label{fig_cs137_run234_timevariation} \textbf{Time stability:} 
This run that collected 40,000 events in 3 hours shows a peak-to-peak gain drift of 0.6\%. No correction for this effect is applied for  this run since its contribution is about 0.2\% in quadrature to the resolution. Trends in the measured 
gain (as the slow one seen in this run) can be caused by changes in temperature, pressure, 
PMT high voltages, grids high voltages and electronics. In situ and continuous calibration sources are 
important to control this type of systematic shifts as well as monitoring systems. }
\end{figure}

We investigated the stability of the data by looking at the peak energy as a function of time. Figure 
\ref{fig_cs137_run234_timevariation} shows a small trend of gain loss during the 3 hour duration of the run. However, 
the effect is small and its contribution would amount to 0.2\% in quadrature to the resolution. As such, we did not include an explicit time correction to the data. 
Continuous monitoring of the voltages, currents, pressure, temperatures, flows are being implemented and are important to control and track this type of systematic 
effect. In addition, a set of LEDs will be installed inside the TPC volume to provide end-to-end electronics and PMT calibration as well as data stability information. 

\begin{figure}[!tbp]
  \centering
  \includegraphics[scale=0.5]{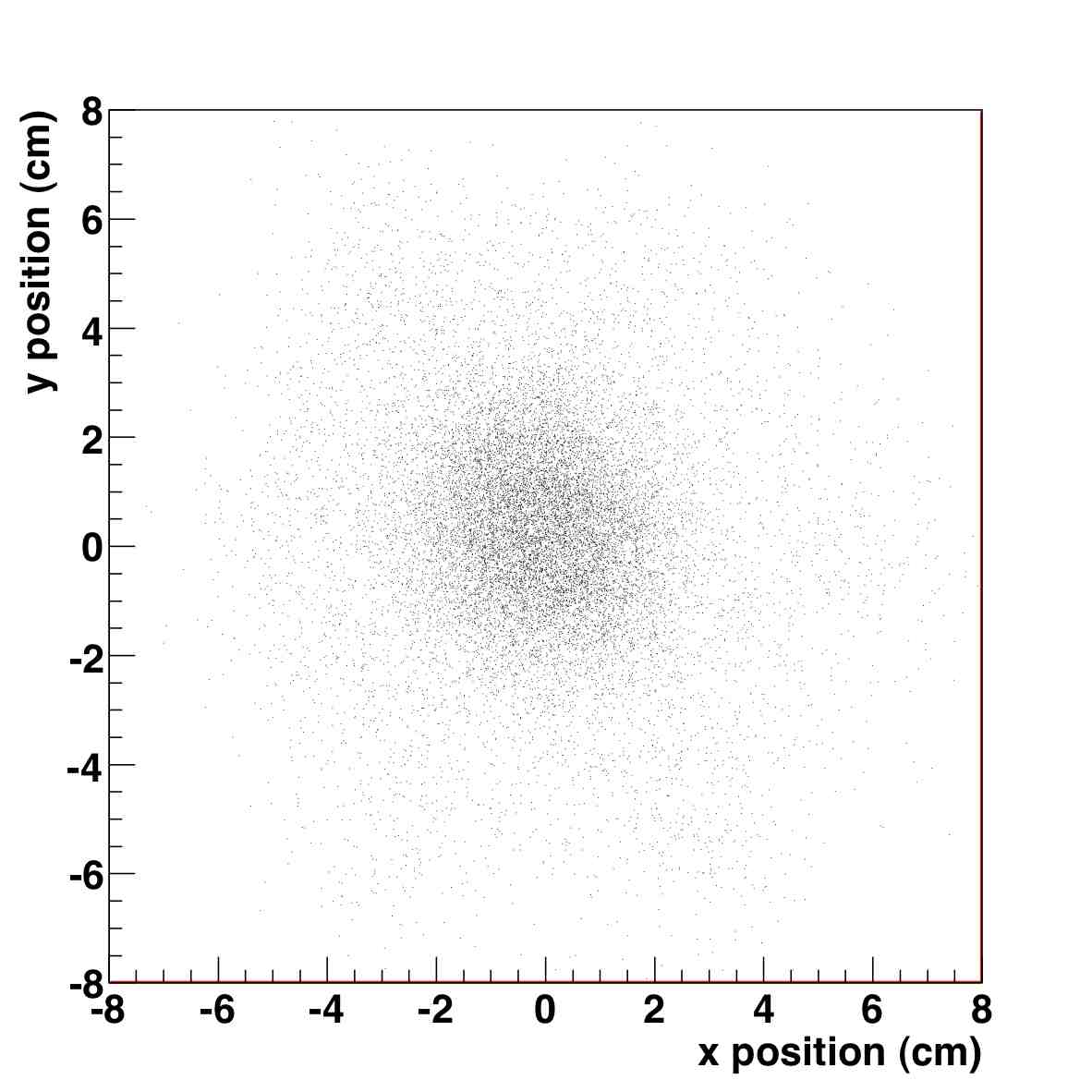}
  \caption{\label{fig_cs137_run234_y_vs_x} \textbf{Reconstructed average position of events:} 
The charges from the 19 PMTs are used to calculate a weighted x and y average position for each event.
The energy measurement PMTs are 13 cm away from the EL region and thus, for a point source, the light 
pattern is relatively uniform on the PMT plane. However, the finite reflectivity of the PTFE wall 
(about 50\% reflectivity) together with the solid angle differences and the large number of photoelectrons
collected give substantial position information. The edge points of the figure show the hexagonal shape 
of the EL region. The bright region in the center is due to gammas from the collimated on-axis Cs radioactive source
that deposit most of their energy near the TPC axis.}
\end{figure}

\begin{figure}[!ptb]
  \centering
  \includegraphics[scale=0.5]{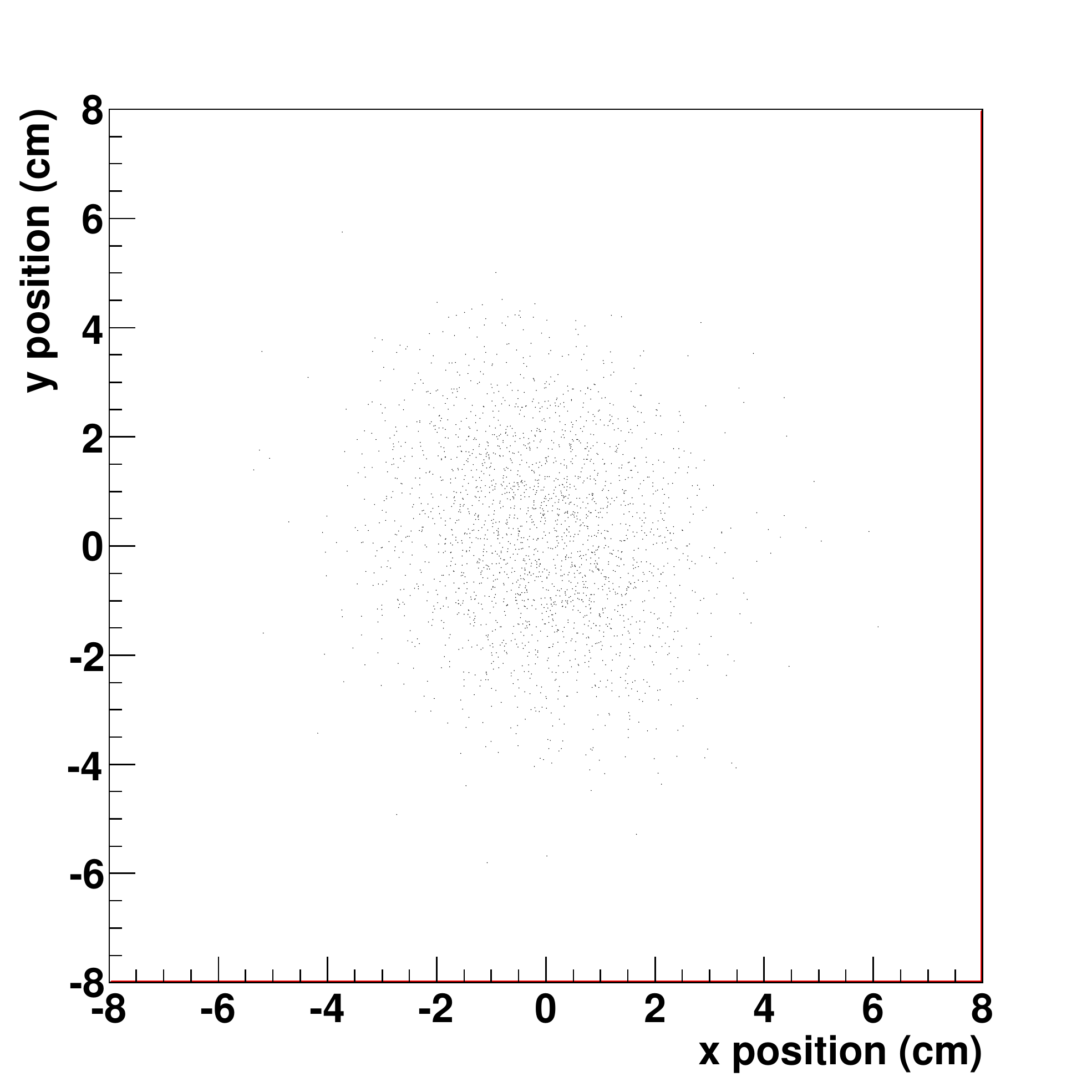}
  \caption{\label{fig_cs137_run234_y_vs_x_peak} \textbf{Reconstructed position of full energy events:} 
Shown are the positions of the subset of events that deposited energies within 30 keV of the peak. 
The spot size is approximately 2 cm in radius, and is primarily due to the source and collimator geometry 
and distance from the TPC active region.}
\end{figure}

Even though the PMT plane is 13 cm away from the EL region, the TPC with only energy side sensors has position sensitivity and 
thus tracking capabilities. For each valid event an average x and y position is found by weighting the PMT positions by the charge observed in each. Because the light 
is relatively uniform at the PMT plane (20-30\% variations) a scaling factor (of ~30) is required to transform the calculated average positions to true x-y TPC 
coordinates.  Figure \ref{fig_cs137_run234_y_vs_x} shows the position of valid events with one S2 pulse. The hexagonal boundaries of the drift region can be clearly 
identified. The dense region in the middle is the spot size of the source after the collimation (a 3 mm hole on a 8 cm long lead block placed about 70 cm from the TPC 
active region).  Figure \ref{fig_cs137_run234_y_vs_x_peak} shows the subset of events that were in the full energy peak forming a spot size of about 2 cm radius.

\begin{figure}[!ptb]
  \centering
  \includegraphics[scale=0.5]{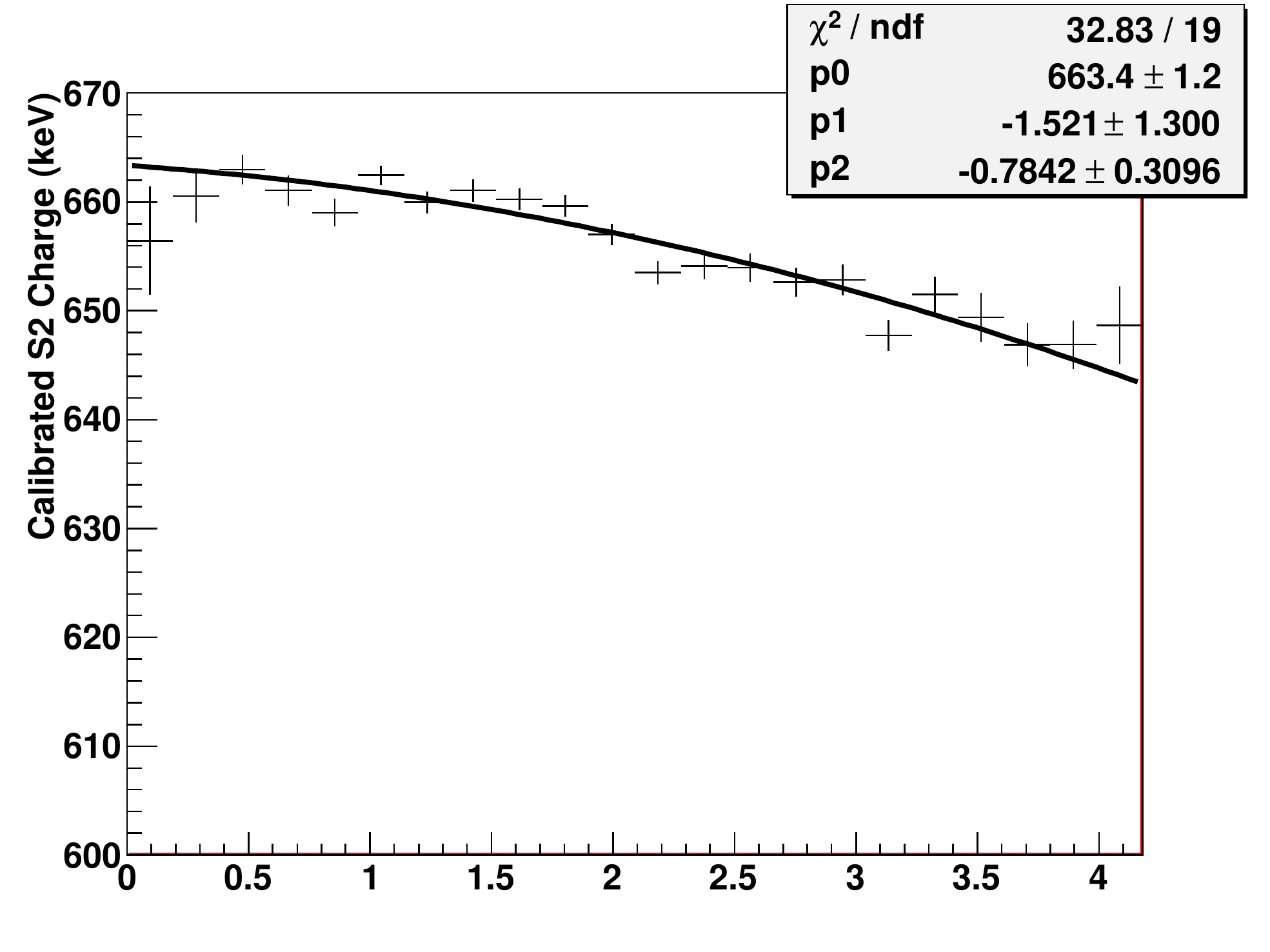}
  \caption{\label{fig_cs137_run234_radialcorr} \textbf{Full 
Energy peak charge vs radial position (in cm):} 
Shown is the full energy peak charge as a function radial position (in cm) of the event. A parabolic fit 
describes the radial gain dependence well. A peak-to-peak variation of 2\% is observed. For this analysis 
only events with radius less than 1.5 cm are used where the gain shows good uniformity without a 
correction. Possible sources of this radial dependence to the gain include: 
forward scatters of gammas in the collimator or window, optical response non-uniformity from finite 
reflectivity of the walls, mesh deformation due to electrostatic force. }
\end{figure}

Systematic gain variations are expected as a function of the event location from forward scatters of gammas in the collimator or window, from optical response 
non-uniformity due to finite reflectivity of the walls and to mesh deformation due to the electrostatic force amongst others. In particular radial systematic gain 
effects need to be measured and understood. Figure \ref{fig_cs137_run234_radialcorr} shows the full energy peak position as a function of the reconstructed average 
radius of the event. The systematic effect observed is small  (or absent) in the region with radius smaller than 1.5 cm. Since most of the full energy events are the 
central region, a cut was applied to the data to select only central events thus avoiding the larger radius gain variations.   

\begin{figure}[!ptb]
  \centering
  \includegraphics[scale=0.5]{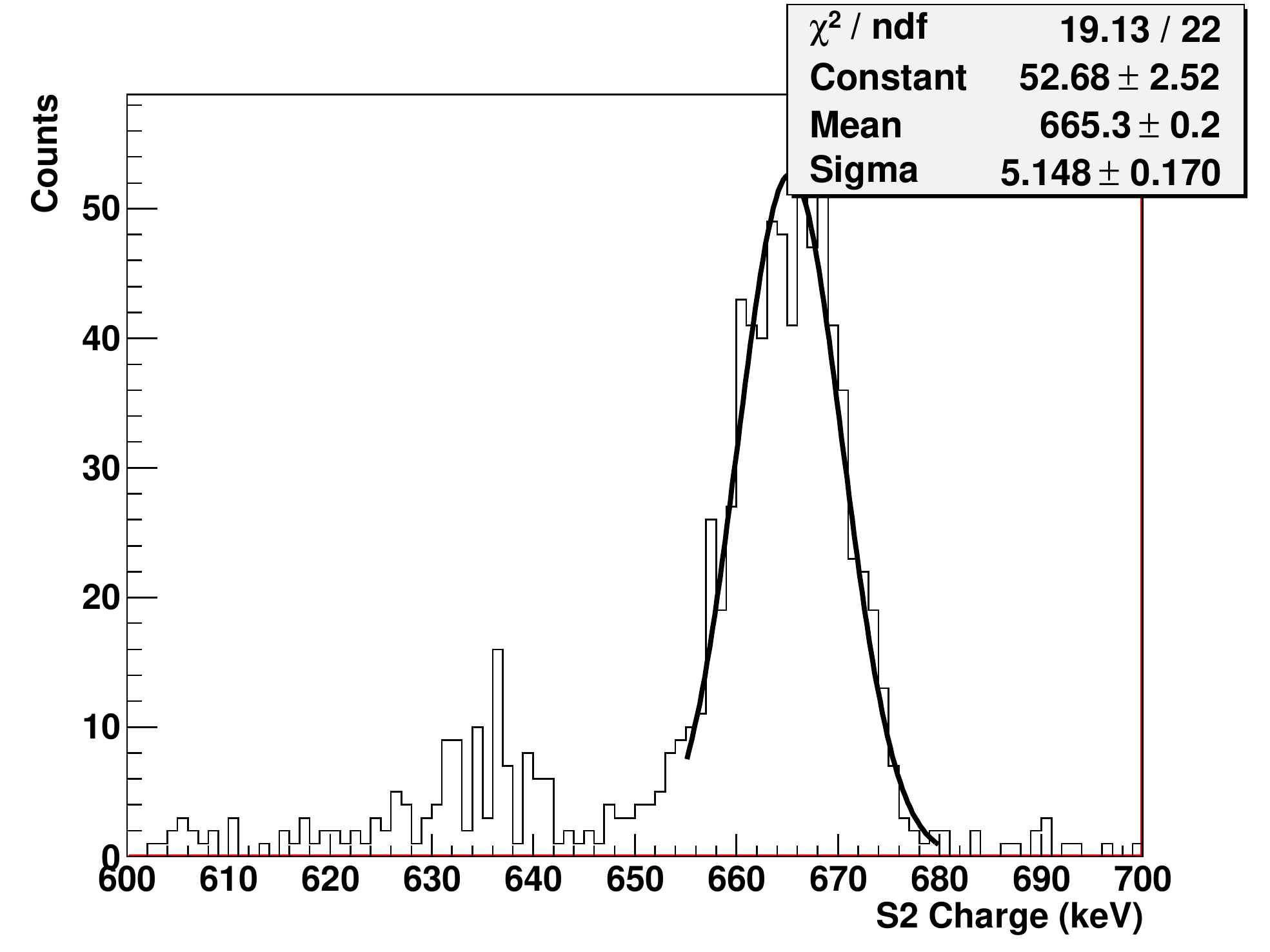}
  \caption{\label{fig_cs137_run234_peak_central_corrected} \textbf{Energy Resolution for central events:} 
Calibrated and attachment corrected spectrum for events with reconstructed radius less than 1.5 cm. 
The FWHM is 1.8\% at 662 keV. This was obtained at 10 bar, with 1.7 $kV/cm$ drift field and $E/P$ of 2.56 
kV/(bar*cm) in the EL region. The smaller peak at 635 keV is due to events where the xenon x-ray/s 
(29-36 keV) were not absorbed in the drift region.}
\end{figure}

After selecting central events the spectrum in figure \ref{fig_cs137_run234_peak_central_corrected} is obtained.
It shows a FWHM of 1.8\% at 662 keV and a well formed gaussian response shape. A smaller peak 30 keV lower is due to xenon x-rays escaping from the active region. 
This results is representative of our good energy resolution configurations and runs.
In the coming months we expect to explore the systematic effects that may be limiting the resolution from the 0.9\%
expected from the Fano factor and total PE statistics.

\subsection{Other analysis}

\begin{figure}[!tbp]
  \centering
  \includegraphics[scale=0.5]{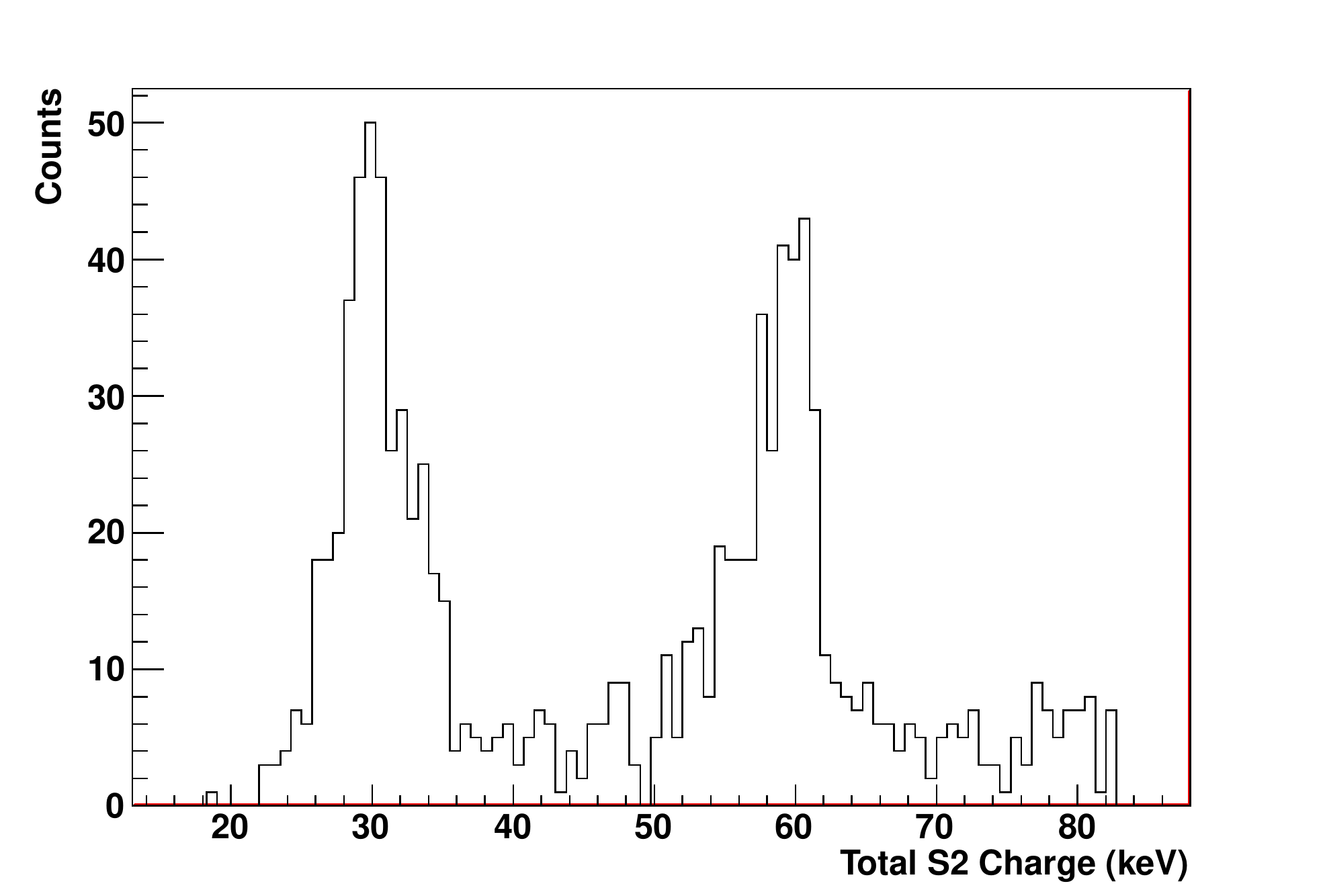}
  \caption{\label{am241_run255_spectrum} \textbf{$^{241}Am$ spectrum:} 
Calibrated and attachment-corrected spectrum for an $^{241}Am$ source placed outside the chamber on the 
reentrance port window (2mm thick), on-axis and about 7.7 cm from the drift region. The peak around 
60 keV (calibrated using the 662 keV Cs line) corresponds the full energy depositions of the 59.4 keV 
gamma rays. The peak at 30 keV is due to photoelectric interactions of the gamma ray with either the 
photoelectric electron or the xenon x-ray photons. }
\end{figure}

\begin{figure}[!tbp]
  \centering
  \includegraphics[scale=0.5]{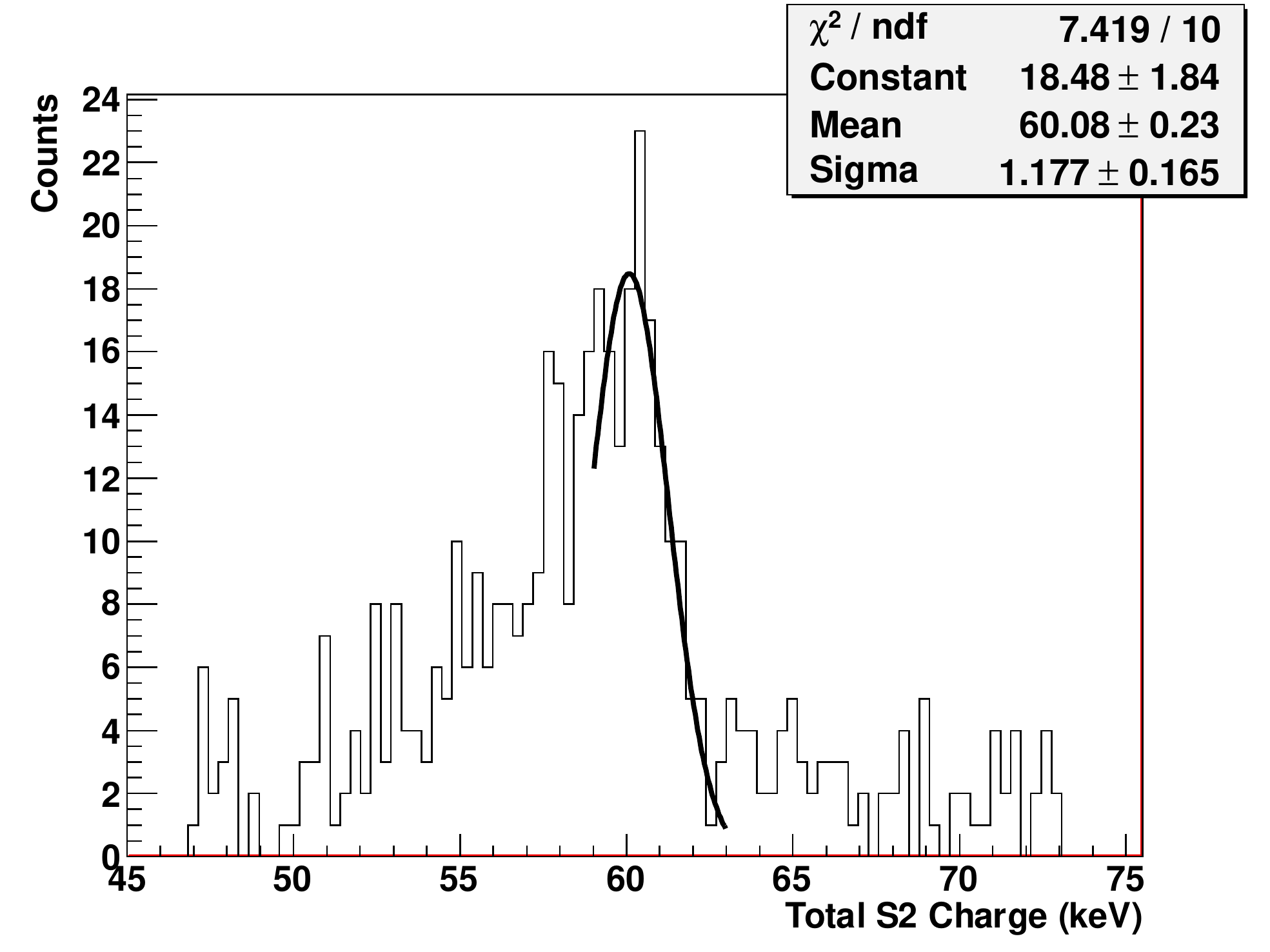}
  \caption{\label{fig_am241_run255_60kevfit} \textbf{Energy resolution at 60 keV:} 
The $^{241}Am$ full energy peak shows a sharp high edge. A fit to the upper edge shows a 4.6\% FWHM at 60 keV. 
The expected resolution from the Fano factor and the photoelectron statistics is 2.6\%. }
\end{figure}

Besides the workhorse $^{137}$Cs calibration source, we analyzed events with 511 keV photons from positron annihilation and 59.4 keV from an $^{241}$Am source. 
The energy spectrum (with corrections) is shown in figure \ref{am241_run255_spectrum}. A fit to the upper edge of the full energy peak shown in figure 
\ref{fig_am241_run255_60kevfit} yielded a  4.6\% FWHM resolution at the 59.4 keV peak. More investigations are planned to understand the asymmetric nature of this 
peak. A naive  $1/\sqrt{E}$ of this resolution to 662 keV would render 1.4\% (0.7\% to \Qbb).  

\begin{figure}[!ptb]
  \centering
  \includegraphics[width=\textwidth]{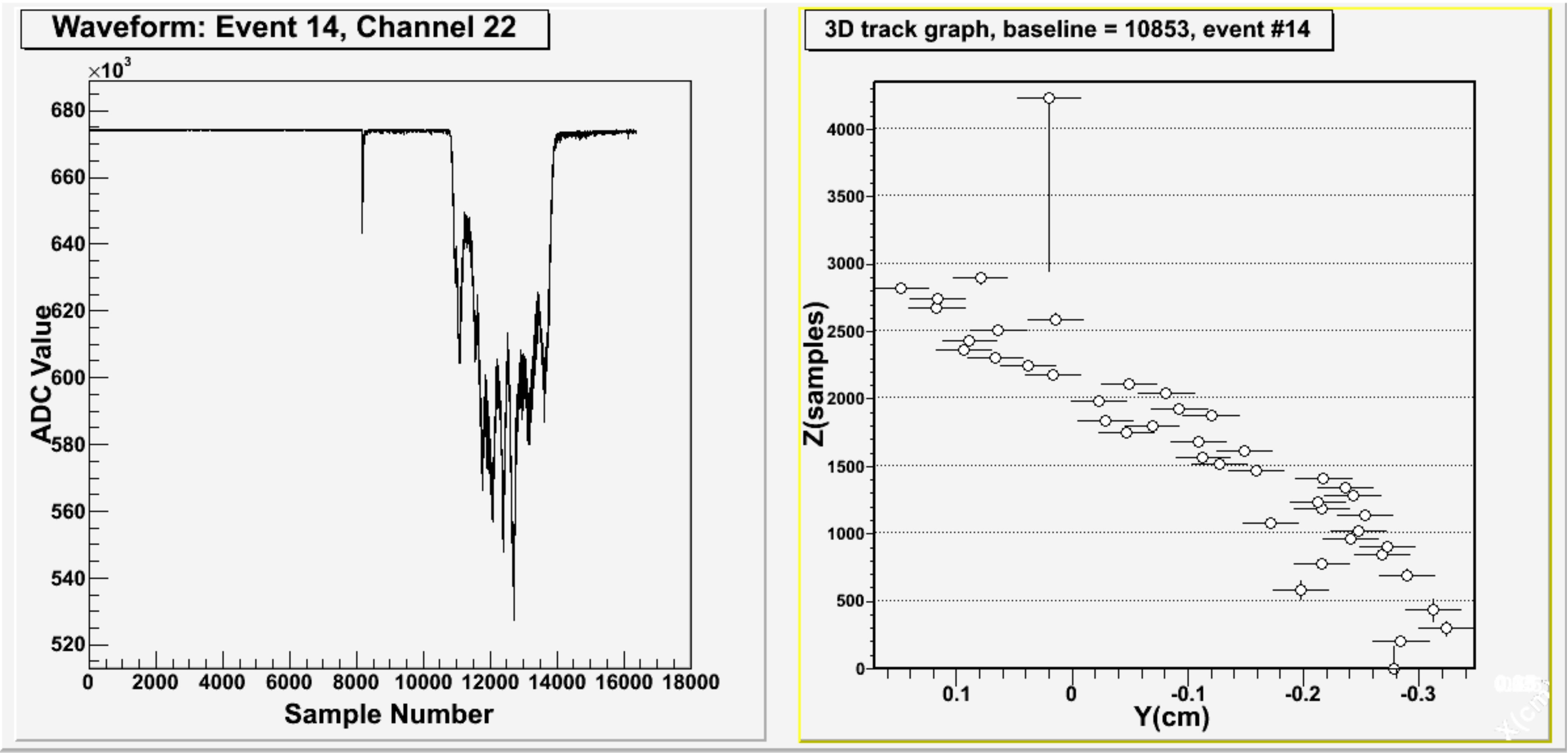}
  \caption{\label{fig_14y} \textbf{Cosmic ray muon tracking:}
Cosmic ray muon measured in the xenon TPC prototype at 10 bar. The left panel shows the summed waveform of the 19 PMTs 
(x axis is proportional to time). The right panel shows track points reconstruction: on the y-axis is the sample number (drift time) 
of a charge in a small time slice and the x-axis is the reconstructed position in the vertical direction perpendicular to the drift. The errors in the reconstruction 
are likely underestimated, but the straight line from the muon track is evident. Individual points represent  about 40keV energy depositions and the full track is about 
20 cm long with a total of approximately 1.2 MeV energy deposition. The trigger was provided by two scintillator pads above and below the TPC that defined muons that 
traverse the TPC diagonally, as reconstructed.}
\end{figure}

\begin{figure}[!ptb]
  \centering
  \includegraphics[width=\textwidth]{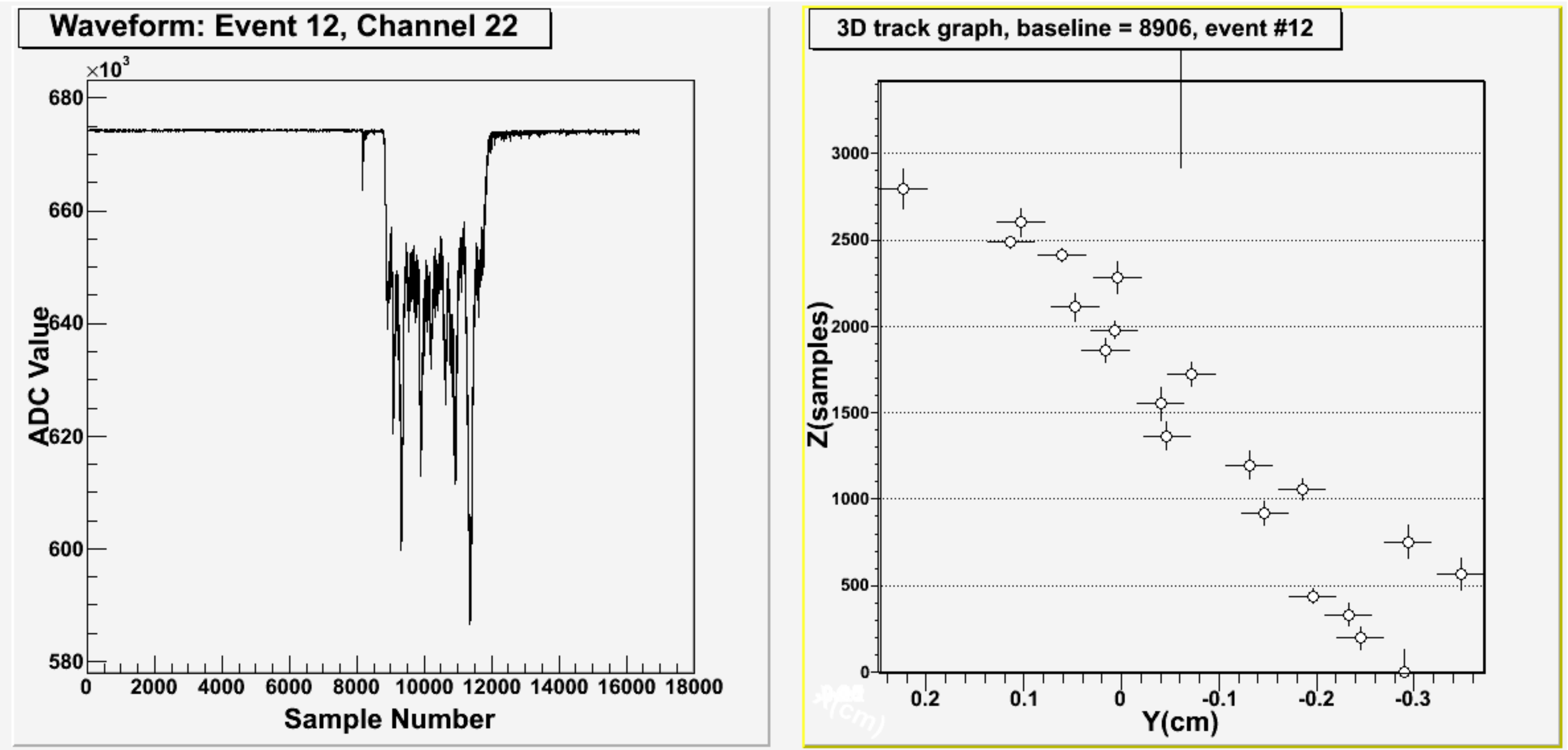}
  \caption{\label{fig_12y} \textbf{Another example of tagged cosmic ray muon imaging/tracking:}For a typical muon tracks such as this there are large variations (the amplitude variations in the left panel) in dE/dx along the track as expected from the Landau distribution of energy loss.}
\end{figure}

Given the proven position sensitivity of the TPC even in the absence of a tracking sensor array, a study of cosmic ray muon tracking was performed. Two small 
scintillators were mounted above and below the TPC pressure vessel, such that muons tagged in coincidence would traverse the TPC at about 45 degrees from the 
vertical in the direction of the drift.  The summed waveforms for the externally triggered muon events were divided in equal energy slices of about 40 keV each. 
For each slice the x an y average position was calculated, yielding a set of (x,y,z) points. The right panels of figures \ref{fig_14y} and \ref{fig_12y} show the z-y 
projection of the track points (the tracks were parallel to the x axis by virtue of the positioning of the scintillators). Clearly, straight tracks can be seen. 
This result is particularly promising in light of the small amount of energy in each slice. In future work we will explore tracking of the gamma ray events that is 
further complicated by the large multiple Coulomb scattering and the possibility of mulitple x-y track points in a given time slice.   

\subsection{Drift velocity and nitrogen}

\begin{figure}[!tbp]
 \centering
 \includegraphics[scale=0.5, angle=270]{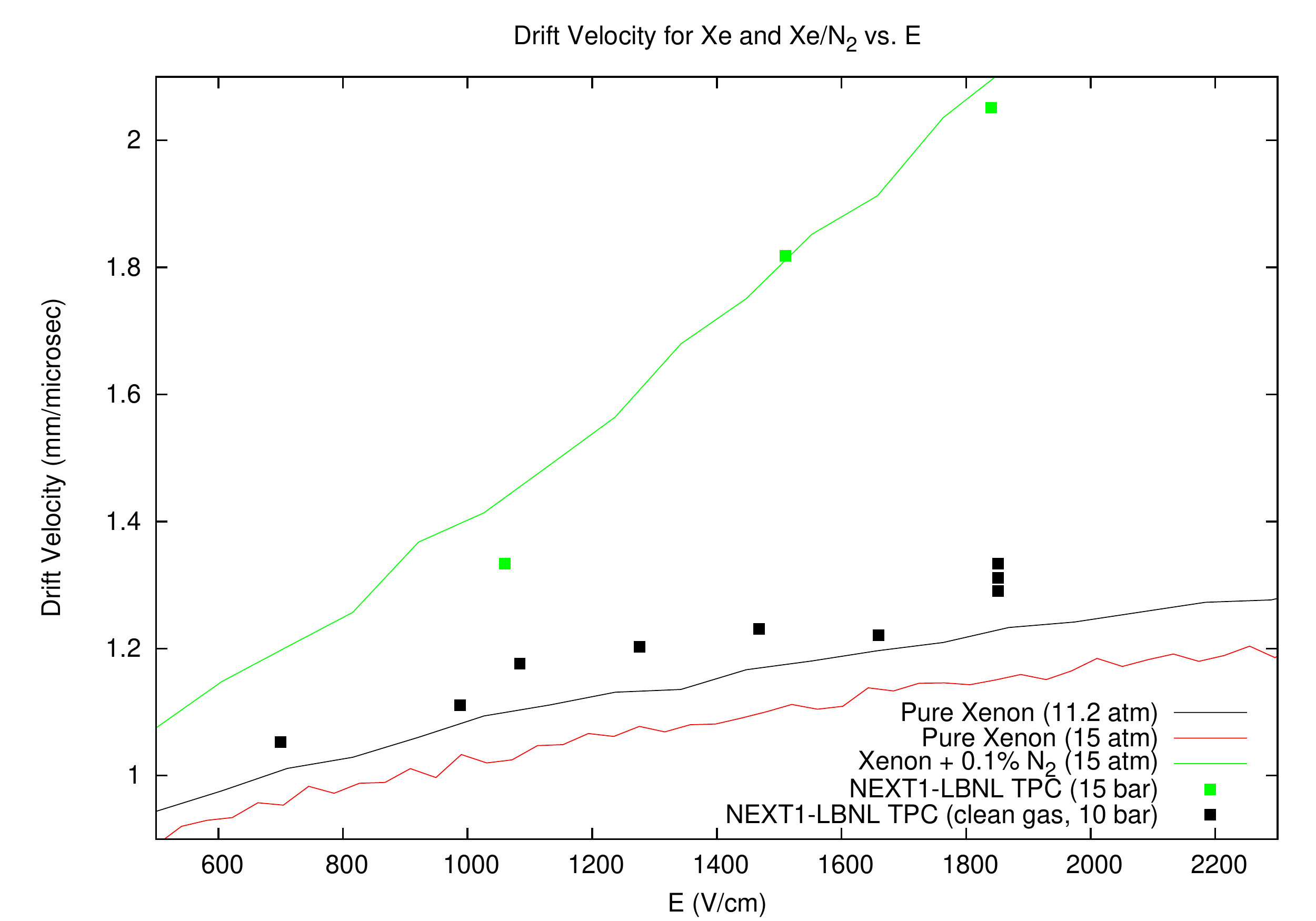}
  \caption{\label{fig_drift_velocity} \textbf{Drift Velocity vs. Electric Field:} Black squares are the measured drift velocities with varying drift electric field at 10 bar.  The black line is the Magboltz calculation of drift velocity for 11.2 bar with pure xenon for comparison. The green squares are drift velocities for an earlier run at 15 bar that deviate significantly from the pure xenon simulation at 15 bar (lowest line in red). The simulation of  Xe+0.1\%N$_2$ at 15 bar, shown as the green line, 
agree with the older data points demonstrating the presence of N$_2$ contamination in the chamber when these data were taken.}
\end{figure}

As mentioned above, the drift velocity is easily measured by looking at the events with the longest drift time. In earlier data taking runs at 11 and 15 bar the drift 
velocities were measured consistently higher than expected for pure xenon. This was later traced (first hints came from RGA scans of working gas samples) to be due to 
the presence on N$_2$ in the TPC gas. The cold getters currently in use in the LBNL prototype do not capture nitrogen (they actually clean N$_2$), so any outgassing in 
the chamber will build up nitrogen in the working gas indefinitely. Newer runs, with xenon fresh from the supplier bottle and with less outgassing of the TPC parts, show 
drift velocities much closer to those expected from pure xenon as can be seen in figure \ref{fig_drift_velocity} where the measurements of the drift velocities of the  
first and second batches are shown along with Magboltz calculations for pure xenon and xenon+N$_2$ mix. The drift velocity for the fresh xenon data are consistent with  
a small (but non-zero) N$_2$ content as expected from the 5-9s quality of the supplier's research grade xenon. A hot getter capable of extracting nitrogen will soon be 
added to the gas system for the LBNL prototype.   

\subsection{EL Yield}

\begin{figure}[!ptb]
  \centering
  \includegraphics[scale=0.5]{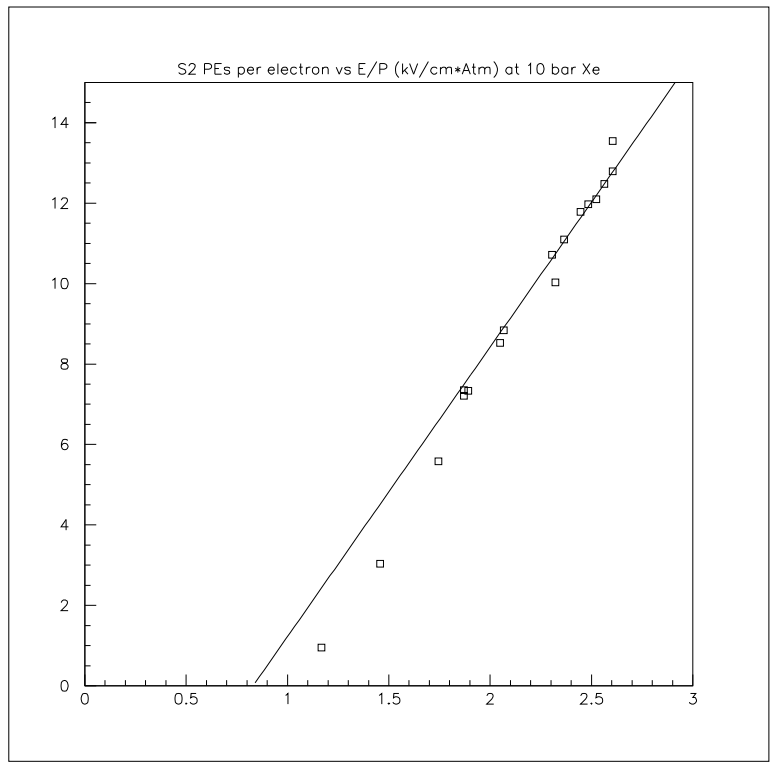}
  \caption{\label{pes_per_e_vs_eop_clean_xe_10bar} \textbf{Photoelctron per ionization electron yield vs. E/P:} The x-axis shows the E/P in the EL region in kV/(bar*cm) for series of runs at 10 bar pressure with the $^{137}$Cs source. The y-axis shows the approximate number of detected photoelectrons per secondary ionization electron assuming a W$_I$ value 24.9 eV and using 300 ADC samples as the average single photoelectron charge. A line crossing the x axis at the nominal E/P EL threshold value of 0.83 is also shown with slope to match the high E/P points. The lower light yield of the low E/P points is due to the presence on N$_2$ at the 1/100,000 level, as demonstrated by Magboltz simulations for the fields relevant in the EL region. }
\end{figure}

Last, we show in figure \ref{pes_per_e_vs_eop_clean_xe_10bar} a measurement of the photoelectron yield per ionization electron as a function of E/P in the EL region. 
The high E/P points extrapolate to the nominal 0.83 kV/(cm*bar) threshold. The deviations from the expected behaviour are again due to the presence on N$_2$, although 
this time in the EL region. This effect was further confirmed with Magboltz calculations showing how much energy was surrendered to inelastic collisions with N$_2$ as a 
function of E/P.  At low E/Ps the electrons in the EL region have energies near the peak of the inelastic cross section (2-3 eV) between electrons and N$_2$  while for 
high E/P electrons are typically higher in energy and the loss to nitrogen molecules is small.

\subsection{Outlook}
After only a few months of operation, the NEXT1-LBNL prototype is already producing high quality data. The analysis presented here shows that reaching, and possibly improving the target resolution of the NEXT experiment with an EASY and SOFT TPC appears well within our capabilities. It also shows the stability and robustness of a HPGXe operated in EL mode. 

In the next few months, the prototype will systematically explore the range of parameters (pressure, drift voltage) relevant for NEXT and will look for ways to  further improve the energy resolution.

\section{A first look at alpha events in NEXT1-IFIC}
The NEXT1-IFIC prototype has been commissioned a few months after NEXT1-LBNL. 
The current configuration corresponds to Run-I (2PMT plane). For the runs discussed here, all of the 19 PMTs placed behind the EL grid (anode) were read out. Because of the 32 channel limitation of the temporary DAQ system employed for the commissioning (the final DAQ has been tested prior to submitting this document and is already in place), only 13 out of the 19 cathode PMTs were read out. Since the beginning of operations we have run at different pressures between 4 and 11 bar. The anode voltage was set to an E/P of near 3.5, while the cathode voltage
was set at a drift voltage of about 330 V/cm. 

The join LBNL-IFIC analysis group is starting to analyze the data. As a first example of the on-going effort we describe here a preliminary analysis made using alpha particles.

\subsection{General features}

\begin{figure}[ptbh!]
\centering
\includegraphics[width=0.4\textwidth]{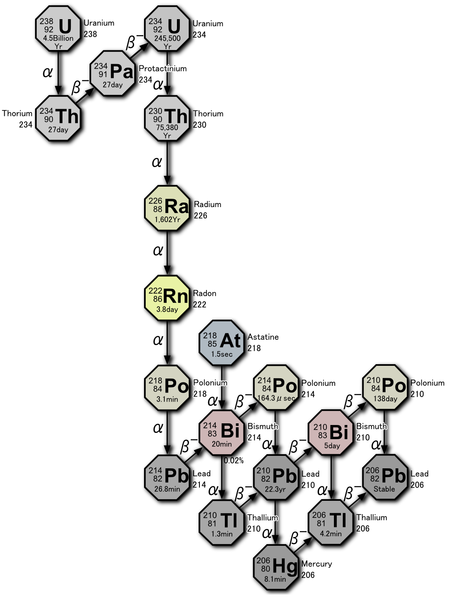} 
\caption{Uranium chain.} 
\label{fig:uc}
\end{figure}
Alpha particles are present in our data, as a byproduct of the radon decay chains. For example, \RN\ appears in the
uranium natural chain (Figure \ref{fig:uc}) and decays with a half life of 3.8 days emitting an alpha particle of 5.6 MeV. They leave a very clear signal that is useful for the initial studies of the NEXT1-IFIC chamber. 

\begin{figure}[ptbh!]
\centering
\includegraphics[width=0.9\textwidth]{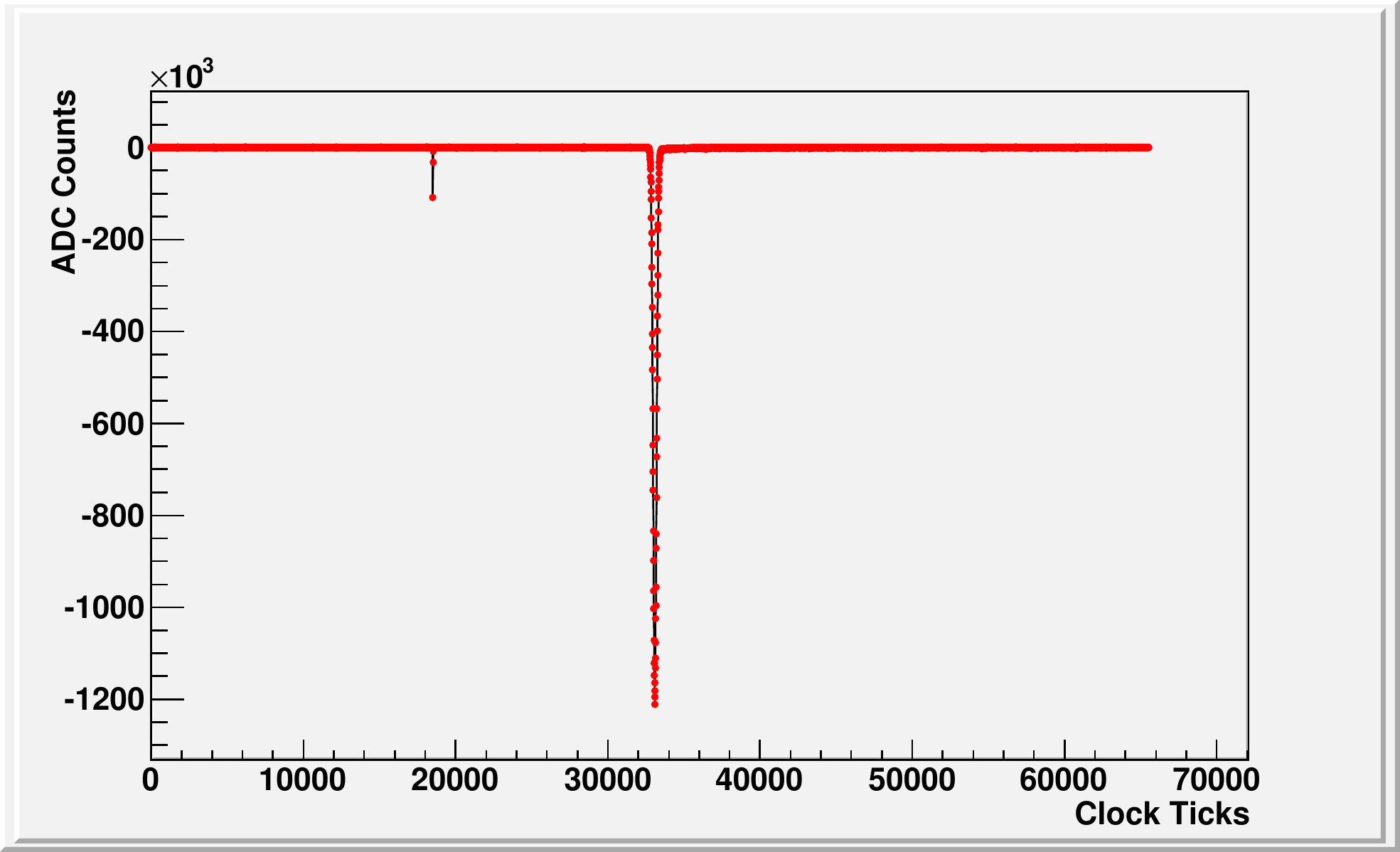} \vspace{0.5cm} \\
\includegraphics[width=0.4\textwidth]{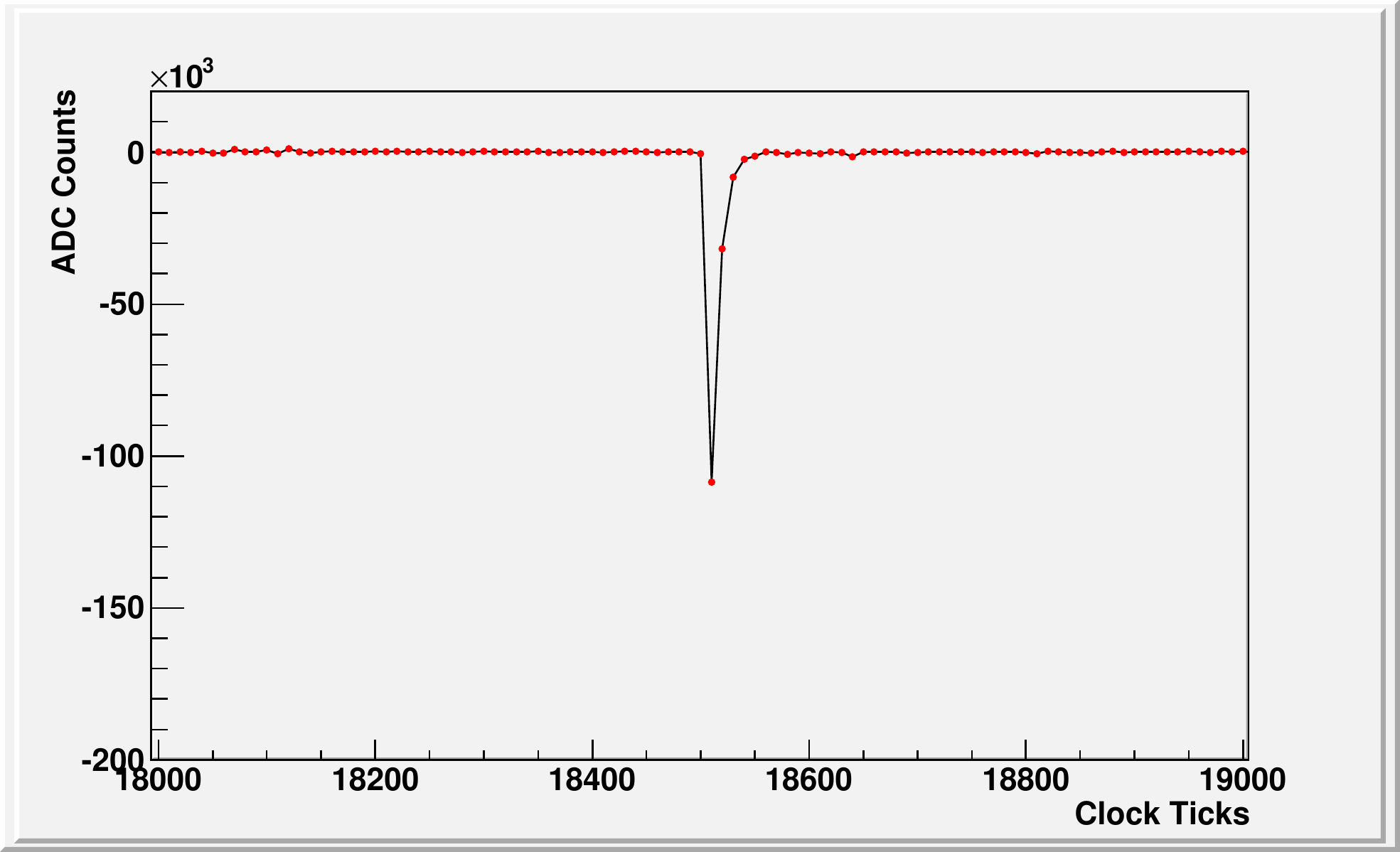} \hspace{0.1\textwidth}
\includegraphics[width=0.4\textwidth]{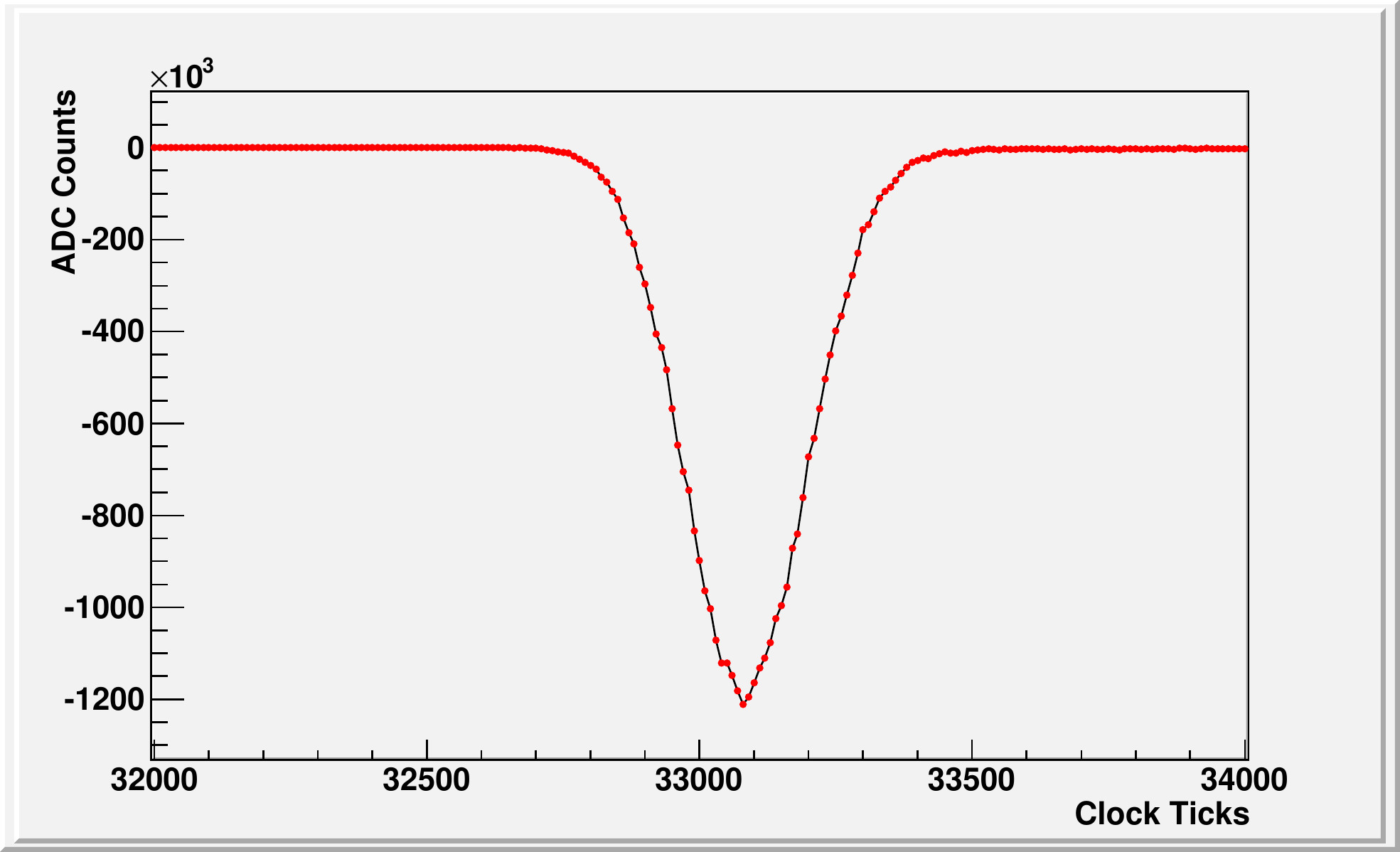}
\caption{Top: cathode sum waveform for a typical alpha candidate in NEXT-1-EL. The S1 and S2 signals are clearly visible. Bottom: zoom on the S1 (left) and S2 (right) peaks, for the same event.} 
\label{fig:next1el_alpha_waveform}
\end{figure}
Figure \ref{fig:next1el_alpha_waveform} shows a typical waveform for a alpha candidate event. The waveform is obtained by summing the 13 cathode PMT, pedestal-subtracted, waveforms. The $\sim$66,000 clock ticks span a $\sim$660 $\mu$s time (10 ns sampling). The trigger occurs in the middle of the DAQ time window, at $\sim$33,000 clock ticks. The S1 and S2 signals are clearly visible. For this event, the S2 signal occurs at the trigger location, and it is preceded by the S1 signal by about 140 $\mu$s. The S1 peak is only a few 100 ns wide, while the S2 peak extends over a few $\mu$s, as expected. The area of the S1 peak is about two orders of magnitude smaller than the S2 area. 

\begin{figure}[ptbh!]
\centering
\includegraphics[width=0.7\textwidth]{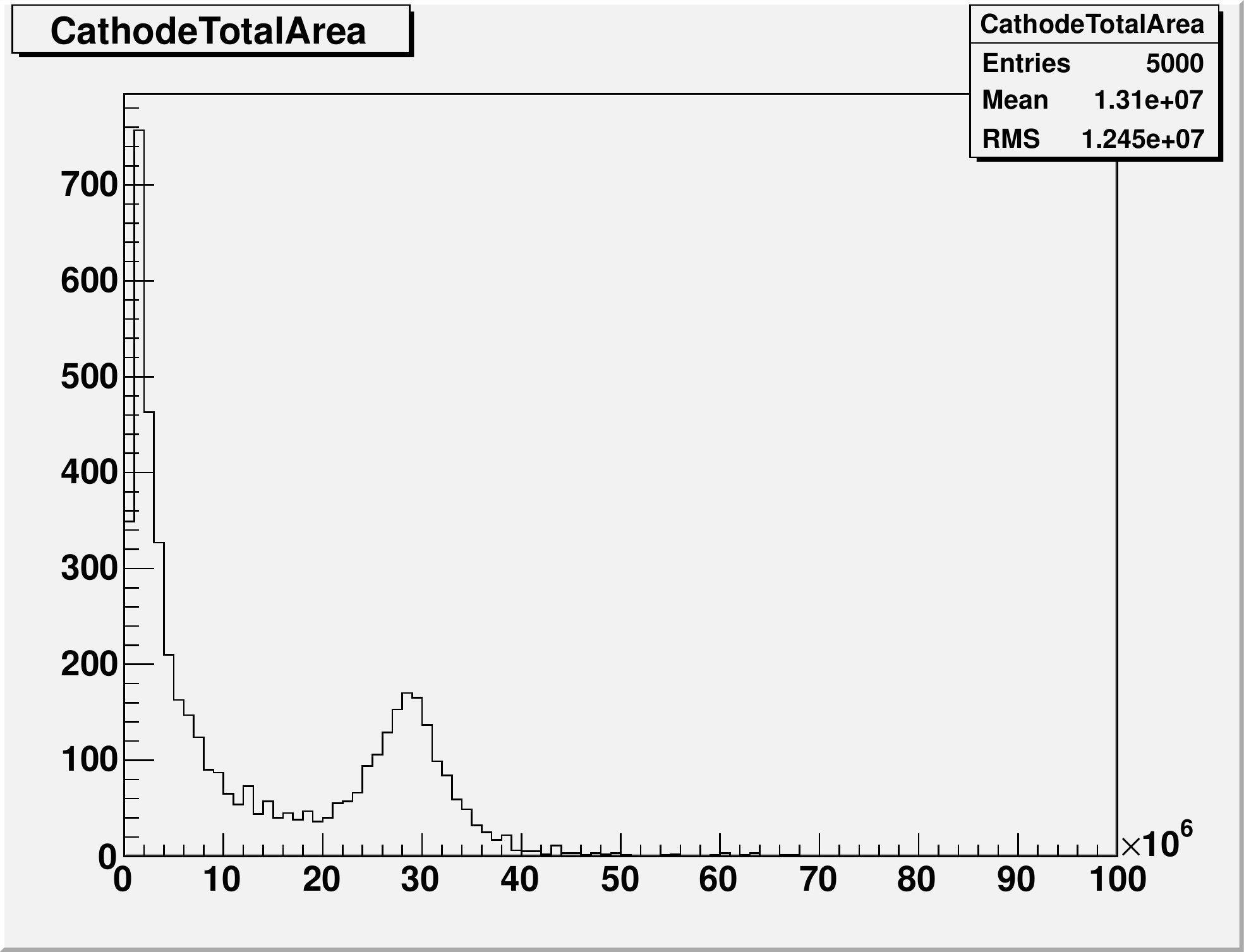} 
\caption{Raw area in ADC counts of the summed waveforms in the cathode.} 
\label{fig:cathodeArea}
\end{figure}
\begin{figure}[ptbh!]
\centering
\includegraphics[width=0.7\textwidth]{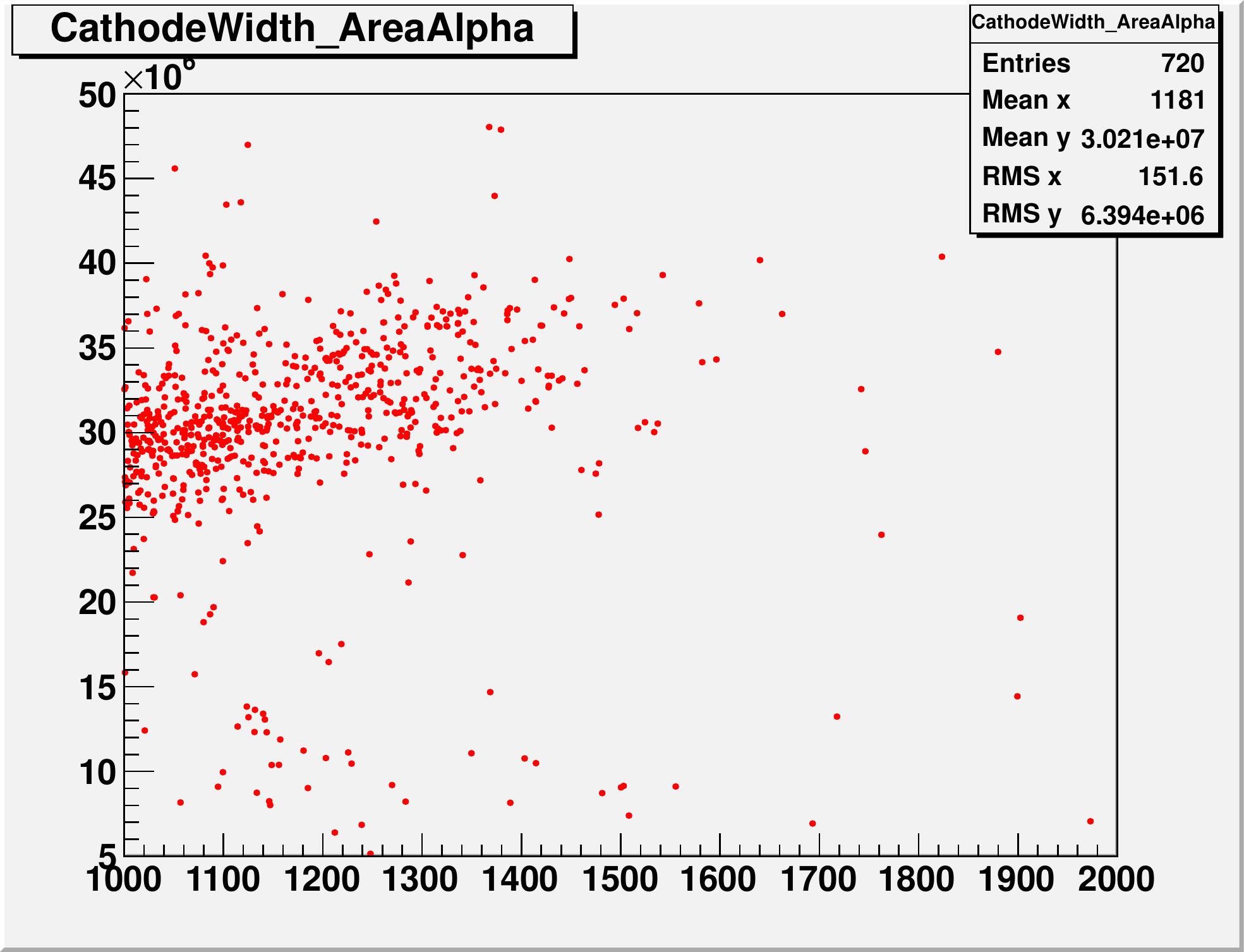} 
\caption{Area versus width of the summed waveforms in the cathode. The area is in ADC counts. 
Multiply by 0.01 to read the area in microseconds.} 
\label{fig:avsw}
\end{figure}

Figure \ref{fig:cathodeArea} shows the uncorrected distribution of the summed area of the cathode PMTs. A wide peak, corresponding to alpha particles appears in the right of the distribution, well separated from the low energy background on the left. Figure \ref{fig:avsw} shows a scatter plot of the area versus the width of the signal. As expected the
total area and the signal width are correlated. Notice however, that the distribution widens in area as it moves toward higher values of the width. This shows the effect of the drift field in alpha particles moving at angles with respect 
to the field lines.   

\begin{figure}[ptbh!]
\centering
\includegraphics[width=0.7\textwidth]{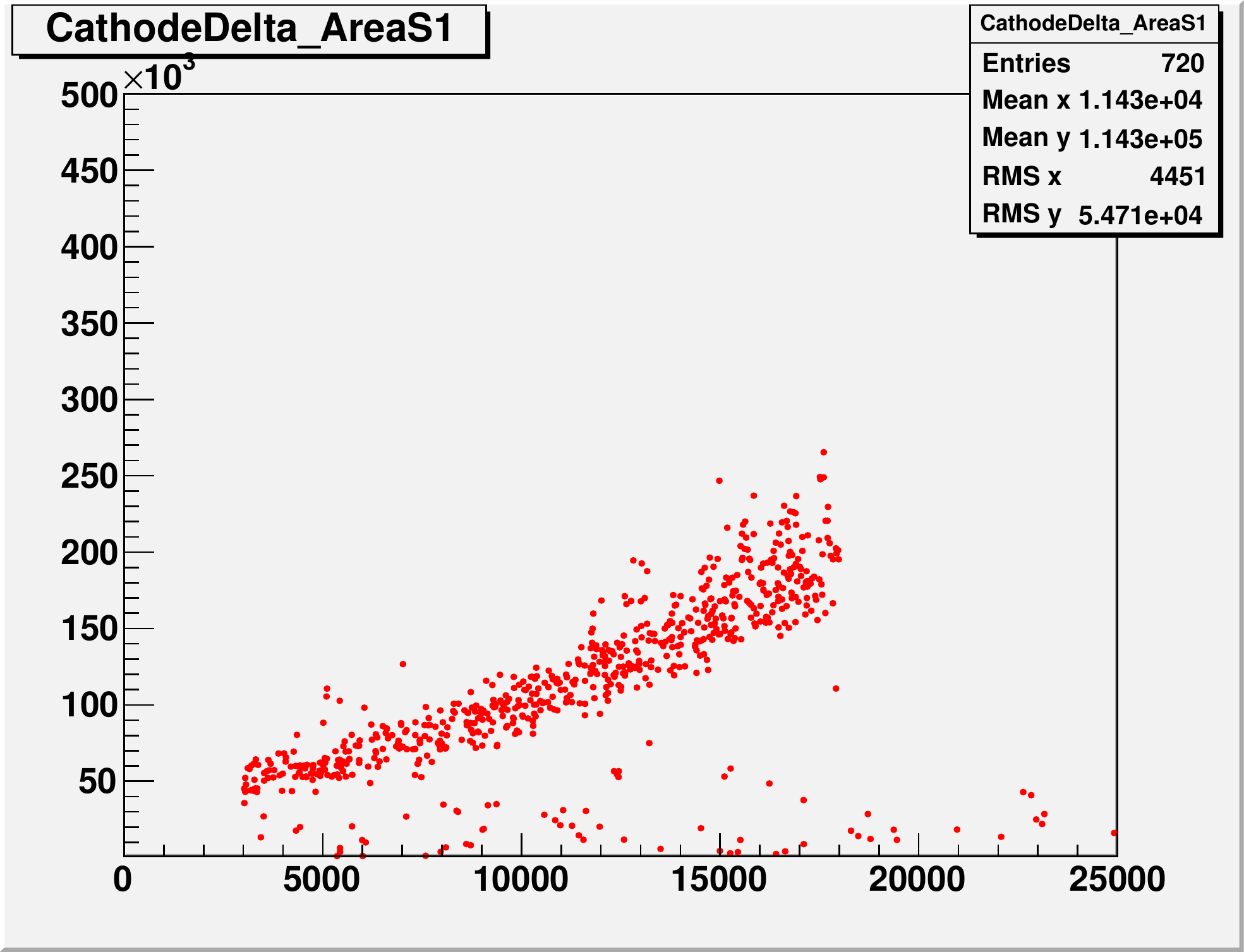} \\
\includegraphics[width=0.7\textwidth]{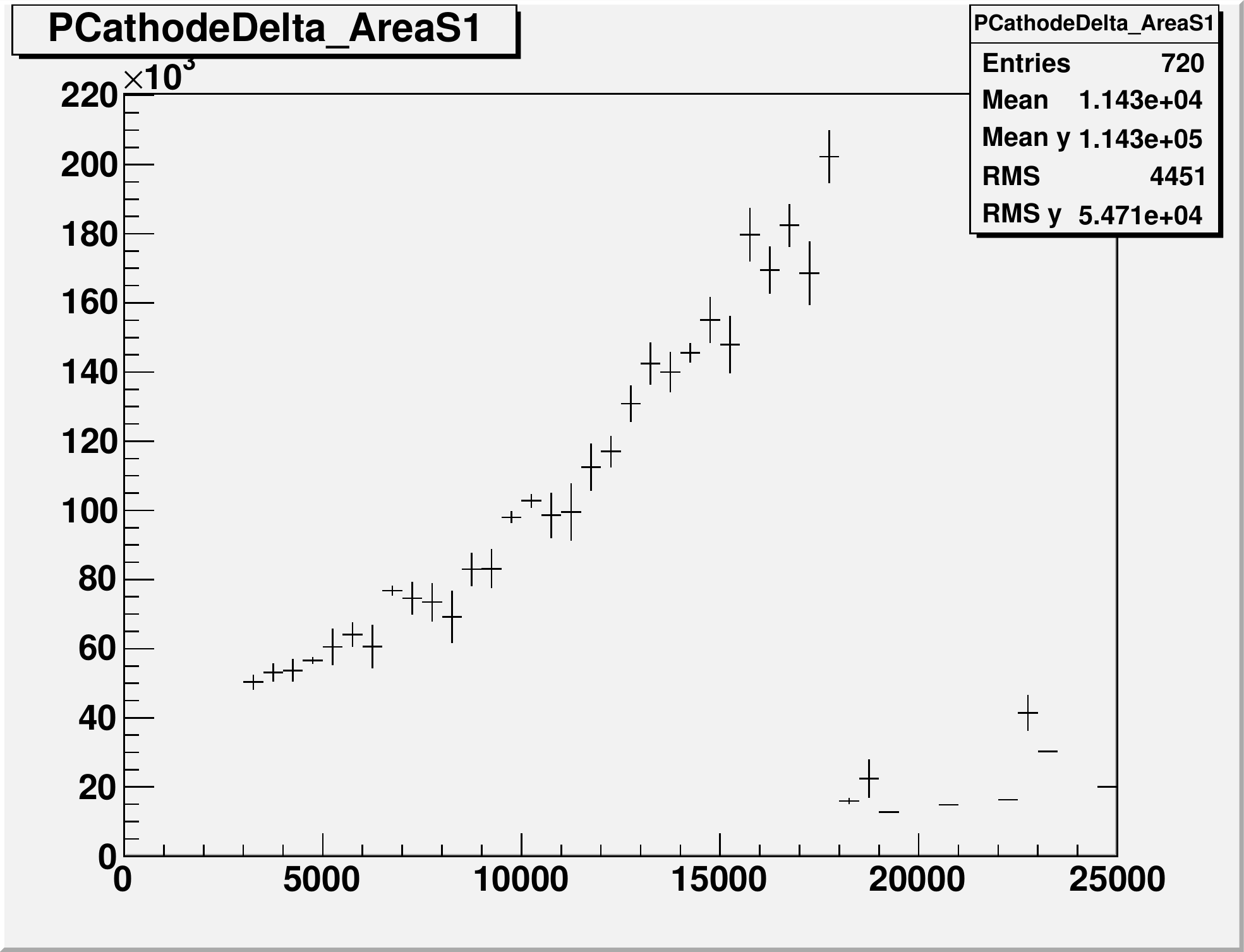} 
\caption{(top): Area of the S1 signal versus width of the summed waveforms in the cathode. The area is in ADC counts. 
Multiply by 0.01 to read the time in microseconds; (bottom) a profile of the same distribution, showing the light attenuation as a function of the distance.} 
\label{fig:s1vsd}
\end{figure}

Figure \ref{fig:s1vsd} (upper) shows a scatter plot of the S1 area versus the drift time, obtained from the difference in
position between S1 and S2. The same distribution is shown as a profile in the lower plot. The light read by the cathode PMTs int the end closer to them is about a factor 5 larger than the light in events happening near the anode. Since our current light tube is made of uncoated PTFE we expect a rather poor reflectivity, of about 50\%, consistent with the observed effect. 

\subsection{Selection of alpha particles}
\begin{figure}[ptb]
\centering
\includegraphics[width=0.46\textwidth]{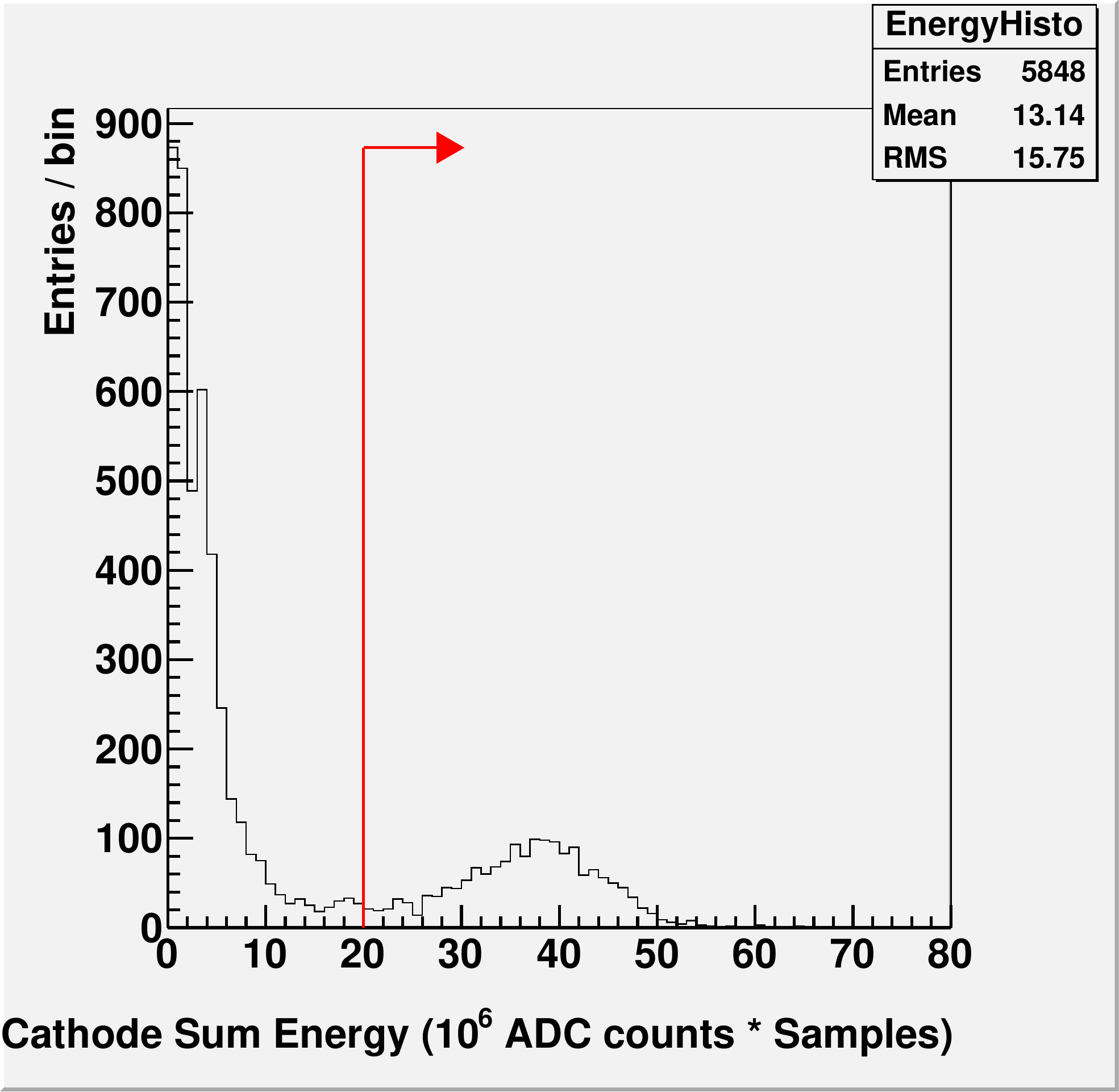} \hspace{0.05\textwidth}
\includegraphics[width=0.46\textwidth]{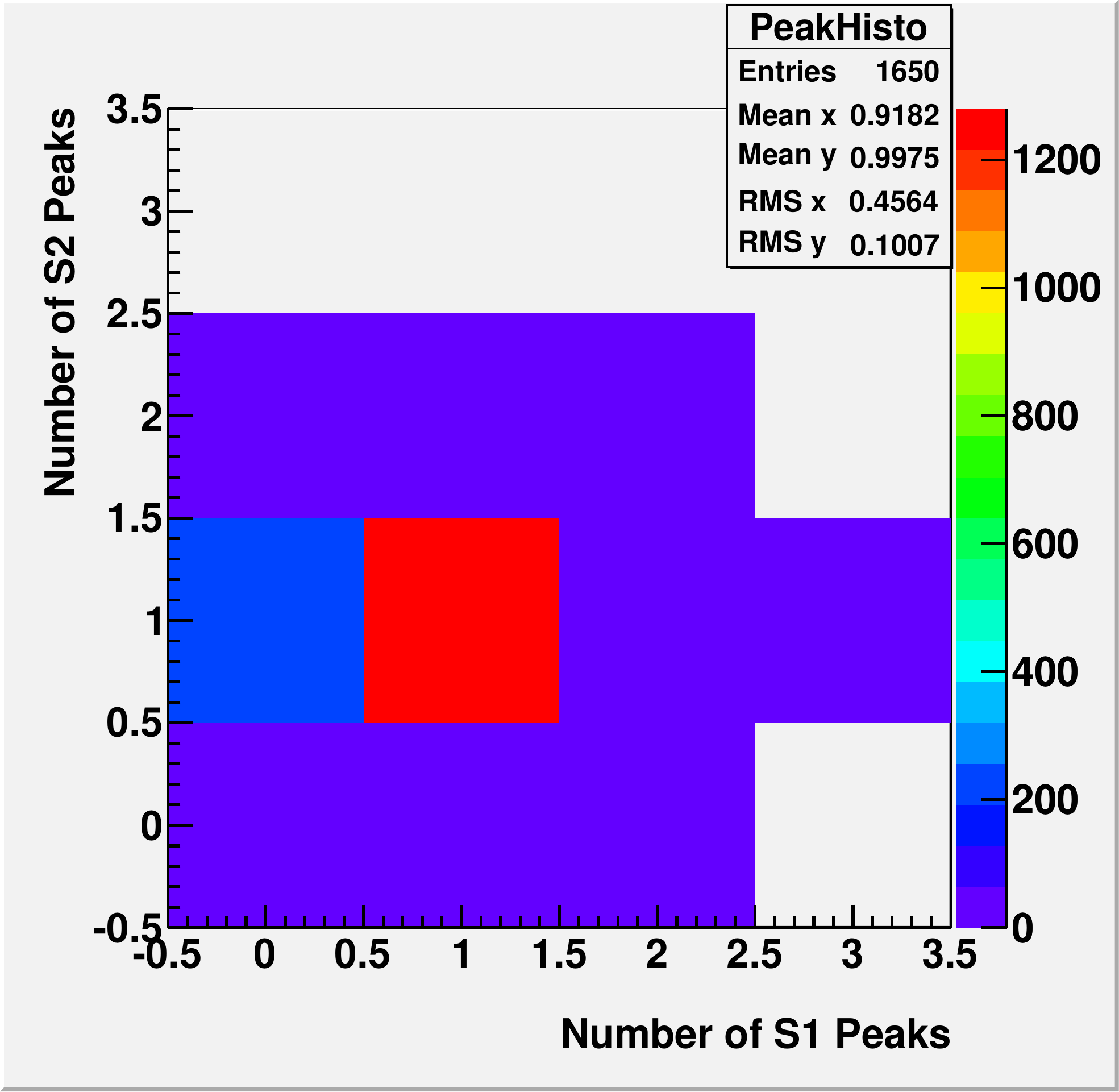} \vspace{0.5cm} \\
\includegraphics[width=0.46\textwidth]{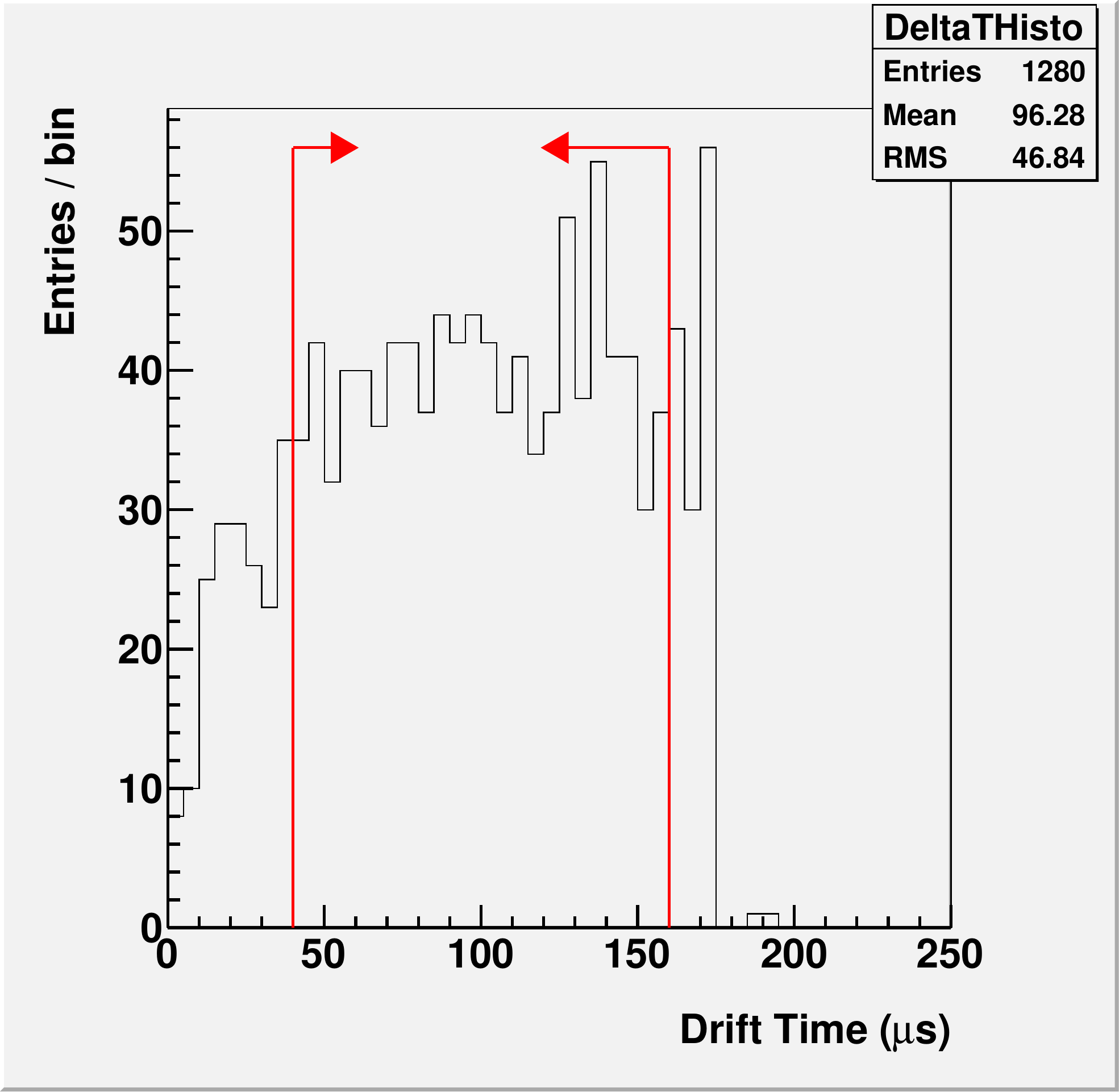} \hspace{0.05\textwidth}
\includegraphics[width=0.46\textwidth]{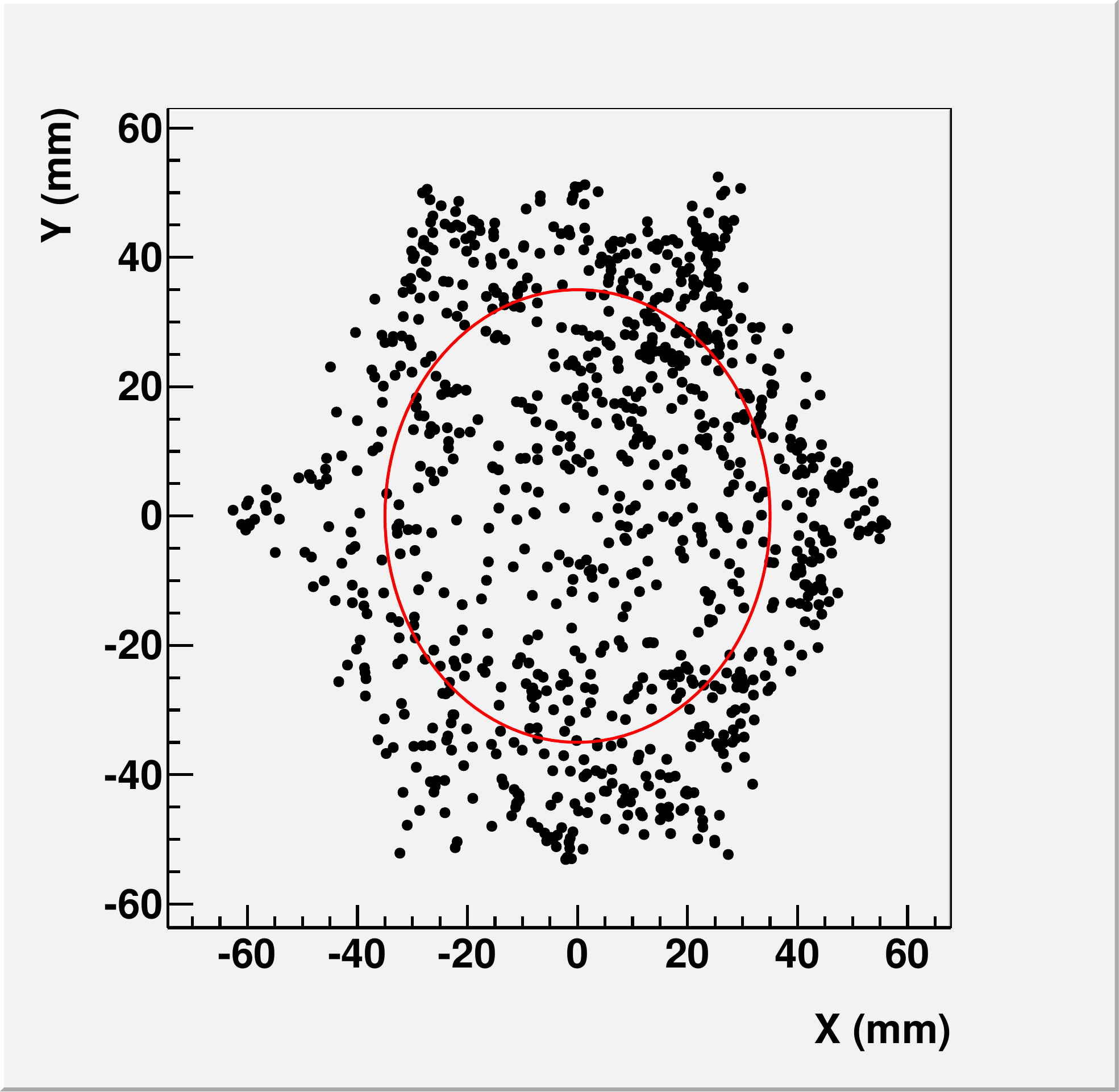} 
\caption{\small Selection of alpha candidates in NEXT-1-EL. See text for details.} 
\label{fig:next1el_alpha_selection}
\end{figure}
Figure \ref{fig:next1el_alpha_selection} summarizes the selection of alpha candidates in NEXT-1-EL. The following cuts are applied:
\begin{description}
\item[Energy cut (Fig.~\ref{fig:next1el_alpha_selection}, top left): ] Given that alpha events are expected to be highly ionizing, only events with a cathode sum $>20\cdot 10^6$ ADC counts are selected.
\item[Waveform peaks cut (Fig.~\ref{fig:next1el_alpha_selection}, top right): ] Alpha events are expected to yield a nearly point-like energy deposition within the chamber. For this reason, only events with a single S2 candidate peak in the cathode PMT summed waveform are selected. In order to obtain a full 3-dimensional reconstruction of the event, the S1 signal needs to be detected. For this reason, the analysis also requires a single S1 candidate peak for the event. As can be seen from the figure, once the energy cut is applied, this (1 S1 $+$ 1 S2) signature is the predominant one.   
\item[Drift time cut (Fig.~\ref{fig:next1el_alpha_selection}, bottom left): ] Once (1 S1 $+$ 1 S2) events have been selected, the drift time can be estimated by taking the time difference between the S1 and S2 signals. As can be seen from the figure, events extend up to $\sim$ 170$\mu$s drift time for this run. Assigning this maximum drift time to the full drift length of the chamber (30 cm) yields a drift velocity of about 1.8 mm/$\mu$s. We select only events with a drift time comprised between 40 and 160 $\mu$s. This cut ensures that only events that are well separated spatially from the chamber metallic grids, and with a unambiguous (1 S1 $+$ 1 S2) signature, are selected. 
\item[Radial cut: (Fig.~\ref{fig:next1el_alpha_selection}, bottom right):] Information from the anode PMT plane is used for a coarse and very preliminary reconstruction of the $(x,y)$ position of the event. This is possible because the face of the anode PMTs is only 3-4 mm away from the EL region. While the $(x,y)$ reconstruction algorithm will certainly be improved in the future, its current implementation is sufficient to clearly reconstruct the hexagonal shape of the NEXT-1-EL light tube, as can be seen in the figure. To ensure that the energy deposition occurs entirely within the xenon gas, a strict requirement on the reconstructed radius of $<35$ mm is used.
\end{description}
%


\subsection{Energy Reconstruction}

In the SOFT concept, the basic observable we use for energy reconstruction is the sum of all (13, in this case) cathode PMTs. While the supply voltage for each PMT has been separately adjusted to provide a uniform PMT gain of $2\cdot 10^6$, a small (about 20\% RMS spread) channel-to-channel variation is seen and corrected for offline. In the following, we discuss how the energy reconstruction changes as a function of spatial location within the chamber, and provide a first energy resolution measurement with NEXT-1-IFIC. 

\subsection{Dependence on Spatial Location}

\begin{figure}[ptb]
\centering
\includegraphics[width=0.46\textwidth]{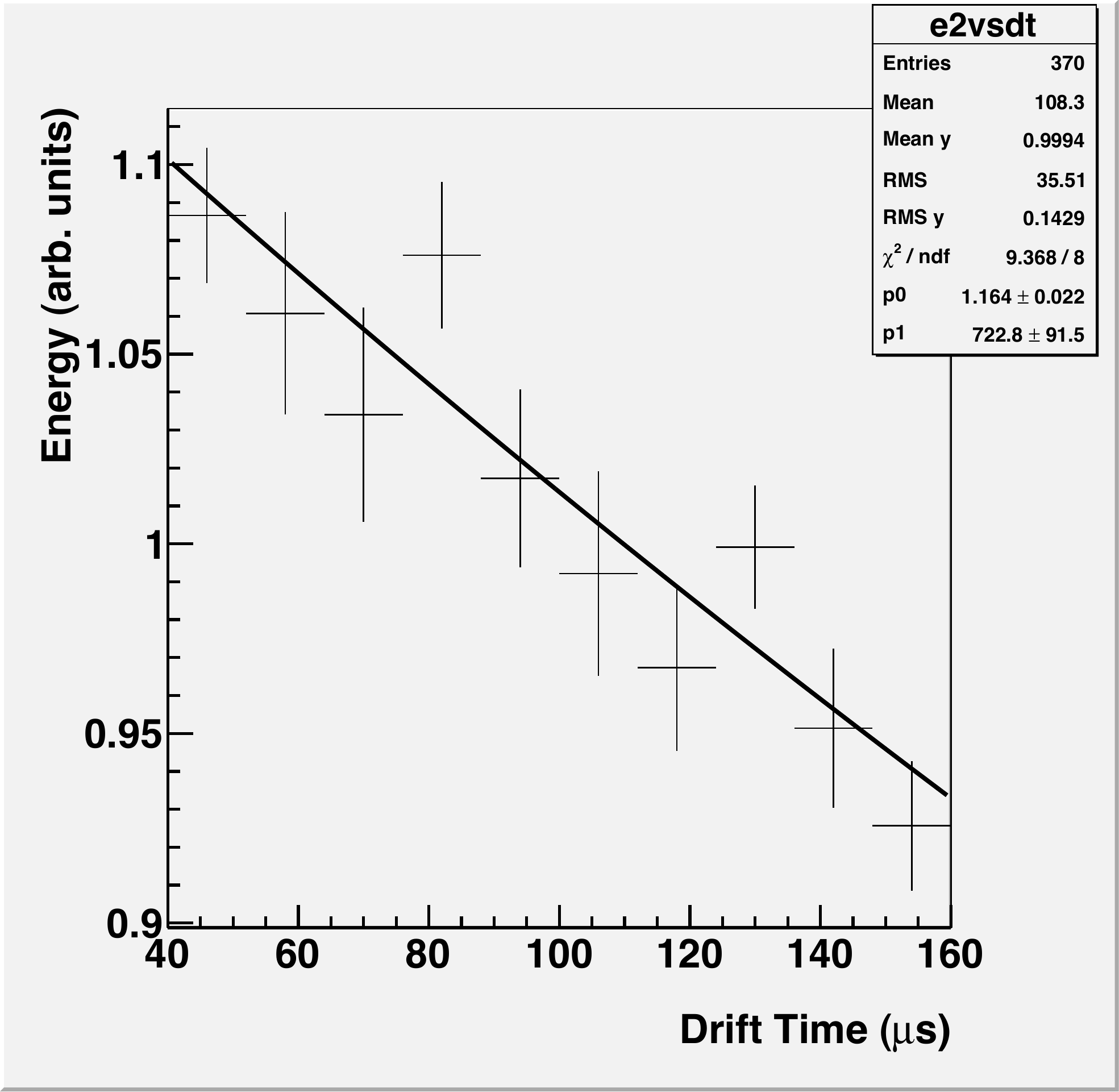} \hspace{0.05\textwidth}
\includegraphics[width=0.46\textwidth]{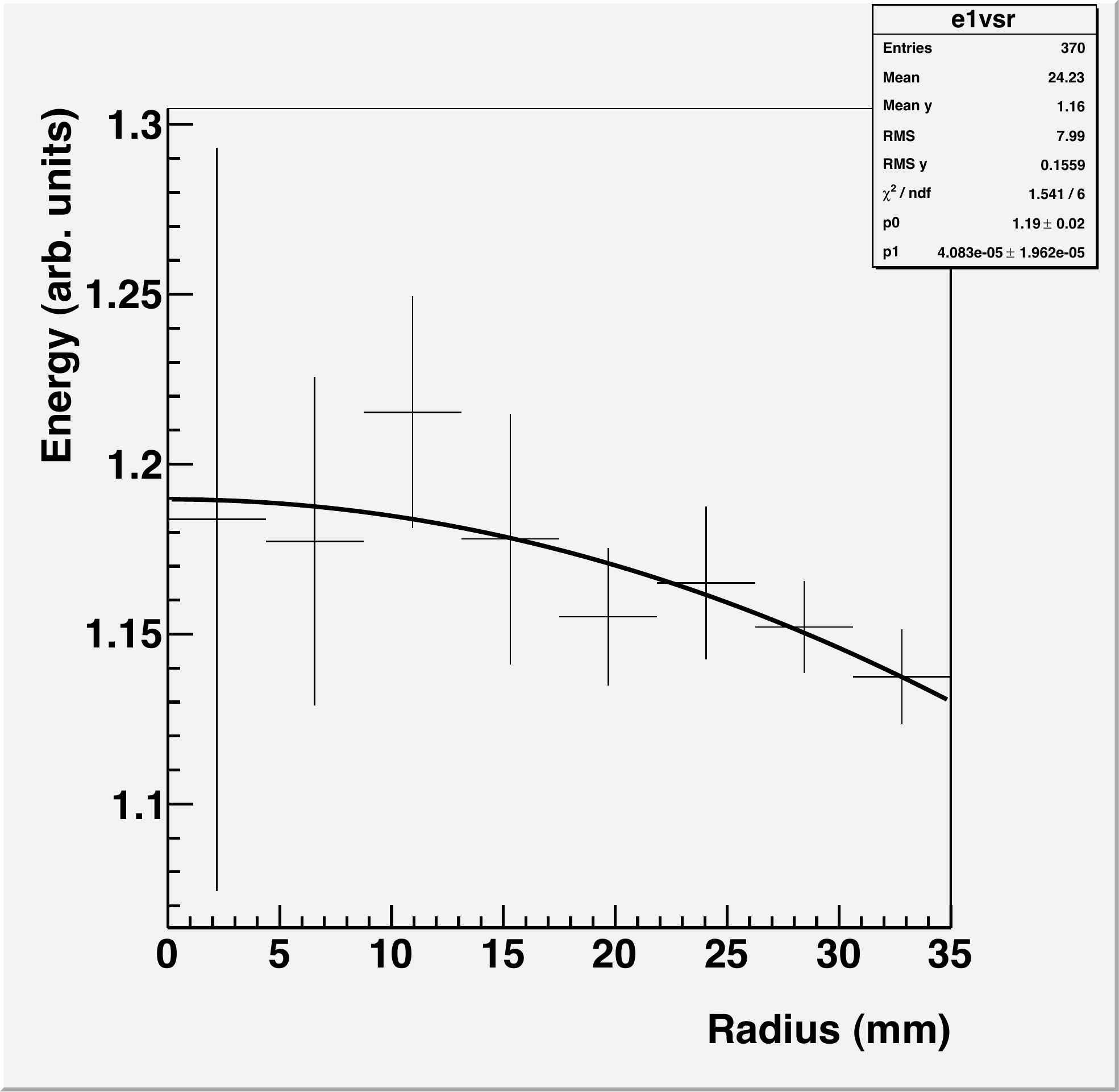} 
\caption{\small Drift time (left) and radial (right) energy corrections in NEXT-1-IFIC.} 
\label{fig:next1el_energy_corrections}
\end{figure}
\indent Once alpha candidate events have been selected, we can study how the amount of light detected by the sum of all cathode PMTs varies, on average, as a function of the event localization within the chamber. The profile histogram in Fig.~\ref{fig:next1el_energy_corrections} (left) shows how the reconstructed energy varies with drift time. As expected from electron attachment, the reconstructed energy decreases with increasing drift time. Nevertheless, for this current run the variation is mild and well understood. Performing an exponential fit to this profile histogram yields a $(723\pm 92)\ \mu$s electron lifetime. 

\indent Once the correction for electron attachment has been applied, we can study how the reconstructed energy varies with reconstructed radial position within the chamber. Given the isotropic nature of the EL light, and the fact that the light guide panels in NEXT-1-IFIC are far from being perfect VUV light reflectors, we expect the reconstructed energy to decrease with reconstructed radius. This is what has been observed on average, as can be seen from the profile histogram in  Fig.~\ref{fig:next1el_energy_corrections} (right). While the effect is sizable as one extends past 35 mm reconstructed radius, the variation is less than 10\% up to 35 mm. A sufficiently good description of this dependence is obtained by fitting the average reconstructed energy to a $(1-c\cdot r^2)$ functional form, with $c=(4.1\pm 2.0)\cdot 10^{-5}\ \mbox{mm}^{-2}$. 

\subsection{Energy Resolution}

\begin{figure}[ptb]
\centering
\includegraphics[width=0.75\textwidth]{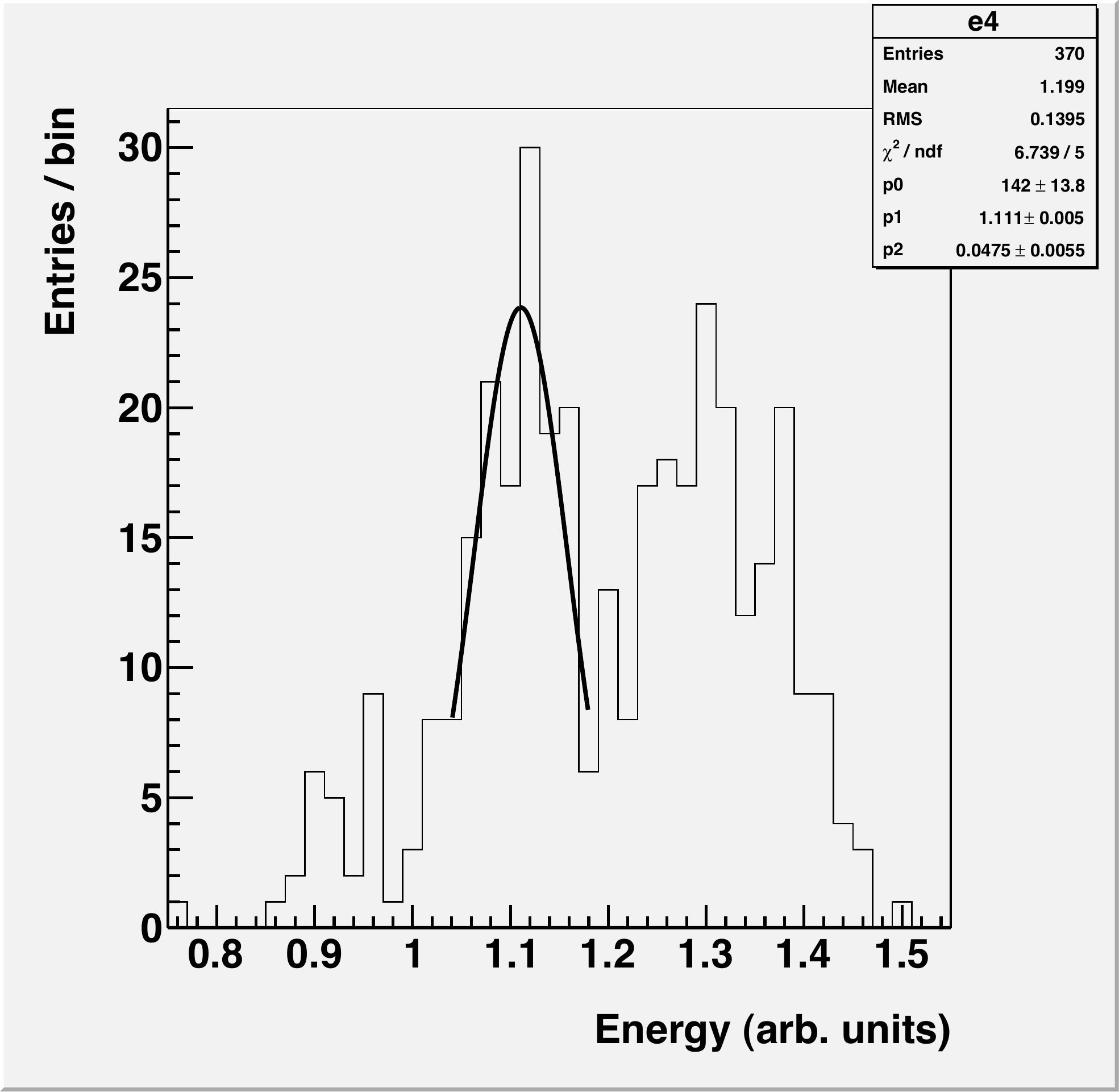} 
\caption{\small Energy spectrum for alpha candidates. PMT relative gain, drift time and radial corrections to the energy have been applied.} 
\label{fig:alphaenergyspectrum}
\end{figure}

Figure \ref{fig:alphaenergyspectrum} shows the energy spectrum of the alpha candidate events, having already applied the PMT relative gain, drift time and radial corrections to the energy. The identification of alpha lines is difficult because of the poor statistics and the poor energy resolution. From these data, we estimate an energy resolution of about 10\% FWHM (see gaussian fit in Fig.~\ref{fig:alphaenergyspectrum}). We are now studying why this first energy resolution measurement is poor. Recombination, in particular given the low drift field for this measurement may be among the explanations. Needless to say, the current results shown here need not be taken as a projection of the expected ultimate performance for NEXT-1-IFIC, but rather as a status report on our current understanding of the chamber after only a few weeks of commissioning.


\section{Other Measurements}

In addition to measurements related to energy reconstruction (discussed above), the alpha analysis provides a few other ancillary measurements, discussed in the following.

\subsection{Drift Velocity}

As can be seen from Fig.~\ref{fig:next1el_alpha_selection} (bottom left), events extend up to $\sim$ 170$\mu$s drift time for this run. Assigning this maximum drift time to the full drift length of the chamber (30 cm) yields a drift velocity of about 1.75 mm/$\mu$s.

\begin{figure}[ptb]
\centering
\includegraphics[width=0.50\textwidth,angle=270]{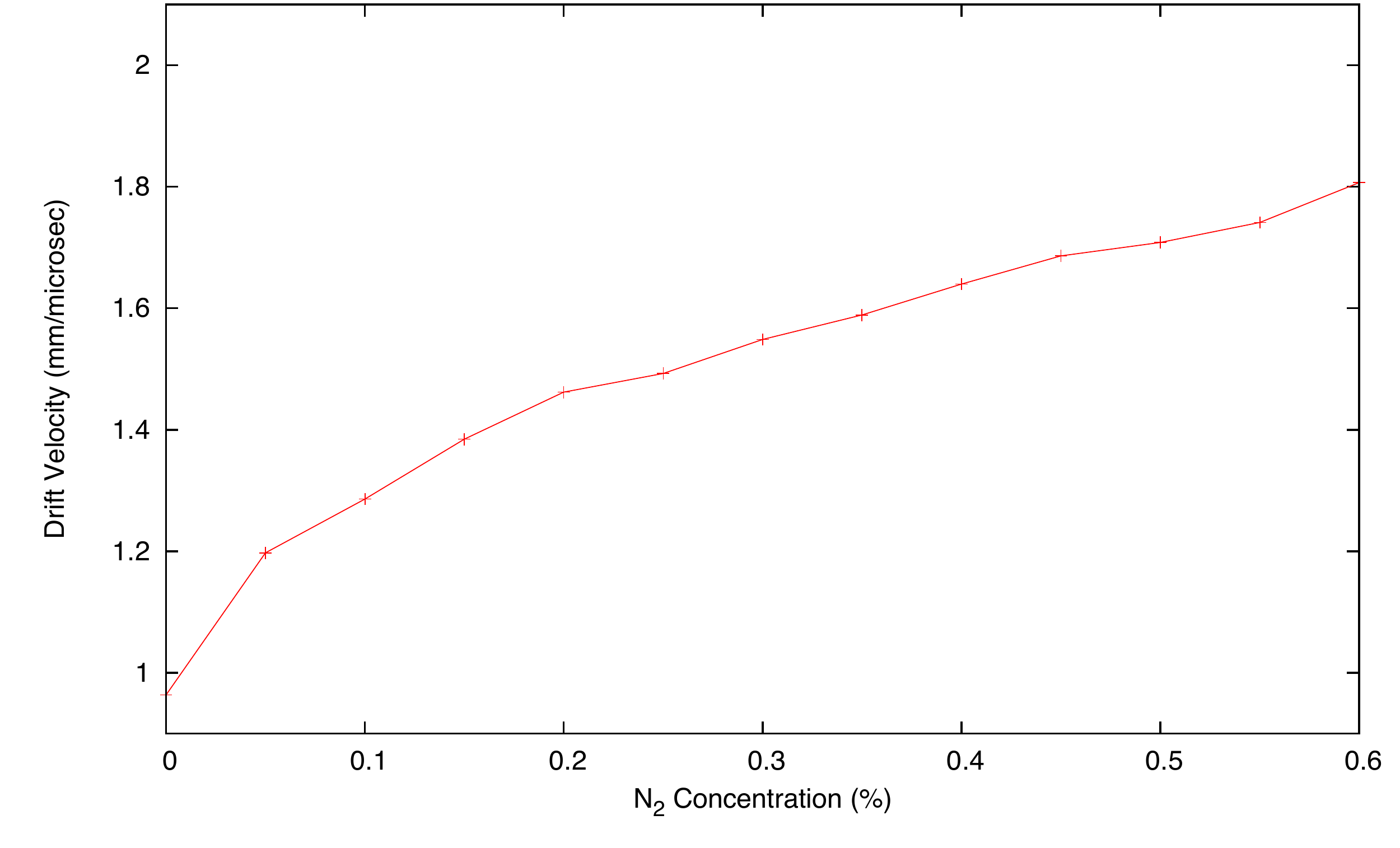} 
\caption{\small Drift velocity in xenon as a function of N$_2$ contamination, as expected from Garfield/Magboltz simulations. The expectation is for xenon gas at 6 bar pressure, and a 0.33 kV/cm drift field. Vertical errors on each point lie in the range of approximately $\pm$2-5\% of their values.} 
\label{fig:driftvelocityvsn2}
\end{figure}

This measured value is significantly higher than expectations for pure xenon at 6 bar and for a 0.33 kV/cm drift field. With ideal gas quality we would in fact have expected a drift velocity of about 1.0 mm/$\mu$s. A possible explanation for such a large drift velocity would be the presence of N$_2$ contamination in the gas in large (0.5\%) amounts, as can be seen from the Garfield/Magboltz simulation results of Fig.~\ref{fig:driftvelocityvsn2}. This is possible, as the current configuration for the NEXT-1-IFIC gas system is not designed to filter N$_2$ out of the gas.

\subsection{S2-to-S1 Ratio}

\begin{figure}[ptb]
\centering
\includegraphics[width=0.75\textwidth]{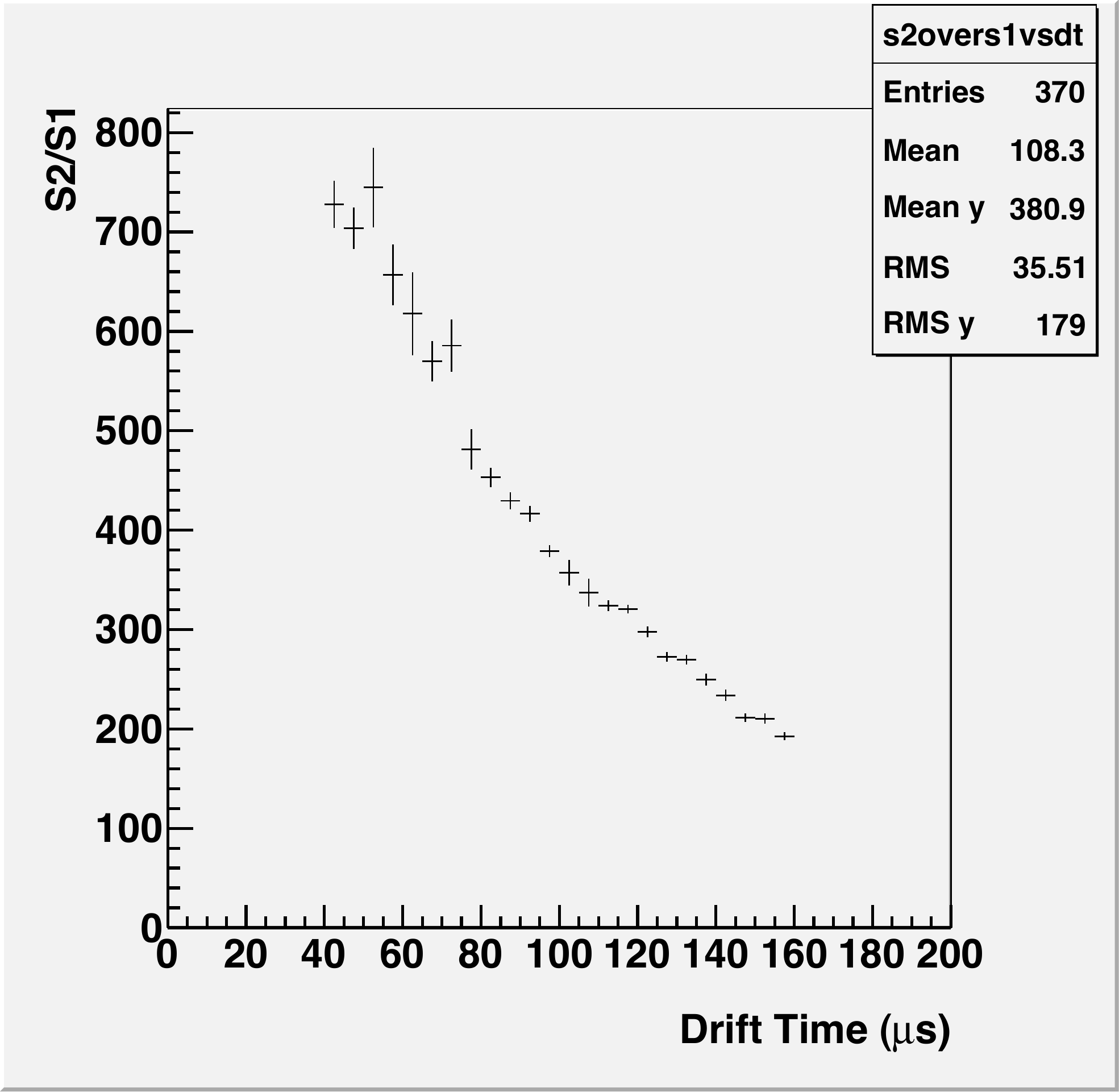} 
\caption{\small Profile histogram of average S2-to-S1 ratio measured in NEXT-1-IFIC as a function of drift time. The ratio extrapolates to S2/S1 $\sim$ 1000 for no drift, that is for S1 and S2 light emitted from the same spatial location.} 
\label{fig:s2overs1}
\end{figure}

Figure \ref{fig:s2overs1} shows the measured average S2-to-S1 ratio as a function of drift time. As discussed above, the strong dependence on drift time is mostly due to S1 signal variations with drift time, caused by the significant VUV light absorption of the NEXT-1-IFIC light tube.

\indent If one extrapolates the measured S2-to-S1 ratio towards zero drift time, one is able to estimate an S2-to-S1 ratio that is free of spatial dependencies, as the S1 and S2 light are, in this case, emitted from essentially the same location. Data of Fig.~\ref{fig:s2overs1} indicate that such an extrapolation would yield a corrected S2-to-S1 ratio of about $10^3$. This is to be compared with an expectation of about $3.3\cdot 10^3$ for such a ratio. This expectation is obtained by considering an expected optical gain of $1.07\cdot 10^3$ at the current chamber conditions, and the fact that it takes about 3 times more energy to produce a primary scintillation photon (about 76 eV) compared to a ionization electron (about 24.8 eV). The deficit of the measured S2-to-S1 ratio compared to the expected value points towards a alpha ionization quenching factor of about 3 in the current chamber conditions.

\subsection{Longitudinal Diffusion and Range of Alpha Particles}

\begin{figure}[ptb]
\centering
\includegraphics[width=0.75\textwidth]{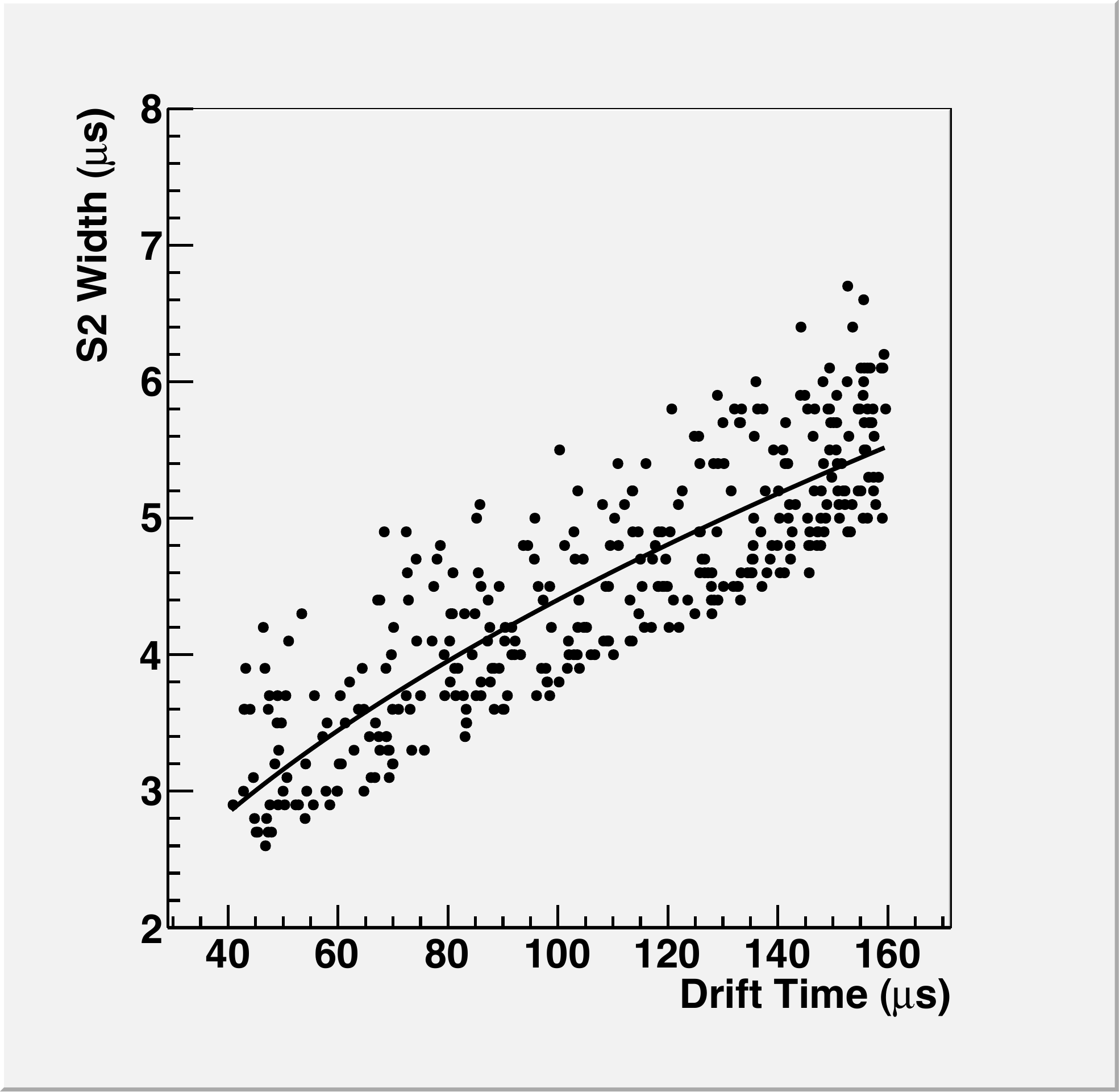} 
\caption{\small Time width of the S2 signal (full width at 10\% of maximum) versus drift time. The effects of longitudinal diffusion and alpha particle range can be shown.} 
\label{fig:s2widthvsdrifttime}
\end{figure}

\begin{figure}[ptb]
\centering
\includegraphics[width=0.75\textwidth]{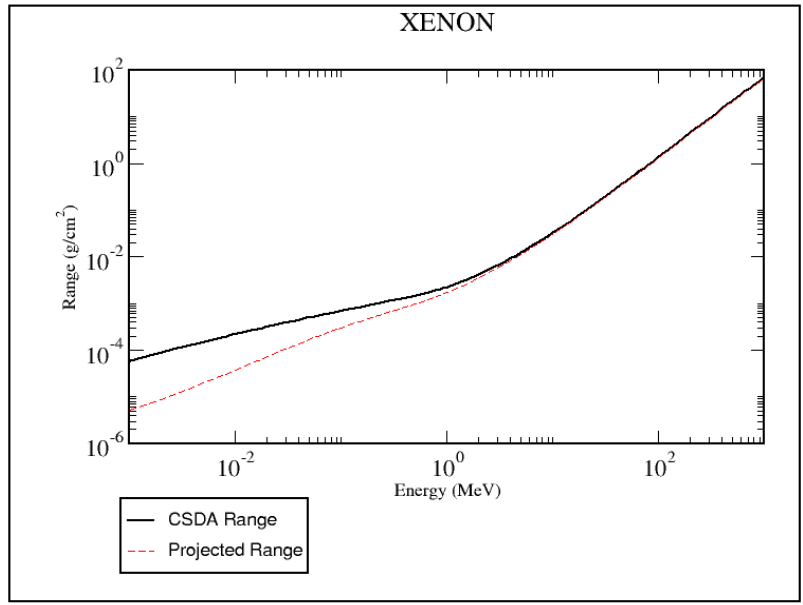} 
\caption{\small Range of alpha particles as a function of particle's kinetic energy. Both the total path length traversed, as well as the path length projected along the initial particle direction, are shown.} 
\label{fig:alpharange}
\end{figure}

It is also possible to perform S2 pulse shape analysis on the selected alpha candidates, and study whether the measured S2 time spreads are consistent with expectations. As indicator of the S2 time width we use the full S2 peak width at 10\% of its maximum. A scatter plot of the S2 width as a function of the drift time for all selected alpha candidates is shown in Fig.~\ref{fig:s2widthvsdrifttime}. The increase in the average S2 width with increasing drift time can be explained via longitudinal diffusion of the ionization electrons during drift. The expected longitudinal diffusion constant for xenon at 6 bar and for 0.33 kV/cm drift field is about 0.6 mm/$\sqrt{\rm{cm}}$, qualitatively in agreement with observations. The spread of about 2~$\mu$s in the S2 width, approximately constant for all drift times, is instead due to the expected range of alpha paticles. As can be inferred from Fig.~\ref{fig:alpharange}, we expect a 3.4 mm projected range at 6 bar for alpha particles at the typical energies of 5 MeV. Considering the measured 1.75 mm/$\mu$s drift velocity, this corresponds to about a 2 $\mu$s S2 width spread due to the orientation of the alpha particle direction with respect to the drift field direction, as observed.

\chapter{Sensitivity of NEXT-100} \label{sec.sensi}
\section{Sources of background in NEXT}
\subsection{\BI\ and \TL}
The \bbonu\ peak of \XE\ is located in the energy region of the naturally-occurring radioactive processes. The half-life of the parents of the natural decay chains, of the order of the age of the universe, is, however, very short compared to the desired half-life sensitivity of the new \bbonu\ experiments ($\sim10^{26}$ years). For that reason, even small traces of these nuclides create notable event rates.

\begin{figure}[tb]
\centering
\includegraphics[width=0.7\textwidth]{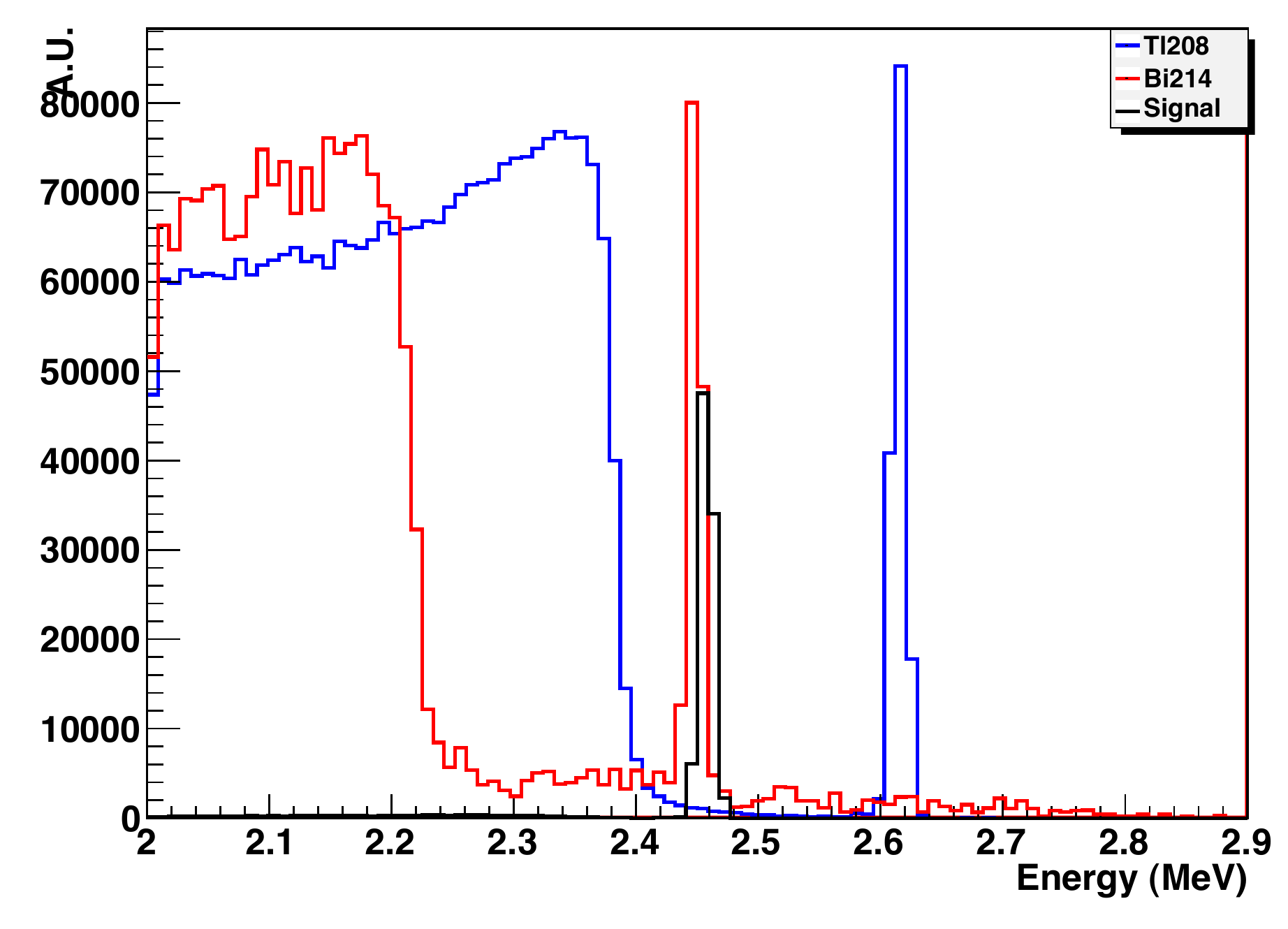}
\caption{\small The landscape near the end-point of \XE , as a function of deposited energy. 
The normalization of the different peaks is arbitrary. The assumed resolution
is 0.5\%. The purpose of the plot is to show how 
the \bbonu\ signal lies between the dominant \BI\ and \TL\ backgrounds.}
\label{fig:landscape2} 
\end{figure}

The only significant backgrounds for NEXT are the high energy gammas produced in the
$\beta$-decays of the isotopes \TL\ and \BI, found in the thorium and uranium series, respectively (Figure \ref{fig:landscape2}). 

The daughter of \TL, \Pb, emits a de-excitation photon of 2614 keV with a 100\% intensity. The Compton edge of this gamma is at 2382 keV, well below \qbb. However, the scattered gamma can interact and produce other electron tracks close enough to the initial Compton electron so they are reconstructed as a single object falling in the energy region of interest (ROI). Photoelectric electrons are produced above the ROI but can loose energy via bremsstrahlung and populate the window, in case the emitted photons escape out of the detector. Pair-creation events are not able to produce single-track events in the ROI. 

After the decay of \BI, \Po\ emits a number of de-excitation gammas with energies above 2.3 MeV.
The gamma line at 2447 keV (intensity: 1.57\%) is very close to \qbb. The photoelectric peak
infiltrates into the ROI for resolutions worse than 0.5\%. The gamma lines above \qbb\
have low intensity and their contribution is negligible. The contribution of pair-creation events is also insignificant.

All materials contain impurities of \TL\ and \BI\ in a given amount.
The dominant source of background in NEXT is the pressure vessel (PV). As discussed in chapter \ref{sec.Angel},
the mass of the PV is circa 2 tons if titanium is used as construction material and more than 8 tons if copper is used. In this
CDR and for the calculations that follow we will assume {\em conservative} specific activities of 200 $\mu$Bq/kg and
50 $\mu$Bq/kg for this two materials respectively. We believe that there is ample room to build a more
radiopure PV, either by using very radiopure batches of titanium or by combining it with very resistant and light composites, such as carbon fiber.

The next source of background in NEXT are PMTs. We know of at least two radipoure PMTs suitable for our
needs, the Hamamatsu models R8520 (1'') and R11410-MOD (3''). The \TL\ and \BI\ activity of the R8520 has been measured to be between 0.5 and 1 mBq  and that of R11410-MOD, which seems better adapted for our needs, given its larger size,
between 2 and 3 mBq. As we discuss later, with a coverage of 25\% the contribution of PMT to the radioactive budget is small compared with that of the vessel. 

All the other components of the detector can be made of PTFE, copper, peek, etc. Careful selection and screening should guarantee that they do not contribute significantly to the radioactive budget. The electronics for the SiPMs will be shielded by a thick plate of very radiopure copper and located in the periphery of the fiducial volume. Its contribution, therefore, is also small compared with that emanating from the bulk of the vessel. 

However, the Zaragoza group has performed an independent sensitivity study which includes a very detailed background model. This study has been performed using as a reference detector a TPC based in micromegas and is not further discussed in this report, although their results are fully compatible with those presented here. We plan to introduce such a detailed background model in our Monte Carlo in the near future.

\subsection{Radon}
Radon constitutes a dangerous source of background due to the radioactive isotopes $^{222}$Rn (half-life of 3.8\,d) from the 
$^{238}$U chain and $^{220}$Rn (half-life of 55\,s) from the $^{232}$Th chain. As a gas, 
it diffuses into the air and can enter the detector. 

\BI\ is a decay product of $^{222}$Rn, and 
\TL\ a decay product of $^{220}$Rn. In both cases, the radon suffers from an alpha decay into polonium, producing a negative ion which 
is drifted towards the anode by the electric field of the TPC.  As a consequence, $^{214}$Bi and $^{208}$Tl contaminations can be assumed to be deposited on the anode surface. Radon may be eliminated 
from the TPC gas mixture by recirculation through appropriate filters. There are also ways to suppress radon in the volume defined by the shielding, whether this is a water tank or a lead castle. 

Radon control is a major task for a \bbonu\ experiment, and will be of uppermost importance for NEXT.

\subsection{Cosmic rays and laboratory rock backgrounds}

Cosmic particles can also affect our experiment by producing high energy photons or activating materials. This is the reason why double beta decay experiments are conducted deep underground. At these depths, muons are the only surviving cosmic ray particles, but 
their interactions with the rock produce neutrons and electromagnetic showers. Furthermore, the rock of the
laboratory itself is a rather intense source of \TL\ and \BI\ backgrounds as well as neutrons.

The above backgrounds can be reduced below those intrinsic to the detector (PV, PMTs, etc) by shielding. A water tank providing 3 meters of ultrapure water shield will reduce the flux of gammas to negligible levels and will also
suppress neutron and muon background. The same shielding capabilities can be provided by a lead-copper housing (``castle'') made of 25 cm radiopure lead and 10 cm radiopure copper.

Given the topological capabilities of NEXT the residual muon and neutron background do not appear to be significant
for our experiment. However, there is the possibility of instrumenting the tank for a detector upgrade. An active veto will further suppress backgrounds other than those emanating from the detector itself.

\section{Signal and background characterization in NEXT}

\subsection{The topological signature}

\begin{figure}[htbp!]
\centering
\includegraphics[width=0.9\textwidth]{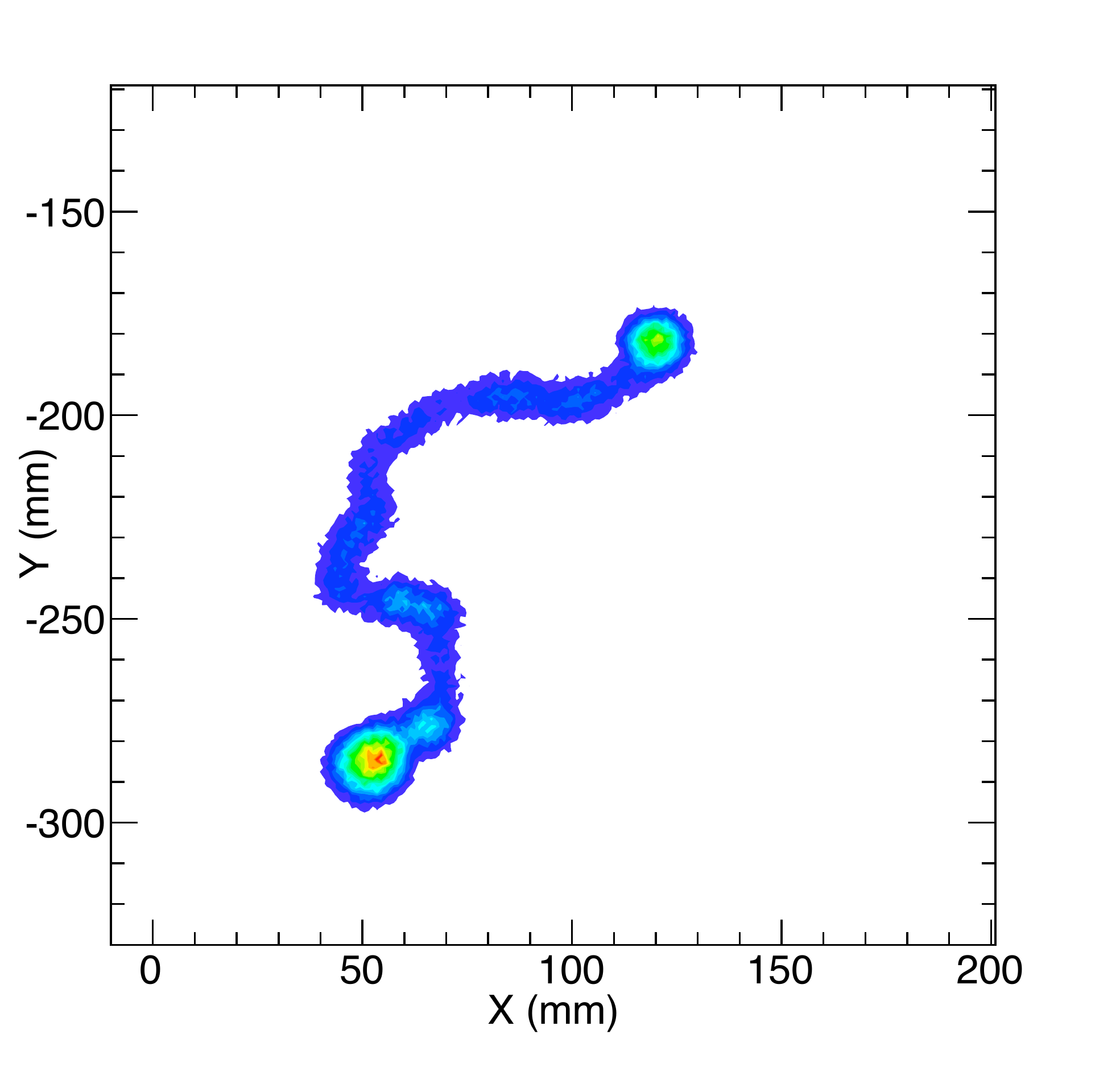}
\caption{\small Double beta decay events have a distinctive topological signature in HPGXe: a ionization track, 
of about 20 cm length at 15 bar, tortuous because of multiple scattering, and with larger depositions 
or \emph{blobs} in both ends.}
\label{fig:track} 
\end{figure}

\begin{table}[tb]
\begin{center}
\begin{tabular}{cccc}
\hline \hline
E$_e$ (keV) & Probability (\%) & E$_{\gamma}$ (keV) & Mean path (cm) \\
\hline
500 &  4 & 12 & 0.14 \\
800 &  6 & 27 &  0.8 \\
1240 & 8 & 58 & 1.8 \\
1680 &10 & 95 & 5.5 \\
2000 & 11 &133 & 11\\
2480 &14 &198 & 33\\
\hline \hline
\end{tabular}
\end{center}
\caption{\small Radiation probability for electrons in xenon at 10 bar. Average energy of the emitted photons and their mean free path in the HPXe is also shown.}
\label{tab:brems}
\end{table}

Double beta decay events have a distinctive topological signature in HPGXe (Figure \ref{fig:track}) : a ionization track, 
of about 20 cm length at 15 bar, tortuous because of multiple scattering, and with larger depositions 
or \emph{blobs} in both ends. 

As the track propagates in the dense gas it emits $\delta$-rays and bremsstrahlung radiation. Those are typically low-energy gammas (Table \ref{tab:brems}) with a mean free path below 1 cm. The convolution of emission of electromagnetic energy with the effect of diffusion (about 1 cm for a drift of 1 m) results in a track for \bbonu\ events that looks more like a wiggly, wide "stripe" of energy deposition, about 1-2 cm wide, than like a well defined ``wire'', as usual when tracking high energy muons, for example.  The implications are quite clear:
\begin{enumerate}
\item Space resolution is not an issue in NEXT. A pitch of about 1 cm is sufficient, given the combined effect of
radiation and diffusion, that blur the track into a stripe. Monte Carlo calculations indicate that even a larger pitch
1.5 to 2 cm may be acceptable. 
\item Instead, identification of low energy gammas nearby the track (at distances of a few cm) is important. This is due to the difference between signal (two electrons of average energy \Qbb/2) and the
dominant background (one electron of energy \Qbb). In the second case the probability of radiation is higher and
the mean free path of the gamma is longer. Topologies with one or more isolated clusters of energy, corresponding to
relatively high energy gammas flying distances of 2 or more cm are therefore more likely for the background than 
for the signal. Therefore, identifying low-energy satellite clusters is important in NEXT. This requires a sensor with low energy threshold, as the SiPMs. 
\end{enumerate}

\subsection{Selection criteria}
\begin{figure}[hptb!]
\centering
\includegraphics[width=15cm]{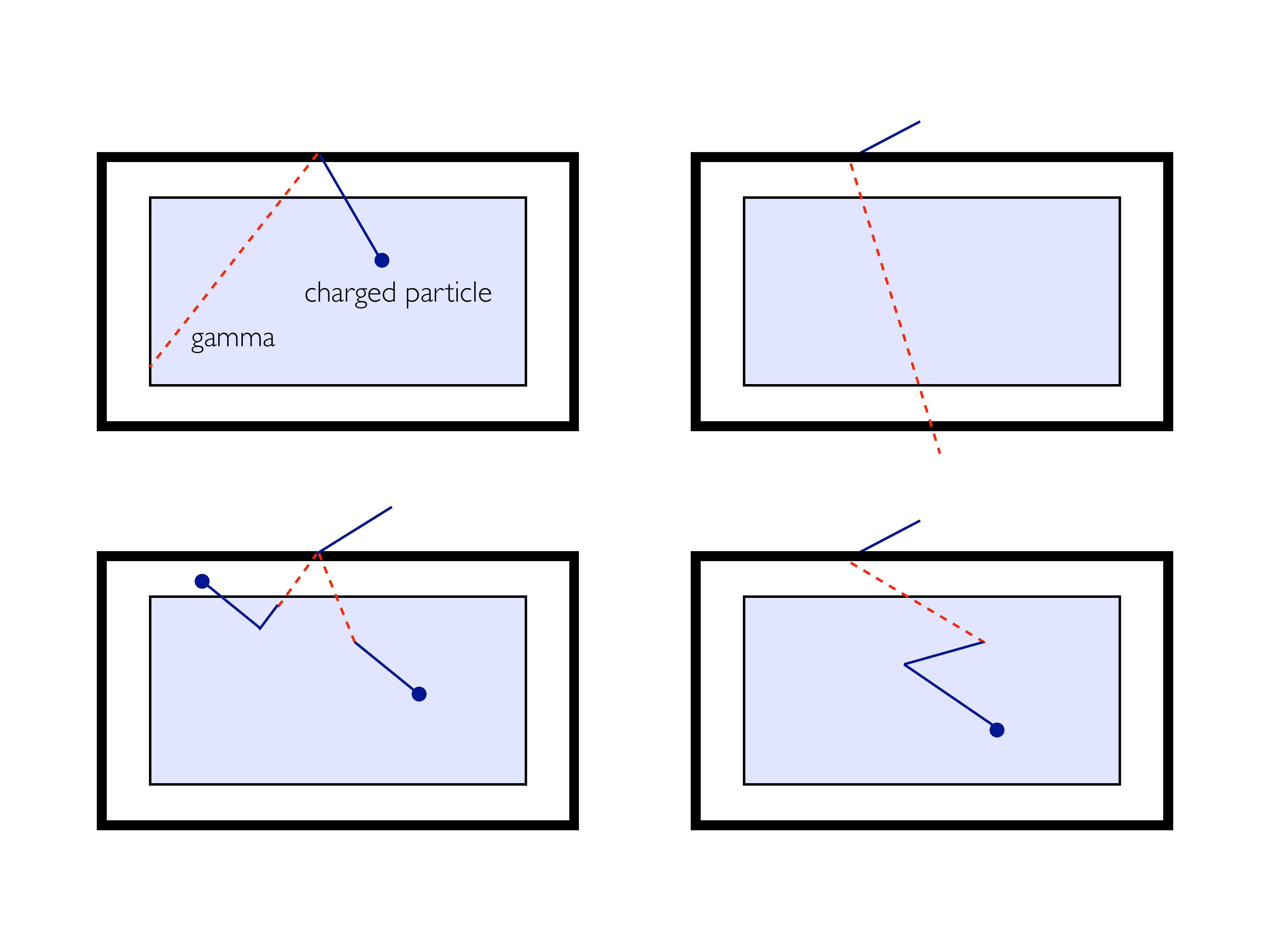}
\caption{\small Charged particle backgrounds entering the detector active volume
		can be rejected with complete 3D-reconstruction (top left). The mean
		free path of xenon for the high-energy gammas emitted in \Bi\ and \Tl\
		decays is $>3$ m, and thus many of them cross the detector without interacting
		(top right). Also, \Bi\ and \Tl\ decay products include low-energy gammas
		which interact in the vetoed region close to the chamber walls (bottom left).
		Only those background events with tracks fully-contained within the fiducial volume
		may mimic the signal (bottom right).
\label{fig:geom_rejection}}
\end{figure}

%

NEXT has three powerful handles to separate \bbonu\ events from backgrounds. These are:
\begin{enumerate}
\item Signal events appear with equal probability in the target, that is, the gas that fills the PV. Defining a fiducial volume, separated from the PV walls by a few cm of active target permits eliminating events in which a high energy gamma is accompanied by charged particles that exit the PV walls. In the ANGEL design, 4 cm of gas are left out between field cage and PV. These are necessary for electric insulation but with proper instrumentation can double up as active veto. In any case, the requirement that the events are strictly contained in the active fiducial volume guarantees that any event with charged activity is eliminated (see Figure \ref{fig:geom_rejection}). Notice that, from the point of view of rejecting backgrounds, an HPGXe behaves in a complementary way than a LXe. While liquid xenon has excellent self-shielding properties, xenon gas, even at (moderately) high pressures, is very transparent to gammas.
\item Signal events have all the same energy. NEXT target resolution is 1\% FWHM and better resolutions can be
achieved (our measurements with small detectors point out to 0.4\%, close to the intrinsic resolution in xenon). Imposing that the events are in the ROI (taken as 1\% FWHM) eliminates substantially the backgrounds.
\item Signal events have a distinctive topological signature that can be exploited to further suppress the 
background.  Our calculations yield a rejection factor of 20 at moderate efficiency cost. This is one of the strongest points offered by NEXT
technology.
\end{enumerate}

%
%

\section{The topological signature} 
\label{sec:sel}

\subsection{Voxelization}

The initial processing of the event allows the ``voxelization'' of the track, which is formed in terms
of connected  3D ``voxels'' formed using the 2D position given by the tracking plane and the
third dimension given by time information. The initial voxels are cubes of 1 cm$^3$~in volume, corresponding 
to the pitch between the SiPMs and twice the EL grid distance. The size of isolated photons is typically one voxel, while a \bbonu\ ``track'' will consist of about 20 voxels. 

Dedicated software (e.g, the RESET algorithm, described in chapter 5) is being developed to voxelize the event. REST also obtains the energy of the event, after correcting for all relevant effects (position of the event, attachment, etc.). The list of (unconnected) voxels is then passed to the pattern recognition algorithms. 

\subsection{Rudiments of graph theory}

For each event, the set of voxels with an energy deposition different from zero can be regarded as a graph: that is, a set of vertices and edges that connect pairs of vertices. Two voxels ($1$~ and $2$~) belonging to a specific set (and regarded as cubes) are connected if they have either one side, one edge or one vertex in common. Mathematically:
\begin{eqnarray*}
x_1 &=& x_2\, (\pm  1) \\
y_1 &=& y_2\,  (\pm  1)\\
z_1 &=& z_2\,  (\pm 1) \\
\end{eqnarray*}

Every group of voxels with energy deposition different from zero is characterized by a square matrix $a$~ of dimension equal to the number of voxels belonging to that group. It is called adjacency matrix, because it describes the closeness of each pair of voxels. The voxels are ordered  and numbered, 1, 2, 3, $\ldots$ Then, the matrix element $a(i,j)$ is built in the following way:
\begin{eqnarray*}
a(i,j) =  \left\{\begin{array}{cccc} 1& \textrm{ if } \left\{\begin{array}{c} x_i = x_j \pm 1\\
y_i  =  y_j \\
z_i  = z_j\\
\end{array}\right.\nonumber &  \textrm{ or } \left\{\begin{array}{c} x_i = x_j \\
y_i  =  y_j \pm 1 \\
z_i  = z_j\\
\end{array}\right.\nonumber &  \textrm{ or } \left\{\begin{array}{c} x_i = x_j \\
y_i  =  y_j \\
z_i  = z_j \pm 1 \\
\end{array}\right.\nonumber \\
\\
\sqrt{2} & \textrm{ if } \left\{\begin{array}{c} x_i = x_j \pm 1\\
y_i  =  y_j \pm 1 \\
z_i  = z_j\\
\end{array}\right.\nonumber &  \textrm{ or }  \left\{\begin{array}{c} x_i = x_j \pm 1\\
y_i  =  y_j  \\
z_i  = z_j \pm 1\\
\end{array}\right.\nonumber &  \textrm{ or }  \left\{\begin{array}{c} x_i = x_j \\
y_i  =  y_j  \pm 1 \\
z_i  = z_j \pm 1\\
\end{array}\right.\nonumber \\
\\
\sqrt{3} & \textrm{ if } \left\{\begin{array}{c} x_i = x_j \pm 1\\
y_i  =  y_j \pm 1 \\
z_i  = z_j \pm 1\\
\end{array}\right.\nonumber &  & \\
\\
10^9 & \textrm{ if } \left\{\begin{array}{c} x_i = x_j\\
y_i  =  y_j  \\
z_i  = z_j \\
\end{array}\right.\nonumber &  & \\
\\
0 & \textrm{ elsewhere } &  & 
\end{array}
\right.\nonumber
\end{eqnarray*}
where $x(y,z)_i$ is the $x(y,z)$ coordinate of the voxel $i$. A matrix element different from zero means that the corresponding voxels are close (and the number represents the distance in cm between the centers of the voxels), while if it is zero, the two voxels are not close. The diagonal elements are conventionally set to an arbitrary high number.

\subsection{The Breadth First Search algorithm}
 
With the adjacency matrix it is possible to ``walk'' across connected voxels, therefore to identify connected subsets (the tracks) and follow paths along tracks. In order to do this the so called BFS (Breadth First Search) algorithm is exploited. The BFS starts from a voxel and calculates the (shortest) path to any other voxel. In this way it is possible to identify the number of tracks for each event and to calculate the length of the path between any pair of voxels of each track. 

\begin{figure}[htb]
\begin{center}
\includegraphics[width=0.7\textwidth]{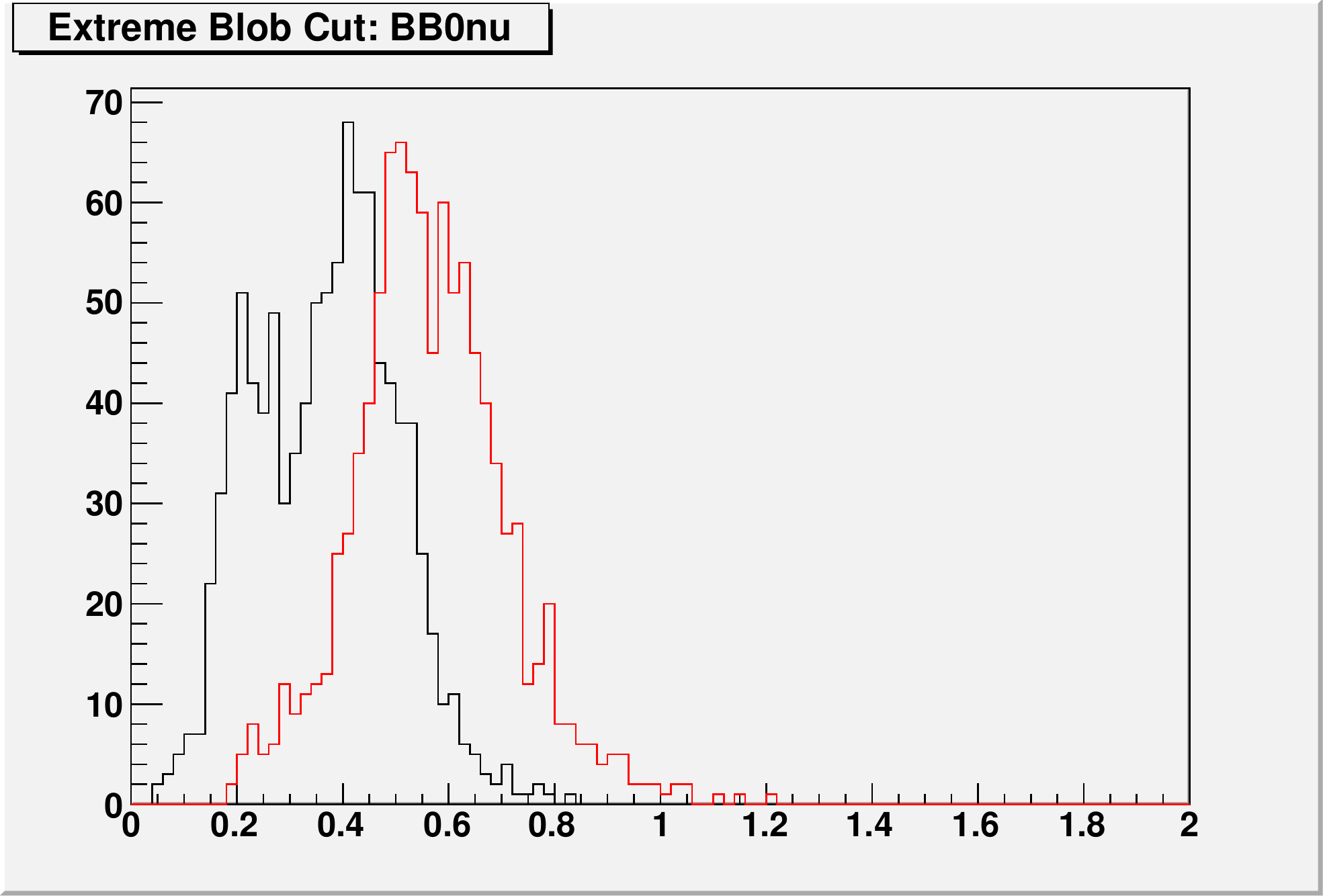}\\
\includegraphics[width=0.7\textwidth]{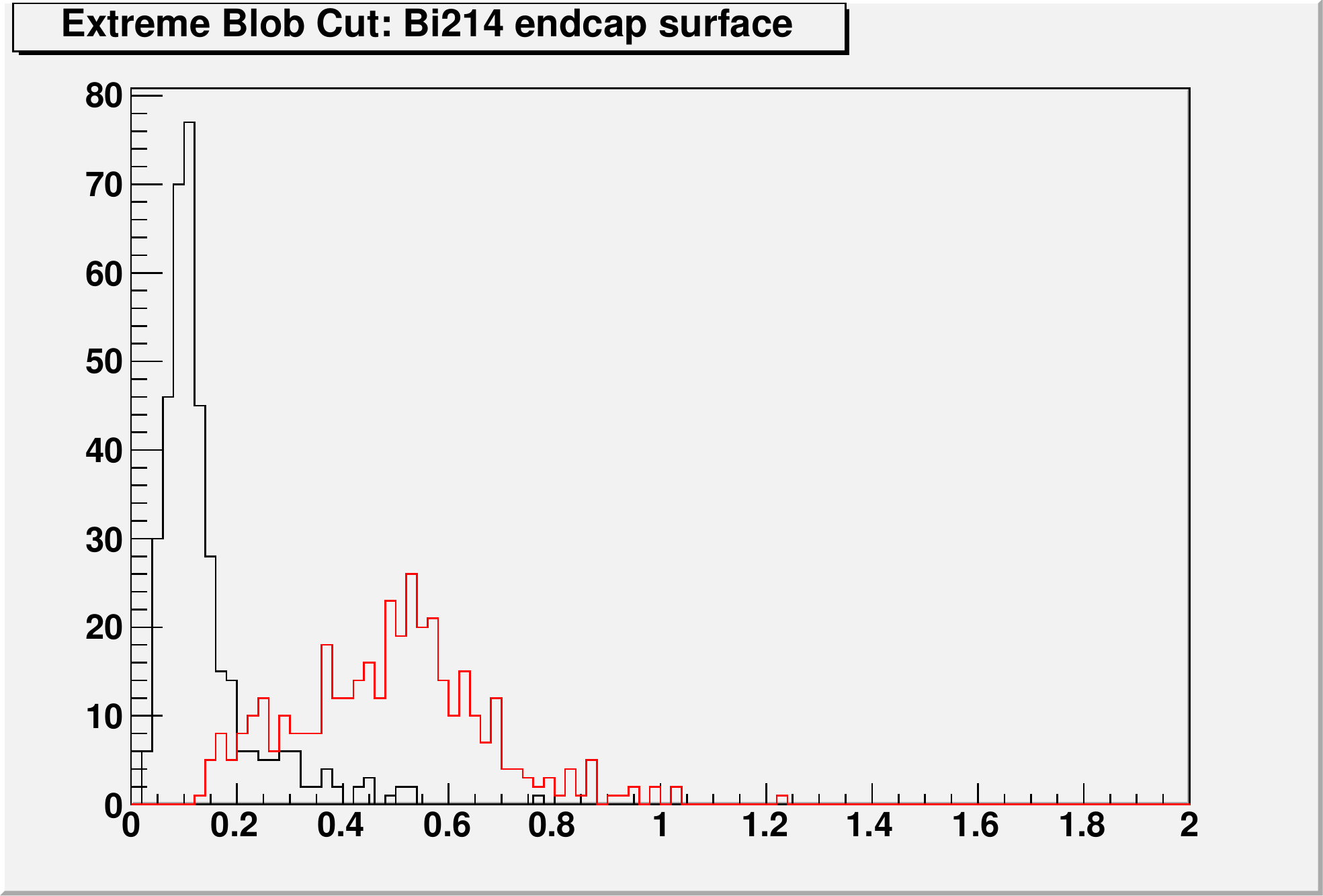}
\caption{\small Distributions of the energy inside a sphere of $R=2$ cm in both extremes, for \bbonu\ events (top) and for \BI\ backgrounds. The higher energy blob is shown in red.}\label{fig:extreme}
\end{center}
\end{figure}

Using the BFS we calculate the paths between any pair of voxels and pick up the longest of such paths. The corresponding voxels of the pair are called the extremes of the track. Afterwards, for each extreme, the voxels whose distance from it is smaller than a radius $R$ are considered and their energies summed up to give $E^R_{1(2)}$, where label 1(2) is given to the extreme with higher (lower) energy. In Fig.~\ref{fig:extreme} the distributions of the energy contained in the sphere of radius $R$ around the extremes of the track are shown.  
Fig.~\ref{fig:extreme} shows the energy inside $R$ for signal and the \BI\ background. In the case of the signal the energy of both extremes is quite similar, as expected, while the background shows very different shapes for the track extremes. A good cut to exploit this difference is to impose that an event have both $E^R_1$ and $E^R_2$ greater than a threshold value $E_{th}$.  

According to our Monte Carlo calculations a cut $E_{th}=0.4$ MeV and $R=2.5$ cm keeps 78\% of the signal while suppressing the background by more than one order of magnitude. The background suppression can be increased to a factor 20 at the expense of less efficiency (55\%).   

\section{Sensitivity}
\subsection{Event selection}

An event is accepted as a \bbonu\ candidate if: 

\begin{enumerate}
\item	{\em The event is fully contained in the fiducial volume}
\item	 {\em Only one reconstructed track}: That is the BFS algorithm finds only one connected set of voxels. Events with more than one object (tracks or disconnected voxels) are rejected. 
 
\item	 {\em Energy in the ROI}: The event is required to be within  1 FWHM (1\%) of \Qbb.

\item	{\em \bbonu\ signature}:  the unique track ends in two blobs of high energy, as expected for a \bb\ event. This is translated into a cut $E_{th}=0.4-0.55$~ MeV and $R=2.5-3.0$ cm, that allows a suppression factor between 10 and 20 for the background at the expense of selection efficiency between 78\% and 55\%.
\end{enumerate}

\subsection{Signal  efficiency and background rejection power} 
\label{sec:rejpower}

To estimate the performance of the NEXT ANGEL design \bbonu\ and background data samples have been generated with the NEXUS Monte Carlo simulation. Signal events are simulated in the volume inside the vessel. Backgrounds (\TL\ and \BI\ decays) are simulated in the vessel and in the readout planes. The events thus generated are passed through the selection procedure.  

\begin{table}[pbht!]
\begin{center}
\begin{tabular}{|c|c|}
\hline \hline
Selection criteria & Fraction of events surviving cuts \\
\hline 
Events analyzed & $10^8$  \\ 
Fiducial, 1 track & $6 \times 10^{-5}$  \\
ROI & $2.2 \times 10^{-6}$  \\ 
Topology & $1.9 \times 10^{-7}$\\ 
\hline \hline
\end{tabular}
\caption{\small Suppression of the \BI\ events by the selection cuts.}
\end{center}
\label{tab:BI}
\end{table}%

\begin{table}[pbht!]
\begin{center}
\begin{tabular}{|c|c|}
\hline \hline
Selection criteria & Fraction of events surviving cuts \\
\hline 
Events analyzed & $10^8$   \\ 
Fiducial, 1 track & $2.4 \times 10^{-3}$  \\
ROI & $1.9 \times 10^{-6}$  \\ 
Topology & $1.8 \times 10^{-7}$\\ 
\hline \hline
\end{tabular}
\caption{\small Suppression of the \TL\ events by the selection cuts.}
\end{center}
\label{tab:TL}
\end{table}%

\begin{table}[pbht!]
\begin{center}
\begin{tabular}{|c|c|}
\hline \hline
Selection criteria & Fraction of events surviving cuts \\
\hline 
Events analyzed & $10^6$   \\ 
Fiducial, 1 track & 0.48  \\
ROI (1 FWHM) & 0.33  \\ 
Topology & 0.25 \\ 
\hline \hline
\end{tabular}
\caption{\small Signal efficiency.}
\end{center}
\label{tab:eff}
\end{table}%

Tables \ref{tab:BI}, \ref{tab:TL} and \ref{tab:eff} summarize the rejection factors for \BI\ and \TL\ backgrounds as well as the signal efficiency. The background rate of the experiment is obtained multiplying by the activity emanating from the pressure vessel, the PMTs and the rest of
the components of NEXT.

As an example, let's consider the background illuminating the fiducial volume from the PV. This is about $2 \times 10^6$~ events per year. Since the rejection factors for \BI\ and \TL\ are about the same ($\sim 2 \times 10^{-7}$),
we can simply multiply one quantity by the other to obtain $4 \times 10^{-1}$ counts/year. Divide now by the 
fiducial mass (100 kg) and by the ROI width (25 keV) to obtain $1.6 \times 10^{-4} \ckky$. Adding the contribution of the PMTs results in $2 \times 10^{-4}$~ counts/keV/kg/year. 

This background rate is one of the lowest in the field. It has improved with respect to the analysis presented in the LOI thanks to the titanium vessel, which has
one order of magnitude less background than the steel vessel considered in the ROI. 

\section{Sensitivity of the NEXT experiment to a light Majorana neutrino}

\begin{figure}[t!b!]
\centering
\includegraphics[width=0.7\textwidth]{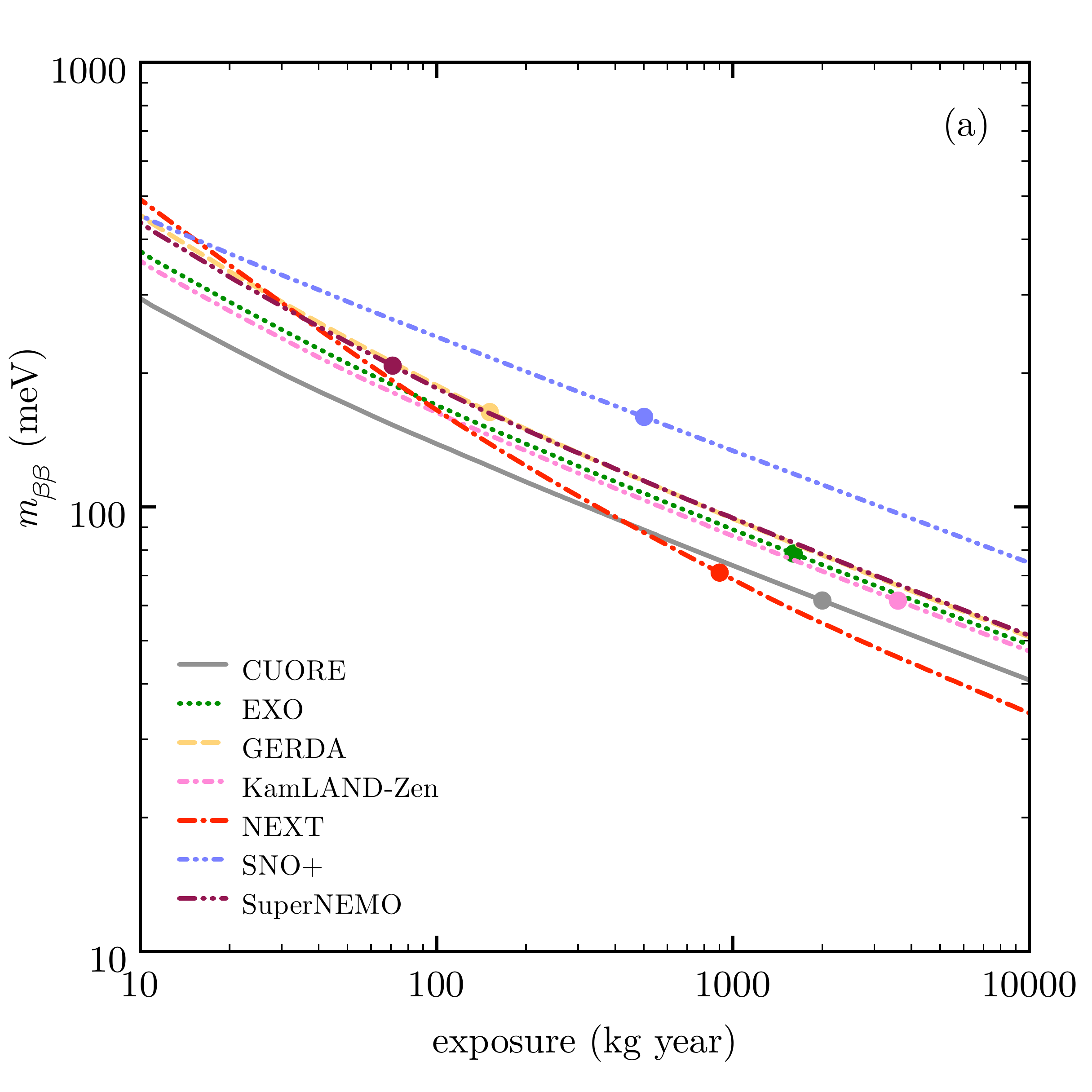}
\caption{The \mbb\ sensitivity (at 90\% CL) as a function of exposure of the seven different \bbonu\ proposals considered. The expected performance of the experiments other than NEXT has been discussed 
in chapter 1 and in \cite{GomezCadenas:2010gs}.  For illustrative purposes, the filled circles indicate 10 years of run-time.} \label{fig.SensiF}
\end{figure}

Figure \ref{fig.SensiF} shows the \mbb\ sensitivity (at 90\% CL) as a function of exposure for  seven \bbonu\ proposal (see chapter 1 and \cite{GomezCadenas:2010gs}). As it can be seen the expected performance of NEXT is excellent.


\chapter{Shielding options for the NEXT detector}

\section{Shielding requirements}

As extensively discussed elsewhere in this report, an exposure of the order of the ton--year of $^{136}$Xe, and a neutrino effective mass of 50\,meV, result in only a few counts per year at the $Q_{\beta\beta}$ energy of 2.457\,MeV. To be sensitive to such low rates NEXT projects a background of the order of $10^{-4} ~\ckky$. This forces a very effective shielding to protect the experiments against the laboratory walls contamination, mainly the 2.615\,MeV  and 2.448\, MeV photons coming from the $^{208}$Tl and \BI\ isotopes. 

The NEXT collaboration has studied two possibilities for shielding. One of them is based on housing the detector in a lead-copper structure (henceforth called the Lead Castle), the other one in the use of a water tank. 

Radon in air, water and from material emanation will affect the experiment and the shielding scheme must be designed to suppress sufficiently this contribution. Consider for example the case of a Lead Castle. The  
radon contamination at the LSC is of the order of 100\,Bq/m$^3$. Assume that the volume of air around the detector is 2.55\,m$^3$ (this corresponds to around 20\,cm of distance between vessel and shielding). Then, more than 100 counts would be registered after cuts in the ROI. On the other hand, air-borne radon can effectively be suppressed by enclosing the detector in a ``plastic bag'' in which a continuous flux or nitrogen is established, or working in a radon free atmosphere (the air in the inner volume passes through radon scrubbers). 

Radon in the water is also a concern in the case of a water tank (for example, GERDA measures around 7-19\,mBq/m$^3$).  Radon suppression in this case is achieved by a slightly over-pressurized nitrogen
blanket which will be kept between the water level and the roof of the tank. In addition the water will be bubbled through a column of nitrogen to eliminate radon degassing from the tank walls. 

\section{The Lead-Copper Castle option}

A lead castle, similar to the XENON100 one,  has been investigated  for the shielding  of NEXT100 against the external gamma radiation. The goal is  to develop a simple and compact design that allows an easy access to the detector.  Rectangular and a round geometries  have  been studied also to minimize the mass of the lead and  the dimensions of the supporting structure; additionally different lead providers have been contacted for a preliminary investigation of the price.

This shielding design offers several advantages:
\begin{itemize}
\item[-] More compact design even replacing a few cm's of lead by water \emph{bricks}.
\item[-] Once it has been mounted is \emph{easy} to maintain.
\item[-] This design could grow up with the experiment.
\item [-] It is flexible enough to face up to any unexpected change (extra space for electronics, extra inner shielding ...).
\item [-] Nitrogen flux not only prevents radon from approaching the detector outer surface, but also offers a certain degree of thermic bath.

\end{itemize}

\subsection{The XENON concept}

\begin{figure}[bthp!]
  \centering{
  \includegraphics[width=10cm]{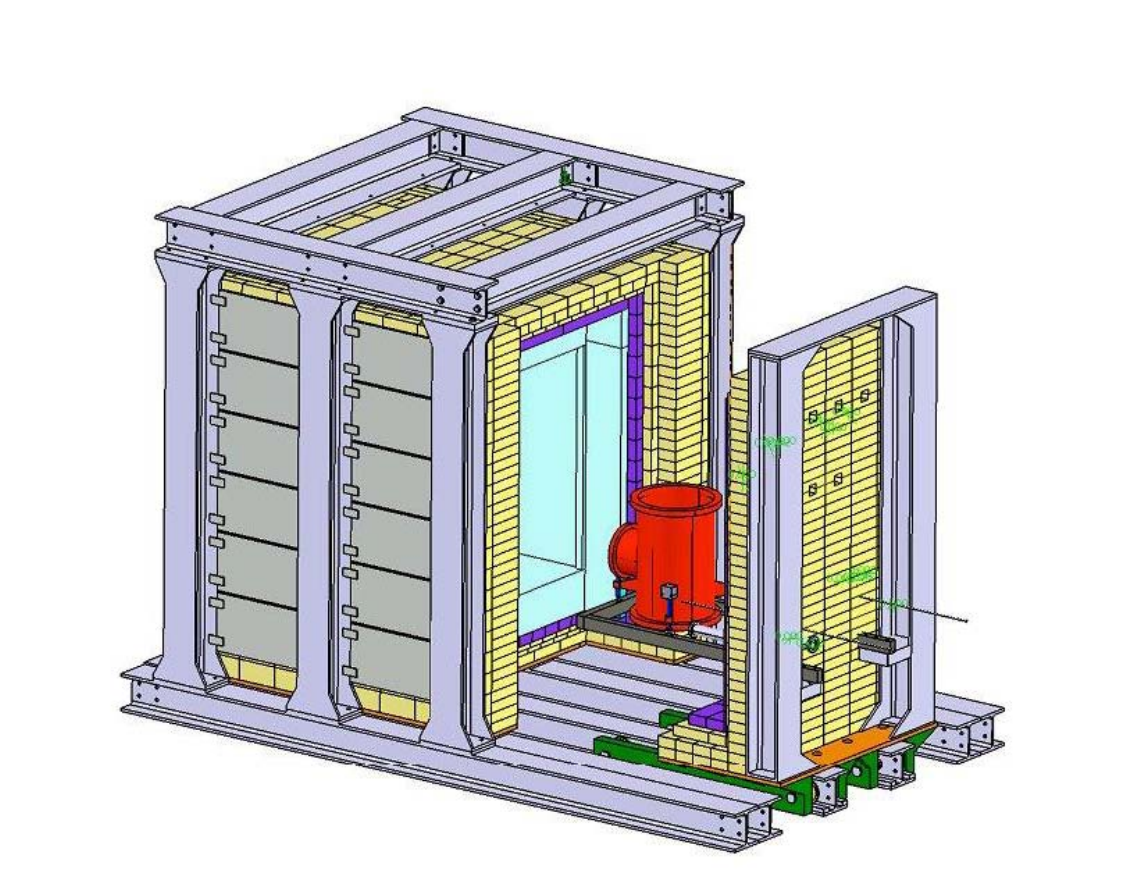} \\
  \caption{Picture showing the XENON10 shielding and the steel moving structure.}\label{figura0}}
\end{figure}

A shield made out of lead bricks has been build by the XENON collaboration for the 10\,kg detector (see figure \ref{figura0}), operated at Laboratori Nazionali del Gran Sasso (LNGS) during the period 2006-2007. Due to its good performances, the same lead castle has been re-used (after minor modifications) for the XENON100 experiment that is currently operating in the same site. The design foresees a cubic steel-framed structure consisting of 20\,cm of Pb with an additional layer of high density polyethylene. The shield structure is secured by steel panels along the outer walls. The detector is attached to a movable door placed on  two rails, and it can be easily put in and out by  one person.
Lead was supplied in standard bricks (5$\times$10$\times$20) by two different companies: Plombum (Poland) for the outer 15\,cm and Fondery de Gentilly (France) for more radiopure  inner 5\,cm layer. An additional inner 5\,cm layer of high-purity copper has been added during the upgrade for XENON100.
All the holes of the shield have been sealed with low--radioactivity silicon gel thus the cubic internal cavity (~1m$^3$) can sustain a slight over pressure. In order to reduce the Rn concentration inside the shield, N$_2$ is continuously flushed at a rate of 1.5\,l/min.  Due to the need of additional shielding, water tanks have been placed around this shield.

\subsection{Water bricks}

This are just containers filled up with water to further reduce the external radiation. The final size of these containers have not yet been defined as there are many different sizes and geometries of water containers available in the market. Even costumed made containers adapted to the inner shielding could be used. In our design, we have used 240\,mm $\times$ 290\,mm $\times$ 380\,mm as an arbitrary choice.

One of the advantages of including water containers in the shielding is the possibility to add PMTs  to use the system as an active veto against muons and external radiation. 

\subsection{Proposed design for the NEXT100 shielding}

This proposal is based on the XENON-like option. The pre-requisitess for our design are:
\begin{itemize}
\item[-] The vessel dimensions, adjusted to the ANGEL design plus a safety margin.
\item[-] Vertical position of vessel.
\item[-] Max. length of cables 5\,m.
\end{itemize}

Some basic points for the shielding design are:

\begin{enumerate}

\item	A shield made up of several different layers (a \emph{sandwich} or \emph{onion}style shield), with:
    \begin{itemize}
        \item[-]inner level: 5\,cm of copper, with the possibility of upgrading to 10 cm or more\footnote{The standard sheet thickness supplied by LUVATA is 2.5\,cm. Two or four plates will be used},
    \item[-]mid level: 15\,cm of lead,
    \item[-]outer level: 1\,m of water, placed into \emph{water bricks} (plastic containers filled up with water).
    \end{itemize}
\item	The shield geometry is composed of straight horizontal and vertical walls.
\item		On the ground level a thicker layer of lead is foreseen (30\,cm) as no water bricks can be used under such a heavy structure.
\item	Cabling and piping require spaces between shield and vessel: at least 100--150\,mm between the side wall of the vessel and the shield and at least 300\,mm  between the vessel endcaps and the shield.
\item	The service plate of the vessel is on the upper part. A covered opening at the shield roof allows cables and pipes to get the vessel.
\item Once installed, the detector should stay in place, motionless, due mostly to its weight, but also to avoid small displacements of inner part of the detector.
\item	After installation, the shielding needs a mechanism to allow an easy access to the immobile detector. This will be done through sliding walls at the front of the shield. Then, cables and pipes will not be affected by the opening of the shield and won't need to be disconnected.
\item	To prevent radon contamination, a plastic bag will be place surrounding the structure and nitrogen flux will be flushed inside.
    \item Finally, standard properties or supplier specifications have been considered for material involved in the design (steel, copper and lead). 
\end{enumerate}

The building method we propose consists of copper panels and chevroned lead bricks. These lead bricks are  commonly used for radiation isolation, have standard geometries and, as they fit one into the other, the built wall are stronger and with less spaces for the radiation to traverse them. They have been placed in an upright position for the walls and lying on their side for the floor and roof, water containers have been piled, and supported steel structures hold the whole assembly. Two different structures are needed: one for the inner copper plus lead shield and another one for  water containers. Catia V5 R19 has been used for the design and Ansys 11.0 for mechanical calculations.

\subsection{Basic copper and lead layers with supporting structure}

The basic elements of the lead shield are:
\begin{itemize}
\item[-] an U-shaped wall (thickness: 150\,mm)
\item[-] the front wall (thickness: 150\,mm) divided into two halves due to its more than 9.5 Tons weight.
\item[-] the floor (thickness: 300\,mm)
\item[-] the roof (thickness: 150\,mm)
\end{itemize}

\begin{figure}[bthp!]
  \centering{
  \scalebox{0.6}{\includegraphics{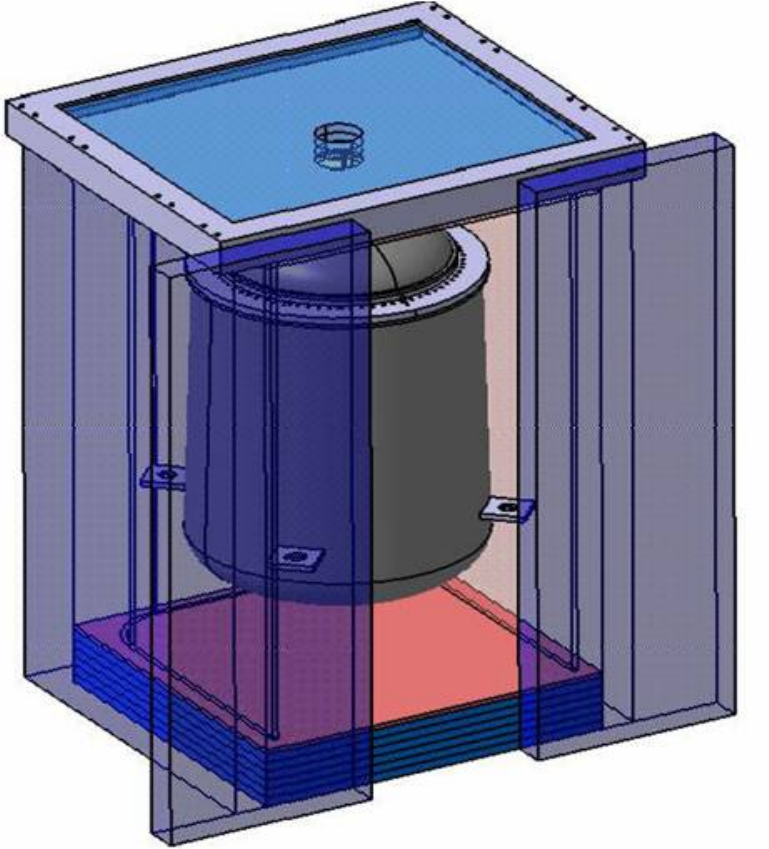} } \hspace{1 cm}
  \scalebox{0.6}{\includegraphics{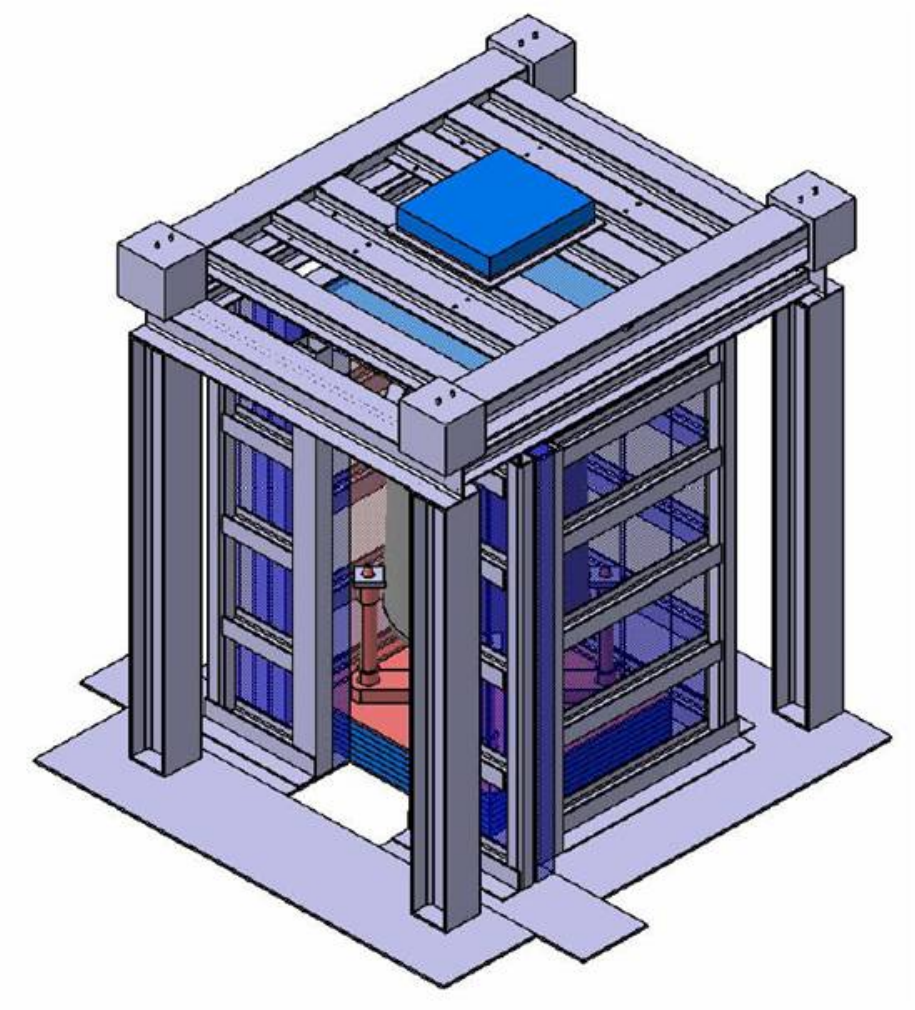}}\\
  \caption{Inner copper and lead shield with doors half open (left).  View of the structure, supporting the copper and lead shielding, and the door opening mechanism (right).}\label{lfig1}}
\end{figure}

\begin{figure}[bthp!]
  \centering{
  \includegraphics[width=12 cm]{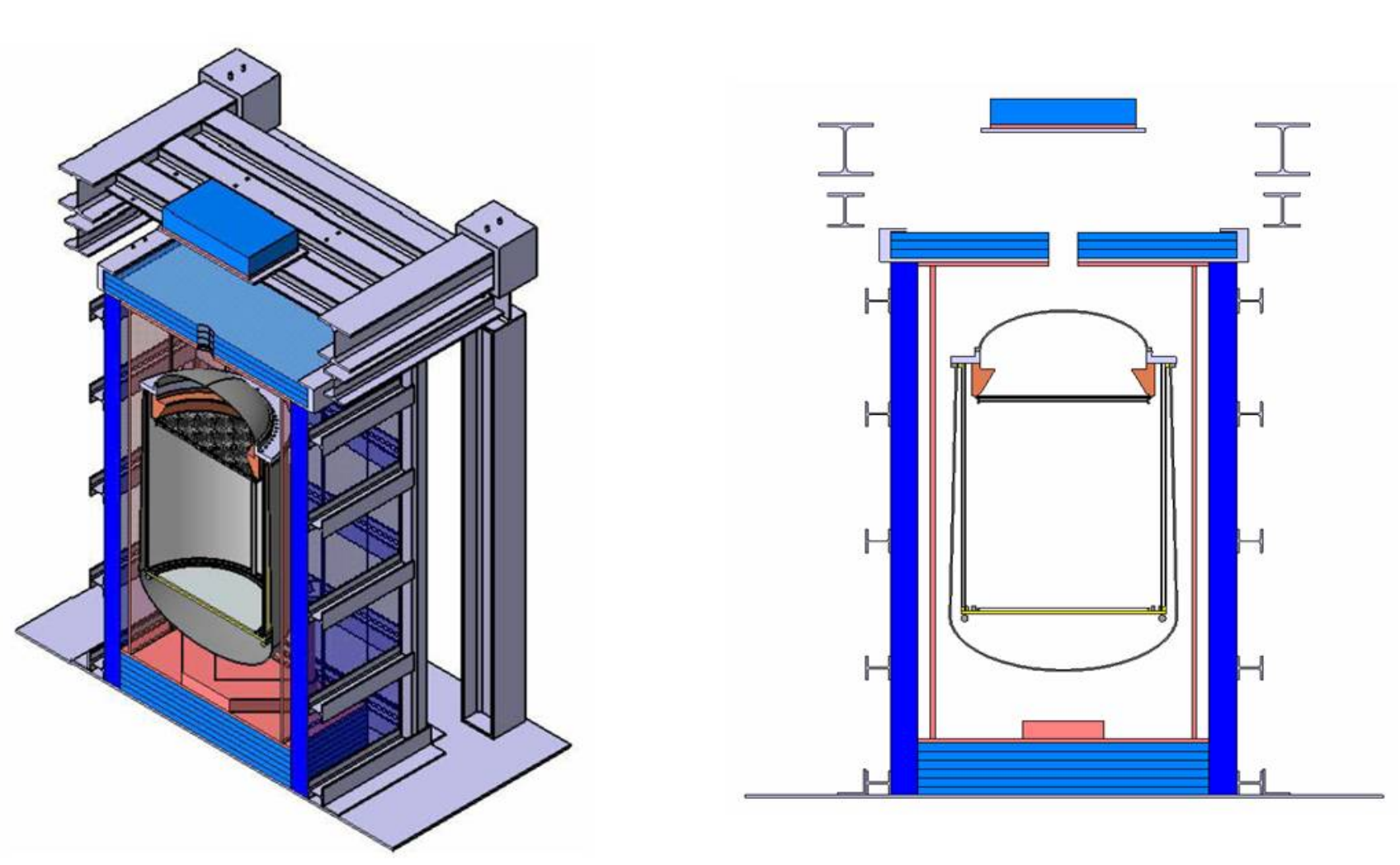}\\
  \caption{Isometric view of the assembly cut by a parallel to the front wall plane.}\label{lfig2}}
\end{figure}

The roof (6 Ton weight) has to be supported by an independent steel structure instead to lie upon the lead walls to avoid the collapse of these vertical walls.
In addition, some structural elements will prevent buckling of the vertical walls. A supplementary element with a layer of copper and a layer of lead has been added over the roof to cover the hole through which cables and pipes exit the inner shield (details in figures \ref{lfig1} and \ref{lfig2})

For the design we have performed static calculations, which take into account the effect of large displacements, and buckling calculations, considering standard properties for lead and steel, and Luvata datasheet's for copper.
Each wall has been considered as a continuous whole body for the calculations, independently of the construction basic elements used for building them. These elements have been thought to fit together properly and eventually be fixed together (glued). The chosen global solution should not have horizontal solicitations or displacements too high on the vertical walls, to maintain this continuity condition.

Under these assumptions we got as results:
\begin{itemize}

\item[-] {\bf Front walls  or U-shaped wall alone}

The general load case doesn't originate problems since deformations produce a maximum total displacement of less that 0.3\,mm and a maximum stress of less than 0.57\,MPa (well below 1.4\,MPa, the elastic limit for lead).

Buckling (or compressive instability) is  dangerous. A little pressure on the upper side of the walls would make them fall down. Without reinforcement elements, the first buckling mode appears at P=87.7\,Pa in front walls and at P=755.8\,Pa in U-shaped wall, well below the pressure caused by a single lead brick (1134\,Pa).

\item[-] {\bf Roof alone}

The weight of the whole roof is near 6\,Tm. Taking into account the buckling problems of the walls, the roof should not lie upon the lead walls of the castle because it would cause the collapse of these walls. Instead of this, it has to be supported by an independent outer structure. In our design, the roof has been placed on a 20\,cm thick steel plate held from 4 anchor points placed at each lateral side of the plate by an outer structure from the top.

 According to calculations, there are no major problems for the general load case. First, we have calculated the reaction forces by considering the anchor points as hard points (we need these forces to apply them on the outer structure that should hold the roof). Then we have considered these anchor points as cylindrical holes with one of their borders fixed (to be closer to a realistic situation). Finally, we have considered a more realistic case: the anchor points are cylindrical holes on the plate and a cylindrical crown around the border of the hole is fixed. For this last case, the maximum displacement that appears on the plate is around 0.5\,mm (that we consider acceptable) and the maximum stress is around 45\,MPa (well below the elastic limit of 220\,MPa for steel).

\item[-] {\bf Assembly with the front walls and the U-shaped wall}

We have performed several calculations of the buckling behavior of this assembly and
some structural elements have to be added to prevent buckling of the vertical walls.

In the assembly without any reinforcement the first buckling mode appears at a pressure level of 754\,Pa (below the level of 1134\,Pa that is the effect of the addition of a single layer of bricks). However, after an iterative process of reinforcement, we have arrived to a final configuration where the first buckling mode appears at a pressure level of 1457\,Pa (near 30\% above 1134\,Pa). Further investigations can be done to improve this result.

Coming back to the independent elements (the U-shaped wall and the front walls), no problems are expected for the U-shaped wall due the reinforcement structure defined for the assembly, but we also need to define a narrow support element for the inner side not compromising radiopurity (shielded by the copper layer) in the case of the front walls.

\item[-] {\bf Support structure for the lead roof}

We have started the definition of the structural elements needed to support the roof. Initial studies show that the designed structure built with standard steel beams do not suffer any deformation or buckling, even if we constrain all the degrees of freedom at the joining points.

\end{itemize}

\subsection{Water bricks and supporting structure}

The water bricks will surround the rest of the shield with upper water bricks (around 2400 kg) placed on a platform above the rest thanks to an independent structure. The design presented here is preliminary and should be refined in future design stages.
Here, we have divided the water bricks into different groups:
\begin{itemize}
\item[-] upper water bricks over the roof of the lead structure,
\item[-] side water bricks, beside the arms of the U-shaped lead wall,
\item[-] back water bricks, beside the back of the U-shaped wall,
\item[-]front water bricks, beside the front lead wall,
\item[-]diagonal water bricks at each one of the 4 corners of the shield.
\end{itemize}
Vertical panels could be included at inner sides of each group of water bricks, to prevent them from falling down. Calculations of the structure needed to hold the water tanks will be done in future stages of development of the castle. In any case, we think that this structure will be less problematic than the one needed for the lead.

\begin{figure}[bthp!]
  \centering{
  \scalebox{0.5}{\includegraphics{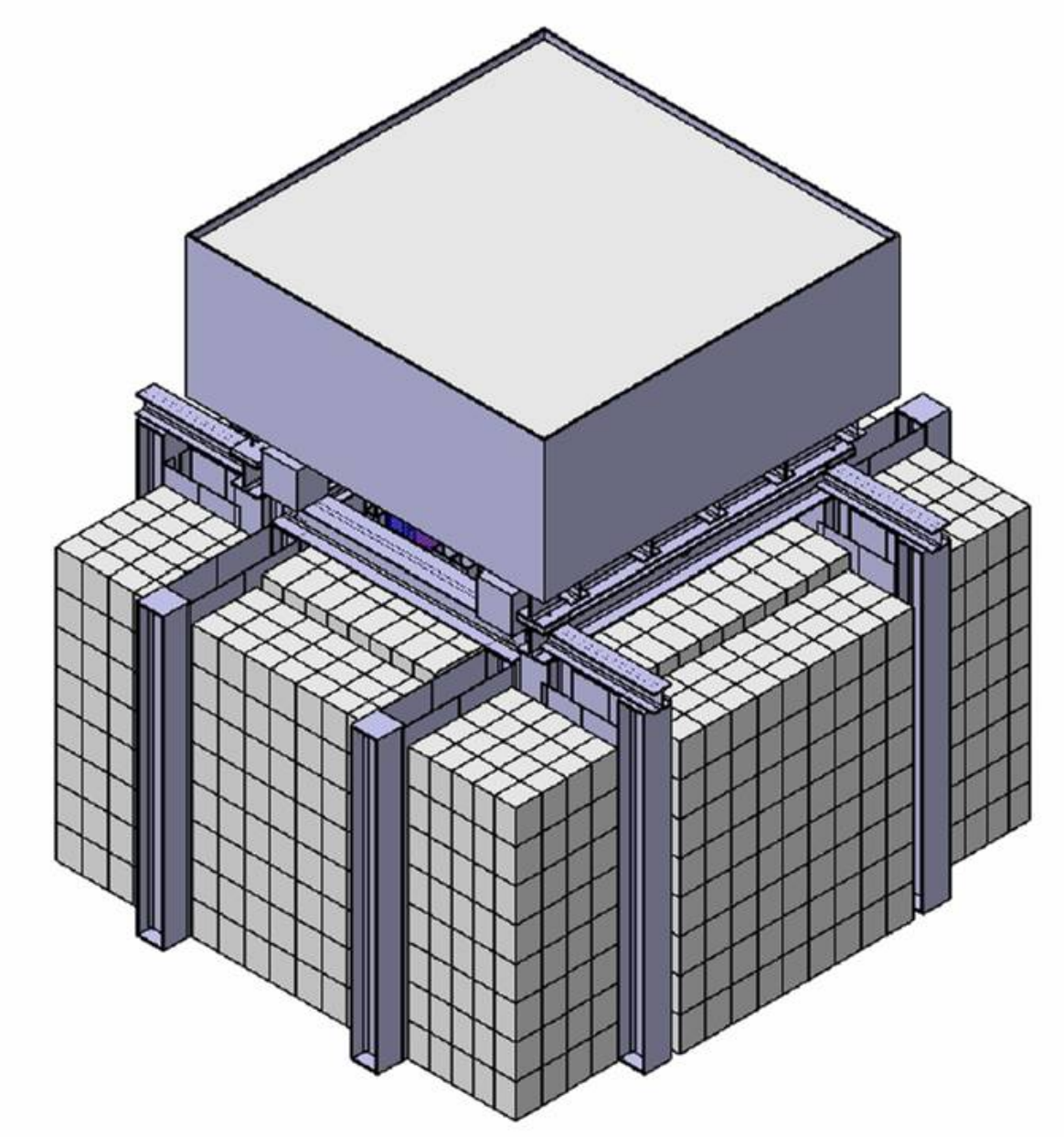} } \hspace{1 cm}
  \scalebox{0.5}{\includegraphics{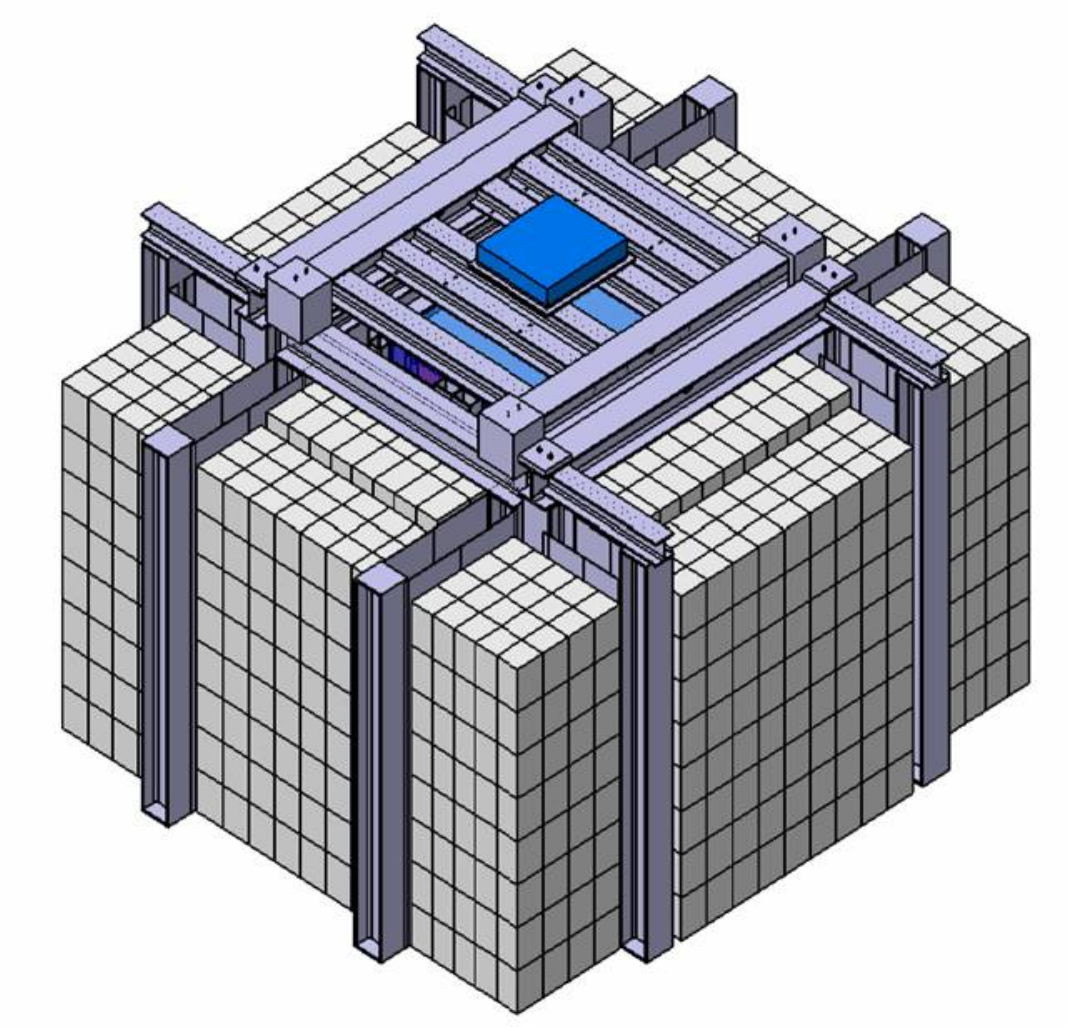}}\\
  \caption{Global view of the whole shield with (left) and without (right) upper box. The top box is designed to contain the water bricks and eventually to act as platform for the electronics.}\label{lfig4}}
\end{figure}

A schematic representation of the final assembly can be observed in figure \ref{lfig4}.

\subsection{Opening sequence}

The present \emph{onion setup} foresees the option of running with the complete shield or with just some layers. An intermediate configuration, with only part of the shield in place, can be considered during the preliminary  underground tests when frequent hardware operation on the detector (feedthrough, gas system, etc) will be required.

\begin{figure}[bthp!]
  \centering{
  \includegraphics[width=15 cm]{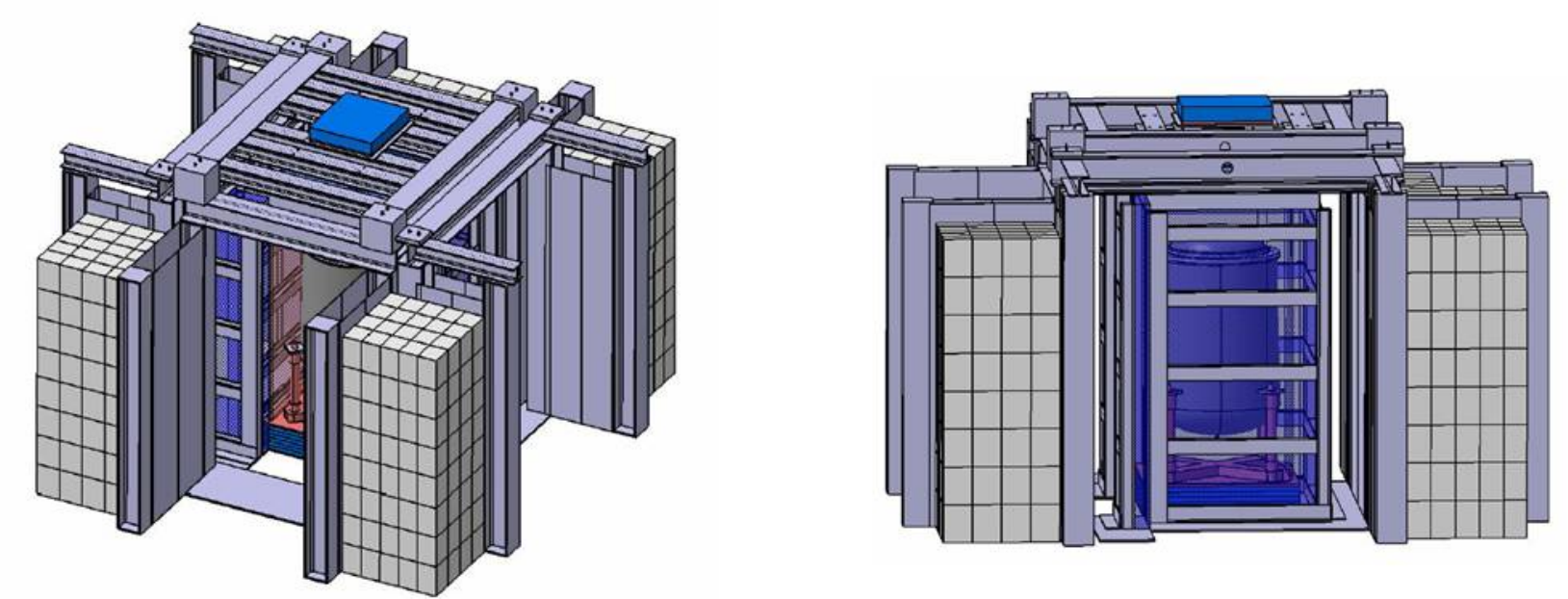}\\
  \caption{Castle partially open, without water roof elements for a better visualization.}\label{lfig5}}
\end{figure}

Once the complete shield is installed, the opening sequence would proceed as follows:
\begin{enumerate}
	\item Front and side water containers removed
	\item Front and side panels removed
	\item Front sliding walls opened. They slide to the sides.
	\item Front Cu removed: this will only be necessary as an independent operation if the Cu front panels are not attached to the sliding doors
\end{enumerate}
Figure \ref{lfig5} shows the shielding after the opening has been completed. Current design can be easily improved, placing the lateral water bricks in such a way that would leave free the necessary space to allow opening the front lead walls without needing to remove these bricks or their supporting panels.

\begin{figure}[bthp!]
  \centering{
  \includegraphics[width=8 cm]{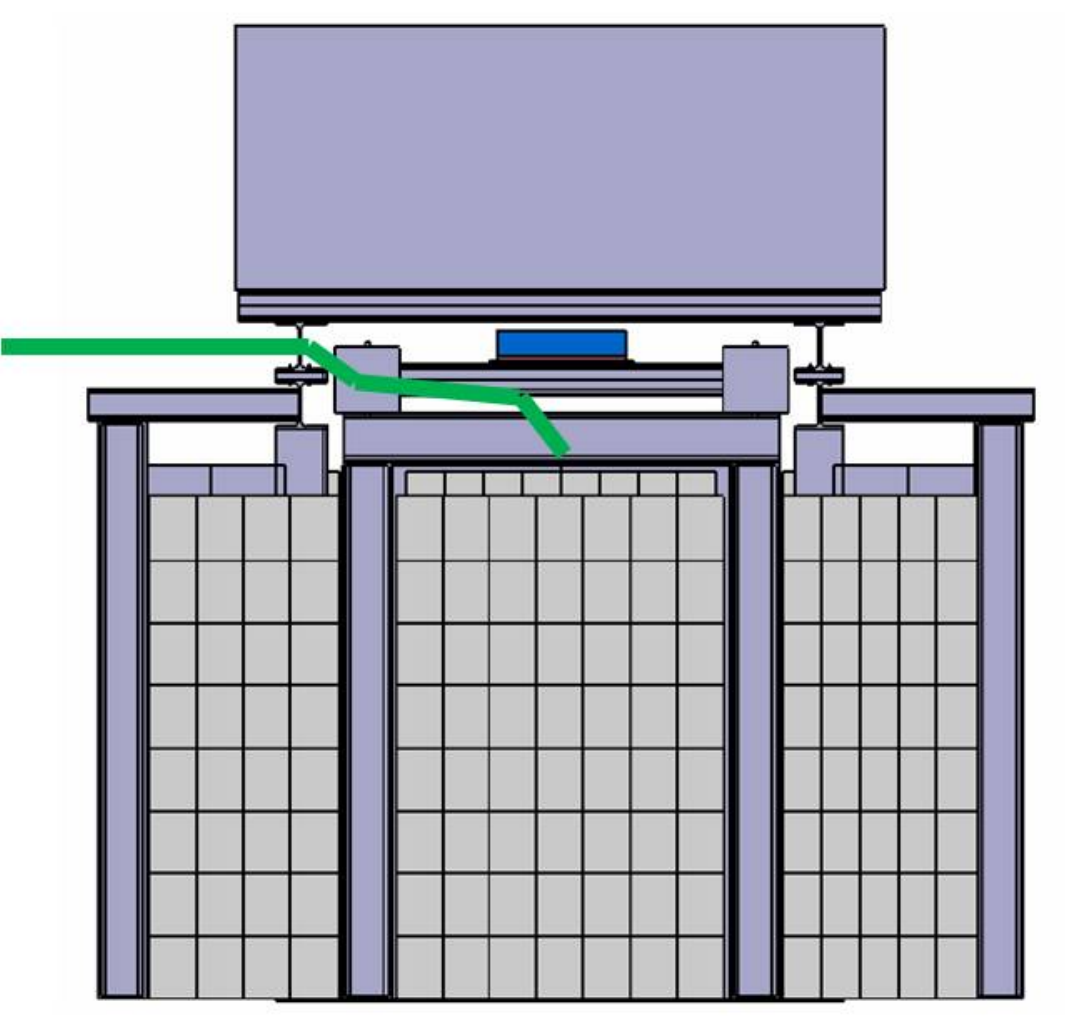}\\
  \caption{Castle front view where one of the possibilities for the pipes and cables to go through the shielding is shown.}\label{lfig6}}
\end{figure}

\subsubsection{Cables and Pipes}

The cables and pipes will go into the shield vertically through the hole at the roof. Then, they will turn horizontal to pass under the supplementary lid over this hole and the steel beams. Finally, a hole in the lead roof will allow cables and pipes to get the vessel. This way, no unprotected lines of pipe or cable are left for radiation to straight penetrate the shielding. Figure \ref{lfig6} illustrates this possibility.

However other possibilities enabling to locate all or part of the electronics closer to the detector can be thought off. The modular design of the Castle Lead makes easier to find a place for the electronics inside the shielding structure.

\subsection{Lead and copper}
\subsubsection{Providers}

Several companies have been contacted as lead and copper providers. An Italian company, COMETA, and a Spanish company, TECNIBUSA, are being considered for two reasons: their competitive prices and their expertise in shielding materials. Possible optimizations of the melting techniques (use of steel matrices and melting in argon atmosphere) are currently studied with both companies in order to get a lower level of radioactive contamination. In case of need of an inner layer of copper, LUVATA company could provide the copper with a high level of radiopurity. In this subsection we will describe the radiopurity screening of materials and some provider offers.

\subsubsection{Radioactivity screening of materials.}

The radioactive contamination of materials such as copper and lead has been measured by Shiva company (EAGlabs) using Glow Discharge Mass Spectrometry (GDMS).
These measurements are quite promising concerning the COMETA lead sample (370\,$\mu$Bq/kg in $^{238}$U --or ${214}$Bi -- and 72\,$\mu$Bq in $^{232}$Th --or 26\,$\mu$Bq/kg in $^{208}$Tl)\footnote{Estimates done considering secular equilibrium in radioactive chain}, and also to a sample from MIFER's lead provider(0,34\,mBq/kg in  $^{238}$U --or ${214}$Bi -- and 100\,$\mu$Bq in  $^{232}$Th --or 36\,$\mu$Bq/kg in $^{208}$Tl), which is also TECNIBUSA's provider. Improvement in melting technique would decrease radioactivity levels. Concerning copper LUVATA company could offer a level of radiopurity of less than 12.4\,$\mu$Bq/kg from $^{238}$U  and less than 4\,$\mu$Bq/kg from $^{232}$Th  (less than 1\,$\mu$Bq/kg from $^{208}$Tl). In case, we do not manage to get more radiopure lead (~10\,$\mu$Bq/kg) we could include 10\,cm of copper inside the lead castle (this would attenuate the$^{208}$Tl 2615\,keV photon coming from lead  in a factor 10).

\begin{table}[bthp!]
  \centering
  \caption{Summary of recent material screening results compare to XENON experiment data. Contaminations are expressed in $\mu$Bq/kg.}\label{radiopurity}
\vskip 0.5 cm
\begin{tabular}{lr r r l }
  \hline\hline
  Material& U238 & Th232 & Tl208& Exp. or Provider \\[1ex] \hline
  Copper & <70& <30 & <11& XENON \\
   Copper &12.4 &4 &1 & LUVATA \\[1ex] \hline
   Lead & <660& <550 & <198& XENON\\
   Lead & 330 & 100& 36 & MIFER \\
  Lead & 370 & 72 & 26 & COMETA \\
  Lead & 730 &140 &50.4 & TECNIBUSA\\ [1ex] \hline
  Poly & <230 & <94 &<34 & XENON\\ [1ex]\hline
\end{tabular}

\end{table}

\subsubsection{Provider offers.}

\begin{table}
  \centering
  \caption{Summary of lead provider offers in \euro/kg. Transportation is included in the case of TECNIBUSA while it costs 2k\euro/truck (a truck can transport up to 24 tons) in the case of COMETA.}\label{offers}
\vskip 0.5 cm
\begin{tabular}[t]{lrrr}
  \hline\hline
  Company & small bricks &  large pieces & Transportation\\ [1ex]\hline
  TECNIBUSA     & 3.78 & 3.96 &no \\
  COMETA & 2.70 & 3.00 & yes \\
   [1ex] \hline
\end{tabular}

\end{table}

Provider offers have been requested to TECNIBUSA and COMETA providers and a summary of these offers is presented in table \ref{offers} where a difference or around 1\euro/kg can be observed between both, being the COMETA offer more competitive.

\subsection{Overall dimensions and final Cost}\label{cost}

For a lead shielding width of 570\,cm, a length of 641\,cm and a height of 584\,cm, the estimated amount of material needed for the castle is: 50800\,kg of lead, 4600\,kg of copper,  15300\,kg of steel as support and 1085\,kg  as base for lead roof, and more of 6400\,kg as \emph{container}for upper water bricks.

The cost of the shield is dominated by the price of the lead necessary for a minimum wall thickness of 15\,cm, though the price of the copper could be important. The impact of the transportation on the total cost is minimal, thus non-Spanish company can in principle be considered as possible lead supplier.  Preliminary estimations foresee 8-10 k\euro for transportation from Rome to Canfranc in case  of the COMETA lead (2 k\euro/truck - 24 ton maximum load).
The external structure, necessary to secure the lead castle, and movable mechanism to move the doors can be built by a machine shop at some of the participating institutions, still to be defined.

These structures are wholly dependent both on the thicknesses of the different material layers making part of the shield and on the precise dimensions of this shield, so its definite configuration will come at the end of the design. A summary of the cost is presented in table \ref{cost2} where around 4\,k\euro/year have to be added for nitrogen.

\begin{table}[bthp!]
  \centering
  \caption{Summary of the total cost for the proposed design.}\label{cost2}
\vskip 0.5 cm
\begin{tabular}{l l}
  \hline\hline[0.2ex]
 Lead & 153\,k\euro  \\ [1ex]
Transportation & 8\,k\euro\\ [1ex]
Copper Layer & 50\,k\euro \\ [1ex]
Water bricks & 10\,k\euro\\ [1ex]
Steel structure (Material) & 20\,k\euro\
  \\ [1ex]
Clean room & 10\,k\euro\
  \\ [1ex]  \hline
  \bf{Total} & \bf{$\sim$251\,k\euro}\\  [1ex] \hline
\end{tabular}

\end{table}

The total price can fluctuate according to the variation of the costs of lead and copper on the market. Consistent variations have been observed during last year. Considering security margins for these basic prices we can set an upper limit of 300\,k\euro for the overall cost of the shield.

%
%
%

\section{The Water Tank option}
 
Water is an excellent shielding system for several reasons: a) once demineralized and fully purified, water is extremely radioclean, b) water is dense; a few meters are sufficient to suppress the gamma background to negligible levels; c) a water tank can be instrumented and turned into an active veto system (against muons and neutrons), and d) water is cheap and the engineering of a water tank straight-forward. Furthermore, one can use the experience of experiments such as GERDA and the synergy with experiments such as XENON-1t that will be building their water tank at about the same time than NEXT.

\begin{figure}[hbtp!]
\centering
\scalebox{0.5}{\includegraphics{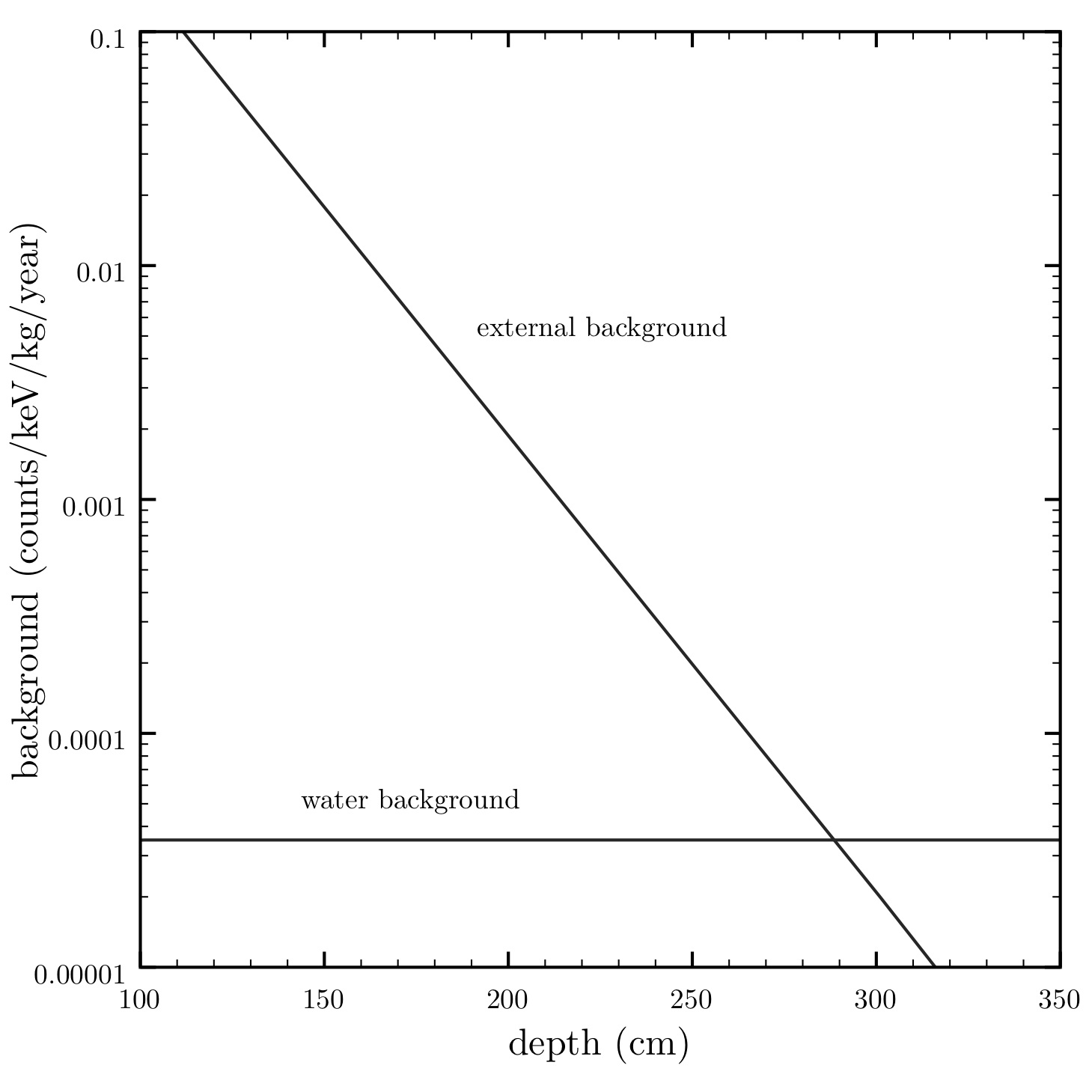}}
\caption{Background rate vs. water tank depth}
\label{fig0}
\end{figure}

As shown in Chapter \ref{sec.sensi}, the intrinsic background rate in NEXT is estimated to be of the order of $2 \times 10^{-4}$ \ckky. Further improvements, in particular in the radiopurity of the PV titanium could allow to reduce this background to  $\times 10^{-4}$~ \ckky or less. The shielding, therefore, must attenuate the background (in particular from the
\BI\ and \TL\ gammas emanating from the LSC rocks) below that level.

Figure \ref{fig0} shows the background rate expected in the detector due to external backgrounds as a function of the
water tank depth. At about 3 m one is reaching the background rate due to the residual contamination of the
purified water, which contributes about $3 \times 10^{-5}$ \ckky, almost one order of magnitude less than the
intrinsic background. 

\subsection{Design and construction of the water tank}

\begin{figure}[h]
\centering
\scalebox{0.5}{\includegraphics{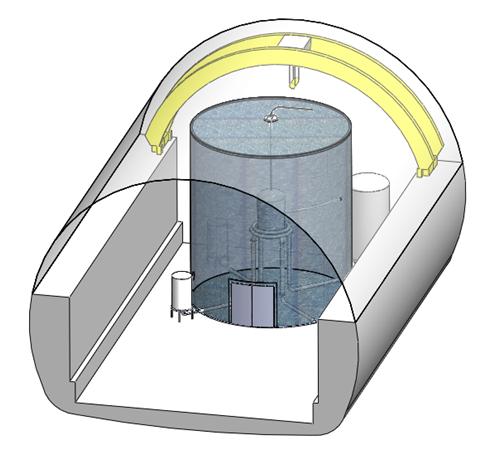}}
\scalebox{0.5}{\includegraphics{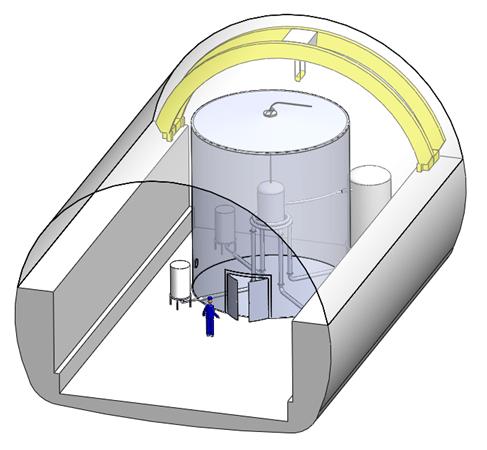}}
\scalebox{0.5}{\includegraphics{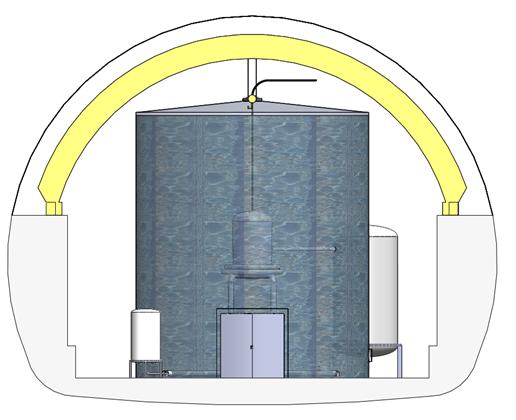}}
\caption{Canfranc underground laboratory settings.}
\label{fig1}
\end{figure}

The water tank has been designed by the Mechanics group of UPV and UdG. It consists of a large cylinder, itself made of welded rings, with a diameter of 8 m and a height of
8 m. The 
floor is made
out of 14 mm thick stainless steel plates, welded together, and placed on a leveled concrete
floor. The lowest ring is made out of 7 mm stainless steel plates. The second ring is made out of 6 mm stainless steel plates while the upper rings and
the roof are all made out of 5 mm plates. 

Radon suppression within the tank will be achieved by a slightly over-pressurized nitrogen
blanket which will be kept between the water level and the roof of the tank. In addition the water will be bubbled through a column of nitrogen to eliminate radon degassing from the tank walls. 

The water tank will be built according to industrial standards. In particular, the API 650 code will be adopted together with Eurocodice 8 for seismic activities. In addition, the Spanish
safety regulations  will be followed. A certified commercial company
will build the water tank. The construction is expected to start in fall 2011. The GERDA
water tank serves as a reference concerning the safety aspects and the quality of the final
product.

Due to its dimensions the tank will be built on-site. Construction materials and technical
gases for welding will be procured by the contracted company. To build the tank we will make use of the 10 t crane for lifting the parts of the tank as it gets built. After construction, the welds will be tested by radiography, and the tightness of the tank will be tested with normal water.

The TPC will be built in place, inside the water tank. After testing with normal water, the tank will be emptied and dried, then transformed in a clean (dust-free). The detector will be assembled and tested inside the tent before filling with clean water.

To fill the water tank one needs extremely clean water. The simplest way to produce such water is to rent a water plant to a commercial company, unless the LSC decides to incorporate a clean water plant to its infrastructures. In order to keep the water as pure as possible, a
continuous water purification will have to be done. Approximately 4 m$^3/\mathrm{h}$ will be pumped
from the top of the water tank, filtered, and returned to the bottom of the tank. The
purification loop will include:
\begin{enumerate}
\item particulate removal
\item radon stripping
\item deionization
\end{enumerate}

Each one of these tasks can be accomplished by commercially available off-the-shelf equipment. 

\subsection{Connecting the TPC to the water tank}

\begin{figure}[h]
\centering
\scalebox{0.5}{\includegraphics{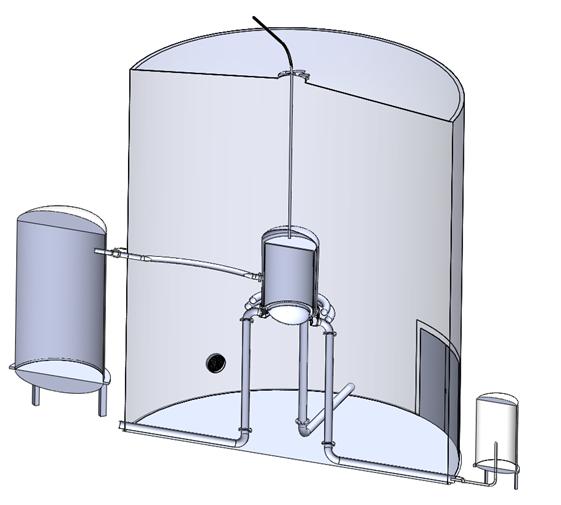}}
\scalebox{0.5}{\includegraphics{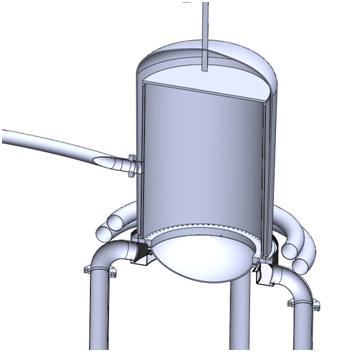}}
\caption{Sections of tank, auxiliary systems and vessel}
\label{fig2}
\end{figure}

The construction of the tank itself is a relatively 
straight forward task. From the engineering point of view the main challenge is the pipe structure that plays a double role: a) supports the TPC, which has to be well centered in the water tank, and b) drives the signals, power and gases in/out the water tank. 

The following connections are needed between tank and TPC vessel.

\begin{itemize}
\item 2 Xe circulation 1/2'' flexible hoses 
\item 1 Xe chamber vacuum  CF100 rigid tube
\item 1 vacuum tubes CF100 rigid tube, including a manifold
\item 1 HV 100 kV flexible hose
\item 1 HV 30 kV flexible hose
\item 1 CF100 rigid tube for emergency cryogenic deposit
\item 1 in / out inlet for hot air with HEPA filter
\item 4 rigid tubes for signal and power electronic PMT and Si-PM (also vessel support beams)
\end{itemize}

All service tubes entering the vessel will include special radiation shield connectors in the vessel side. Many other special requirements will have to be taken into account, such as the xenon vacuum pump being as close as possible to the vessel, mobile platforms to maintain and repair the vessel etc.

\subsection{Water Purification System}

A purification system to provide {\em ultra--pure} water is needed for the experiment. The requirements of this system depend on the desired water quality and, in general, use a combination of purification technologies, which must be used in an appropriate sequence to optimize their particular removal capabilities. In essence, it will consist of a reverse osmosis system to filter most of the impurities and of a re--gasification system with bubbling $\textrm{N}_2$ to continuously eliminate radon.

Next we describe the major components of the water purification system.

\subsubsection*{Raw water}

The raw water is obtained from a spring and pumped to the laboratory. Hall A has two water taps, one inside and the other one in the entrance. In both cases, the pipe is DN20 and the flow is 0.5 l/s, approximately. 

The analysis of the raw water has been performed by the {\em Analaqua} ({\em http://www.laboratorio-medioambiental.com/index.htm}) company. The results are shown in Table \ref{t1}.

\begin{table} [!h]
\centering                  
\begin{tabular}{||c|c||}                                                                                                \hline
{\bf Ions}             & {\bf Total concentration} mg/l  
\\ \hline
\hline
Ammonium (NH$_4$)      & $<$0.05
\\ \hline
Potassium (K)          & 0.2
\\ \hline
Sodium (Na)            & 1.3
\\ \hline
Magnesium (Mg)         & 4.1
\\ \hline
Calcium (Ca)           & 36.1
\\ \hline
Strontium (Sr)         & $<$2 $\mu$g/l
\\ \hline
Barium (Ba)            & $<$0.2
\\ \hline
Carbonate (CO$_3$)     & $<$0.5
\\ \hline
Bicarbonate (HCO$_3$)  & 117
\\ \hline
Nitrate (NO$_3$)       & 1.0
\\ \hline
Chloride (Cl)          & 4.6
\\ \hline
Fluoride (F)           & $<$0.1
\\ \hline
Sulfate (SO$_4$)       & 6.8
\\ \hline
Silica (SiO$_2$)       & 4.1
\\ \hline 
Boron (B)              & 0.12     
\\ \hline 
Conductivity           & 186 $\mu$S/cm      
\\ \hline 
pH                     & 7.6 U.     
\\ 
\hline 
\end{tabular}
\caption{Raw water analysis.}
\label{t1}
\end{table}

\subsubsection*{Water requirements}

Table \ref{t2} compares the water quality parameters to be used in NEXT to those used in GERDA and LUX. Notice that symbol ? denotes that the information has not been provided and - that the parameters are not checked.

Two particular features of GERDA are: i) the water tank temperature is in equilibrium with the laboratory temperature, about 10$^\circ$C, and ii) only the electric resistivity is monitorized.

\begin{table}[t]
\centering                  
\begin{tabular}{||c|c|c|c|c||}                                                                                                \hline
{\bf Parameters}                & {\bf Units}   & {\bf LUX} & {\bf GERDA}  & {\bf NEXT} 
\\ \hline \hline
Elect. Resist at 25$^\circ$C    & M$\Omega$/cm  & $<$18     & 17.5         & $<$18      
\\ \hline
Coliform bact. concent.         & \% ml         & $<$10     & -            & $<$10/100  
\\ \hline
Total Organic Carbon            & ppb           & $<$50     & -            & $<$50      
\\ \hline
U                               & ppb           & $<$0.002  & $<$0.00023   & $<$0.0001  
\\ \hline
Th                              & ppb           & $<$0.004  & $<$0.0001    & $<$0.0001  
\\ \hline
$^{226}$Ra                      & mBq/m$^3$     & ?         & $<$2         & $<$2       
\\ \hline
$^{222}$Rn                      & mBq/m$^3$     & -         & $<$0.1       & $<$0.1     
\\ \hline
Dissolved O$_2$                 & ppm           & $<$0.5    & -            & $<$0.5     
\\ \hline
Ultrafiltration                 & $\mu$m        & 0.2       & 0.1          & 0.2        
\\ \hline
Temperature                     & $^\circ$C     & ?         & 8            & 10         
\\ \hline                               
Recirculation Rate              & lit. per min. & 26.46     & ?            & 26.5       
\\ \hline 
Fill Rate                       & lit. per min. & 26.46     & ?            & 26.5       
\\ \hline         
\end{tabular}
\caption{Water requirements. ? denotes information not provided, and - parameter not checked}
\label{t2}
\end{table}

\subsubsection*{Purification system}

In addition to the initial fill of ultra-clean water, the purification system has to guarantee that water
quality is maintained during operations (this requires recirculation of the water) and that air quality (dust free)
is maintained in operations with dry tank. 

%
%
\begin{figure}[h]
\centering
\scalebox{0.8}{\includegraphics{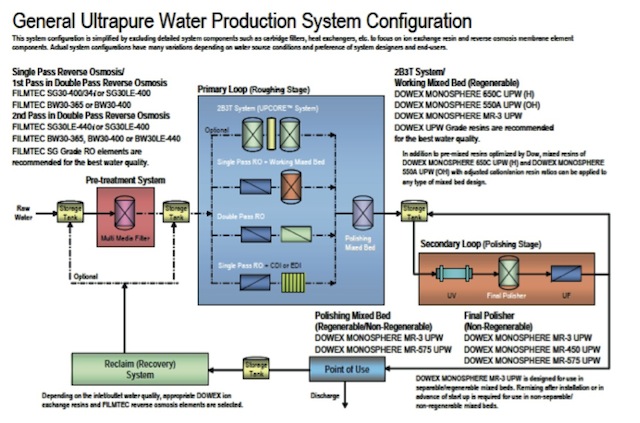}}
\caption{Systems to produce ultra--pure water according to DOW ({\em http://www.dow.com/})}
\label{fig2}
\end{figure}

There are several procedures to obtain Ultra--Pure Water (UPW)  
which must be applied in an appropriate sequence (pretreatment, reverse osmosis and refining systems) to optimize performance.

Figure \ref{fig2} shows a scheme of the various procedures:

\begin{itemize}
\item The pretreatment equipments such as water softening and carbon filters are designed to remove contaminants that may affect purification equipment located after.
\item Reverse osmosis removes from 90\% to 99\% of all contaminants: dissolved solids, turbidity, asbestos, lead and other toxic heavy metals, radium, bacteria and many dissolved organics.
\item Filter system removes the impurities from water already pretreated. For example, ultraviolet radiation to reduce the TOC; ion exchanger to remove ions; vacuum de--gasser to remove about 99\% of the oxygen and 96\% of the radon.
\end{itemize}
%
%
%
%
%

\begin{figure}[hbtp!]
\centering
\scalebox{0.4}{\includegraphics{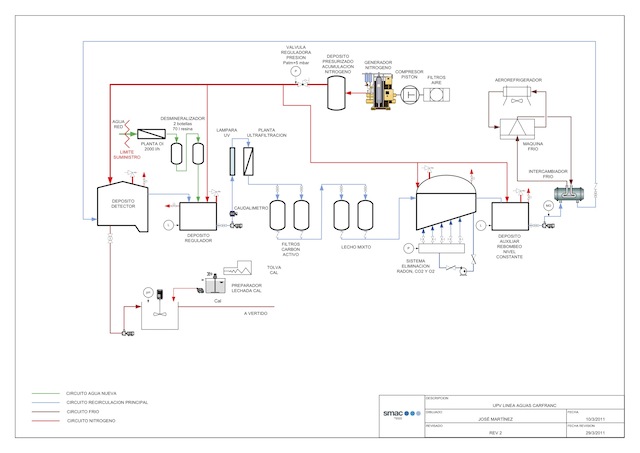}}
\caption{Purification system design developed by SMAC.}
\label{fig3}
\end{figure}
Figure \ref{fig3} shows the design of the purification system developed by the SMAC company. Notice that the water flows through all components of the system for filling, while through only a few of them for the circulation. We discuss next the most important systems.

\subsubsection*{Pretreatment equipments}

\begin{center}
\sc Coarse Filter
\end{center}

50 micron filter with a pleated cartridge filter (not winding) of 20 inches, mounted on a opaque polypropylene container.

The system has pressure gauges to check the status of the filter. Filter suitable for a throughput of up to 4,000 liters / hour.

\begin{center}
\sc Anti--fouling dosage
\end{center}

Product dispensing equipment to prevent osmosis membranes of lime or other salts present in the raw water. It is composed by:
\begin{itemize}
\item Water meter with pulse generator, maintenance--free for maximum pressure of 7 bar and temperature of 60 $^\circ$C
\item Dispensing pump B--02, SEKO trade mark, series TEKNA, DPR type, model pH 602 with integrated control and proportional dosing
\item Injector and delivery pipe
\item Tank for dispensing anti--fouling, D-02
\item Minimum level probe that prevents the pump runs out of product
\end{itemize}

\begin{center}
\sc Medium filter
\end{center}

5 micron filter with a pleated cartridge filter (not winding) of 20 inches, mounted on a opaque polypropylene container.

\begin{center}
\sc Polishing filter
\end{center}

1 micron filter with a pleated cartridge filter (not winding) of 20 inches, mounted on a opaque polypropylene container.

\subsubsection{*Reverse osmosis}

\begin{center}
\sc Reverse osmosis plant
\end{center}

The plant consist of:
\begin{itemize}
\item Set input instrumentation osmosis, consisting of conductivity with temperature correction, pressure switch and temperature transmitter
\item High pressure pump feeding a vertical multistage reverse osmosis. Parts in contact with the fluid and stainless steel rotors technopolymer. It is a type multistage pump Mercabomba mark, model V816F75T with a power of 7.5 kW and maintenance free. Pump has flanged connections and all parts in contact with water are, at least in stainless steel AISI-304
\item Reverse Osmosis composed of a fiber container with a capacity of two layers of 8 inch. Osmosis polyamide cartridges incorporate two 8 "x40" Koch trade mark, model 8040-HR TFC-375 to ensure ion rejection of at least 99.5\%, which will leave a water with about 7 microSiemens / cm. before entering the column deionization resins. Osmosis system has a total hydraulic performance of 65\% in the configuration for a final plant flow of 48,000 liters / day.
\end{itemize}

\begin{center}
\sc Demineralization plant
\end{center}

The plant consist of:
\begin{itemize}
\item Set input instrumentation deionization columns, conductivity transmitter consists of a valve sampler to take samples properly and gauge. Data from these sensors are sent to the PLC that determines the actions to be performed
\item Resin Column System de--ionizers anion/cationic regeneration system with semi-automatic. Columns with 70 liters each, and volumes up to 2,000 l/h
\end{itemize}

\subsubsection{Filtering--polishing systems}

\begin{center}
\sc UV lamp
\end{center}

To maintain water quality conditions in the microbiological aspect, an UV lamp will be installed. 

%
%
\begin{figure}[h]
\centering
\scalebox{0.35}{\includegraphics{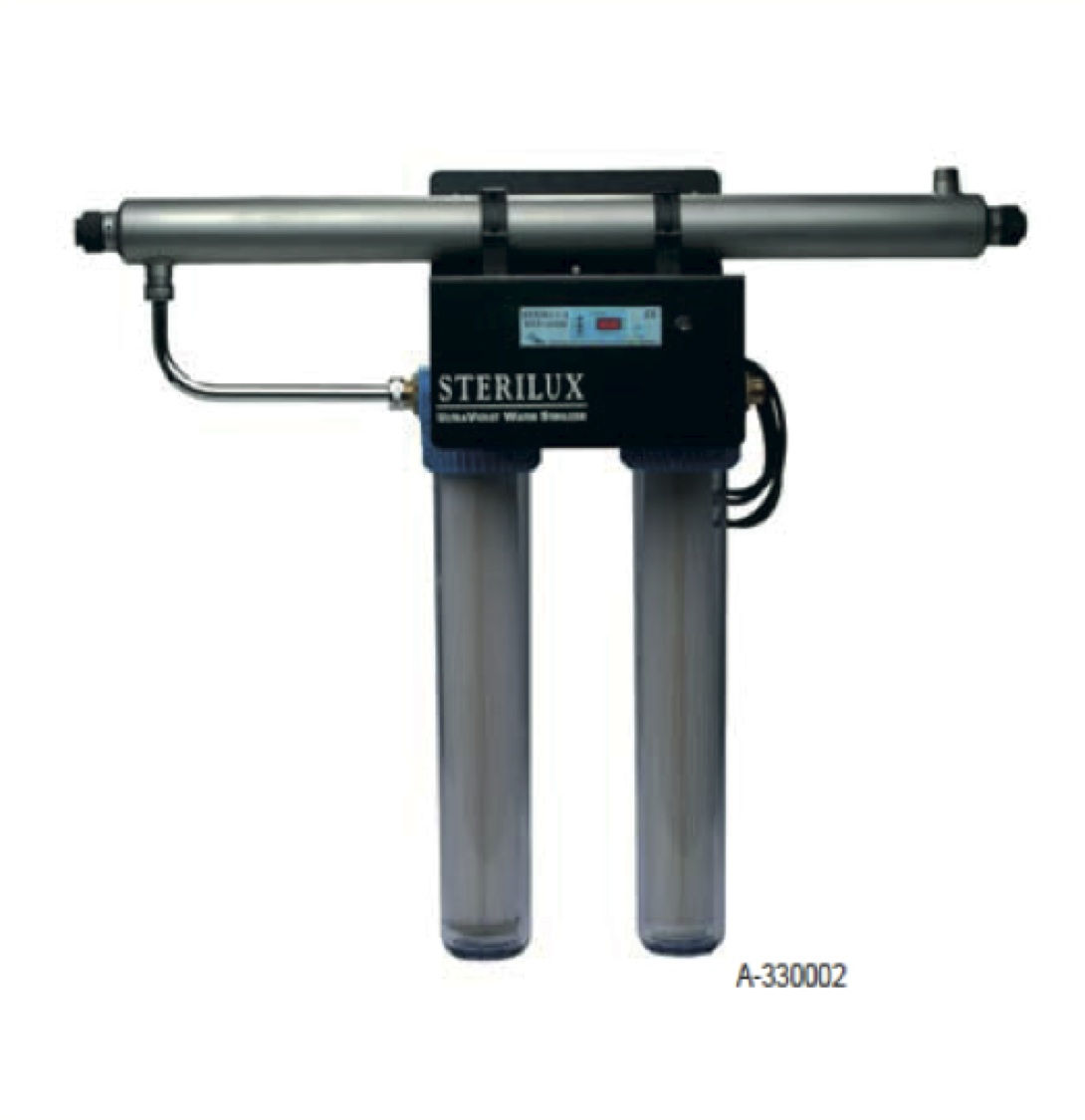}}
\caption{Ultraviolet (UV) lamp}
\label{fig4}
\end{figure}

\begin{center}
\sc Active carbon filtration
\end{center}

For adsorption of particles and organic nature pollutants will be installed two active carbon filters MG1050 of 486 mm in diameter and 150 l of carbon. Only one will be on duty, changing it manually when you reach the limit of service. With this system we obtain a filtration rate of 10.8 m / h, slow enough to allow adsorption of the pollutants. The bottles will be equipped with a manual valve KERAMIS K56.

\begin{center}
\sc Polishing mixed bed
\end{center}

Two bottles will be installed with 150 l of exchange resin of mixed bed refining unit to achieve and to maintain the desired conductivity. 

\begin{center}
\sc Conductivity
\end{center}

To control the conductivity in the recirculation will be installed a probe  CONDUMAX ENDRES + HAUSER W CLS 15D. For the transmission and control of signal will be installed a transmitter LIQUILINE CM442.

\begin{center}
\sc Ultrafiltration
\end{center}

Compact installation of a filtration plant Likuid-URA -20.

\begin{center}
\sc Rn elimination
\end{center}

{\em Henry} law determines the solubility of a gas in water is directly proportional to the partial pressure of gas to pressurize in a nitrogen atmosphere. The objective is to make the rest o the gases (oxygen, carbon dioxide, radon, ...) insoluble in water. For this purpose, a pump with a pressurizing and depressurizing area with ultra--pure nitrogen atmosphere and positive pressure will be installed. 

All equipment will be built in AISI 314 L, with an area of 1 m$^2$ depressurization and recirculation flow rate of 1 m$^3$/h.

\subsubsection*{Auxiliary components}

\begin{center}
\sc Flowmeter
\end{center}

All systems work at a constant flow regulated by a PID and a flowmeter. It will be installed a flowmeter ENDRES + HAUSER PROWIRL PROLINE 72 F, which is Vortex type and it has no moving parts. In addition, it is calibrated to measure liquid volumes of very low conductivity.

%
%
\begin{figure}[h]
\centering
\scalebox{0.35}{\includegraphics{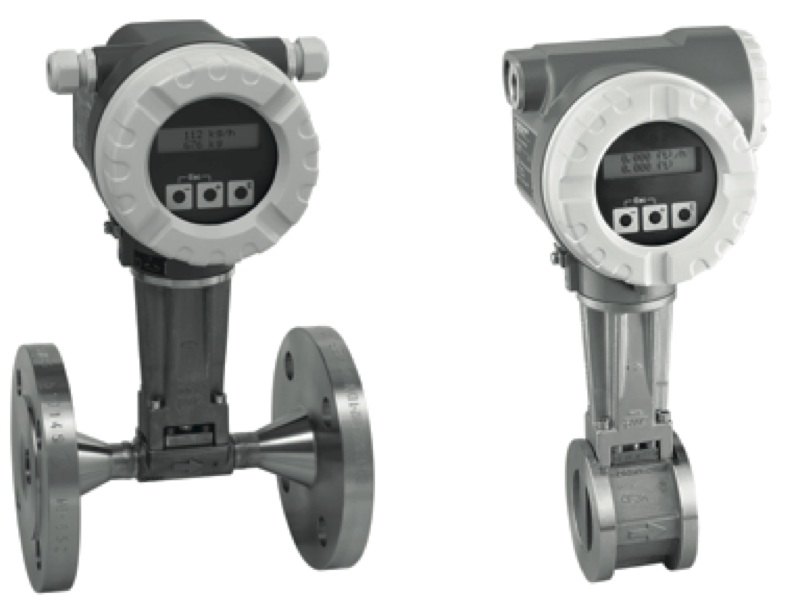}}
\caption{Flowmeter}
\label{fig4}
\end{figure}

\begin{center}
\sc Auxiliary tanks
\end{center}

To maintain the overall system and ensure the correct booster two auxiliary tanks will be installed PP, closed and pressurized ultrapure nitrogen atmosphere, with corresponding levels ENDRES + HAUSER trade mark PTP model 31.
These deposits will be lined with an insulation of total thickness 0.6 mm.

\begin{center}
\sc Pumps
\end{center}

Installation of a pump PROLAC SWFI 20 OD KIT T100/112 CL in each pumping deposit.

\begin{center}
\sc Cooling system
\end{center}

Two refrigeration/evaporator machines with forced ventilation and a heat exchanger will be installed to keep down the temperature of the water reservoir.

\begin{center}
\sc Gas line
\end{center}

For the removal of Rn, it is needed to work on atmosphere of nitrogen and to prevent the dissolution of atmospheric gases in water, especially carbon dioxide would increase the conductivity and aggressiveness of water. For this purpose, a nitrogen generation system with a wealth of 99.999\% and a capacity of 600 l/h for "inert" all deposits will be used. 

\subsubsection*{Air quality}

In order to maintain the air quality during periods of maintenance (empty water tank), an air conditioning (TECNIVEL, CHF-4-AE) will be installed. This devices has two operating modes:
\begin{itemize}
\item Mode recirculating air drying before starting maintenance work and after draining the tank
\item Normal mode, with complete air
\end{itemize}

This system keeps humidity and air temperature in the tank and the installed filters can match the quality of air to a cleanroom.

\chapter{Toward the NEXT-100 technical design}
\section{The NEXT-100 detector}
The NEXT-100 detector consists of:

\begin{enumerate}
\item The shielding system, described in the previous chapter.
\item The NEXT-100 TPC, whose design is based in the ANGEL concept, described in chapter 3.
\item The gas system.
\item Other systems (DAQ, slow controls, monitoring).
\end{enumerate}

In addition, the detector requires a data analysis system (DAS) to process the data.

In this chapter we describe some engineering aspects of the NEXT-100 TPC, as well as the gas system, DAQ and DAS. We also describe the general techniques mastered by our collaboration for TPB coating. Last but not least, we sketch an outline of the NEXT project, that will be further developed in the near future.

\section{Engineering design of the pressure vessel}\label{sec.PV}
The entire detector including the pressure vessel and internals are all supported on a support flange of the pressure vessel. There is a large copper shield ring having extension "fingers" around its periphery which rest in a groove of this support flange. This shield ring  provides a base for mounting the SiPM plane, EL mesh and field cage. The SiPM front end electronics boards have some activity and are located behind this shield ring. The copper shield provides a heatsink for the F.E. electronics and will conduct heat to the support flange which is in contact with both water and the other pressure vessel parts, providing passive cooling.

\subsection{Layout}
The pressure vessel comprises four parts, all of pure titanium, ASTM grade 3 (or 2) which are bolted together:
\begin{enumerate}
\item An upper head which contains the PMTs,
\item A main cylindrical vessel, which incorporates at its base a thick section which allows penetrations for services (Xe gas flows, power and signal feedthroughs, EL HV cable, etc.), 
\item A support flange, upon which the detector is built, and to which floor supports connect to,  
\item A lower inverted elliptical head, designed to withstand pressure on its convex surface; the inverted design minimizes wasted Xe volume. This lower head is removed to access the SiPM plane and front-end electronics.
\end{enumerate}
All flange pairs are bolted together with grade 3 titanium bolts and nuts, and have O-ring pairs for sealing. PCTFE is the preferred O-ring material, having the lowest leakage of any polymer. A small pumpout port samples the annulus between the O-rings to check for either Xe (out) or water (in) leakage.
\subsection{PMT Head}
 The choices for this head shape (in axial cross section) are elliptical, torispheric or flat. The torispheric head has a spherical center section called a "crown" with a toroidal edge section ("knuckle"), to transition to a short cylindrical section that is welded to the flange. It was chosen over an elliptical shape because the spherical shape of the crown allows all edges of radial openings drilled into it (normal to the surface) to be circular; this simplifies the machining of these openings and any flanges that mate to them. The torispheric (and elliptic) shape is far more mass efficient than a flat plate for holding pressure; a thickness of 15mm is required, whereas a flat plate requires close to 100 cm or more. It is only somewhat harder to machine the openings than for a flat head, CNC machining is not required. 

\begin{figure}[h]
	\centering
	\includegraphics[width=150mm]{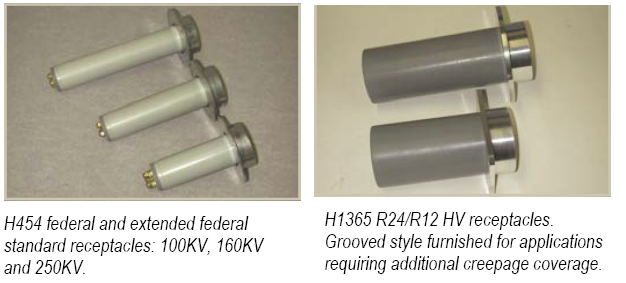}
	\caption{Hydraulic Bolt Tensioner}
	\label{HV-FT}
\end{figure}

The crown has 3 concentric rings of openings that accommodate pure titanium CF (NW100 size, 15 cm O.D.) flanges with diffusion bonded sapphire windows, for a total of 36 PMTs.  A central opening having a large inside radius is used for the high voltage feedthrough, either the TAMU designed feedthrough, or a standard cable receptacle for a 100 kV medical X-ray machine (e.g. Federal Standard or Euro RXX designs) as pictured here: \ref{HV-FT} 
which can be made in radiopure materials, and rated for pressure. CF flanges are also used on the outer surface of the crown (with a welded-in housing) to both seal the PMT's against water, and to provide a secondary containment for Xenon, should a sapphire window break or gasket leak. This design is chosen as the nominal design since, apart from the-beam welding of the head shell to the flange, no further welding is required and all the openings for PMTs and CF flanges can be machined into the crown.

An optional head having 60 PMT heads is also feasible, but will require the e-beam welding of custom machined PMT housings into bored holes in the torispheric head, since the close spacing disallows the use of CF flanges. Polymer gaskets , such as PCTFE or PEEK would be used to seal the sapphire windows instead of CF metal seals. Windows or gaskets are replaceable without removing the head in this design. This head has a further advantage in that the PMT faces need not be flush with the crown, and can be inserted down into the cathode buffer space, up to where the faces are roughly in a plane, thus reducing wasted Xenon volume. This head will be significantly more expensive to fabricate and require significant prototyping. 

\subsection{Design Requirements}

The pressure vessel is designed to safely maintain a maximum operating internal pressure of 15 bar (absolute), with atmospheric pressure on the outside, and also to withstand an external pressure of 1.5 bar (5 m H$_2$O hydrostatic pressure on outside, with full vacuum on inside). It also serves as ground potential, on its inside surface for the field cage, and must be compatible with the insulation gas, here Xenon, and withstand any breakdown. There is over 2 MJ of stored energy, which would be catastrophic if suddenly released by vessel rupture. It is designed in full accordance with American Society of Mechanical Engineering (ASME) Pressure Vessel (PV) Code, section VIII- Division 2 (or equivalent European Standard), and will be pressure tested to a minimum of 1.25x the maximum allowable pressure. 

The Pressure Vessel will necessarily be a welded construction due to the size. Shells will be roll formed and seam  welded, machined, then welded to machined flanges. There may be a final machining of each flange face after welding. Welds will be exclusively electron beam, made in high vacuum, and all welds will be fully radiograph inspected afterwards; this allows the welds to have full strength, reducing the wall thickness needed.  Titanium is very easily e-beam welded, even in thick sections, with excellent resulting properties.  Laser or diffusion bonding (welding) methods are also possible, however laser welding typically does not have the penetrating power of e-beam, and diffusion bonding is not mentioned as an allowable process in the ASME PV code. 

The pressure vessel must also be corrosion resistant in ultrapure water. Titanium has excellent corrosion resistance in pure water; copper corrodes in ultrapure water, and would need some sort of radiopure barrier coating. 

The pressure vessel interior surface also serves as a ground boundary for the field cage and must withstand potential sparkovers.  It is not known whether titanium has better spark damage resistance than copper, it has a higher melting point, but lower thermal conductivity; testing is warranted.

\subsection{Pressure Vessel Construction Issues}
The final design of the Pressure Vessel will only be released after close consultation with the fabricator, to assure that no unexpected difficulties arise.  A detailed specification will be drafted and potential manufacturers will be required to submit a full fabrication plan. Crucial details must be worked out with fabricators, material suppliers  etc. to not only assure that the vessel is constructed in full accordance withe the PV code but remains radiopure when finished. Many important details must be worked out based on practical realities of supplier and manufacturer capability, such as where to place flange welds, what prototyping and testing must be done to assure the vessel is made correctly, what material samples are to be taken, what are  permissible cleaning and preparation processes, etc. Below are some initial concerns with fabrication and assembly of the various sections:  
\paragraph{CF Flange 36 PMT Torispheric Head}
This head requires no welding of PMT housings into the crown, but is limited to 36 PMTs due to the large CF flange OD. The shell is formed by metal spinning, the crown is thicker than the knuckle and cylindrical section, so these sections my be formed separately and e-beam welded together. Due to the flange stud holes, which limit the reinforcing area around the PMT opening, a thicker shell is required, 20mm instead of 13mm for the welded 60 PMT head. It still requires welding the formed shell to the flange; this will be e-beam welded. However flange welds should not require a CNC controlled e-gun (as in the 60 PMT head), a fixed gun will work, with a rotary table holding and spinning the head. If the knuckle cannot be formed integral to the crown, it can be e-beam welded to it in the same manner. Machining of the CF flange ports should be straightforward, and could be done with a mill having a tilting head, with the PV head. The ports can be machined using milling cutters, for less  chatter, but the final surface of the knife-edge will need be machined, ground, or burnished using a single axis (spindle) rotating tool such as a boring/facing head, to avoid scratches across the  knife edge. Gaskets for the outer CF flanges should be chosen to be as soft as possible, as the knife edges may be more delicate in titanium than in stainless steel. Silver may be a good choice, especially for the outer CF flange gasket which must not create galvanic corrosion in either the Ti or gasket. Copper will corrode in ultrapure water. Tin may be an option as well, as it can be made radiopure. An option is to use O-ring seals; these may even be used in CF flanges. The inside surface of the head may need machining after welding to assure that a smooth ground surface results. 
\paragraph{Welded-in 60 PMT Torispheric Head (option)}
This head will likely require a CNC controlled e-gun to make the PMT housing to crown welds. The crown will be bored for the PMT holes; and then accurately measured in a coordinate measuring machine (CMM) afterwards as most of the crown metal is being removed, and nominal dimensions will change once all the holes are bored (Swiss cheese effect). The PMT housings will be machined to fit snugly in the bores, and they will need to be held in place with individual fixtures.  Some provision for the e-beam to be absorbed without damaging these holding fixtures may need to be incorporated. There will be some degree of shrinkage around the welds, as the metal cools; this will require several intermediate CMM measurements of the remaining bores to assure that the welds are being made in the right place (unless welder can otherwise assure precise positioning). Thus the fixturing of PMT housings in bores will need to be staged, and the sequence of welding will need to be carefully thought through to assure accurate welding. It is likely that each ring of PMT's will need to be fixtured and welded separately, and welds should be done around each ring in a staggered fashion, as is typically done in tightening bolted flanges uniformly.
The PMT housings themselves will need to be machined from thick pipe or perhaps even solid bar, since the reinforcement section (large fillets on weld flange) must have a substantial radial dimension and cannot be formed as part of the crown. 
\paragraph{Main cylindrical Vessel}
There will be  a longitudinal seam weld made after roll forming the cylinder which may need to be ground smooth. Edges will need machining  to allow precise welding to flanges and to the lower thick section, which  will likely be made separately.   
\paragraph{Support Flange}
This is a relatively straightforward machined plate, though required thickness may be high as this plate supports the entire detector, including the large internal copper shield, transferring weight to the outside legs. There will be a significant edge moment in the plate, and we desire to maximize the internal radius so as to provide clear access to the feedthroughs.
\paragraph{Lower Flat/Elliptical Head}
This is also a relatively straightforward to fabricate. Shell thickness is nominally 6mm so this should be easy to spin. Carrying an external pressure requires tight tolerances on out of roundness or ellipticity, so CMM measurements should be made before welding. A method for safely lowering this head and moving it out of the way for SiPM service access is needed, as crane access is difficult. 

\subsection{Assembly of Pressure Vessel}
\begin{figure}[h]
	\centering
	\includegraphics[width=90mm]{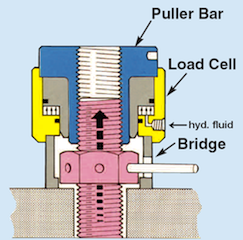}
	\caption{Hydraulic Bolt Tensioner}
	\label{hyd-tens}
\end{figure}

There are currently 84 M20 bolts on each set of flanges which must be tightened uniformly around each flange. All bolts and nuts are also titanium which has a very high propensity for galling, which must be avoided. Silver plating should be considered on all nuts, and testing should be performed to see how many repeated bolt torqueings are allowable. Polymer coatings may be possible, but can have a friction coefficient that is too low to prevent nuts from backing off. No coating prove acceptable however, and it may be necessary to avoid all wrench torquing of large bolts by using a hydraulic tensioning/nut runner system as shown in fig. \ref{hyd-tens}. Such systems work by screwing small hollow hydraulic cylinders (load cells) onto each bolt, above the nut (which is surrounded by the bridge), then pressurizing them all simultaneously through a common manifold, thus tensioning the bolts, then running the nuts down to the flange to maintain the bolt tension when the hydraulic pressure is released and the hydraulic cylinders are removed. This will also have the advantage of being much faster, far less strenuous, and give better results for bolt tension. It is envisioned that groups of these hydraulic cylinders will be loosely mounted to sections of a ring, so as to be easy to handle and install.

\section{The NEXT-100 gas system} \label{sec.GasSystem}
The gas system performs one of the pivotal roles in NEXT. It is responsible for removing electronegative impurities that would degrade the performance of the detector by reducing the drift length of charged tracks as well as causing opacity to UV light. The expected level of outgassing from the materials that make up the field cage and the reflector demand a constant purification of the gas in a re-circulation loop to a level of a few parts per billion (ppb) of unwanted contaminants. Additionally removal of the radon from the gas is highly important as the decay daughters would add an unacceptable level of background inside the active volume of the detector. The nominal re-circulation flow rate of 100 slpm has been identified as the required goal. \mbox{NEXT-100} will operate at 15 bar, thus leading to the appropriate volumetric reduction from the standard temperature and pressure flow rate stated previously. During the physics exploitation runs the detector will operate with gas enriched to 90\% level of \XE\ isotope. The enriched gas has already been acquired by the LSC and delivered to the site. The high cost of the enriched gas places additional demands on the gas system to minimize accidental leakages and achieve full gas recovery in case of accidents. 

\begin{figure}[p]
\begin{center}
\includegraphics[width=1.25\textwidth,angle=90]{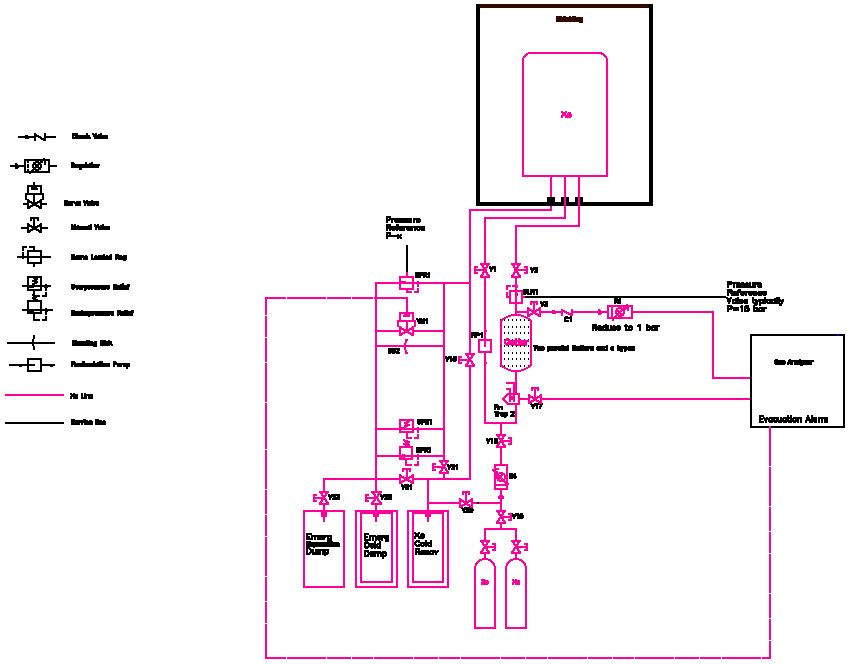} 
\end{center}
\caption{Schematic of the NEXT-100 gas system.}\label{fig:SG}
\end{figure}

\begin{figure}[tbh!]
\begin{center}
\includegraphics[width=0.75\textwidth]{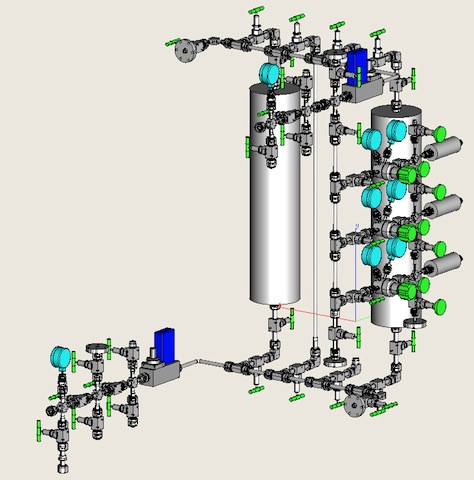} 
\end{center}
\caption{The engineering drawing of the gas system.}
\label{fig:Gas_Sys_Eng}
\end{figure}

The design of the system responsible for the filtration of the xenon follows closely the experienced gained at University of Coimbra, LBNL, UNIZAR and IFIC from constructing and operating smaller Xe purification systems. A CDR for the NEXT-100 gas system \cite{Next10:Internal} intended for installation at the LSC has been approved by the Collaboration and is presently being constructed. System schematic and engineering drawing are shown in Figure~\ref{fig:SG} and Figure~\ref{fig:Gas_Sys_Eng}, respectively. 

\begin{figure}[t!b!]
\begin{center}
\includegraphics[width=0.25\textwidth]{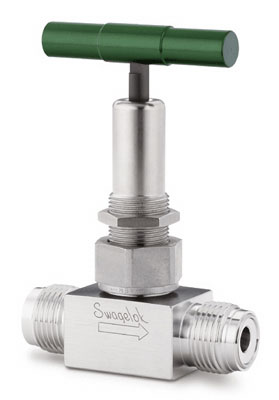} 
\end{center}
\caption{Example of all metal 0.5'' VCR gas valve used in the gas system.} \label{fig:SS-8BK-VCR}
\end{figure}

To insure the gas tightness and purity of the system, all piping and all the fittings have been chosen as 1/2'' VCR stainless steel. All the gas valves are metal-to-metal seals with all the wetted surfaces also manufactured from stainless steel. An example of such valve is shown in Figure~\ref{fig:SS-8BK-VCR}. These components were chosen to insure the cleanliness of the gas system as well as to allow the plumbing to be baked at high temperature ($~200^\circ$C) to remove water contamination prior to use.

\subsection{Xenon purification}
\begin{figure}[t!b!]
\begin{center}
\includegraphics[height=10cm]{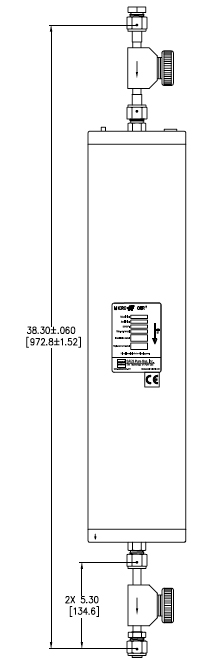} \hspace{2.5cm}
\includegraphics[width=5cm]{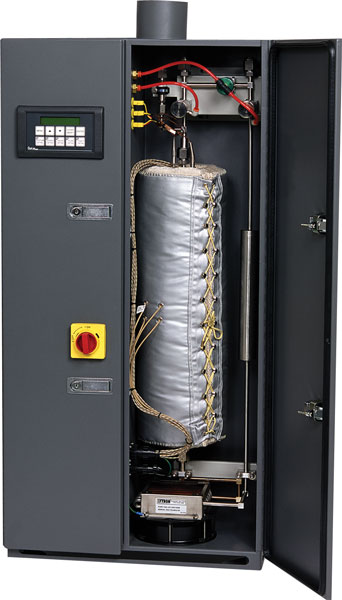}
\end{center}
\caption{Left: Drawing of the MC4500-902FV SAES Pure Gas getter, with VCR valves, to be used for the purification of the Xe. The gas system will contain two such parallel getters. Right: Example of a heated SAES getter model number PS4-MT50 with nominal flow of 150 slpm. This is a possible future upgrade to the gas system. The hot getter has been reported to be better at filtering out \RN\ and therefore can be preferred purification technology to achieve ultra low radiation background rates.}\label{fig:Getter}
\end{figure}

MicroTorr cold getter model number MC4500-902FV has been chosen as the purification filter for the Xe gas. Capable of removing electron negative impurities to less
than 1 ppb, the model chosen has a nominal flow rate of 200 slpm, well in excess of the required flow rates for NEXT 100, offering sufficient spare capacity. The gas system
will contain two such getters in parallel with a bypass. This configuration has been developed and used by the smaller gas systems operating at UNIZAR and IFIC. The second
spare getter is placed in parallel in the event of accidental contamination of one of the getters, allowing uninterrupted running. The ability to bypass the getters will
allow the testing of the purification of the gas and aid in diagnostic and monitoring of the gas system. A drawing of the MC4500-902FV getter is shown in Figure~\ref{fig:Getter} (left).

While cold getter technology is capable of reaching the required purity levels in water and oxygen, a hot getter, like the one shown in Figure~\ref{fig:Getter} (right), can also
remove nitrogen and methane. Furthermore, 362 g of cold getter material has been measured to emit 1.64 atoms of Rn \cite{RnImperial}. In that regard a future upgrade
to a hot getter technology has been considered. Alternatively (or in addition to the use of hot getters) an effective means of removing Rn can be used. We discuss Rn trapping
below.

%
%

\subsection{Rn Trapping}\label{subsubsection:RnTrap}

\begin{figure}[tbh!]
\centering
\includegraphics[width=0.99\textwidth]{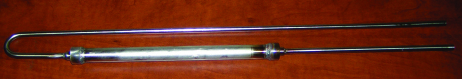} 
\caption{\RN\ trap operated at SNO lab.The trap uses activated carbon spheres to slow transmission of radon atoms. The trap tested on removing Rn from Ar gas and operates at a $-110^\circ$C.}
\label{fig:SNO_RnTrap}
\end{figure}

Radon is a natural by product in the decay chain of Uranium. Therefore, most materials will emanate Rn adding to the radioactive background.
Several research groups in the world have been interested in developing methods of trapping Rn and preventing its entry into ultra low background detectors.
A group at Queens University, Kingston Ontario, Canada has recently produced such a device  \cite{RN_ODwyer:2011}. Figure~\ref{fig:SNO_RnTrap} shows such a trap.
The trapping mechanism operates by passing the gas being filtered through a cooled column of activated carbon spheres. The carbon slows the diffusion
of Rn and the low temperature retains the atoms in the trap. 

\begin{figure}[tbh!]
\centering
\includegraphics[width=0.99\textwidth]{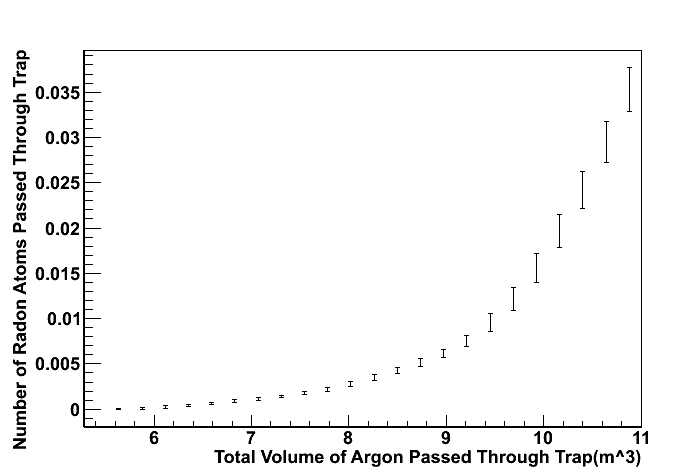} 
\caption{Total number of \RN\ atoms that got through the trap as a function
of the amount of argon gas for a 1 mBq/m$^3$ assumed radon concentration in the
argon supply.}
\label{fig:RnTrapEff}
\end{figure}

The trap was tested with Ar gas containing 1 mBq/m$^3$ activity due to \RN\ gas. Figure~\ref{fig:RnTrapEff} illustrates the efficiency of this setup for \RN\ trapping.

In the case of NEXT 100 the collaboration aims to develop and test a \RN\ trap similar to one describe above for use with Xe. The operating
temperature of this trap needs to be higher than the one developed by the Queens University as Xe will also freeze at $-110^\circ$C.
This can prove a very cost effective way of reducing radioactive background due to \RN\ in the NEXT 100 detector.

\subsection{The Re-Circulation pump}

\begin{figure}[tbh!]
\centering
\includegraphics[width=0.6\textwidth]{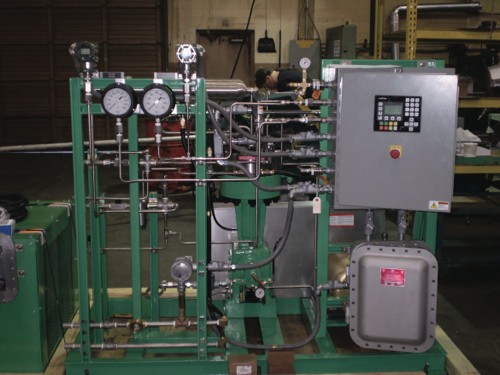} 
\caption{Example of the re-circulation pump chosen for the Xe purification gas system. This is a triple stainless steel
diaphragm pump capable of 100 slpm flow rates.}
\label{fig:Re-circulationPump}
\end{figure}

The enriched \Xe\ used in NEXT 100 project is very expensive and therefore the pump to move the gas through the re-circulation loop
must have sufficient redundancy to minimize the probability of failure and leakage. 
Furthermore, to preserve the purity of the gas all metal to metal seals must be used. We have chosen a pump manufactured by Pressure Products Industries
\footnote{\texttt http://www.pressureproductsindustries.com/compressors/diaphragm\_compressors.html}.
This pump is made with metal-to-metal seals on all the wetted surfaces. The gas is moved through the system by a triple stainless steel diaphragm. Between each
of the diaphragms there is a sniffer port to monitor for gas leakages. In the event of a leakage automatic emergency shutdown can be initiated Figure~\ref{fig:Re-circulationPump}.

\begin{figure}[tbh!]
\centering
\includegraphics[width=0.85\textwidth]{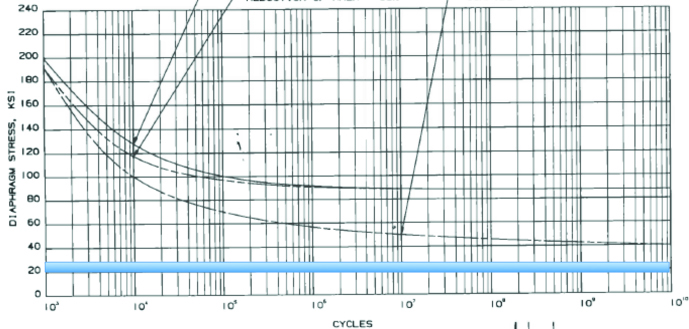} 
\caption{Failure curve for the single stainless steel diaphragm as a function of operating cycles. The pump is designed to operate
in the shaded band. This translates into an expected time to failure of continuos operation in excess of 40 years.}
\label{fig:FalureCurve}
\end{figure}

In terms of redundancy and reliability Figure~\ref{fig:FalureCurve} is showing failure curve for a single diaphragm. Under normal condition
the time between failures during continuos operation per diaphragm is in excess of 40 years. Therefore this is a very safe, clean and reliable pump with a nominal
flow rate of 100 slpm as is the goal of the xenon gas system.

\subsection{Xenon recovery system}
\label{subsubsection:XeRecovery}

In an unlikely event of an emergency condition the valuable Xe must be automatically evacuated into a safe container. This is best achieved with an emergency, stainless steel reservoir, permanently kept cold with liquid nitrogen. In an emergency condition an isolating valve connecting the detector chamber and the emergency reservoir will open and the enriched Xe will be cryo-pumped into a safe container.

Two primary conditions emergency conditions have been identified to trigger automatic evacuation.
The first is an overpressure that could potentially cause
an explosion. Because the gas system for NEXT-100 will be operated in a
closed mode the overpressure condition could occur only under two possible
conditions. The first is during the filling stage of the operation, when the
system is being filled with gas and the second in the case of thermal expansion of the gas due to laboratory fire.
In the case of overpressure we will
have an electromechanical valve, activated by a pressure switch, that will
open a pipe from the NEXT 100 chamber to a permanently cold recovery
vessel. This will then cryo-pump xenon into the recovery vessel, causing the gas
to freeze in the recovery tank. In the event of the electromechanical valve
failing, a mechanical spring-loaded relief valve, mounted in parallel to the
electromechanical valve, would open and allow the xenon to be collected in the
recovery vessel. As a final safety feature, in case both the electromechanical
and spring-loaded valves fail, a bursting disk, also mounted in parallel to the
electromechanical and spring-loaded valves, will burst connecting the NEXT 100 chamber with the recovery vessel causing the Xe to be thus evacuated.

The second emergency condition would be detected as under-pressure
indicating a leak in the system which would require evacuation of the NEXT 100 chamber to prevent Xe loss. If this is detected, an electromechanical valve
sensing under-pressure will open and evacuate the Xe into the recovery vessel.
As the recovery vessel is to be used for safety, there cannot be a valve
to close once the emergency recovery has been completed. Therefore, in the
event an automatic system is triggered, a signal will be sent to an operator
to arrive at the laboratory to supervise post recovery procedures. The aim
of post recovery is to move the collected Xe from the recovery vessel to a
closable gas bottle for future purification and use once the problems that
caused the emergency conditions have been rectified. This operation must
be conducted by the responsible person and would involve cryo-pumping the
gas collected in the recovery vessel into a stainless steel bottle which will be
valved off post transfer.
The same system can be used for controlled recovery of Xe if work needs
to be done on the NEXT-100 chamber.

\begin{figure}[h]
\begin{minipage}{18pc}
\centering
\includegraphics[width=12pc]{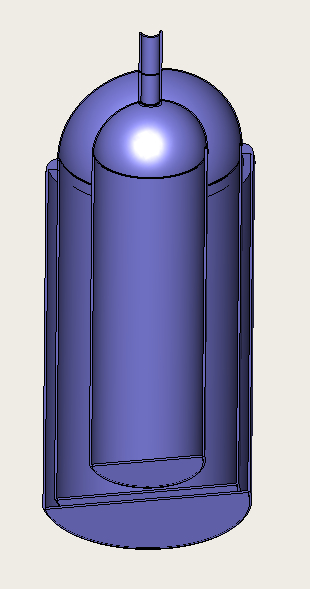}
\caption{\label{fig:Cold_dump}Cutaway view of the conceptual design for a permanently Cold Emergence Xe recovery tank. This is a triple wall vessel with liquid nitrogen cooling the inner cylinder. In an an emergence condition the Xe will be cryo-evacuated into the inner vessel where the temperature of the wall will cause it to condense and freeze on the walls.}
\end{minipage}\hspace{0.3pc}
\begin{minipage}{19pc}
\vspace{11.5pc}
\centering
\includegraphics[width=19pc]{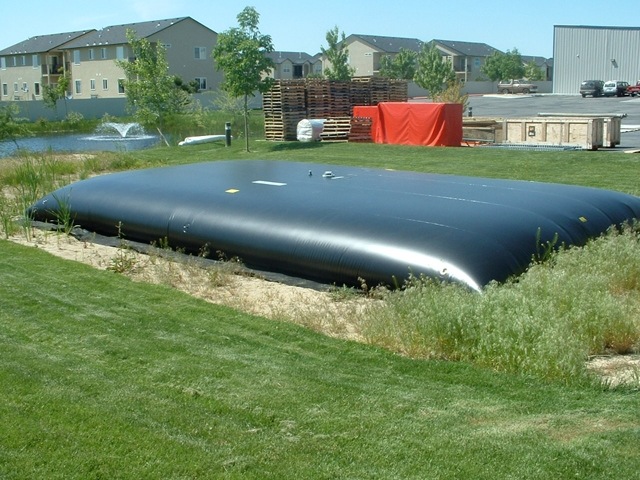}
\caption{\label{fig:tank_10000}Flexible fuel bladder like this one can be used as a last level of security in the highly unlikely event that all the other safety measures to prevent the loss of enriched \XE\
to the atmosphere fail.}
\end{minipage}
\end{figure}

A conceptual drawing of the cold reservoir is shown in Figure~\ref{fig:Cold_dump}. This sort of system has been already developed and built for the LUX collaboration by our TAMU collaborators. As a final line of defense in the extremely unlikely event that cryo-evacuation has failed the xenon will be evacuated into a flexible fuel bladder like one shown
in Figure~\ref{fig:tank_10000}. In that way loss of Xe to the atmosphere will be held at an absolute minimal risk level.

\subsection{Monitoring and Control}

\begin{figure}[h]
\begin{minipage}{17pc}
\centering
\vspace{3.55pc}
\includegraphics[width=15pc]{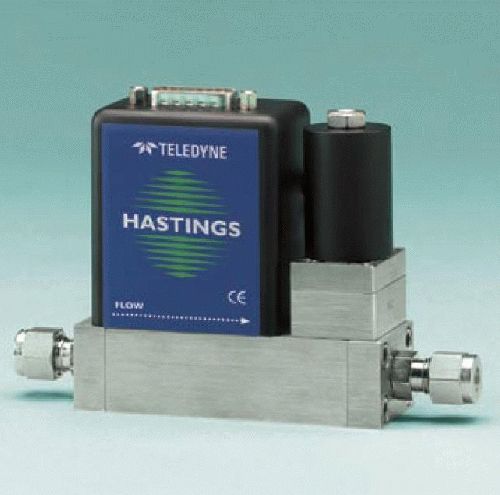}
\caption{\label{fig:HFC302}Example of a Teledyne-HASTINGS flow controller being considered for the flow control application of the NEXT 100 gas system.}
\end{minipage}\hspace{1pc}
\begin{minipage}{19.5pc}
\centering
\includegraphics[width=13pc]{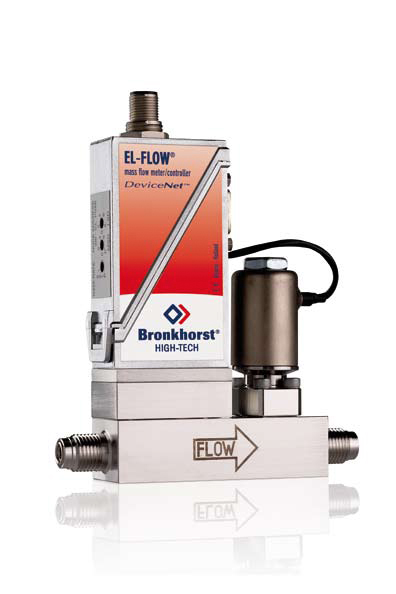}
\caption{\label{fig:Bronkhorst}Example of a Bronkhorst flow controller that can also be used for flow cotrol application in the NEXT 100 gas system.}
\end{minipage}
\end{figure}

The Monitoring and Control can be divided into passive and active roles. On the passive side the logging of temperature and pressure of the gas
at different point in the re-circulation circuit will provide information about the health of the system. On the active side the monitoring of parameters that require
automatic action from the system will be implemented. These would involve the control of pressure and and flow through the use of flow controllers of the type
shown in Figures~\ref{fig:HFC302}~and~\ref{fig:Bronkhorst}. The monitoring for Xe leaks and automatic decision to evacuate the Xe also forms part of this system.
All this monitors and controllers will be integrated into the slow control of the experiment.

\subsection{Electron Lifetime Monitor}

The drift length of the NEXT 100 detector is envisioned to be over 1 m. This places a very high requirement on the absence of electro-negative imputes in the
gas that would recombine with the charged tracks or absorb UV light.
A lifetime monitor would able to determine the level of electro-negative impurities in the gas by measuring the mean drift length of changes. Such a device would be a very useful
addition to the diagnostic arsenal of the Xe purification system. The development of such a device is a separate project and a number of the NEXT collaborators are interested
in this. 

\begin{figure}[h]
\begin{minipage}{22 pc}
\vspace{2.5 pc}
\centering
\includegraphics[width=22pc]{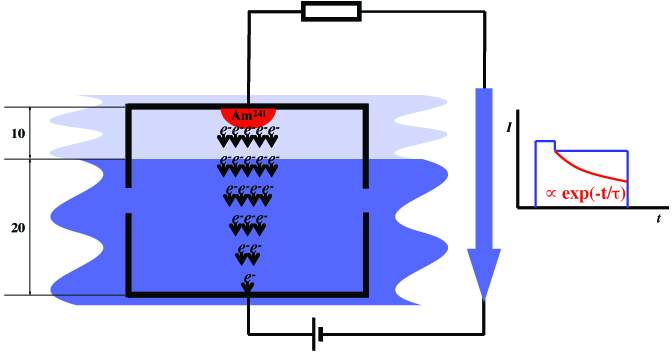}
\caption{\label{fig:LTM_Z3_schem}Principle of operation of the electron lifetime monitor developed by the ZEPLIN III collaboration to measure the purity of the liquid Xe used for
Dark Matter Searches. An $^{241}$Am source produces ionization in the top section of the chamber where there is a layer of gaseous Xe. Under the action of an electric field the charges are then drifted through the liquid Xe. The measured current flowing through the chamber is proportional to $exp(-t/\tau)$ where $\tau$ is the lifetime. }
\end{minipage}\hspace{0.65 pc}
\begin{minipage}{15pc}
\centering
\includegraphics[width=13pc]{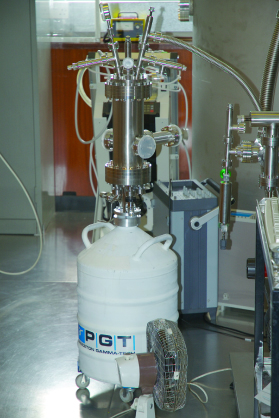}
\caption{\label{fig:BLTM_Z3}Photograph of the ZEPLIN III electron lifetime monitor positioned on the liquid nitrogen cryostat.}
\end{minipage}
\end{figure}

Figure~\ref{fig:LTM_Z3_schem} shows principle of operation of a liquid Xe electron lifetime monitor developed my the ZEPLIN III collaboration and Figure~\ref{fig:BLTM_Z3}
shows this devise prepared for operation \cite{Walker:LTM}. This is a very simple piece of kit. However, because NEXT 100 will be operating using gaseous Xe liquifying
the gas to measure the purity may not give accurate information if the impurities present in gas had been frozen by liquefaction. Thus a gaseous lifetime monitor
will be developed by the NEXT collaboration in the future.

\subsection{Vacuum Evacuation}

\begin{figure}[tbh!]
\centering
\includegraphics[width=0.5\textwidth]{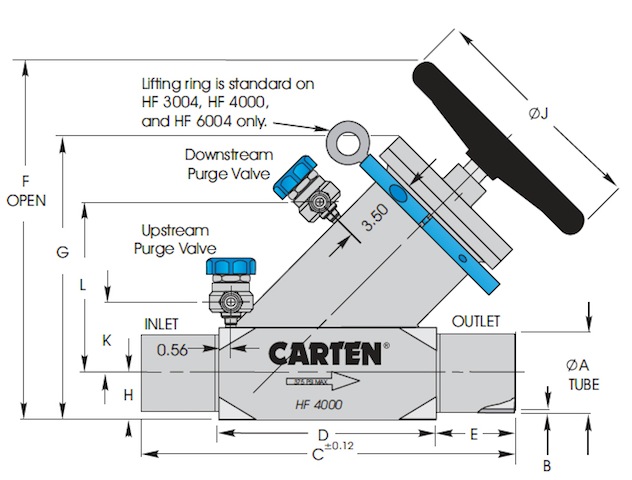} 
\caption{The HF4000 CARTEN valve. }
\label{fig:HF4000}
\end{figure}

To insure the cleanliness of the chamber and the Xe gas system prior to the introduction of Xe both the chamber and the Xe gas system need to be vacuum evacuated
to as low pressure as possible. A reasonably good vacuum is in the range of \mbox{$10^{-5}$ to $10^{-6}$ mbar}. To achieve this the turbo-molecular pump needs to be positioned
as close as possible to the vessel being evacuated. For that reason the turbo-molecular pump station will be directly connected to the NEXT 100 vessel through a
large conductance valve rated for vacuum and pressure as close as possible to the vessel inside the shielding. The valve satisfying this requirement is shown in
Figure~\ref{fig:HF4000}, it is rated down to high vacuum and up to 25 bar pressure. It has a 100 mm orifice and thus well suited to the vacuum needs of the NEXT 100.
A smaller valve of the same type has been extensively used by out Laurence Berkeley Laboratory collaborators. After reaching desired vacuum the valve will be closed, the pumping station removed and the valve enclosed in a radio pure cap to shield the detector
from the radioactive impurities in the materials that make up the valve but also to protect the valve from the corrosive action of the purified water. The details of the water
shielding is discussed elsewhere.

However, many internal structures of the NEXT 100 detector, such as the light pipe surrounding the active volume will now allow good conductance for vacuum evacuation.
Therefore, the chamber will be flushed with Ar and again evacuated a number of times. Furthermore, continuos re-circulation of the Xe gas will further aid it cleaning the system
of impurities.

The Xe gas system constructed from 1/2'' pipe will have low conductance for vacuum evacuation and therefore instead of evacuating the system from a single
pint the vacuum manifold will be connected to several points simultaneously and the system heated to  $~200^\circ$C to remove water. Also flushing with Ar
several times with air in the cleaning process. Finally, as in the case of the main detector, continuous gas re-circulation will clean the Xe gas system.


\section{Data Acquisition}
	
\begin{figure}[phbt!]
\centering
\includegraphics[width=0.99\textwidth]{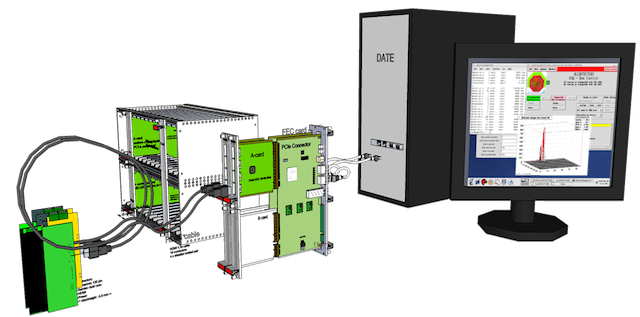}
\caption{Physical overview of the SRS concept.}\label{fig.srs}
\end{figure}

The DAQ has been developed in close collaboration with the CERN RD-51 collaboration \cite{RD51}. RD-51 defines a scalable (target applications ranging from small setups to large scale experiments) and configurable (based on programmable logic, allowing different trigger and DAQ schemes) DAQ architecture that has been developed by CERN-PH-AID and UPV-ELEC under the name SRS (Scalable Readout System) \cite{Mueller}. 
Figure \ref{fig.srs} shows the main SRS components that are the base for NEXT-100 electronics.

\begin{figure}[phbt!]
  \centering
  \includegraphics[width=0.8\textwidth]{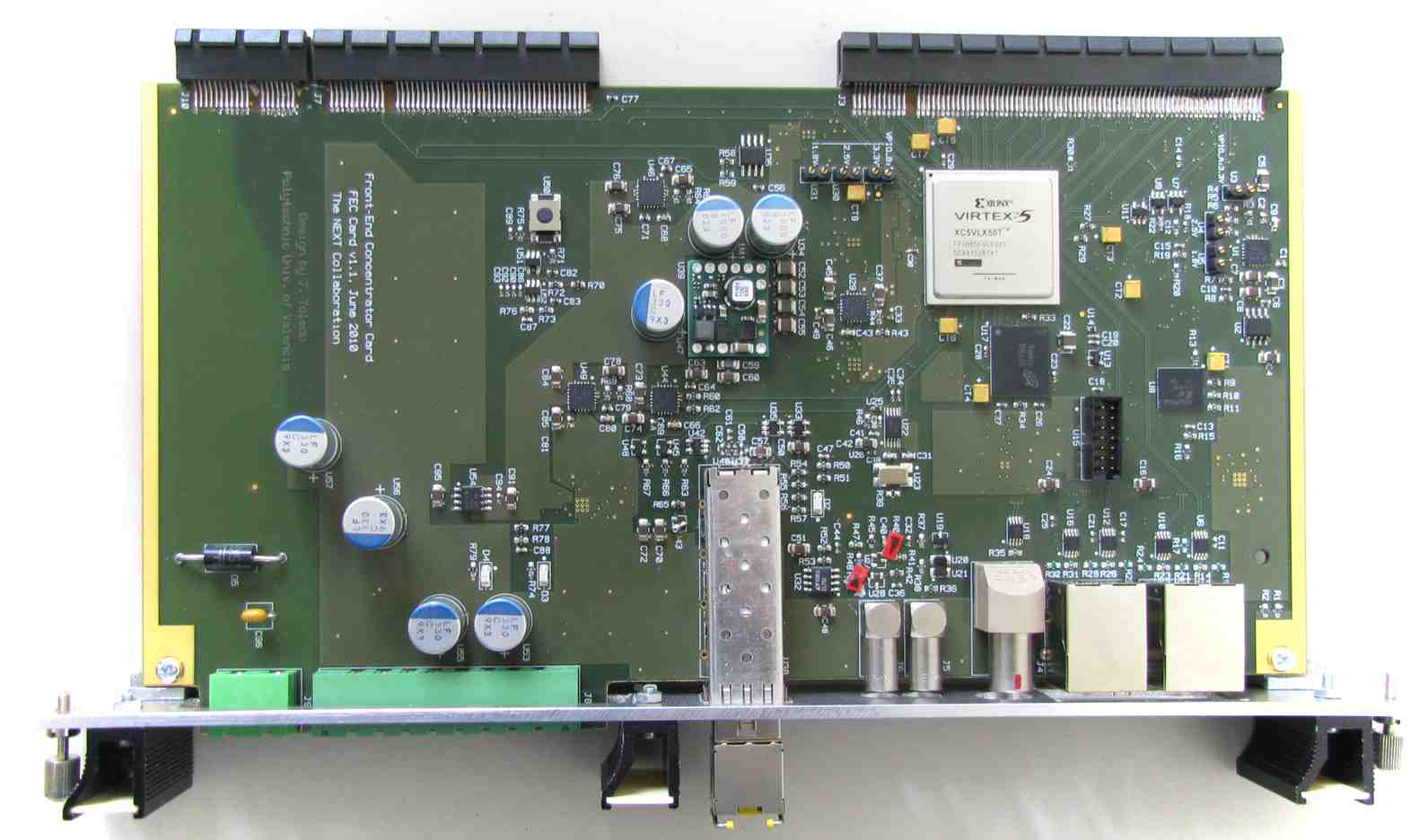}
  \caption{The FEC card. Top edge: connectors for the add-in cards. Bottom edge, from left to right: power connectors, SFP module for gigabit ethernet, NIM input, NIM output, LVDS input and two RJ-45 connectors carrying each one 4 LVDS signal pairs.}\label{fig.fec}
\end{figure}

\begin{figure}[phbt!]
  \centering
  \includegraphics[width=0.5\textwidth]{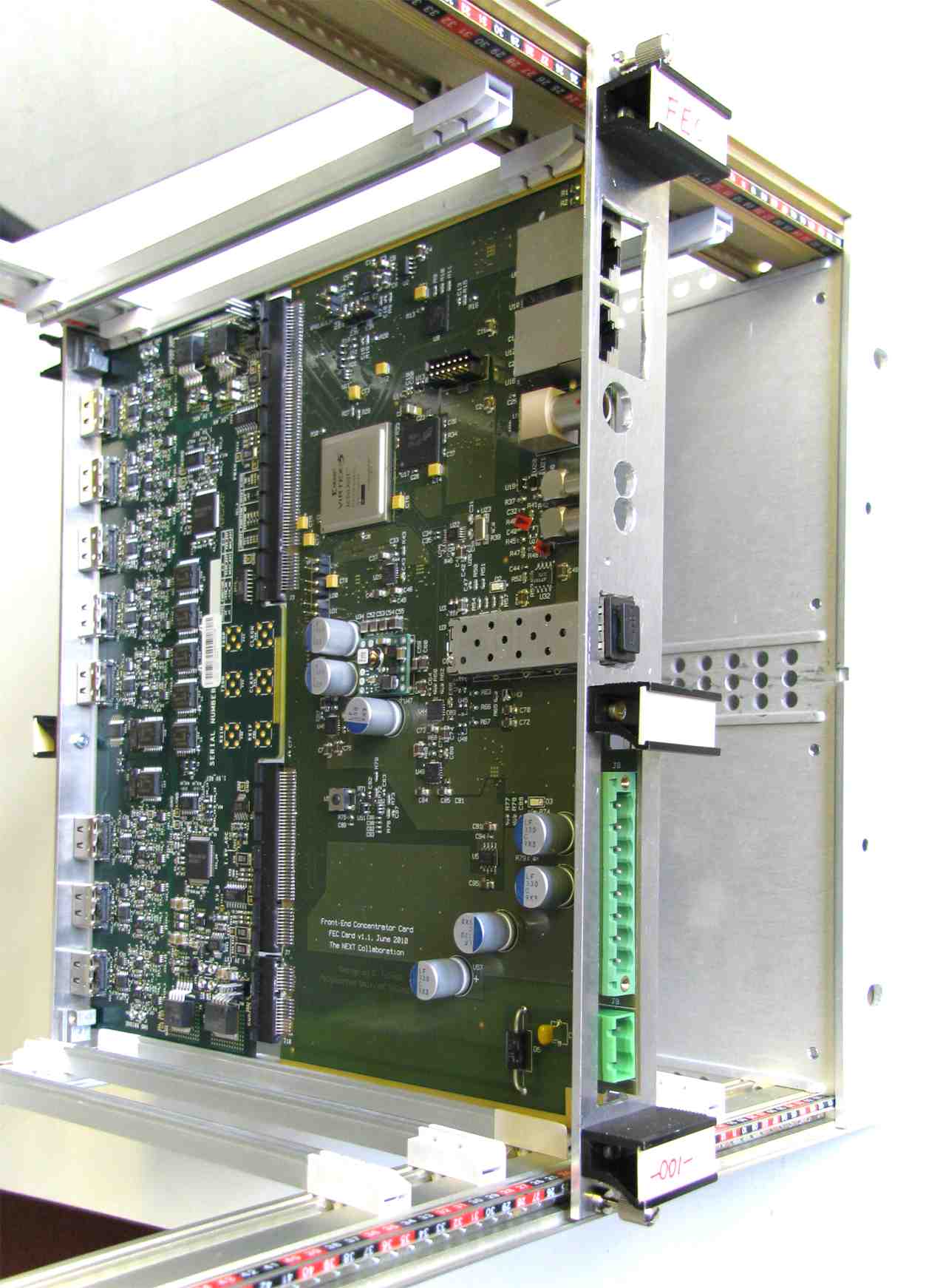}
  \caption{FEC (designed by NEXT) and the ADC digitizer card (designed at CERN) form a $6U \times 220$ mm Eurocard. Used to readout the PMT plane.}\label{fig.fecadc}
\end{figure}

The three main hardware components in the proposed architecture for the DAQ and Trigger systems are:
\begin{itemize}
\item {The FEC card}: The Front-End Concentrator card (FEC card) (Figure \ref{fig.fec}) is a generic interface between a wide range of front-ends and the Online system. This is accomplished by using interface add-in cards between the FEC module and the application-specific front-ends. The ensemble (FEC + add-in card) form a $6U \times 220$ mm Eurocard that fits on a 6U 19'' subchassis. Designed, produced and tested by UPV for RD51 and NEXT, it is currently used in several RD-51 applications other than NEXT, like a muon scanner application in the Florida Institute of Technology \cite{Gnanvo}, the readout of a THGEM in Coimbra/Aveiro/Israel, the characterization of GEM foils at the Helsinki Institute of Physics and the readout of resistive strip MM at CERN (ATLAS group). Coming applications include BEETLE and TIMEPIX ASICs readout, as well as support for the future CERN GBT link. The FEC card design is currently being upgraded to accommodate a larger and more powerful FPGA, a 4x larger and 4x faster on-board event buffer, on-board power supply filters for enhanced noise immunity and extended I/O connectivity with the addition of three SFP+ sockets. This will allow Ethernet-based slow controls.

\item {The CERN/RD-51 ADC add-in card}: The CERN/RD-51 ADC add-in card is used to digitize energy and trigger outputs from up to 16 PMTs using HDMI cables between the front-end boards and the digitizers for enhanced signal integrity. The ADCs are 12 bits and the digitization rate for PMTs will be 50 MHz. Data are processed in the FEC module and are sent upstream to the Online system via gigabit ethernet links. Designed at CERN, there are currently 6 ADC add-in cards available for NEXT-1 operation and other 4 cards will be received from CERN in May 2011. This card is also successfully being used in other RD-51 applications. As shown in Figure \ref{fig.fecadc}, the FEC and the ADC card fit together to form a powerful 16-channel 12-bit 40-to-50 MHz digitizer with strong data processing capabilities in the FEC FPGA. Processed data are sent to the Online system or optionally to an upper DAQ stage (SRU modules) via gigabit ethernet (GbE) links (copper or optical) or dedicated LVDS links over standard RJ-45 connector and CAT6 network cable (400 Mb/s sustained throughput tested by UPV-ELEC in the NEXT-1 setup).
\item {The CDTC16}, a digital interface card for the tracking plane readout. This is a FEC add-in card and a NEXT design. It receives digitized data from up to sixteen SiPM front-end boards via standard network cables. Data are passed to the FEC module, where sub-events are built and then sent to the online system via gigabit Ethernet links. 

\item {The SRU (Scable Readout Unit)} module is a derivative of an existing CERN module (named LCU, LED Control Unit). It accepts LVDS connections via standard CAT5 or CAT6 network cable from up to 36 FECs and is intended for large-scale setups. Though GbE is the chosen DAQ link technology for NEXT-1, 10-GbE will be implemented and tested in the SRU module in 2011, with the aim to evaluate 10-GbE for use in NEXT-100. The SRU module can be used for distributing trigger and precise timing to the readout modules, as well as to provide to the slow controls system with access to the readout modules and indirectly to the front-end electronics. 
\end{itemize}


The Online System is a PC farm running the proven and supported CERN ALICE DAQ software (named DATE after Data Acquisition and Test Environment). DATE has been modified to support GbE technology in request of RD-51 and also as candidate technology for the ALICE DAQ upgrade. NEXT signed a memorandum of understanding that granted us the right to use DATE. 

Figure \ref{fig.ndaq} shows the DAQ system currently operating at NEXT-1-IFIC. Each module can take up to 8 PMT channels in raw data mode in the current FEC version, but will be able to accommodate 16 channels with the revised FEC available in Winter 2011. We need 5 modules to run the current configuration of NEXT-1-IFIC (38 PMTs) and 8 modules to
run the ANGEL detector (60 PMTs). The system is essentially ready for NEXT-100 operation.

The proposed DAQ system can work either in external trigger or in self-triggered mode. In the external trigger mode, either a single SRU or two dedicated FEC+CDTC16 ensembles act as clock+trigger fanout to all DAQ FECs using the available RJ-45 connectors. This scheme has already been proven in the NEXT-1 prototype in Valencia. In self-triggered mode, the upstream channels in the RJ-45 connectors carry trigger candidates from the DAQ FECs to the trigger SRU module, where a centralized low-latency trigger algorithm is run.

\subsection{FE electronics and Data Acquisition for the tracking plane}

The tracking plane poses a greater challenge due to the much higher number of channels. Although having ten thousand signal wires across the TPC vessel is in principle feasible, in-chamber electronics appears as a more practical solution. This implies installing SiPM front-end electronics, digitizers and data multiplexers inside the vessel in order to reduce the number of feedthroughs.

Currently we are studying two options: a)  using 64-ch ASICs that include the analog chain (amplifier + integrator), the digitizers and the data multiplexer, and b) small scale discreet electronics mounted in 64 PC boards.

The possibility of adapting one of the existing ASICs for our application arises from the expertise existing in the UPV group concerning the development of chips such as the ALTRO, widely used in ALICE and in many other HEP experiments. On the other hand, we need a solution fully operative in less than 2 years. Therefore we will also explore the use small PCBs with discreet components. Each PCb would also serve 64 channels, corresponding to a
Daughter Board.

In both cases a commercial serializer/deserializer chip interfaces the ASIC/PCb to an optical transceiver, using a single optical fiber for a full-duplex data connection to the DAQ stage (FECs equipped with specific add-in cards). With this multiplexing factor, 10k channels require just 157 optical links across the vessel. 

The group's expertise and know-how on ASIC design for PMT and SiPM readout is shown in \cite{Herrero, Herrero:2009}. Moreover, R.~Esteve is one of the co-designers of the popular ALTRO chip 

A modification of the existing CDTC16 card in which RJ-45 connectors are replaced by optical transceivers has to be designed. This card will interface the readout of the SiPMs to the FEC cards. The multiplexing factor in the FECs is determined by the trigger rate and the channel occupancy, taking into account that the FEC throughput is limited to a nominal 1 Gb/s, but a factor of 8 is a conservative one and will result in the need of 157/8=20 FECs with its optical-to-electrical add-in cards. 

We must also design and test a full-duplex optical link for 1.25 Gbps that includes the following elements:
\begin{itemize}
\item Serializer/deserializer: a commercial IC that interfaces the ASIC via 8 or 10 bit buses and, on the other side, an optical transceiver. As an example, the TLK1221 part from Texas Instruments could be used. It has a power dissipation of 200 mW.
\item Optical transceiver in a reduced footprint like 1$\times$9, low power, preferably carrying the upstream and downstream channels in a single fiber. 
\item A fiber optic vacuum feedthrough and 4 or 8 optical fibers in order to reduce as much as possible the number of feedthroughs in the vessel.
\item An air-side fiber of approx. 25m, a mating optical transceiver and a receiving FPGA.
\end{itemize}

With a total of 30 FECs, a single SRU unit (3k\euro) can be used to receive trigger candidates and distribute trigger, slow controls commands and a common clock to all the approx. 30 DAQ FECs in the experiment, resulting in a compact and cost-effective trigger system.

\section{The NEXT offline software} \label{sec.Software}
NEXT software includes the NEXUS detector simulation framework, the nextoffline software framework for reconstruction and data analysis, and the nextonline framework which has been summarized in the DAQ section.

\begin{figure}[t!b!]
\begin{center}
\includegraphics[width=0.65\textwidth]{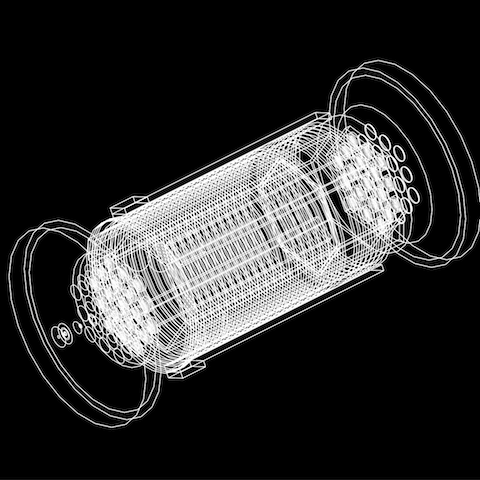}
\end{center}
\caption{Geometry of the NEXT-1-IFIC prototype as implemented in NEXUS.}\label{fig.nexus-next1}
\end{figure}

NEXUS includes:
\begin{itemize}
\item  A simulation package using the latest version of Geant4 (version 4.9.3 at the time of preparing this CDR). This version contains important, non-backward compatible, improvements in the low-energy electromagnetic physics processes, an essential feature for our simulations.
\item Charge drift and electroluminescence packages.
\item Microscopic models of EL based on Garfield.
\item Detailed geometries for the different prototypes (NEXT-0, NEXT-1) as well as for  NEXT-100, including  the water tank (see, for instance, Figure \ref{fig.nexus-next1}).
\item A preliminary version of an event generator for the gamma particles reaching the LSC Hall-A has been developed. This uses measurements taken in the ``old'' laboratory. As soon as new measurements are available, they will be incorporated into the imulation.
\end{itemize}

\begin{figure}[t!b!]
  \centering{
  \includegraphics[width=8cm]{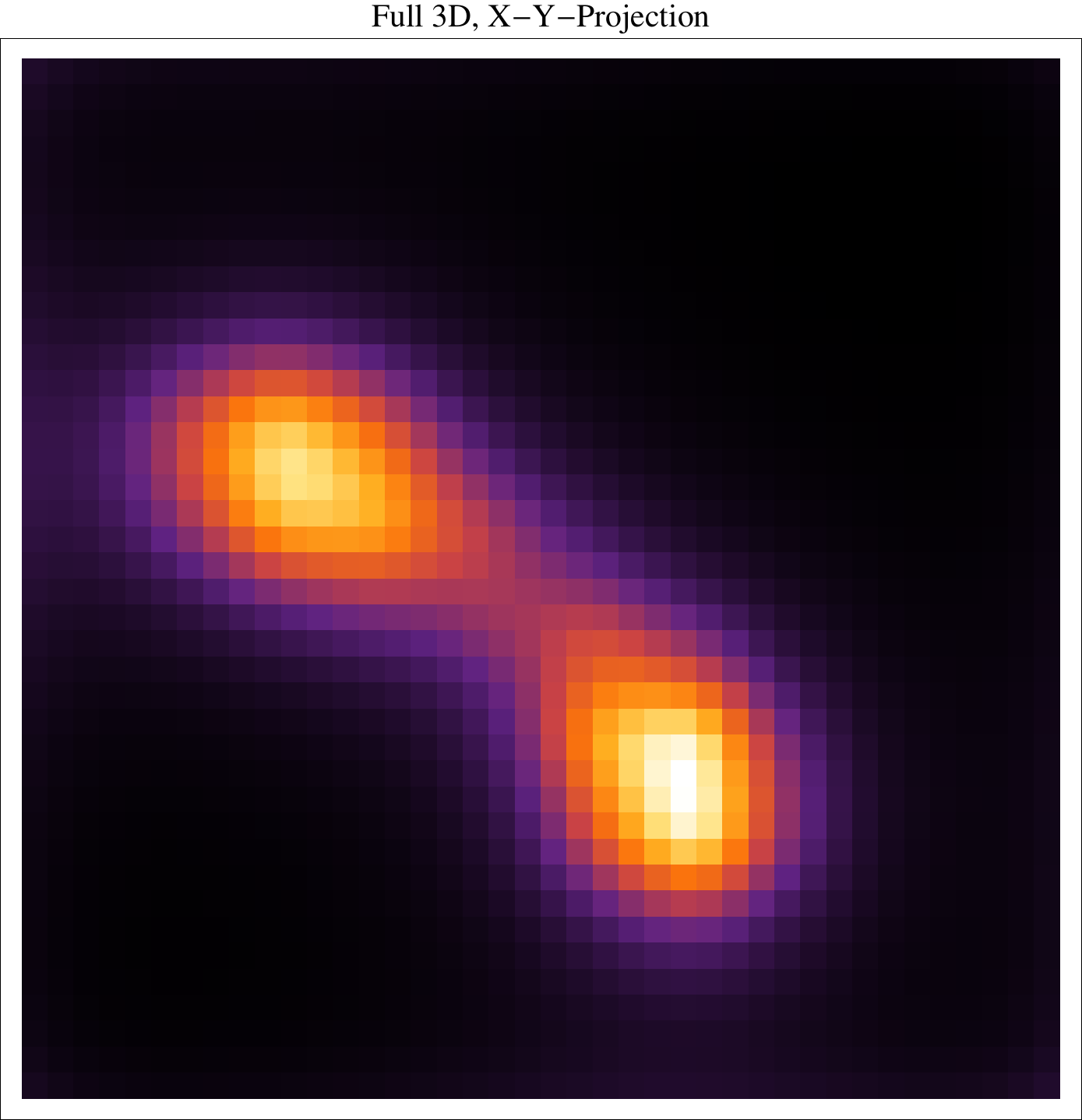}\\
  \caption{Example of a 3D reconstruction of a track in NEXT. The track energy is measured in the PMT plane applying corrections from the positions measured by the SIPMt pixels. The blobs can be clearly seen.}\label{fig.reco}}
\end{figure}

Concerning the nextoffline software framework:
\begin{itemize}
\item It can read and process both real and simulated data. 
\item In particular, simulation of electroluminescent light production, propagation and detection can now be made also in nextoffline and not just in NEXUS, using look-up tables. This approach is motivated by the need to speed up the NEXUS full simulation.  
\item A simulation of the photosensor (both PMTs and SiPMs), front-end electronics, and digitization response properties has been implemented. This allows the study of effects related to dark current and electronics noise, photodetector gain spread, analog/digital signal non-linearity, deadtime, and thresholds.
\item Significant efforts were also devoted to the reconstruction code. Algorithms to reconstruct the total energy deposited within the chamber per event, as well as the imaging of the energy deposition within the chamber, were designed, developed, and tested. The imaging provides a pixelised 3-dimensional picture in terms of chamber volume elements (or voxels). These algorithms are based on the Maximum Likelihood Expectation Maximization method, typically used in medical imaging, and it assumes an EL TPC, although it could be easily adapted to other technologies. An example of a reconstructed track is shown in Figure \ref{fig.reco}. 
\item An improved and more realistic algorithm to reconstruct tracks (for the topological signature) has been developed, starting from the imaging of the energy deposition within the chamber discussed above. Such algorithm is based on graph theory: for each event, the group of voxels with an energy deposition different from zero can be regarded as a graph, that is a set of vertices and edges that connect pairs of vertices. Notice that the algorithm is used to search for the two-electron (``spaghetti with two meat balls'') signature, but not to compute the total energy of the event, which is found by adding (after geometrical corrections) the contributions of all the PMTs.
\end{itemize}

Much progress has also been made in differentiating signal from background events via topological information, using simulated data, as described in the previous chapter.

\section{Shifting the VUV light for NEXT}
In NEXT detector application, the SiPMs are the photosensors mostly suitable for the pattern reconstruction as argued earlier, with the only drawback of poor sensitivity to the Xenon 175~nm scintillation light.  The high cost of APD's and UV sensitive photomultipliers makes the use of a wavelength shifter (WLS) on SiPMs an essential issue in the development of the tracking readout of NEXT detector.
1,1,4,4-Tetraphenyl-1,3-butadiene (TPB) of $\ge99$\%  
purity grade is an organic WLS widely used in many experiments 
to shift scintillation light produced in the UV by Liquid Argon or Liquid Xenon to the visible, where it can be detected by commercial photomultipliers. 
Among the commonly used organic WLS fluors, TPB is the WLS whose emission spectrum matches best the photo-detection spectrum of the SiPMs  (see Figure~\ref{Fig:PDE}) as it absorbs light in a wide UV range and re-emits it in the blue with an emission peak around 430~nm.  In addition, TPB can be applied by vacuum evaporation from crystalline form 
directly onto surfaces as vessel walls or the front face of photomultipliers. 
The TPB coating obtained that way is reported to be hard and durable with good adherence to the substrate and high resistance to mechanical abrasion. It is not soluble in water but may be degraded by humidity, and can be removed when necessary by the use of  benzene or other organic solvents.  

A TPB coating procedure on SiPMs has therefore been developed at ICMOL and IFIC with special attention to the search of the coating thickness that provides the optimal fluorescence efficiency 
as well as the conditions of long-term stability of the coatings. 
Indeed, the fluorescence efficiency of the coating depends on its thickness as has been shown in previous published studies \cite{Lally}. It decreases with increasing coating thickness, whereas  it is limited by evaporation inhomogeneity at small thicknesses.
This is due to the evaporation pattern of TPB which does not sublime when heated as TPH does for instance, but goes through a liquid phase with appearance of bubbling that can lead to a poor quality coating if not well controled. The protocole of coating on SiPMs has therefore to be investigated in order to meet the requirements of both high fluorescence efficiency and high coating quality.

\subsection{TPB coating technique at ICMOL}
\begin{figure}[h]
\centering
\includegraphics[width= 8cm]{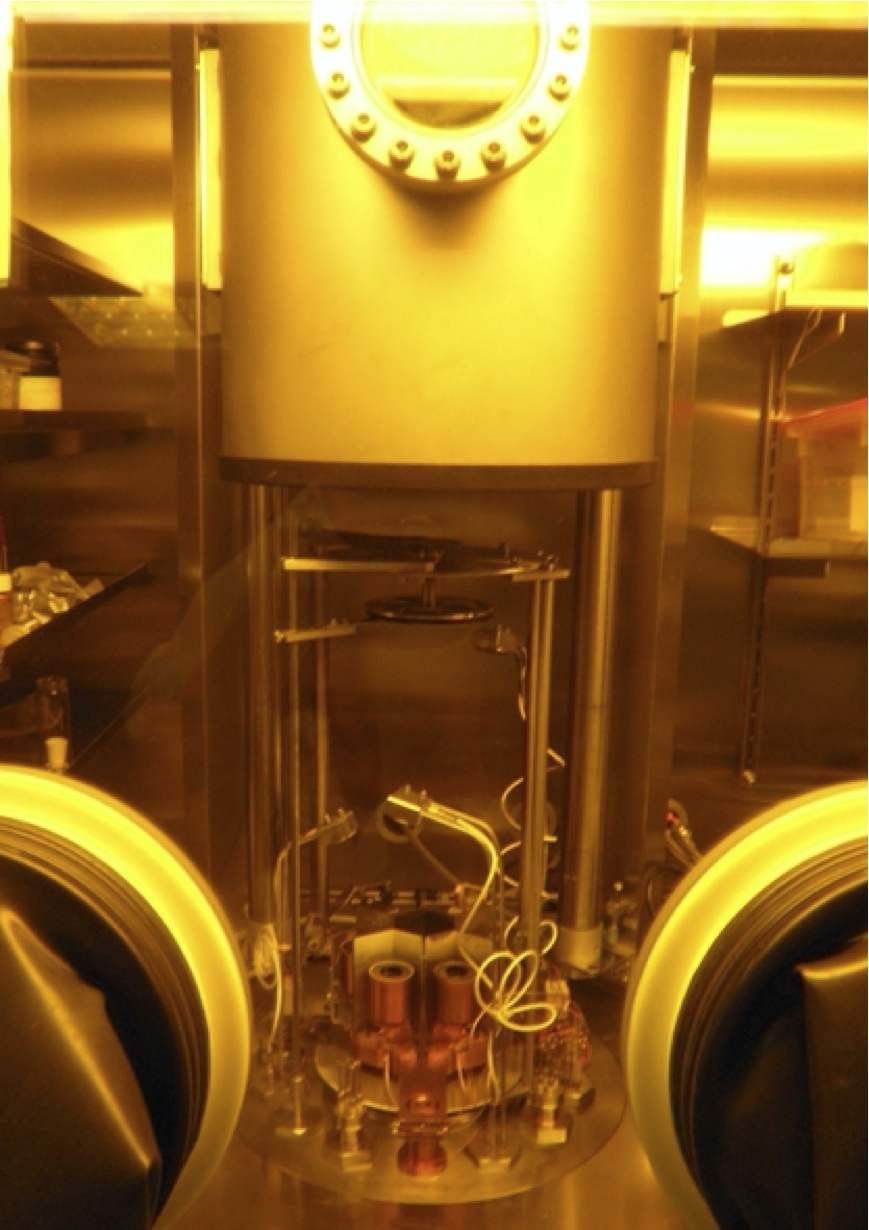}			
\caption{The Evaporator where one can distinguish from bottom to top, the crucibles and the deposition-sensors, the shutter, the sample-holder supported by a spinning disk and the vacuum chamber which is closed down during evacuation and coating.}
\label{Fig:evaporator}
\end{figure}

The coating facility of the Instituto de Ciencias de Materiales (ICMOL) was used. This is located in a class 10.000 clean-room since stringent cleanliness conditions are required for high quality depositions of the Molecules on different substrates. 
 In Figure~\ref{Fig:evaporator} a photograph of the evaporator s shown. This is enclosed in a  glove-chamber filled with with N$_2$  to insures the conservation of the organic compounds by avoiding their oxidation and hydration. 
The coating setup consists in a vacuum chamber enclosing 4 ceramic crucibles used to contain and melt simultaneously up to 4 compounds. During the TPB evaporation campaign for NEXT the whole setup was cleaned to remove any traces of other Molecules. One crucible only was used and filled with TPB pounder. It was heated by a cartridge with an adjustable current for monitoring the temperature
and controling the evaporation rate to avoid bubbling and sputtering of the TPB onto the exposed substrate. 

\begin{figure}[h]
\centering	
\includegraphics[width= 8cm]{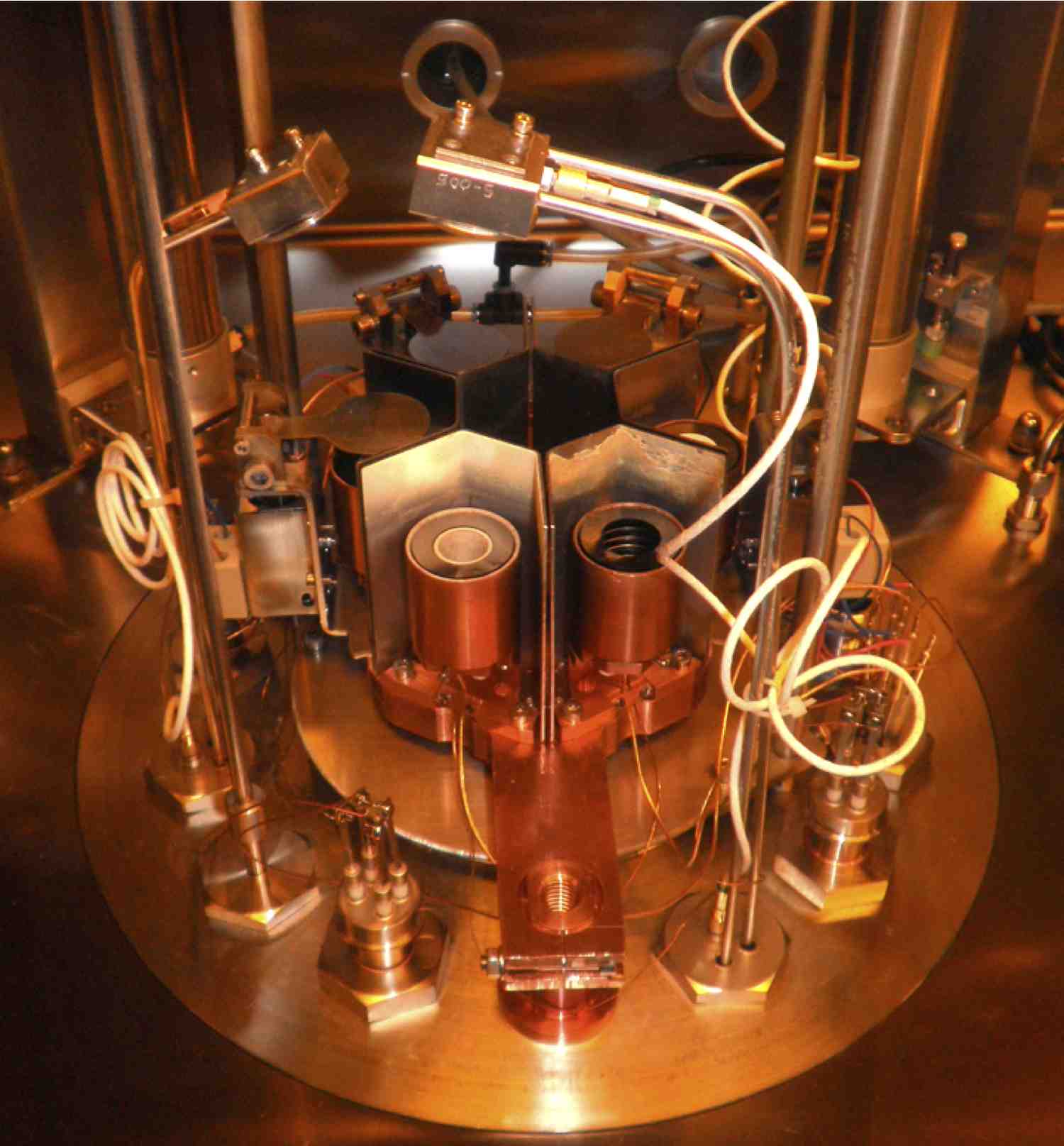}		
\caption{TPB crucible (on the right) and the deposition-sensor positioned on top of  it, half-way before the surface to be coated.}
\label{Fig:deposition_sensor}
\end{figure}
\begin{figure}[h]
\centering
\includegraphics[width= 6.cm]{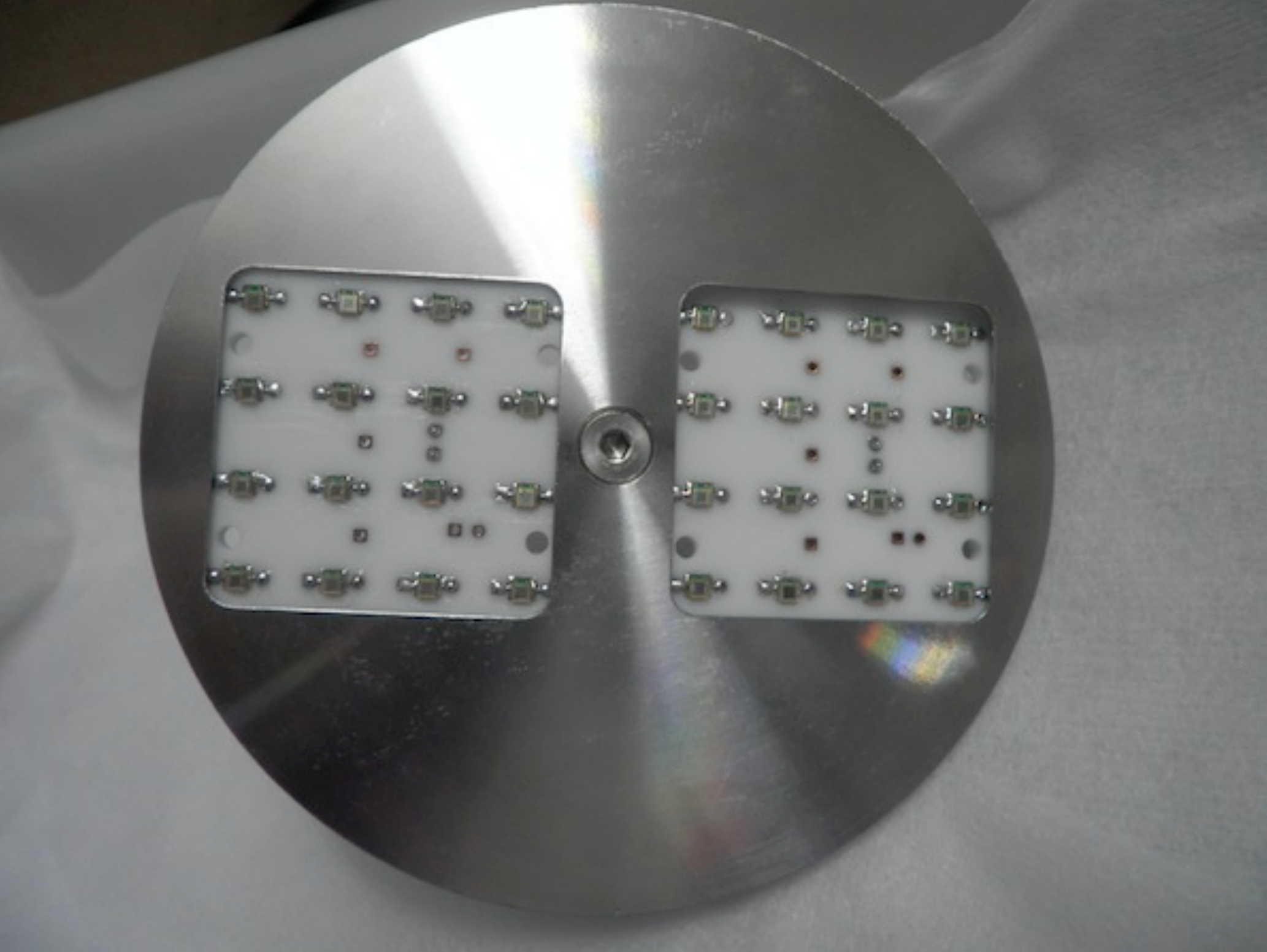}	
\includegraphics[width= 6.cm]{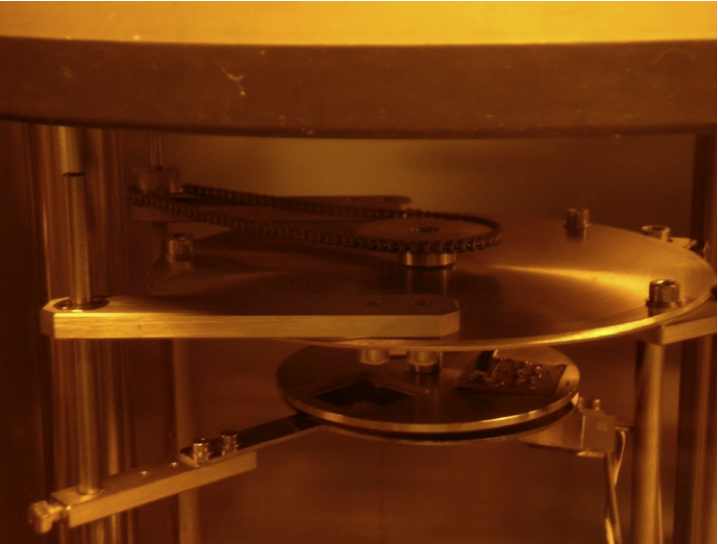}		
\caption{(Left) Two 16-SiPM DB positioned in their lodging in the sample-holder of the evaporator.
(Right) The sample-holder fixed to the spinning disk of the evaporator. The shutter underneath covers the exposed surface of the samples.  }
\label{Fig:DB_support}
\end{figure}

The surface to be coated is positioned on a sample-holder located 15 cm above the crucible. This holder is fixed on a spinning disk which allows the rotation, at a constant velocity, of the samples during evaporation. The spinning in the case of TPB has been proved to be essential to insure the uniformity of the deposition. Below the sample-holder a shutter allows to mask the exposed surfaces when stopping
the TPB deposition is required. 
 Another crucial element of the ICMOL coating system is the deposition-sensor (explain here the principle) to record the TPB deposition thickness in \AA and the deposition rate in \AA/s. This sensor is located half-way between the crucible and the surface to be coated as can be seen in
 Figure~\ref{Fig:deposition_sensor}. 
 
Prior to their introduction in the Evaporator, the samples to be coated are rigorously cleaned with alcohol at 150$^{o}$C and dryed with pressurized N$_2$.  
 After positioning of the clean samples in the Evaporator,  the vacuum-chamber is closed and evacuation is started using a diaphragm and a turbomolecular pumps.  After the optimal vacuum-level is reached, typically  $4\times10^{-7}$~mbar, heating of the crucible is started while the shutter is maintained closed. The relevant parameters as the vacuum level, the temperature in the crucible, the deposition rate and thickness are displayed in co-deposition control units, allowing a constant monitoring of the evaporation process.
 The TPB melting temperature is 203$^{o}$C at atmospheric pressure. At the vacuum level reached, TPB starts evaporation at about 75$^{o}$C.  This is the temperature at which the deposition-sensor starts to record an increasing deposition thickness. The shutter is then opened when the deposition rate stabilizes around a constant value, typically  between 1.8 and 2.4 \AA/s. This is an indication that a steady  evaporation process of the TPB is established.  At the same time, the spinning of the sample-holder is initiated to insure a uniform deposition of the TPB on the exposed surface. When the desired thickness is reached the shutter is closed to avoid additional deposition on the coated surface.  

\begin{figure}[h]
\centering	
\includegraphics[width= 10cm]{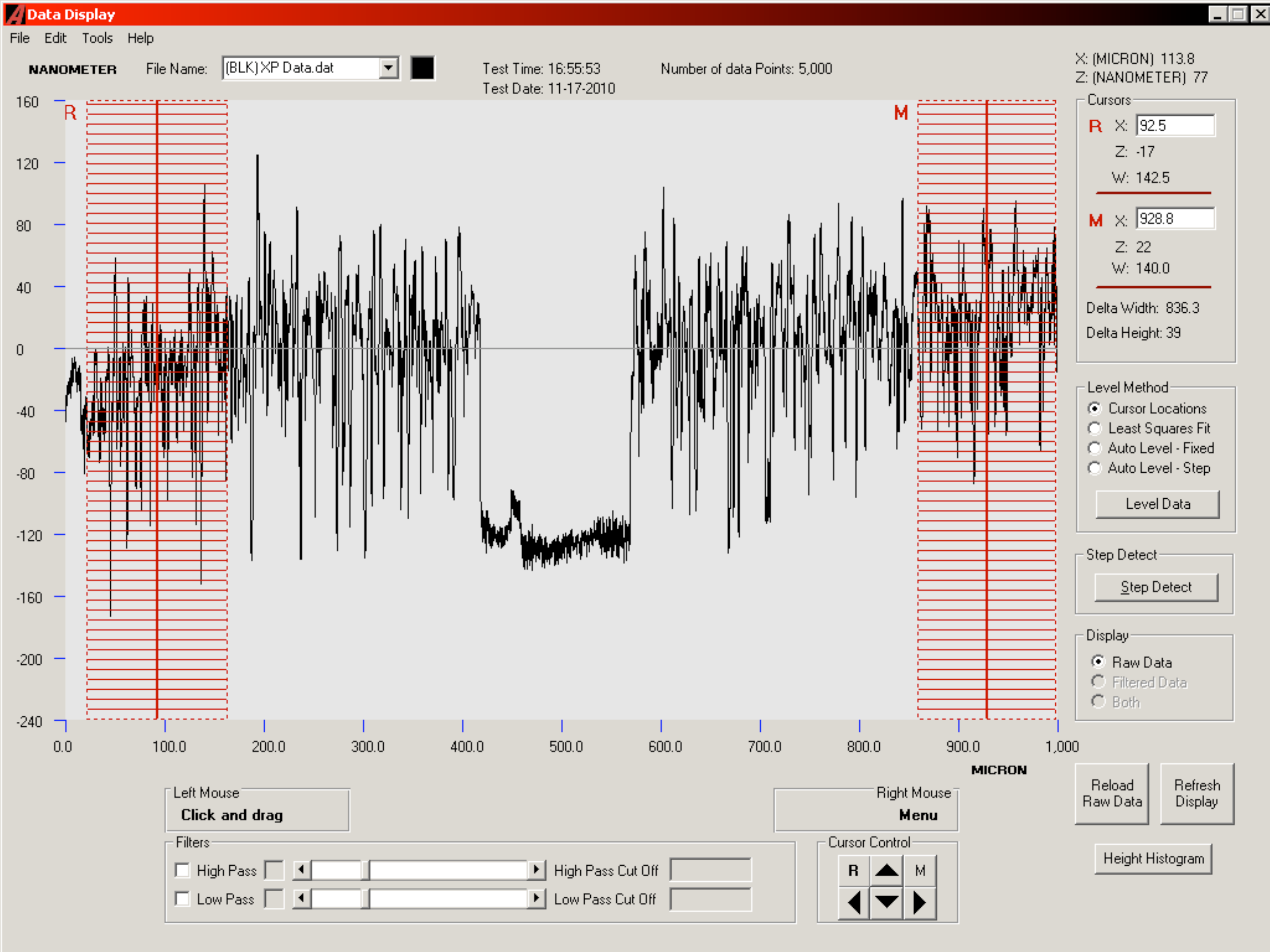}		
\caption{Profile of the TPB deposition thickness obtained on a glass-slice with the profiler.}
\label{Fig:TPB_thickness}
\end{figure}

The relationship between the recorded deposition thickness and the true thickness of the TPB coating on the substrate was determined using a profiler. The coated glass slice was scratched to produce a groove across the coating layer, using a sharp cutter. The thickness profile of the layer is recorded using the needle-sensor of the profiler, made to scan the TPB surface across the groove. A spectrogram of the thickness profile as shown in  Figure~\ref{Fig:TPB_thickness} is produced and the thickness of the TPB layer is measured with a precision of $\pm 4$ Angstrˆm. 
The result from the profiler as compared to the value displayed by the deposition-sensor in the evaporator provides the calibration factor for the coating system. Several successive coating trials are produced on TiO$_2$ glass-slices to adjust the coating parameters for the production of high quality coatings of a chosen thickness.  
\begin{figure}[h]
\centering
\includegraphics[width= 6.5cm]{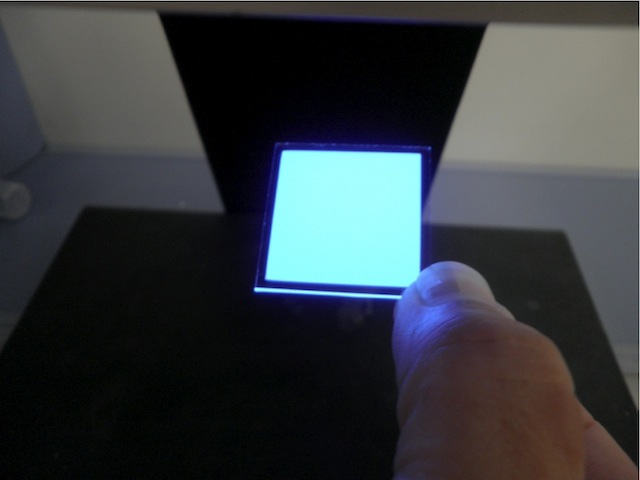}	
\includegraphics[width= 6.cm]{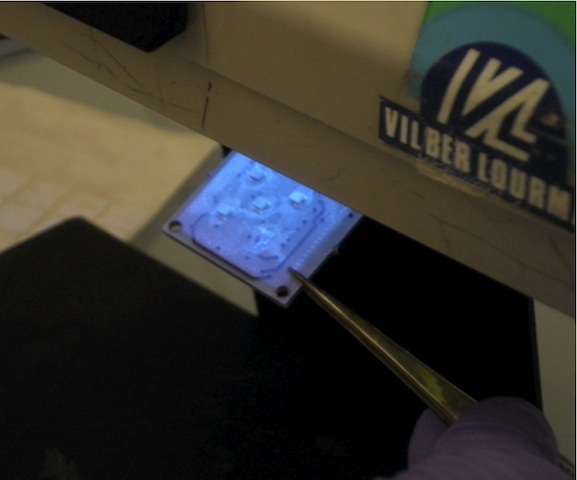}		
\caption{Illumination with 240~nm UV light  of a glass-slice (left) and a 5-SiPM board (right) both coated with TPB.}
\label{Fig:UV_illumination}
\end{figure}

Several successive depositions of a chosen thickness have been successfully produced on TiO$_2$ glass-slices  and on several samples of 5-SiPM  boards. The results have shown to be reproducible provided that the batch of TPB used has been properly conserved.
In Figure~\ref{Fig:UV_illumination} a glass-slice (left) and a 5-SiPM board (right) coated with TPB and illuminated with UV light at 240~nm,  clearly appear re-emitting in the blue.
The coated glass-slices and SiPM samples have been tested and characterized at IFIC and ICMOL with different UV light sources. Main results are described in the following subsections. 

\subsection{TPB fluorescence spectrum} 

\begin{figure}[h]
\centering
\includegraphics[width= 11cm]{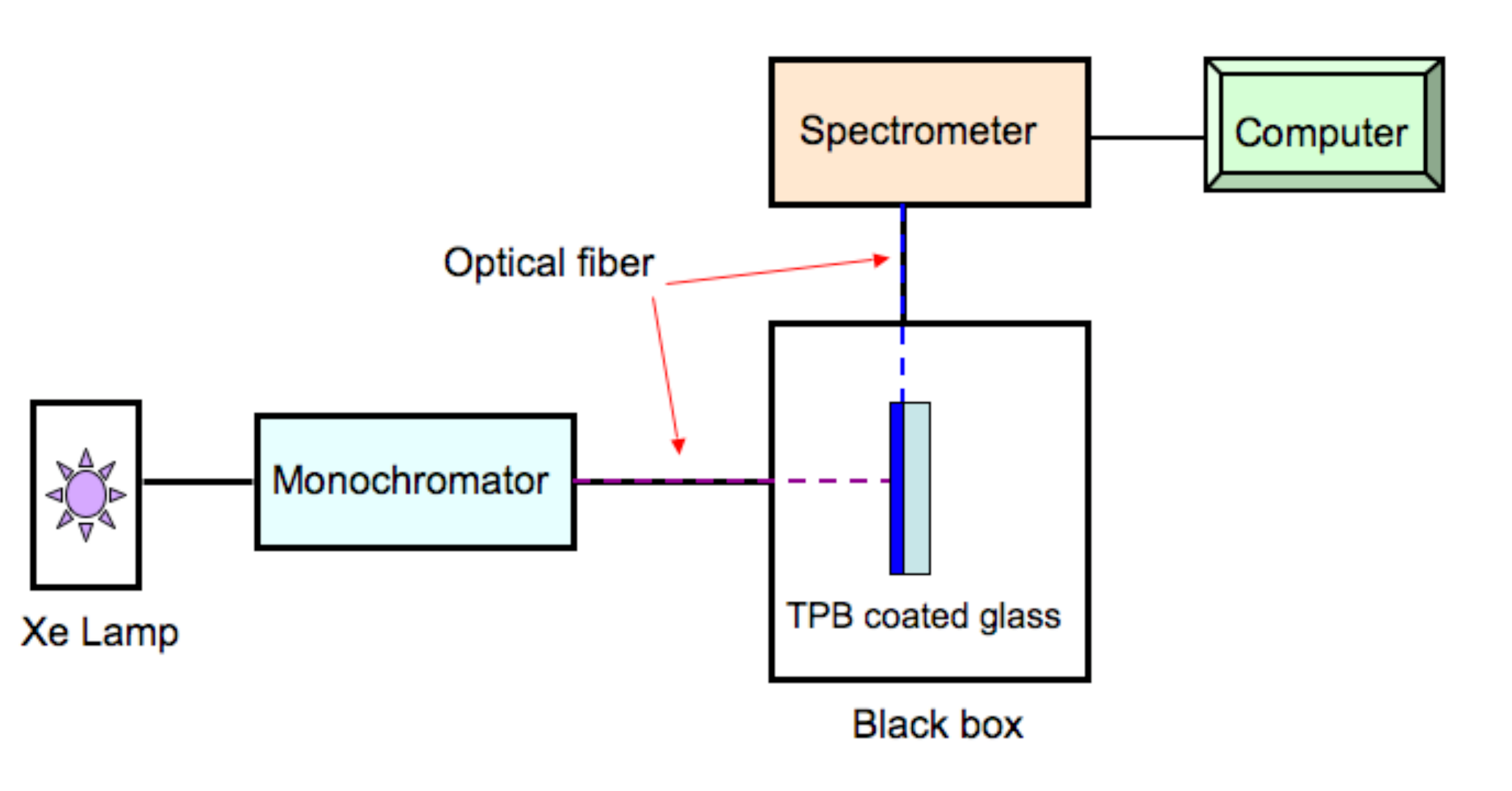}			
\caption{Diagram of the fluorescence measurement setup.}
\label{Fig:setup_diagram}
\end{figure}

\begin{figure}[h]
\centering
\includegraphics[width= 10cm]{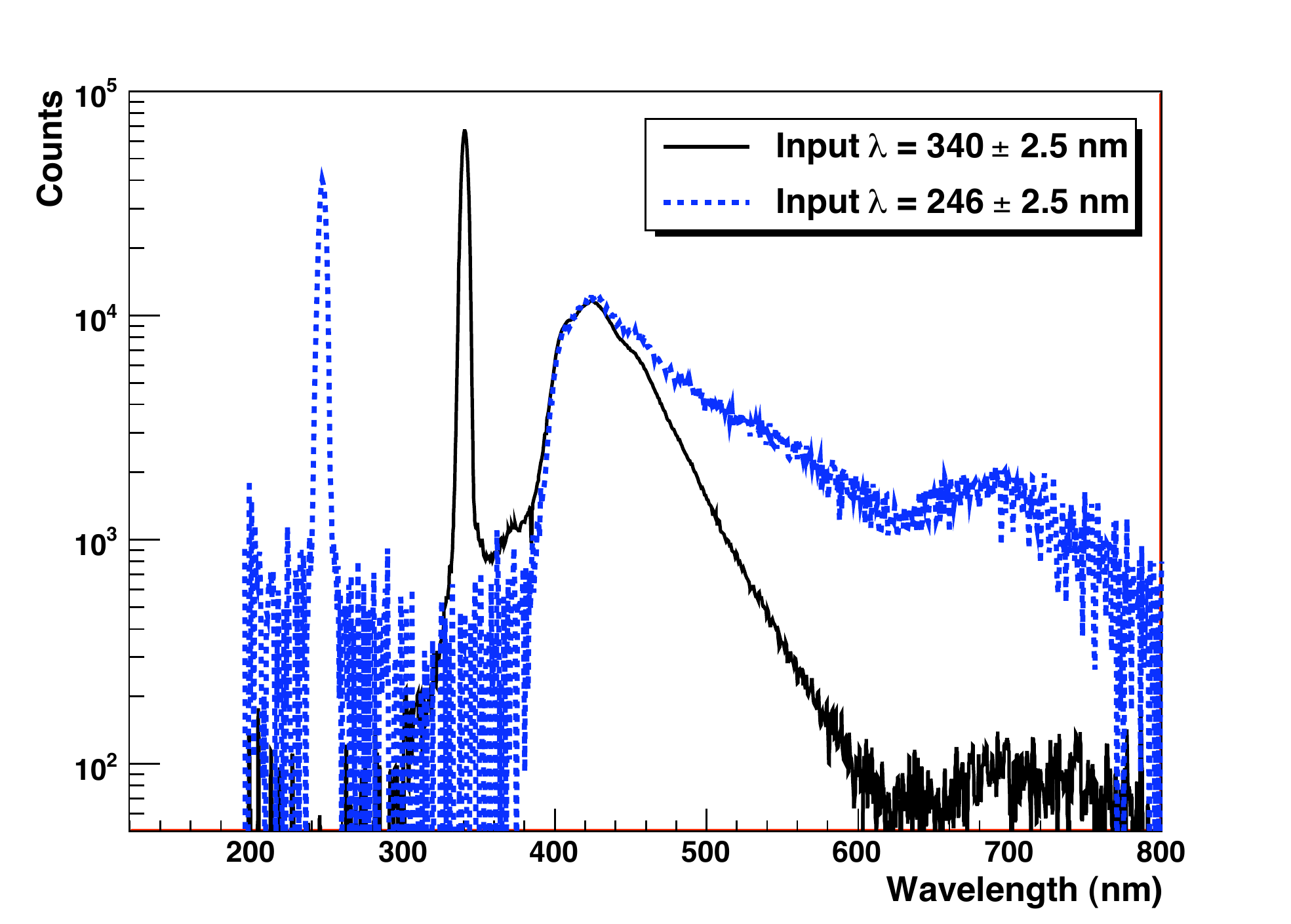}			
\caption{TPB emission spectra measured at ICMOL using a glass-slice coated with  0.1~mg/cm$^2$ of TPB illuminated at 246~nm and 340~nm.}
\label{Fig:TPB_spectrum}
\end{figure}

\begin{figure}[h]
\centering
\includegraphics[width= 10cm]{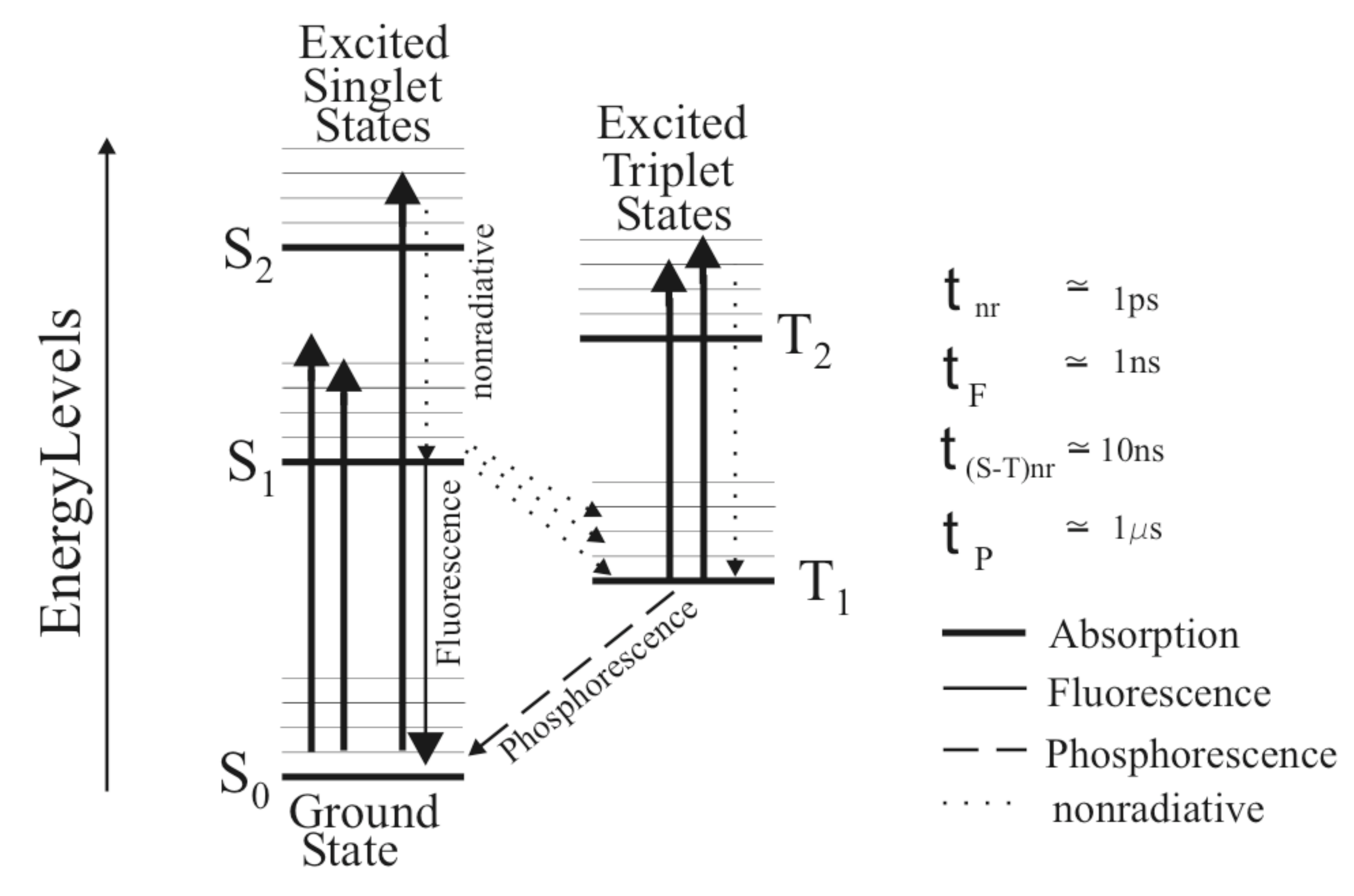}		
\caption{Energy levels diagram  of an organic molecule with $\Pi$-electron structure.}
\label{Fig:fluorescence_scheme}
\end{figure}

A glass-slice coated with 0.1~mg/cm$^2$ of TPB has been used to measure the fluorescence spectrum of the TPB, using a Xenon lamp coupled to a Monochromator for the selection of the input wavelength.
A Spectrometer (Hamamatsu Photonic Multichannel Analyzer C10027) allowed to record the spectrogram of the output light from the TPB layer as shown in
the diagram of Figure~\ref{Fig:setup_diagram}. 
The TPB re-emission spectra measured at $247\pm2.5$~nm and  $340\pm5$~nm  input wavelengths are shown in Figure~\ref{Fig:TPB_spectrum}. 
In these spectra one can see the light source peaks at  
$247\pm2.5$~nm and  $340\pm5$~nm reflected by the glass-slice and the fluorescence peak which lies at $427\pm20$~nm. 
As one can see this fluorescence peak from the S1 Singlet States 
(see Figure~\ref{Fig:fluorescence_scheme})  shows no dependance with the input wavelength in the UV range below 340~nm.  The exposition time of the glass sample in the measurements at the two wavelengths were not the same, the statistics is though different. 
The phosphorescence peak which lies at 680~nm from T1 Triplet States has little contribution to the emission spectra. This however appears to be more significant  at low input wavelength.  

\subsection{Deposition homogeneity}
\label{sec:homogeneity}

\begin{figure}[h]
\centering
\includegraphics[width= 7cm]{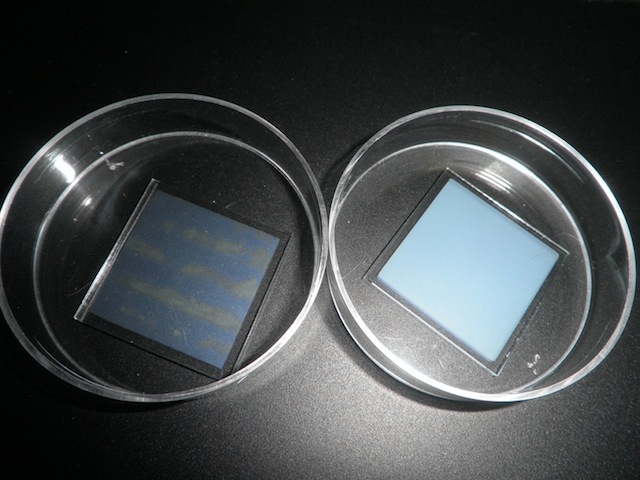}		
\caption{(left) Glass sample coated without spinning. (right) Glass sample coated with spinning.}
\label{Fig:spinning}
\end{figure}

In the first trials the TPB evaporation process was performed without spinning the surface to be coated.  The homogeneity of the deposit was poor in these conditions as one can see in Figure~\ref{Fig:spinning}(left). The process was repeated with spinning at a constant velocity keeping other parameters as the deposition time and the vacuum level unchanged. The result can also be seen in  Figure~\ref{Fig:spinning}(right) where a nicely homogenous deposit
was formed.  This result depends however on the exposition time as the spinning velocity could not be changed and consequently on the deposition thickness. This effect was studied using ICMOL
coated glass slices and a reference non-coated 5-SiPM Board. The Board was placed behind the coated glass slice with its TPB surface in close contact with the SiPMs. 

\begin{figure}[h]
\centering	
\includegraphics[width= 10cm]{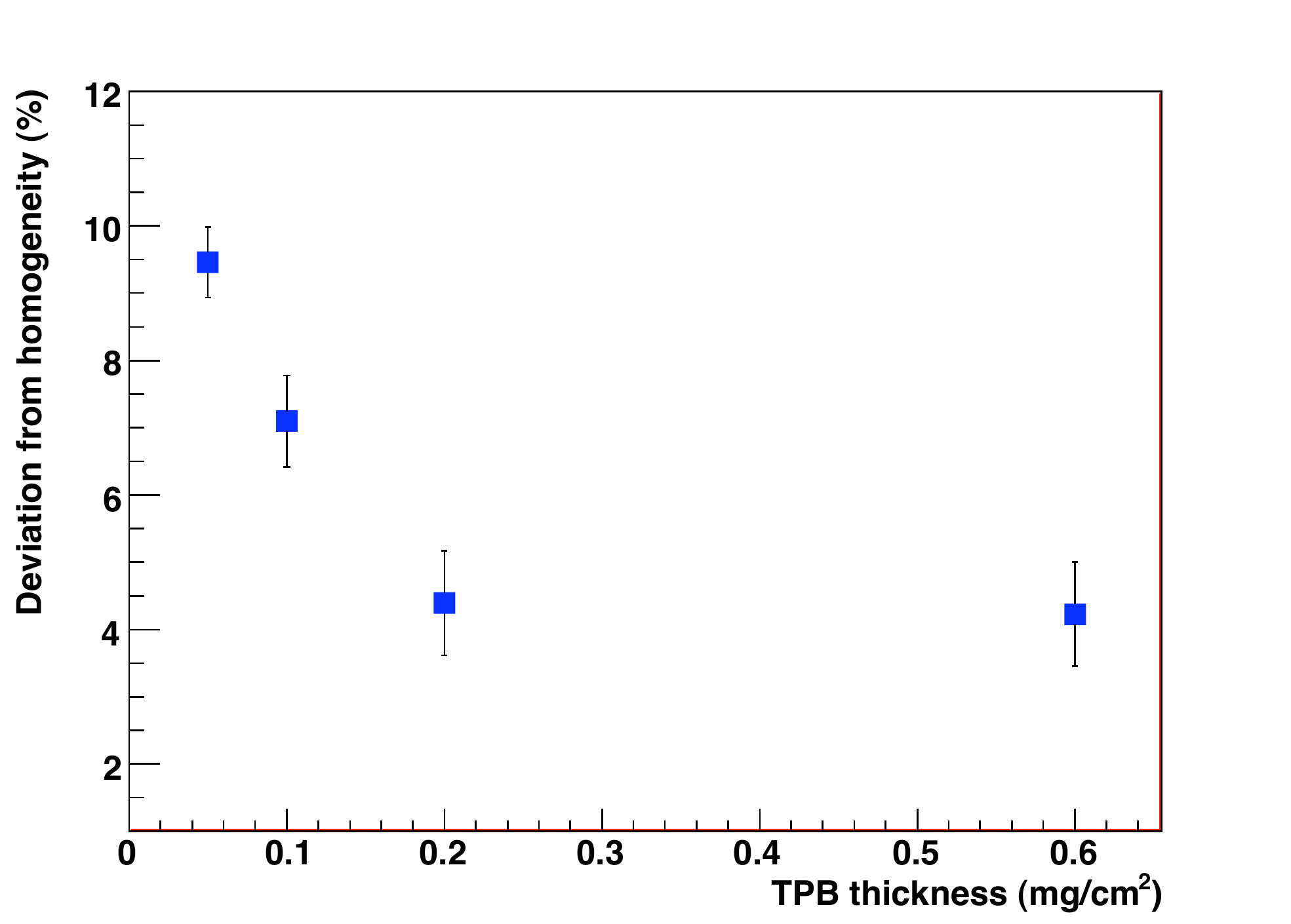}		
\caption{Deviation from homogeneity of the TPB coatings on glass-slices of 3~cm$\times$3~cm as a function of the coating thickness.}
\label{Fig:homogeneity_vs_thickness}
\end{figure}
 In Figure~\ref{Fig:homogeneity_vs_thickness}, this standard deviation is plotted as a function of the coating thickness. 
 The coatings considered in the homogeneity measurement were produced on four identical glass-slices, with the same spinning velocity and at the same vacuum level in the evaporator. As one can see in Figure~\ref{Fig:homogeneity_vs_thickness}, 
the deviation from homogeneity is close to 10\% in the worst case at the lowest coating thickness of 0.05~mg/cm$^2$ and is slightly above 4\% for the thickest coating of 0.6~mg/cm$^2$.  

\subsection{Transmittance of  the TPB at its emission wavelength}

\begin{figure}[h]   
\centering	
\includegraphics[width= 10cm]{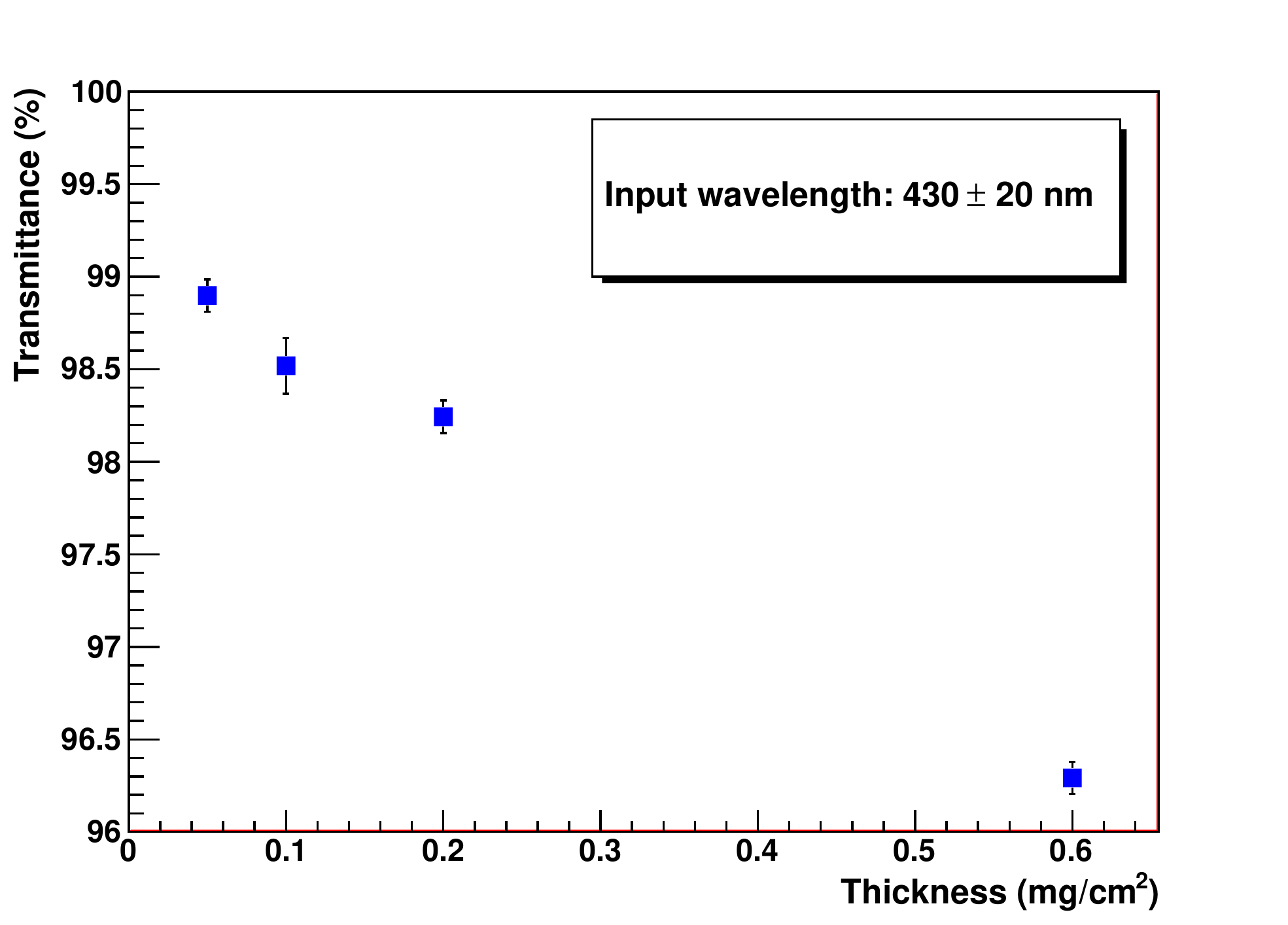}		
\caption{Transmittance of the TPB at its emission wavelength as a function of thickness.}
\label{Fig:self_absorption}
\end{figure}

The transmittance of the TPB at its emission wavelength has been measured as a function of the coating thickness using an LED emitting at $430\pm20$~nm which illuminates successively four glass-slices of 3 cm $\times$ 3 cm coated with 0.05~mg/cm$^2$, 0.1~mg/cm$^2$, 0.2~mg/cm$^2$ and 0.6~mg/cm$^2$  of TPB.  The non-coated side of the glass-slices was coupled to the window of  a 1 inch PMT Hamamatsu R8520. The anode current of the PMT was measured for each TPB thickness and compared to the current measured with a non-coated glass-slice of the same size and thickness as the coated ones. The illumination conditions were kept unchanged  for all the samples measured. The transmittance of the TPB coating shown in Figure~\ref{Fig:self_absorption} as a function of the thickness indicates an increase of the absorption with the thickness as expected. However the amount of absorption keeps below 4\% which is comparable to the uncertainties due the deposition inhomogeneity. 

Several 5-SiPM boards coated with different TPB thickness have been tested with collimated light from LEDs emitting at different wavelengths.  The response of the coated SiPMs to UV light as a function of the coating thickness has been measured and compared to their response to the same illumination prior to coating. The increase in the SiPMs photo-detection efficiency produced by the TPB layer is plotted in Figure~\ref{Fig:fluorescence_vs_thickness} as a function of the TPB thickness.
It is clearly seen that the response of the coated SiPMs increases with decreasing coating thickness and is maximum at 0.05 mg/cm$^2$. This result confirms previous published studies 
on TPB depositions \cite{Lally, Boccone:2009kk}.  
\subsection{Response of coated SiPMs as a function of TPB thickness}

\begin{figure}[h]
\centering
\includegraphics[width= 10cm]{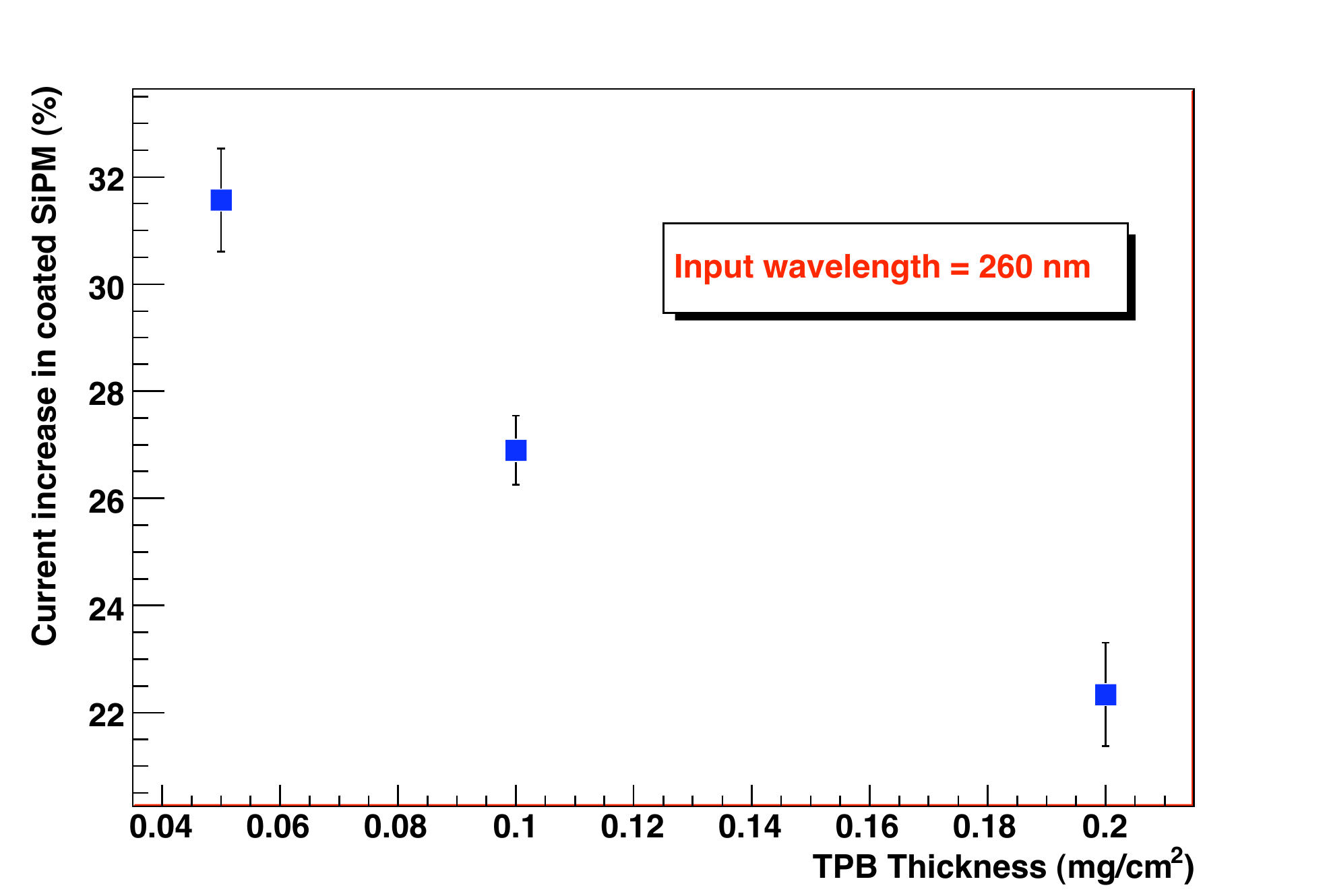}		
\caption{Response of the TPB coated SiPMs to UV light (260 nm) as a function of the coating thickness. For each thickness, the average of the currents measured in each of the 5 SiPMs of a board is represented.}
\label{Fig:fluorescence_vs_thickness}
\end{figure}

An important issue addressed for coating the SiPMs in NEXT is the search of the coating thickness which allows the optimal conversion efficiency of the TPB.  For this purpose, depositions of various thickness of TPB were performed on 5-SiPM Boards which were characterized using a collimated LED emitting at 260~nm. The current output from the coated SiPMs was measured and compared to the current measured prior to coating with the same illumination and temperature conditions.
In Figure~\ref{Fig:fluorescence_vs_thickness} the increase of current of the SiPMs after coating when illuminated at 260~nm is shown as a function of the TPB thickness.  This current increase similar to an increase of the photo-detection efficiency of the SiPMs
depends on the coating thickness and is shown to have maximum value at 0.05 mg/cm$^2$.  This result confirms reported studies on TPB depositions performed on Photomultipliers \cite{Boccone:2009kk}.  However for the coating of the SiPMs of NEXT1-EL DBs and due to the small size of the active area of the SiPMs (1$\times$1 mm$^2$), a compromise between the deposition homogeneity and the fluorescence efficiency of the TPB was considered. Thus a thickness of 0.1 mg/cm$^2$  was chosen as it fulfills both conditions of good homogeneity and high enough fluorescence efficiency.

\subsection{Coating and testing of NEXT1-EL SiPM Daughter-Boards }

Prior to coating the SiPM-DBs were tested with an LED and then cleaned using isopropanol in an ultrasonic bath. They were after dried in an oven at 70$^o$C during 2 hours, then stored in the N$_2$ atmosphere of the ICMOL coating facility until their introduction in the coating chamber. The cleaning protocol was previously agreed with Hamamatsu to be safe for the SiPMs which were tested before and after this process. The 18 SiPM-DBs of NEXT1-EL tracking plane were coated with 0.1 mg/cm$^2$ of TPB and then stored in a vacuum chamber until their mounting on the readout Mother-Board.  A couple of SiPM-DB samples have been dedicated to testing with an LED emitting at 240~nm immediately after coating. The currents produced in the SiPMs measured before and after coating in the same illumination and temperature conditions are compared. In Figure~\ref{Fig:DB_LEDtest} the dispersion in the current increase measured in the 16 SiPMs of one coated DB is shown.  More than 45\% of current increase is measured at 240~nm
which is more than what has been measured with previous samples at 260~nm. The dispersion of this measurement is within the current dispersion obtained in the SiPMs of one DB operated at the same voltage (71.15 V) and below the current dispersion due to the TPB inhomogeneity at 0.1mg/cm$^2$ thickness, which is an indication of an excellent coating quality.   

\begin{figure}[h]
\centering
\includegraphics[width= 6.5cm]{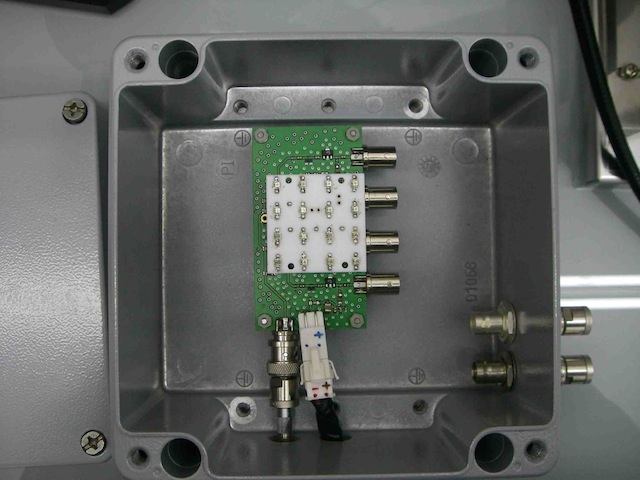}
\includegraphics[width= 6.5cm]{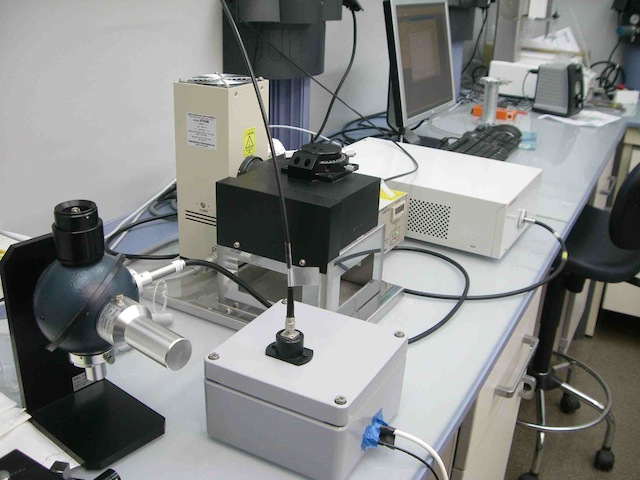}		
\caption{(Right) The testing black-box with one coated SiPM DB plugged on a readout PCB.  
(Left) The closed black-box with the optical fiber from the Monochromator output plugged on its top cover. }
\label{Fig:Test_bbox}
\end{figure}

\begin{figure}[h]
\centering
\includegraphics[width= 10cm]{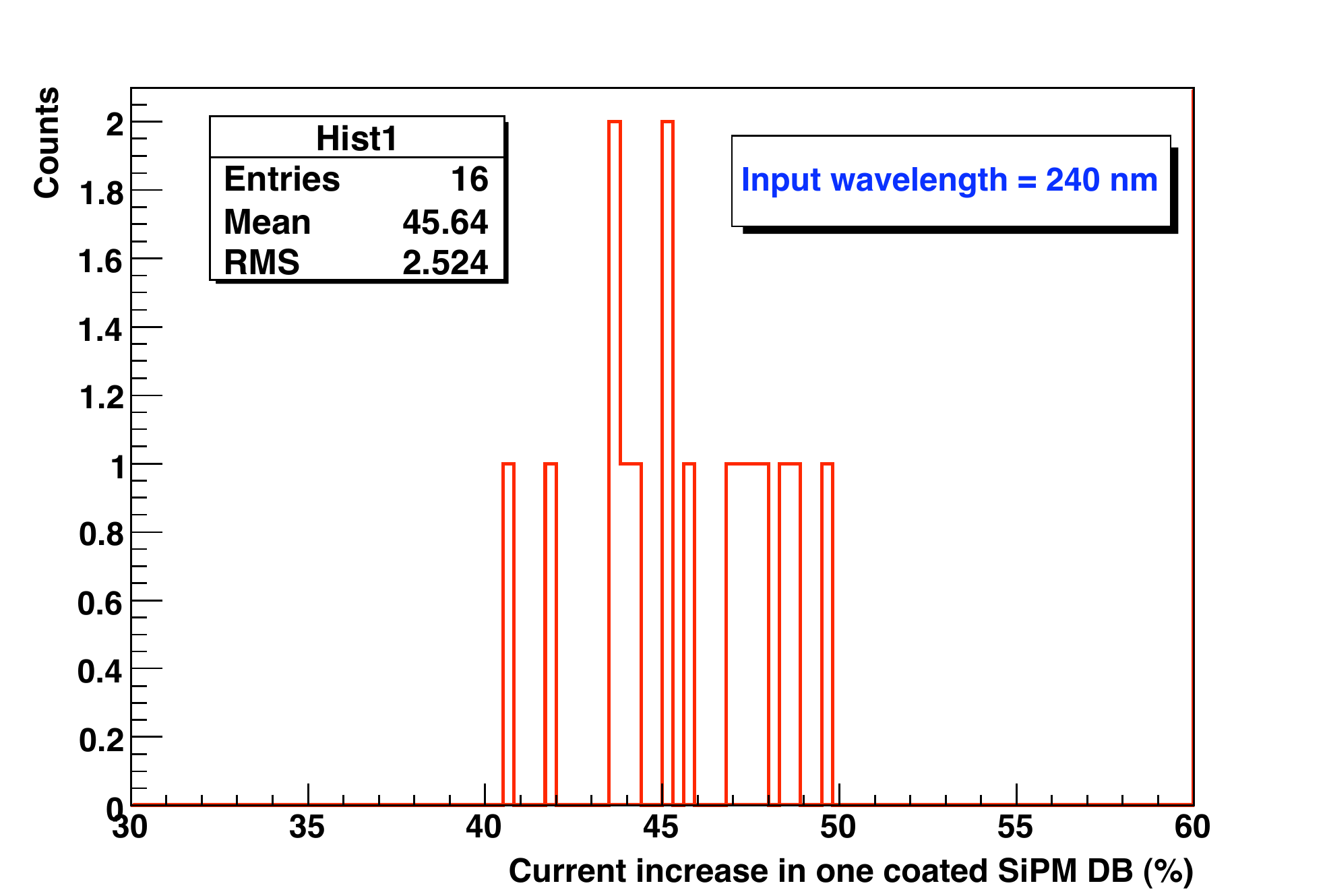}	
\caption{Dispersion of current increase in the TPB coated 16 SiPMs of one DB, illuminated at 245~nm from a LED before and after coating with TPB.}
\label{Fig:DB_LEDtest}
\end{figure}

\begin{figure}[h]
\centering
\includegraphics[width= 10cm]{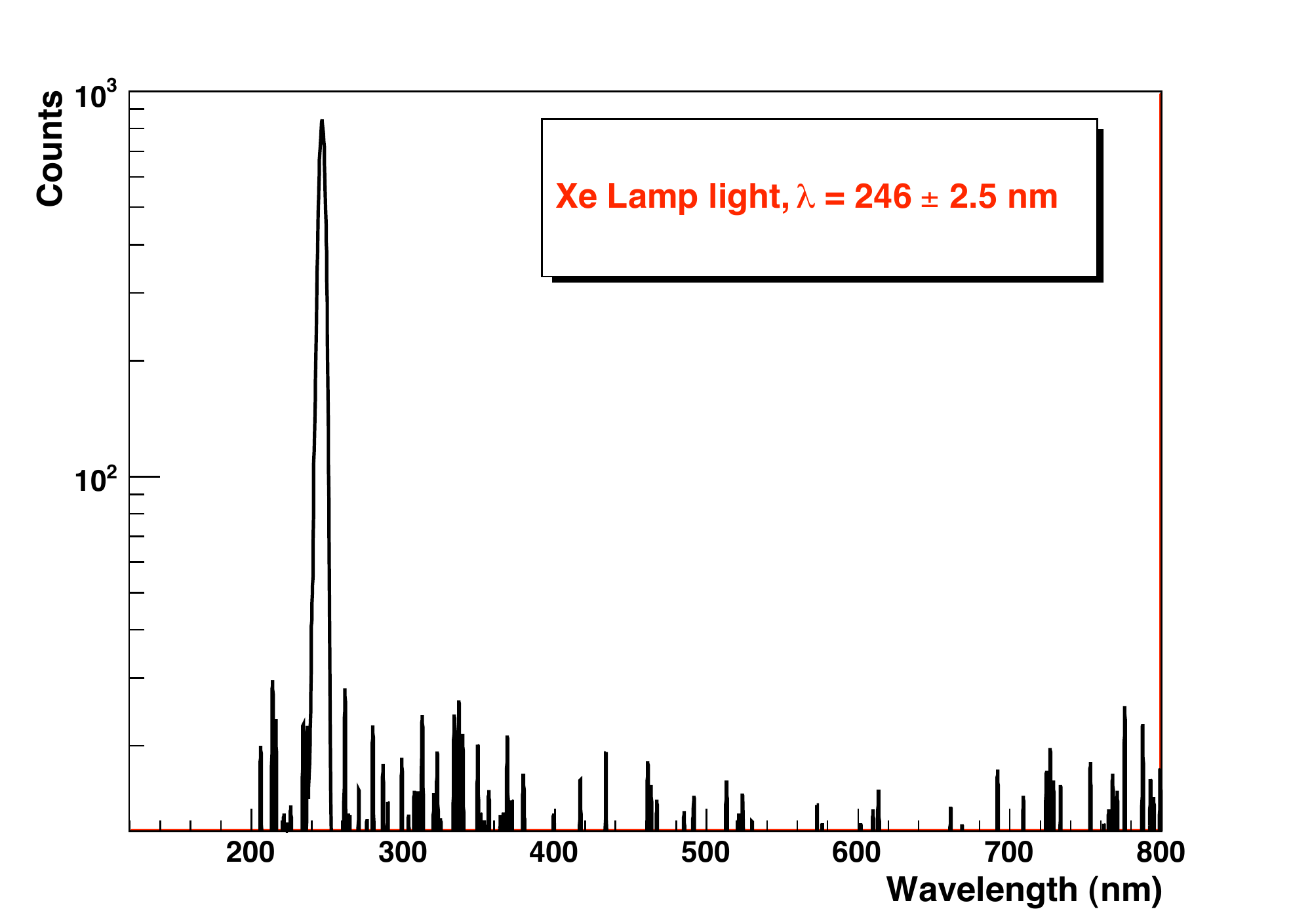}	
\caption{Emission spectrum of the Xenon lamp after the Monochromator with filtering set at  246~nm. }
\label{Fig:monochromator_spectrum}
\end{figure}

The response of the SiPMs of the DB sample as a function of the wavelength was measured using a Xenon lamp coupled to a Monochromator which allowed to select the input wavelength from the Xenon emission spectrum down to 240 nm, with a spread of $\pm5$~nm.  The DB was plugged into a readout PCB placed inside a testing black-box as shown in Figure~\ref{Fig:Test_bbox}.
An SMA feedthrough on the top cover of the black-box allowed to plug the optical fiber from the Monochromator to illuminate a chosen SiPM of the DB.  
A spectrometer was also used to analyze the spectra from the Monochromator output in order to determine the input spectrum used to illuminate the coated SiPM-DBs. An example of such spectrum is shown in Figure~\ref{Fig:monochromator_spectrum} from the Monochromator set at 247~nm.    
The average current from the SiPMs of the DB is recorded at each input wavelength using an Electrometer and compared to the typical average current of the non-coated SiPMs of the DBs obtained in the same illumination conditions. 
In Figure~\ref{Fig:Xenon_lamptest} a significant increase of the current in the coated SiPMs is observed at input wavelengths below 340~nm. 

\begin{figure}[h]
\centering
\includegraphics[width= 10cm]{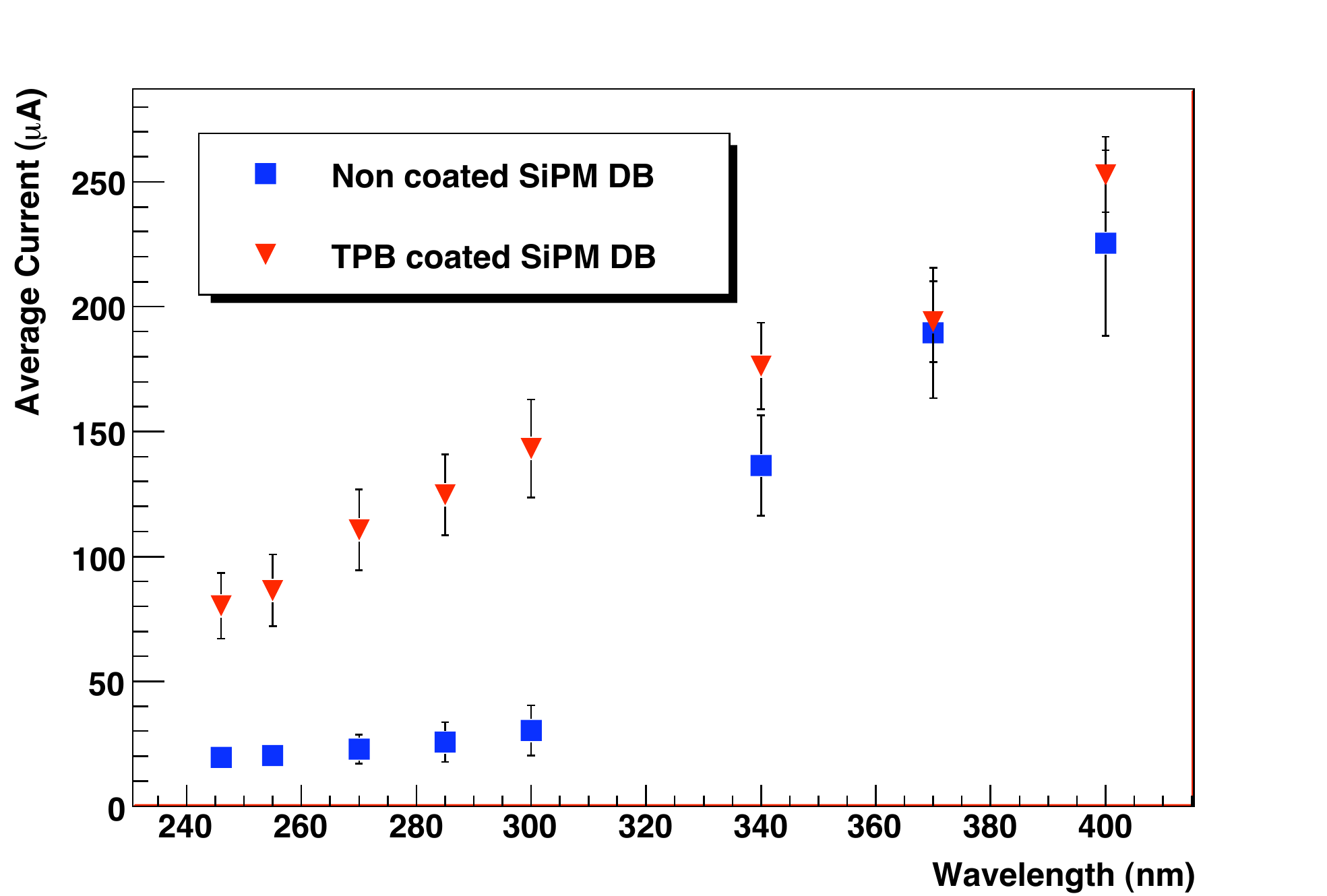}			
\caption{Average current in the TPB coated SiPMs DB compared to the average current of a non coated SIPM DB.  Both SiPM DB are illuminated by the same Xenon Lamp coupled to a Monochromator to select the input wavelength.}
\label{Fig:Xenon_lamptest}
\end{figure}

\section{NEXT Project}

The NEXT project (NP) refers to the sharing, organization, coordination, tracking and resource management of the different tasks related with the construction, commissioning and
operation of the NEXT-100 detector. 

The NP defines the following projects.
\begin{enumerate}
\item {\bf PV:} Design and construction of the NEXT-100 pressure Vessel. 
\item {\bf EL:} Design and construction of the field cage, EL grids and High Voltage feedthroughs. 
\item {\bf LT:} Design and construction of the light tube. 
\item {\bf EP:} Energy plane sensors (PMTs) and front end electronics.
\item {\bf SiPM:} Construction and testing of SiPMs daughter boards (DB).
\item {\bf DB-TPB:} Coating of the DB with TPB prior to mounting in Mother Board (MB). 
\item {\bf MB:} Mother board design and construction.
\item {\bf DB-FE:} FE electronics for the DB (discreet version).
\item {\bf DB-CHIP:}  FE electronics for the DB (ASICs version).
\item {\bf DB-OPT:}  Optical links for SiPMs.
\item {\bf DAQ:} Data acquisition and online monitoring.
\item {\bf SLOW:} Slow controls.
\item {\bf SHL:} Shielding design and construction.
\item {\bf Gas:} Gas system.
\item {\bf RAD:} Radon depuration and monitoring.
\item {\bf INT:} Integration (includes interface between the shielding and the pressure vessel, electrical services and gas services).
\item {\bf SFT:} Safety and risk analysis.
\item {\bf OFS:} Offline software (reconstruction, data processing).
\item {\bf ONS:} Online software (DAQ, monitoring).
\item {\bf QLTY:} Quality control. 
\item {\bf MC:} Monte Carlo simulation.
\item {\bf Calib:}  Calibration.
\item {\bf RPY:} Screening of components, radiopurity measurements.
\item {\bf GM:} Study of gas mixtures to improve the response of pure Xenon.
\end{enumerate}

In the following we briefly describe each project.

\subsection{The Pressure vessel project} 
The PV project includes the design and construction of the vessel for NEXT-100. 
The project will be coordinated by LBNL (design) and IFIC (construction). Other institutions that have confirmed their participation are UPV, UdG (finite element simulations) and UniZar (overall design, together with IFIC).

The design using the ASME norm has been completed at the time of writing this CDR (LBNL, IFIC) and will be extended and further refined using finite elements analysis (UPV, UdG). The construction of the pressure vessel will be carried out by the company TRINOS vacuum system, which has obtained a special grant of the ministry of industry for this purpose.  
 
\subsection{The EL project} 
The EL project refers to the design and construction of the field cage, EL grids and High Voltage feedthroughs. The project will be coordinated by Texas A\&M (TAMU), which has built the same systems for NEXT-1-IFIC. The  participants institutions will be TAMU, and LBNL, as well as UniZar/Saclay (radiopure feedthroughs). 

\subsection{The light tube project} 
The light tube will be constructed along the lines developed by the ArDM collaboration, maximizing the synergy with this experiment. 

A  coating machine suitable for coating large TTX+3M foils has been made available to NEXT by Z. University. The machine will be assembled and tested at IFIC in the next few months.

The project will be coordinated by IFIC, with the participation of JINR. 

%
\subsection{Energy plane sensors}
The procuration and testing of PMTs will be carried out at IFIC. The front-end electronics of the PMTs will be jointly developed by IFIC and UPV. 

\subsection{ Construction and testing of SiPMs daughter boards} 
About 160 SiPM daughter boards, each with 64 SiPMs must be mounted and fully tested. The project will
be coordinated by U. of Aveiro, with the  participation of Coimbra, Aveiro, U. Santiago and UAN.

\subsection{Coating of DB with TPB } 
The coating of NEXT1 DB with TPB will be done at IFIC, using a machine available from ICMOL. 

\subsection{Testing of functionality of DB}
An acrylic chamber with gloves for external manipulation has been prepared at IFIC for NEXT-1. The chamber is being used to test the SiPM MB of NEXT-1. For NEXT-100 we plan to build a mother board "mockup" (MBM) that will have the same functionality than the MB but smaller size. The MBM will be used to test all DB prior to installation in the final MB.  

\subsection{Mother board design and construction}
The MB design and construction will be developed by UPV.

\subsection{SiPM FE electronics }
Three projects will run in parallel. The development of small-scale integrated electronics (LBNL), the development
of a custom-chip (UPV) and the development of the Optical links for SiPMs (UPV, LBNL).

\subsection{Data acquisition and online monitoring} 
The UPV will continue coordinating the DAQ, developed in the RD51 framework. IFIC will contribute with the development of the online software.

\subsection{Slow Controls} 
Slow Controls refers to the system that follow operative parameters such as high voltages for the field cage, PMTs and SiPMTs, gas system, temperature and others. The project will be coordinated by U. Coimbra, with the participation of JINR. 

\subsection{Shielding} 
The shielding project has been coordinated by U. Zaragoza and UPV (mechanics group), and they will continue working in the project to choose a final scheme and proceed to the construction of the system. 

\subsection{The Gas system} 
The construction of the gas system will be a collaboration between IFIC and U. Zaragoza. The system for radon monitoring (including radon cold traps) can be developed by U. Zaragoza and the group of JINR.

\subsection{ Integration} 
The integration of the various subsystems will be coordinated by LSC, IFIC and U. de Zaragoza.

\subsection{Safety} 
Safety will be primarily coordinated by LCS, in close collaboration with the technical coordination of 
NEXT (IFIC and U. Zaragoza). 

\subsection{ Offline software and quality control} 
IFIC will coordinate the software offline and Monte Carlo development. Data Quality control will be 
developed by U. Nariño and U. Santiago.

\subsection{Monte Carlo simulation} 

The Monte Carlo simulation of NEXT has been developed at IFIC, with substantial cooperation of 
U. Aveiro and UAN, who will continue contributing to the project.

\subsection{Calibration}
The development of calibration strategies (and equipment where needed) for NEXT will be coordinated by U. Santiago.
  
\subsection{Radiopurity} 
The screening of components and radiopurity measurements will be jointly coordinated
by LSC and U. Zaragoza. 

\subsection{Gas Mixtures} 
The study of gases that could improve the response of pure Xenon will be coordinated by Saclay and U. Zaragoza. 

\subsection{NEXT Project Management Plan}

The NEXT Project Management Plan (NPMP) coordinates the construction of the NEXT-100 detector. It is under the direct supervision of the Spokesperson (SP) and the Project Manager (PM).

The PMP follows the progress of each NEXT project, monitors deliverables and dead lines and keeps track of invested resources including personnel. It also identifies potential show-stoppers and synergies (as well as possible conflicts) between the different projects and optimizes the sharing of resources.

A draft of the PMP is currently under preparation.

\subsection{Further Developments}
The Next Project and the Next Project Management Plan will be fully developed by the collaboration in the next few months, and will be the focus of our next report to the LSC.  In addition we will submit a full costing of the detector. 

%

\chapter*{Acknowledgements}
This work received support by the Spanish Ministry of Science and Innovation (MICINN) under grants CONSOLIDER-INGENIO 2010 CSD2008-0037 (CUP), FPA2009-13697-C04-04 and RYC-2008-03169. J.~Renner acknowledges the support of a US DOE NNSA Stewardship Science Graduate Fellowship, under contract number DE-FC52-08NA28752.

\bibliographystyle{JHEP}
\bibliography{biblio}

\end{document}